\documentclass[a4paper,11pt]{article}
\pdfoutput=1

\usepackage{jheppub}

\usepackage[T1]{fontenc}
\usepackage{graphicx}
\usepackage{epstopdf}
\usepackage{bm}
\usepackage{epsfig}
\usepackage{graphics}
\usepackage{xspace}
\usepackage{hyperref}
\usepackage{slashed}
\usepackage{xcolor}
\usepackage{longtable}
\usepackage[small]{subfigure}
\usepackage[normalem]{ulem}

\newcommand{\df}{\mathrm{d}}

\newcommand{\ep}{\epsilon}

\newcommand{\tmop}[1]{\text{#1}}
\newcommand{\mathd}{\mathrm{d}}

\newcommand{\MSBar}{{\overline{\text{MS}}}}

\title{\boldmath Toward massless and massive event shapes in the large-$\beta_0$ limit}

\preprint{\begin{flushright} IFT-UAM/CSIC-21-14\end{flushright}\vspace*{-1cm}}

\author[a]{N.\,G.~Gracia}
\author[a,b]{and V.~Mateu}

\affiliation[a]{Departamento de F\'isica Fundamental e IUFFyM, Universidad de Salamanca,\\ Plaza de la Merced s/n, E-37008 Salamanca, Spain}
\affiliation[b]{Instituto de F\'isica Te\'orica UAM-CSIC,\\ C/ Nicol\'as Cabrera 13-15, E-28049 Madrid, Spain}

\emailAdd{ngonzalez@usal.es}
\emailAdd{vmateu@usal.es}

\abstract{We present results for SCET and bHQET matching coefficients and jet functions in the large-$\beta_0$ limit. Our computations exactly predict all terms of the
form $\alpha_s^{n+1} n_f^n$ for any $n\geq 0$, and we find full agreement with the
coefficients computed in the full theory
up to $\mathcal{O}(\alpha_s^4)$. We obtain all-order closed expressions for the cusp and non-cusp anomalous dimensions (which turn out to be unambiguous) as well
as matrix elements (with ambiguities) in this limit, which can be easily expanded to arbitrarily high powers of $\alpha_s$ using recursive algorithms to obtain the
corresponding fixed-order coefficients. Examining the poles laying on the positive real axis of the Borel-transform variable $u$ we quantify the perturbative convergence of
a series and estimate the size of non-perturbative corrections. We find a so far unknown $u=1/2$ renormalon in the bHQET hard factor $H_m$ that affects the normalization
of the peak differential cross section for boosted top quark pair production. For ambiguous series the so-called Borel sum is defined with the principal
value prescription. Furthermore, one can assign an ambiguity based on the arbitrariness of avoiding the poles by contour deformation into the positive or negative
imaginary half-plane.
Finally, we compute the relation between the pole mass and four low-scale short distance masses in the large-$\beta_0$ approximation (MSR, RS and two versions of the
jet mass), work out their $\mu$- and \mbox{R-evolution} in this limit, and study how their implementation improves the convergence of the position-space bHQET jet function,
whose three-loop coefficient in full QCD is numerically estimated.}

\begin{document}
\maketitle
\flushbottom

\section{Introduction}
\label{SectionIntroduction}
In the LHC era, the scientific interest in jet physics has experienced a significant boost. It has been noticed by Cacciari and Salam (see e.g.~\cite{CacciariSalam}) that roughly 60\%
of experimental articles by ATLAS and CMS make use of jets (sprays of energetic particles traveling in nearly the same direction). Some of the theoretical
efforts to better understand their dynamics seek to tame so-called hadronization effects, that is, corrections that arise from the fact that, even at high energies,
what we observe are not quarks and gluons but hadrons. These corrections are enhanced since jet cross sections are not completely inclusive observables. The two
most prominent approaches to pursue such goal are either parametrizing those corrections from first principles, or designing observables that are less afflicted by them.
Since the most relevant non-perturbative effects come from low-energy particles, so far only soft hadronization has been dealt with. In the first approach, one can show
within factorization theorems derived the context of Soft-Collinear Effective Theory (SCET)~\cite{Bauer:2000ew,Bauer:2000yr,Bauer:2001yt,Bauer:2001ct,
Bauer:2002nz} (or the equivalent CSS formalism \cite{Collins:1981uk,Korchemsky:1998ev,Korchemsky:1999kt,Korchemsky:2000kp,Berger:2003iw})
that the soft function (which describes large-angle soft radiation) can be re-factorized in a partonic part and a non-perturbative shape function,
see Refs.~\cite{Korchemsky:1994is,Korchemsky:2000kp,Lee:2006nr} (other theoretical approaches can be found in~\cite{Gardi:1999dq,Gardi:2000yh,Dokshitzer:1995zt}),
whose first moment encompasses in a single parameter the leading hadronization effects in the tail of the distribution (the effect of hadron masses on
power corrections has been investigated in Refs.~\cite{Salam:2001bd,Mateu:2012nk}). In the second approach one defines grooming algorithms that
consistently remove soft radiation in a way in which theoretical computations are not overly complicated, such as Soft-Drop~\cite{Larkoski:2014wba} of its
recursive version~\cite{Dreyer:2018tjj}.

Taking a different perspective, one can investigate power corrections through so-called renormalons, that is, the asymptotic behavior of the associated
perturbative series. This approach was applied to the soft function appearing in the factorized cross section for the doubly differential hemisphere mass
distribution (which can be projected onto thrust and heavy jet mass in a straightforward manner) in Ref.~\cite{Hoang:2007vb}, while a generalization for
any other SCET-I type observable was worked out in \cite{Mateu:2012nk}. The same soft function appears in cross sections for massless or massive
quarks~\cite{Fleming:2007qr}. Earlier studies of renormalons in the context of $e^+e^-$ event shapes can be found in Refs.~\cite{Webber:1994cp,Beneke:1995pq,Gardi:2001ny}.
In this article we investigate non-perturbative corrections coming from the remaining pieces of the SCET and a boosted version of Heavy Quark Effective Theory
(bHQET)~\cite{Eichten:1989zv,Isgur:1989vq,Isgur:1989ed,Grinstein:1990mj,Georgi:1990um} factorization theorems, which are therefore associated to
different physical scales. Power corrections take the generic form $(\Lambda_{\rm QCD}/{\cal Q})^n$ with $n\geq 1$ and $\cal Q$ a magnitude with dimensions of energy
appearing in the corresponding matrix element. As we shall discuss, a double pole implies a logarithmic enhancement in the power correction, which signals the
presence of an anomalous dimension in the associated operator, with $n_f$ dependence at leading order within an operator product expansion (OPE).

While a thorough assessment of the large-order behavior of a series in full QCD is only possible in a few cases (such as the pole mass renormalon), one can carry
out the relevant computations in the large-$\beta_0$ limit of QCD. In order not to violate important properties of the strong interactions such as confinement,
this limit formally corresponds to making the number of active flavors $n_f$ tend to $-\infty$. Since for phenomenological applications one
takes a finite (and positive) number of flavors, in practice one considers the (equivalent) $\beta_0\to \infty$ limit.\footnote{Equivalently one has an
effective number of flavors $n_f^{\rm eff}=3(11-\beta_0)/2$ which tends to $-\infty$ as $\beta_0\to\infty$.} In this setup one reorders the perturbative series
by considering $\alpha_s\beta_0\sim\mathcal{O}(1)$ such that the leading contribution ---\,beyond tree level\,--- contains all terms of the form
$\alpha_s\times (\alpha_s\beta_0)^n\sim\mathcal{O}(\alpha_s)$ for any $n\geq 0$. The series might have an $\mathcal{O}(\alpha_s^0)$ term, such that tree-level
and one-loop terms are exactly reproduced.
The infinitely many terms that contribute at leading order are equally important and should be summed up. While this
sum can be carried out unambiguously for convergent series, for asymptotic ones (that is, with null convergence radius), the sum is ill-defined. Using the Borel
resummation method one can, however, {\it define} the sum of the series through the principal value prescription
and assign an ambiguity to this procedure. It is believed this Borel sum is close to the ``true answer'' within the estimated ambiguity.
For an excellent introduction to renormalons the reader is refereed to Ref.~\cite{Beneke:1998ui}, while a very useful technical review can be found
in \cite{Grozin:2003gf}.

In this article we compute in the large-$\beta_0$ approximation the following renormalized matrix elements: a)~the hard function $H_Q$ for quark-initiated processes
appearing in many SCET factorization theorems, b)~the hemisphere jet function $J_n$ for massless quarks describing collinear radiation in the thrust, heavy-jet-mass and
C-parameter event-shape cross sections, c)~the SCET to bHQET matching condition $C_m$ showing up in factorization theorems for any massive-quark event-shape
distribution in the peak region, in the form of the hard factor $H_m$, and d)~the bHQET hemisphere jet function $B_n$, describing radiation which is soft in the rest frame
of the boosted heavy quark, for the set of event shapes just mentioned. As a byproduct of these
computation we will also obtain closed expressions for the universal cusp anomalous dimensions and their non-cusp counterparts for the SCET hard, jet, and soft
functions, as well as the bHQET matrix element and jet function.\footnote{The large-$\beta_0$ cusp anomalous dimension was computed
in Ref.~\cite{Scimemi:2016ffw}. We find full agreement and provide an efficient recursive algorithm to obtain the corresponding fixed-order coefficients
to any order.} We find that all matrix elements are ambiguous, while the cusp and non-cusp anomalous dimensions have the same non-zero convergence radius. Additionally,
we provide recursive algorithms to expand all matrix elements and anomalous dimensions to any order in perturbation theory.

The jet function(s) calculation's complexity is similar to that of the matching conditions, since the collinear thrust measurement is
inclusive and the computation can be cast as the discontinuity of a 1-loop forward-scattering matrix element with a modified gluon propagator.
This property does not hold for non-boosted momenta, and therefore to determine the large-$\beta_0$ soft-function it is required to compute Feynman diagrams of
two-loop complexity.\footnote{One needs to compute real-radiation phase space integrals with two on-shell particles,
similar in complexity to two-loop virtual-momentum integrals.} We therefore relegate this calculation to the future (note that in
Refs.~\cite{Hoang:2007vb,Mateu:2012nk} only the asymptotic behavior of the soft function was computed, which does not exactly reproduce the terms
leading in $1/\beta_0$ at low orders).

To carry out these computations we generalize the procedure presented in Ref.~\cite{Grozin:2003gf} to series with cusp anomalous dimension [\,that is, for cases with
$1/\varepsilon^2$ poles at $\mathcal{O}(\alpha_s)$ in dimensional regularization\,]. We solve the corresponding RGE equation exactly at $1/\beta_0$ accuracy.
The closed form for the renormalized series can be written as the sum of a \mbox{$\mu$-independent} inverse-Borel integral, where all ambiguities
are confined, plus other $\mu$-dependent terms with finite convergence radius. The former depends explicitly on $\Lambda_{\rm QCD}$, the non-perturbative QCD scale.

We also discuss the relation between the pole and MSR~\cite{Hoang:2008yj,Hoang:2017suc} masses in the \mbox{large-$\beta_0$} approximation, and compute a closed
form for its R-anomalous dimension $\gamma_R$. We show that, even though the Borel transform of the series defining the MSR mass has a singularity at $u=1/2$, the
poles for $\gamma_R$ appear only for $u\geq 3/2$. We analytically solve the fixed-order R-evolution~\cite{Hoang:2009yr} equation in this limit. From our result for the
position-space bHQET jet function we compute the large-$\beta_0$ expression for two versions of the jet mass: as defined in Ref.~\cite{Jain:2008gb} and an alternative scheme which we define in
analogy to the soft-gap subtraction introduced in Ref.~\cite{Bachu:2020nqn}. We compute their $\mu$- (only in the original definition) and R-anomalous dimensions
in a closed form.

The results of our computations are used in three phenomenological applications: i)~the associated leading ambiguity is used to quantify the dominant hadronization
correction,
ii)~the convergence radius of all anomalous dimensions is found, and iii)~we study under which circumstances the various matrix elements converge best.
In particular, we asses the effect of $\pi$-resummation and the various low-scale short-distance masses in the convergence of $H_Q$ and $B_n$, respectively. We present a
tentative procedure to remove the $\mathcal{O}(\Lambda_{\rm QCD})$ renormalon present in $H_m$ by expressing the pole mass in terms of the $\MSBar$ scheme.
Phenomenological studies of this cancellation beyond $1/\beta_0$ are relegated to future work.

This article is organized as follows: In Sec.~\ref{eq:remarks} we provide an introduction to the large-$\beta_0$ expansion and the general treatment of the leading
$1/\beta_0$ correction; in Sec.~\ref{sec:noSCETren} we work out closed expressions for series which do not have a cusp anomalous dimension, providing a solution for
their renormalization group equation, while in Sec.~\ref{sec:SCETren} we adapt the formalism for series with cusp anomalous dimension. Our prescription to compute the
inverse Borel transform when there are poles in the integration path is presented in Sec.~\ref{sec:ppv}. The massive-quark self-energy is dealt with in
Sec.~\ref{sec:MSbarPole}, including the computation of closed forms for the relation between the pole and $\MSBar$ masses, the wave-function renormalization
in the on-shell scheme, the $\MSBar$ mass anomalous dimension and the MSR R-anomalous dimension.
The computation
of the SCET hard and jet functions is carried out in Sec.~\ref{sec:SCET}.
Analogous computations for
bHQET are contained in Sec.~\ref{sec:bHQET}, where it is shown how the leading renormalon found in each one cancels when expressing the pole mass in terms
of a short-distance mass. In Sec.~\ref{sec:JetMasses} we discuss two alternative versions for the so-called jet mass in the \mbox{large-$\beta_0$} approximation,
and analyze how effectively they remove the renormalon of the bHQET jet function. In Sec.~\ref{sec:estimate} we provide a numerical estimate of the unknown 3-loop
position-space jet function for a boosted heavy quark. Our conclusions can be found in Sec.~\ref{sec:conclusions}.

\section{\boldmath General remarks about the large-$\beta_0$ expansion}
\label{eq:remarks}
Let us consider a renormalized perturbative series starting with a non-vanishing tree-level coefficient, which for simplicity is taken to be $1$ (we also assume
that the lowest-order Feynman diagrams contain no gluon propagators), written in terms of the renormalized coupling $\alpha_s$ (for simplicity we do not
show the dependence of $\alpha_s$ on $\mu$, but we mark with a superscript that of the fixer-order coefficients $c^\mu_i$):
\begin{equation}
A = 1 + \sum_{i = 1} \biggl( \frac{\alpha_s}{4 \pi} \biggr)^{\!\! i} \sum_{n = 0}^{i - 1} c^\mu_{i, n} n_f^n\,,
\end{equation}
where the sub-index $i$ counts the loops while $n$ matches the powers to which the number of quark flavors $n_f$ is raised to.
We can now rewrite the series in terms of powers of $\beta_0=11 - 2 n_f/3$ as follows:
\begin{equation}
A = 1 + \sum_{i = 1} \biggl( \frac{\alpha_s}{4 \pi} \biggr)^{\!\! i} \sum_{n = 0}^{i - 1} a^\mu_{i, n} \beta_0^n\,,\qquad\quad
a^\mu_{i, n} = (- 1)^n \sum_{j = n}^{i - 1} \frac{11^{j - n}c^\mu_{i, j}}{\bigl( \frac{2}{3}\bigr)^j} \binom{j}{n} .
\end{equation}
For $a_{n, n - 1}$ (that is, the leading terms in the large-$\beta_0$ expansion) only one term contributes to the sum in the second equality.
In the
limit we are interested in one considers $\alpha_s \beta_0 \sim \mathcal{O}(1)$, and the series is reorganized accordingly:
\begin{align}\label{eq:a-notation}
A = &\, 1 + \sum_{n = 1} \frac{1}{\beta^n_0} \sum_{i = n} a^\mu_{i, i - n}
\biggl( \frac{\beta_0\alpha_s}{4 \pi} \biggr)^{\!\! i} \equiv \, 1 + \sum_{j=1}^{\infty}\frac{1}{\beta_0^j} f^\mu_j(\beta)\, ,\\
f^\mu_j(x) \equiv &\, \sum_{i = j} a^\mu_{i, i - j} x^i\,,\qquad \beta\equiv \frac{\beta_0\alpha_s}{4 \pi}\sim \mathcal{O}(1)\,, \nonumber
\end{align}
where the sum defining $f^\mu_j$
starts contributing at $\mathcal{O}(\alpha_s^j)$. The parameter $\beta$ is order one and should not be mistaken with the QCD $\beta$-function, which we denote
$\beta_{\rm QCD}$. The small parameter upon which we reorganize our perturbative series is obviously $1/\beta_0$. Since all terms that enter
the definition of $f_j$ are equally important, the sum has to be carried out all the way to infinity. In this article we will be concerned with
$f_1^\mu(\beta)\equiv \beta_0\, \delta\! A(\mu)$ only.

In this counting one has
$\beta_n \alpha_s^{n + 1}\sim\mathcal{O} (\alpha_s)$ for
$n\geq 1$.\footnote{This is caused by the fact that the coefficients of $\beta_{\mathrm{QCD}}$ satisfy
$\beta_{n>0}\sim\mathcal{O}(\beta_0^n)$. In particular one has that
both $\beta_0$ and $\beta_1$ have only linear dependence on $n_f$.} Therefore, in $d=4-2\varepsilon$ dimensions the QCD $\beta$-function takes a simple
form, \mbox{$\beta_{\rm QCD} =- 2 \alpha_s (\varepsilon + \beta)$}, such that one can write $\df\! \log(\beta)/\df \!\log(\mu) = -2 (\varepsilon + \beta)$. This result
can be used to write down a closed form for the strong coupling renormalization factor in this approximation.
To derive this result one has to recall the relation between the renormalized strong coupling $\alpha_s(\mu)$ and the (scale-independent)
bare coupling $g_0$: \mbox{$\alpha_s(\mu) =Z^{- 1}_{\alpha} (\mu^2 e^{\gamma_E})^{- \varepsilon} g_0^2/(4 \pi)^{1- \varepsilon}$},
where we already adopt the $\MSBar$ convention for the renormalization scale $\mu^2\to \mu^2 e^{\gamma_E}/(4\pi)$.
Taking logarithms on both sides and applying a derivative with respect to $\log(\mu)$ one gets \mbox{${\rm d}\! \log(Z_\alpha)/{\rm d} \!\log(\mu)=2 \beta$}.
Since the $\mu$ dependence of $Z_\alpha$ comes through $\beta$ only, using the tree-level boundary condition
$Z_\alpha(\beta=0)=1$, we obtain a unique solution for the renormalization factor
\begin{equation}\label{eq:Za}
\frac{{\rm d} \log (Z_\alpha)}{{\rm d} \beta} =
-\frac{1}{\varepsilon + \beta} \quad\longrightarrow \quad Z_{\alpha} = \frac{1}{1 + \frac{\beta}{\varepsilon}}\,.
\end{equation}
The RGE solution for the strong coupling in the large-$\beta_0$ limit simply corresponds to the full QCD leading-logarithm (LL) result, which for the $\beta$ parameter can
be written as
\begin{equation}\label{eq:betaRun}
\beta (\mu) = \frac{\beta (\mu_0)}{1 + 2\beta (\mu_0) \log \Bigl(\frac{\mu}{\mu_0} \Bigr)} = \frac{1}{2\log \Bigl(
\frac{\mu}{\Lambda_{\rm QCD}} \Bigr)}\,,
\end{equation}
where the $\mu$-independent quantity with dimensions of energy $\Lambda_{\rm QCD}=\mu\, e^{-1/[2\beta(\mu)]}$ determines the non-perturbative regime of
QCD and specifies by itself a boundary condition for the strong coupling.

Let us consider an unrenormalized perturbative series at $\mathcal{O}(1/\beta_0)$ expressed in powers of the bare coupling $g_0$, whose ($\mu$-independent) coefficients
$a^{(0)}_{n, n - 1}$ contain $1/\varepsilon^m$ divergences with $m\geq n$ if the series does not have a cusp anomalous dimension, or $m\geq n + 1$ if it has.\footnote{One has
$m \geq n-1$ if the series has no anomalous dimension at all, making the one-loop term finite.} If the series carries an anomalous dimension of either kind, even after
expanding it in powers of the renormalized coupling $\alpha_s$, additional (multiplicative) renormalization will be necessary:
\begin{align}\label{eq:bare}
A_0=\,&1 + \frac{1}{\beta_0} \sum_{i = 1} a^{(0)}_{n, n - 1}\! \biggl[
\frac{g_0^2 \beta_0}{(4 \pi)^{2 - \varepsilon}} \biggr]^n = 1 + \frac{1}{\beta_0} \sum_{n = 1} a^{(0)}_{n, n - 1} (\mu^2 e^{\gamma_E})^{n
\varepsilon} \biggl( \frac{\varepsilon \beta}{\varepsilon + \beta} \biggr)^{\!\! n}\\
=\,&1 + \frac{1}{\beta_0} \sum_{n = 1} a^{(0)}_{n, n - 1}(\mu^2 e^{\gamma_E})^{n
\varepsilon} \sum_{i = 0} \frac{(- 1)^i (n)_i}{i!} \biggl( \frac{\beta}{\varepsilon}
\biggr)^{\!\!i + n} \,,\nonumber
\end{align}
where in the second equality of the first line $g_0$ has been expressed in terms of $\beta$,
and in the second line the Pochhammer symbol \mbox{$(a)_n=\Gamma(a+n)/\Gamma(a)$} has been used to re-expand the series in powers of $\beta$.

It can be shown that when inserting the vacuum polarization function $n$ times in the gluon propagator, and after performing a na\"\i ve
non-abelianization~\cite{Broadhurst:1994se,Beneke:1994qe,Beneke:1998ui} with the replacement $n_f\to -3\beta_0/2$, one gets the following effective gluon propagator:
\begin{align}\label{eq:bareA}
\frac{i g^{\mu\nu}}{-p^2} \to\,& \biggl[ e^{-\varepsilon \gamma_E} \frac{\beta_0 g_0^2}{(4 \pi)^{2 -
\varepsilon}} \frac{D (\varepsilon)}{\varepsilon} \biggr]^n \frac{i}{(-
p^2)^{1 + n \varepsilon}} \biggl( g^{\mu \nu} - \frac{p^{\mu} p^{\nu}}{p^2}\biggr)\,,\\
D (\varepsilon)\equiv\,& \frac{6 e^{\varepsilon \gamma_E} \Gamma (1 + \varepsilon) \Gamma^2 (2 -
\varepsilon)}{\Gamma (4 - 2 \varepsilon)} \,.\nonumber
\end{align}
The transverse part of the effective propagator does not contribute to the computations carried out in this article, therefore in practice we use the Feynman gauge with a modified
gluon propagator.\footnote{If the quantity of interest has no anomalous dimension one can renormalize the one-loop
fermion bubble in the $\MSBar$ scheme, which implies the renormalized coupling is being used. When inserted $n$ times one gets
\begin{equation}
\frac{i g^{\mu\nu}}{-p^2} \to \Biggl[-\beta \log\biggl(e^{\frac{5}{3}}\frac{\mu^2}{-p^2}\biggr)\Biggr]^n
\frac{i}{-p^2} \biggl( g^{\mu \nu} - \frac{p^{\mu} p^{\nu}}{p^2}\biggr).
\end{equation}
When summing up all possible insertions of the fermion bubble one gets an inverse Borel transform
\begin{equation}
\beta \sum_{n=0} \beta^n \log^n\!\biggl(e^{\frac{5}{3}}\frac{\mu^2}{-p^2}\biggr) =\!
\int_0^\infty \!{\df} u\,e^{-\frac{u}{\beta}}\! \sum_{n=0}\frac{1}{n!}\log^n\!\biggl(e^{\frac{5}{3}}\frac{\mu^2}{-p^2}\biggr)^{\!\! u}
=\int_0^\infty \!{\df} u\,e^{-\frac{u}{\beta}}\biggl(e^{\frac{5}{3}}\frac{\mu^2}{-p^2}\biggr)^{\!\! u}\,.
\end{equation}
Therefore, also in this case the gluon propagator has a shifted power $(- p^2)^{1 + u}$. The modified 1-loop computation
is carried out directly in $d=4$ and the following result for the renormalized is found
\begin{equation}\label{eq:Anogam}
A = 1 + \frac{1}{\beta_0}\int_0^\infty \!{\df} u\,e^{-\frac{u}{\beta}}\! \biggl(e^{\frac{5}{3}}\frac{\mu^2}{{\cal Q}^2}\biggr)^{\!\! u} b(0,u)
= 1 + \frac{1}{\beta_0}\int_0^\infty \!{\df} u\biggl(\!e^{\frac{5}{3}}\frac{\Lambda_{\rm QCD}^2}{{\cal Q}^2}\biggr)^{\!\! u} \,b(0,u)
\,.
\end{equation}}
The gauge dependence of course cancels after summing up all diagrams. The effective propagator for $n\geq 0$ is $\mathcal{O}(1)$
in the large-$\beta_0$ counting, and therefore any term contributing at leading order can be obtained from gluon bubbles. In practice it is enough to compute the
one-loop contribution to the series under study with a shifted power $(- p^2)^{1 + h}$ in the gluon's propagator denominator. This computation defines an
$\varepsilon$- and $h$-dependent bare one-loop result which can be written as
\begin{equation}\label{eq:1-loopbare}
A_\text{1-loop}= \frac{g_0^2}{(4 \pi)^{2 - \varepsilon}}\, \frac{b(\varepsilon, h)}{{\cal Q}^{2(h+\varepsilon)}}\,,
\end{equation}
where ${\cal Q}$ represents a parameter with dimensions of energy
that renders the computation non-vanishing in dimensional regularization.
For the computations carried out in this article ${\cal Q}$ will stand for $Q$ (center of mass energy), $m$ (heavy quark mass), $R$ (scale of the MSR, RS and jet masses)
${\cal Q}^2=s$, the (squared) invariant mass of a jet, and $\hat s = (s^2-m^2)/m$. The bare coefficient
appearing in Eq.~\eqref{eq:bareA} can be expressed in terms of the modified 1-loop computation of Eq.~\eqref{eq:1-loopbare},
\begin{equation}
a^{(0)}_{n, n - 1} = \biggl[\! \biggl(\!\frac{\mu^2 e^{\gamma_E}}{{\mathcal Q}^2}\!\biggr)^{\!\!\varepsilon} \frac{D
(\varepsilon)}{\varepsilon} \biggr]^{\!n - 1} \frac{b[\varepsilon, (n - 1) \varepsilon]}{{\cal Q}^{2\varepsilon}} \,.
\end{equation}
When inserted in Eq.~\eqref{eq:bare} the bare series can be expressed as follows:
\begin{align}\label{eq:sumF}
A_0 =\,& 1 + \frac{1}{\beta_0} \sum_{n = 1} \frac{F^\mu(\varepsilon, n \varepsilon)}{n}
\sum_{i = 0} \frac{(- 1)^i (n)_i}{i!} \biggl( \frac{\beta}{\varepsilon}\biggr)^{i + n}\,,\\[-0.1cm]
F(\varepsilon, u) \equiv\,& u D (\varepsilon)^{\frac{u}{\varepsilon} - 1} b (\varepsilon, u - \varepsilon) e^{\varepsilon \gamma_E} \,,\qquad
F^\mu(\varepsilon, u) \equiv F(\varepsilon, u)\!\biggl(\frac{\mu}{\cal Q}\biggr)^{\!\!2 u} . \nonumber
\end{align}
The $F$ function is standard in renormalon calculations, see e.g.\ Refs.~\cite{Beneke:1994qe,Beneke:1995pq,Beneke:1998ui,Grozin:2003gf}.
In the second line, a factor of $u$ is included such that $F(\varepsilon, u)$ is regular (and in general non-zero) when $u\to 0$ for series without
cusp anomalous dimension.
Moreover, $F^\mu(\varepsilon, 0)= F(\varepsilon, 0)$ does not
depend on $\mu$. Computations in the context of effective field theories for jets have $1/\varepsilon^2$
poles at one loop and, when using a shifted gluon denominator, this translates into $F (\varepsilon, u)\sim 1/u$ when $u\to 0$. In Sec.~\ref{sec:SCETren} we shall
adapt the procedure to account for this fact.

\section{\boldmath Series without cusp anomalous dimension}
\label{sec:noSCETren}
In this section we review the standard procedure to find closed expressions for series without cusp anomalous dimension. In Sec.~\ref{sec:MSbarPole}
we apply the formalism to the case of the massive quark self-energy.
The presentation follows closely Ref.~\cite{Grozin:2003gf}, but we provide explicit proofs for all intermediate steps. The regularity condition allows us to expand the function $F(\varepsilon, u)$
in powers of $\beta$ and $\varepsilon$
\begin{equation}
F^\mu(\varepsilon, u) = \sum_{m = 0} u^m \sum_{j = 0} \varepsilon^j F^\mu_{j, m}\,.
\end{equation}
Since the coefficients $F^\mu_{j, 0}$ do not depend on $\mu$ we denote them simply by $F_{j, 0}$. Inserting this expansion in Eq.~\eqref{eq:sumF}, shifting some
indices, manipulating the order of sums and defining $A_0 = 1 + \delta\!A_0$ we obtain:
\begin{align}\label{eq:nocuspDer}
\beta_0\,\delta\!A_0 \equiv &\, \beta_0\Biggl[ \delta A(\mu) + \sum_{j = 1}^\infty \frac{\delta A^{1/\varepsilon^j}}{\varepsilon^j}\Biggr] \!=
\sum_{n = 1} \sum_{i = 0} \frac{(- 1)^i (n)_i}{i!} \beta^{i + n} \!\sum_{m = 0}
n^{m - 1} \!\sum_{j = 0} \frac{F^\mu_{j, m}}{\varepsilon^{i + n - m - j}}\\
=&\sum_{i = 1} (- \beta)^i \!\sum_{m = 0} \sum_{j = 0}
\frac{F^\mu_{j, m}}{\varepsilon^{i - m - j}} \sum_{n = 1}^i (- 1)^n \binom{i - 1}{n - 1} n^{m - 1}\nonumber\\
=& \sum_{i = 1} (- \beta)^i \!\sum_{j = 0}
\frac{1}{\varepsilon^{i - j}}\! \sum_{m = 0}^j F^\mu_{j - m, m} \sum_{n = 1}^i (-1)^n \binom{i - 1}{n - 1} n^{m - 1}\nonumber\\
=& \sum_{j = - \infty}^0 \varepsilon^j\!\! \sum_{i = \max (1,
- j)} (- \beta)^i \!\sum_{m = 0}^{j + i} F^\mu_{j + i - m, m} \!\sum_{n = 1}^i (- 1)^n
\binom{i - 1}{n - 1} n^{m - 1} + \mathcal{O}(\varepsilon)\,,\nonumber
\end{align}
where to get the second line we shift $i\to i-n$ and push the summation over $n$ to the right-most position; to obtain the third line
we shift $j\to j-m$ and swap the order between the sums over $j$ and $m$; the last line takes its form after shifting $j\to j+i$ and interchanging
the resulting sums over $j$ and $i$. Since after renormalization one takes $\varepsilon\to0$, in the last line the $j$ upper summation
limit is set to zero. In Appendix~\ref{sec:AppSum} it is shown that the rightmost sum over $n$ in the last line vanishes unless $m = 0$ or $m \geqslant i$,
and in the former case it equals $-1/i$. Therefore we isolate the $m=0$ contribution and restrict the remaining sum to values of $m$ larger or equal than $i$.
Shifting in this last sum $m\to m + i$ one gets
\begin{equation}\label{eq:intermediate}
\beta_0 \delta\!A_0 = \!\!\sum_{j = - \infty}^0 \varepsilon^j\!\!\! \sum_{i =
\max (1, - j)} \!\!\!\!\!(- \beta)^i \! \Biggl[ \,\sum_{m = 0}^j F^\mu_{j - m, m + i} \sum_{n =
1}^i (- 1)^n \binom{i - 1}{n - 1} n^{m + i - 1} - \frac{F_{i + j, 0}}{i}
\Biggr] .
\end{equation}
For negative values of $j$ the first sum vanishes and the lower limit of the sum over $i$ becomes $-j$. Reversing the sing of $j$ and shifting $i\to i+j$ one gets for the coefficients of $1/\varepsilon^j$, defined in the first line of Eq.~\eqref{eq:nocuspDer}, the result
\begin{equation}
\beta_0 A_0^{1 / \varepsilon^j} = - (- \beta)^j \sum_{i = 0} (- \beta)^i \frac{F_{i, 0}}{i + j} \,.
\end{equation}
For $j = 0$ the lower limit in the sum over $i$ in Eq.~\eqref{eq:intermediate} becomes $1$, and the sum over $m$ has only one term, $m=0$. As shown in
Appendix~\ref{sec:AppSum}, this sum can be carried out explicitly
\begin{equation}\label{eq:sum2}
K_i\equiv \sum_{n = 1}^i (- 1)^n \binom{i - 1}{n - 1} n^{i - 1} = (- 1)^i (i - 1) ! \,,
\end{equation}
such that the second piece of the finite term takes also a simple form. Adding the divergent terms the following form for the bare series in terms of the
renormalized coupling is obtained:
\begin{equation}\label{eq:A0sum}
\beta_0 \,\delta\!A_{0} = \sum_{i = 1} \beta^i \biggl[ \Gamma (i)
F^\mu_{0, i} - \frac{(- 1)^i F_{i, 0}}{i} \biggr] - \sum_{j = - \infty}^{- 1}
\varepsilon^j \! \sum_{i = - j} (- \beta)^i \frac{F_{i + j, 0}}{i} \, .
\end{equation}
From this equation one can read off the expression for the renormalized fixed-order coefficients in the large-$\beta_0$ limit. Setting $\mu={\cal Q}$ the
non-logarithmic coefficients, which are factorially divergent, can be obtained (the logarithmic pieces can be easily obtained,
as discussed next). Following the notation introduced in Eq.~\eqref{eq:a-notation} and defining \mbox{$a_{i,i-1}\equiv a^{\cal Q}_{i,i-1}$} one obtains:
\begin{equation}\label{eq:nologFOnocusp}
a_{i, i-1} = (i-1)!\, F_{0,i} - \frac{(-1)^i F_{i,0}}{i} = (i-1)!\, F_{0,i} - \frac{\hat \gamma_A^i}{2 i}\,,
\end{equation}
where $\hat \gamma_A^i$ will be defined in Eq.~\eqref{eq:nocuspgamma} below.

At this point one can figure out closed forms for the various sums appearing in Eq.~\eqref{eq:A0sum} in terms of the function $F$ defined in the second line of
Eq.~\eqref{eq:sumF}.
The second term of the finite contribution [\,at leading order in $1/\beta_0$ the finite term coincides with the renormalized series $\beta_0\,\delta\!A(\mu)$\,] involves
a UV subtraction related to renormalization,
\begin{equation}
\sum_{i = 1} \frac{F_{i, 0}}{i} (- \beta)^i = - \!\!\int_{- \beta}^0 \frac{{\df} \varepsilon}{\varepsilon}\sum_{i = 1} \varepsilon^i F_{i, 0} = -\!\!
\int_{- \beta}^0 \frac{{\df} \varepsilon}{\varepsilon} \bigl[F (\varepsilon, 0) - F (0,0)\bigr],
\end{equation}
that makes sure the integrand is convergent when $\varepsilon\to 0$. For the terms proportional to $1/\varepsilon^j$ with $j>0$ subtractions are not necessary (as we shall see, the $1/\varepsilon$ term fully determines the anomalous dimension of the series in the large-$\beta_0$ limit),
and all of them can be worked out at once:
\begin{equation}
\beta_0 A_0^{1 / \varepsilon^j} = - \! \sum_{i = 0} (-\beta)^{i + j}
\frac{F_{i, 0}}{i + j} = \!\int_{- \beta}^0 \!{\df}
\varepsilon\, h^{j - 1} \sum_{i = 0} h^i F_{i, 0} = \!\int_{-\beta}^0 \!{\df} h\,
h^{j - 1} F (h, 0)\,.
\end{equation}
The closed expression for the first contribution to the renormalized series takes, as expected, the form of an inverse Borel transform:
\begin{equation}
\sum_{i = 1} \beta^i\, \Gamma(i) F^\mu_{0, i} = \!\int_0^{\infty}\!\! \frac{{\df} u}{u}\, e^{- \frac{u}{\beta}}
\sum_{i = 1} u^i F^\mu_{0, i} = \!\int_0^{\infty} \!\frac{{\df} u}{u} \,e^{- \frac{u}{\beta}}
\bigl[F^{\mu}(0, u) - F (0, 0)\bigr] ,
\end{equation}
where again the UV subtraction makes the integrand converge as $u\to 0$. With this result we can write a closed
form for the renormalized series. Defining $A(\mu) = 1 + \delta A(\mu)$
we get
\begin{equation}\label{eq:FOnocusp}
\beta_0 \,\delta\!A(\mu) = \int_0^{\infty} \!\frac{{\df} u}{u} \,e^{- \frac{u}{\beta}}
\bigl[F^{\mu}(0, u) - F (0, 0)\bigr] + \!\int_{- \beta}^0 \frac{{\df} \varepsilon}{\varepsilon}\bigl[F(\varepsilon, 0) - F (0, 0)\bigr] .
\end{equation}
The ambiguities appear in the first integral [\,the second is not ambiguous because
$F(\varepsilon, 0)$ has poles for $\varepsilon>0$ only\,], and written in this way it seems they might depend on $\mu$. Later in this section it will shown that
they do not depend on any renormalization scale.

The series is renormalized multiplicatively \mbox{$A_0 = Z A=(1+\delta Z)(1+\delta\!A)\simeq 1+\delta Z + \delta\!A$} where $A(\mu)$ is finite in the
$\varepsilon \rightarrow 0$ limit and $\delta Z\equiv Z - 1$ in the $\MSBar$ scheme has only $1/\varepsilon^j$ poles. We have discarded the crossed term
$\delta Z\times \delta\!A$ since it is $\mathcal{O}(1/\beta_0^2)$, making renormalization {\it effectively} additive at $\mathcal{O}(1/\beta_0)$. It is
straightforward to find a closed expression for $\delta Z$
\begin{equation}\label{eq:ZNC}
\beta_0 \,\delta Z = -\!\sum_{j = 1}^\infty \frac{1}{\varepsilon^j} \sum_{i = j}
(- \beta)^{i} \frac{F_{i - j, 0}}{i} = \sum_{j = 1}^\infty \frac{1}{\varepsilon^j} \! \int_{- \beta}^0\!\! {\df} h\, h^{j - 1} F (h, 0)\,.
\end{equation}
For series with no anomalous dimension $F(\varepsilon, 0)=0$ and therefore $Z=1$.
Since $F_{i,0}$ and, consequently $F(-\beta,0)$, are $\mu$ independent, the $Z$ factor has no dependence on $\mu$ beyond that of $\beta$. Since in this limit
one also has $\log(Z)\simeq \delta Z$ and $\log(A)\simeq \delta A$, the anomalous dimension \mbox{$\gamma_A\equiv {\df }\!\log\! A/{\df}\! \log\mu$} for the renormalized quantity $A(\mu)$ can be
computed as the negative of the logarithmic $\mu$-derivative acting on $Z$. We use the chain rule in the large-$\beta_0$ approximation for $d=4-2\varepsilon$ effectively replacing
\begin{equation}
\mu \frac{\mathrm{\text{d}}}{\mathrm{\text{d}} \mu} \to - 2 \beta (\varepsilon + \beta) \frac{\df}{{\df} \beta} \,.
\end{equation}
Taking into account that $\gamma_A$ is finite as $\varepsilon\to0$ it is clear that the anomalous dimension must come entirely from the $\varepsilon$ term in the previous equation
acting on the $1/\varepsilon$ contribution to $Z$. The remaining contributions are individually $\mathcal{O}(1/\varepsilon^k)$ with $k>0$ and therefore must cancel when added up
to render a finite $\gamma_A$. Let us show this fact with an explicit computation:\footnote{The same result can be obtained
applying the derivative directly to the integral form of $\delta Z$:
\begin{align}\label{eq:gammaNC}
2(\varepsilon+\beta)\beta \frac{\df}{{\df} \beta} \!\!\!\sum_{j=-\infty}^{-1}\!\!\varepsilon^j\!\!\int_0^{\beta} \!\!{\df} h\, \frac{F (-h, 0)}{(- h)^{j + 1}} =
2F (-\beta, 0)\!\Biggl[\,\sum_{j = - \infty}^{- 1} \!\!\!\varepsilon^{j + 1} (- \beta)^{- j} -\!\!\!\! \sum_{j = - \infty}^{- 1} \!\!\varepsilon^j (- \beta)^{1 - j}\Biggr]\!
= 2\beta F(-\beta,0)\,,
\end{align}
where one can shift $j\to j-1$ in the first term of the second equality to see that all divergent terms cancel.}
\begin{align}\label{eq:nocuspgamma}
\beta_0\gamma_A(\beta) \equiv&- \!2 \beta (\varepsilon + \beta) \beta \frac{\df}{{\df} \beta} \!\sum_{j = - \infty}^{- 1}
\varepsilon^j \sum_{i = - j} (- \beta)^i \frac{F_{i + j, 0}}{i} \\[-0.2cm]
=& -\!2\!\!\sum_{j = - \infty}^{- 1} \varepsilon^{j + 1} \! \sum_{i = - j} (-\beta)^i F_{i + j, 0} +2\!\! \sum_{j = - \infty}^{- 1} \varepsilon^j \! \sum_{i= - j} (- \beta)^{i + 1} F_{i + j, 0}\nonumber\\
=& -\!2\sum_{i = 1} (- \beta)^i F_{i - 1, 0} = 2\beta\sum_{i = 0} (- \beta)^i F_{i, 0} = 2\beta F(-\beta,0)\,,\nonumber
\end{align}
where one can shift $j\to j-1$ and $i\to i+1$ in the first term of the second line to show that all divergent terms cancel, obtaining the first equality in the last line.
Therefore, the corresponding fixed-order coefficients have the following form:
\begin{equation}
\gamma_A(\beta) \equiv\frac{1}{\beta_0}\sum_{i=0}\hat \gamma_A^i \,\beta^{i+1}\,,\qquad
\hat \gamma_A^i = 2(-1)^iF_{i, 0}\,,\qquad
\gamma_A^i = \hat \gamma_A^i \beta_0^{i} \,.
\end{equation}
By the same reasons exposed above, $\hat \gamma_A^i$
are $\mu$ independent, and the dependence of $\gamma_A$
on the renormalization scale happens through $\beta$ only.
The RGE for $A(\mu)$ can be easily solved, and in the large-$\beta_0$ takes the form [\,$\beta_i\equiv \beta(\mu_i)$\,]:
\begin{align}\label{eq:nocusprun}
\beta_0[A(\mu_2) - A(\mu_1)] = \,&-\!\frac{\beta_0}{2}\! \int_{\beta_1}^{\beta_2}\! \frac{{\df}\beta}{\beta^2} \,\gamma_A(\beta)
= -\!\int_{\beta_{\mu_0}}^{\beta}\!\frac{{\df}\varepsilon}{\varepsilon}F (- \varepsilon, 0)\\
= \,& -\!\frac{\gamma_0}{2} \log\!\biggl(\frac{\beta_2}{\beta_1}\biggr) +\!\int^{\beta_1}_{\beta_2}\!\frac{{\df}\varepsilon}{\varepsilon}\,
[F (- \varepsilon, 0) - F(0,0)]\nonumber\\
= \,& -\biggr[\frac{\hat\gamma_0}{2} \log\!\biggl(\frac{\beta_2}{\beta_1}\biggr)\biggr]_{\rm LL} -
\frac{1}{2}\sum_{i=1}\biggr[\frac{\hat\gamma_A^i}{i}(\beta_2^i - \beta_1^i)\biggr]_{{\rm N}^i{\rm LL}}\,,\nonumber
\end{align}
where in the second line we have pulled out the term which is not regular when either $\beta_i$
tend to zero. In the last equation we use the expanded version of the anomalous dimension, and given the fact that $\gamma_A$ is not ambiguous, provided the values of $\beta$
are not too large, the sum can be carried out to arbitrarily high orders. This approach, employed also in Ref.~\cite{Hoang:2021fhn}, provides an effective way of computing the numerical
integral in the second line to arbitrary precision. Furthermore, the first term in the last line corresponds to LL accuracy, while adding $n$ additional terms from the sum
yields N$^n$LL accuracy. This result shows that the difference of two series with different values of $\mu$ is renormalon free.
Taking the values $\mu_2=\mu$, $\mu_1=\mu_0$ and using Eq.~\eqref{eq:FOnocusp} to obtain the boundary condition $A(\mu_0)$, from Eq.~\eqref{eq:nocusprun}
we obtain the following closed resummed form for $\delta\!A(\mu)$:
\begin{equation}\label{eq:no-cusp-sumup}
\beta_0\,\delta\!A(\mu) =\!\! \int_0^{\infty}\! \frac{{\df} u}{u} \,e^{- \frac{u}{\beta_{\mu_0}}} \!
\Biggl[\!\biggl(\frac{\mu_0}{\cal Q}\biggr)^{\!\!2u}\! F(0, u) - F (0, 0)\Biggr]\!
+
\!\int_{- \beta}^0\!\! \frac{{\df} \varepsilon}{\varepsilon}
\bigl[F (\varepsilon, 0) - F (0, 0)\bigr] - \frac{\gamma_0}{2} \log\!\biggl(\frac{\beta}{\beta_{\mu_0}}\biggr) ,
\end{equation}
with $\beta_{\mu_0} \equiv \beta(\mu_0)$. Written in this way, we have explicitly
singled out the dependence on $\mu$ ---\,as expected, renormalon-free\,--- from the ambiguous piece which now depends on the matching scale $\mu_0$.
However, since the matching scale is arbitrary,
the ambiguities must be also independent of $\mu_0$. To proof this explicitly we need the following relations
\begin{align}\label{eq:mu0workout}
\biggl( \frac{\mu_0}{\cal Q} \biggr)^{\!\!2 u}F(0, u) - F (0, 0) = & \biggl( \frac{\mu_0}{\cal Q} \biggr)^{\!\!2 u} \bigl[F(0, u) - F (0, 0)\bigr]
+ \frac{\gamma_0}{2} \biggl[\! \biggl( \frac{\mu_0}{\cal Q} \biggr)^{\!\!2 u} -1 \biggr], \\
\int_0^{\infty} \frac{{\df} u}{u} e^{- \frac{u}{\beta_{\mu_0}}} \biggl[ \!\biggl(
\frac{\mu_0}{\cal Q} \biggr)^{2 u} - 1 \biggr] = & - \!\log \biggl[ 1 + 2
\beta_{\mu_0} \log \biggl( \frac{\cal Q}{\mu_0} \biggr) \!\biggr] = \log \biggl(
\frac{\beta_{\cal Q}}{\beta_{\mu_0}} \biggr), \nonumber\\
e^{- \frac{u}{\beta_{\mu_0}}} \biggl( \frac{\mu_0}{\cal Q} \biggr)^{\!\!2 u} = & \biggl(
\frac{\Lambda_{\rm QCD}}{\cal Q} \biggr)^{\!\!2 u}\,,\nonumber
\end{align}
where in the first line we have simply added and subtracted $F(0,0)(\mu_0/{\cal Q})^{2u}$, and noted that $F_{0,0}=\gamma_0/2$.
The solution to the RGE equation for $\beta$ in Eq.~\eqref{eq:betaRun} has been used to re-write the argument of the logarithm in the second line
and to show the independence of $\Lambda_{\rm QCD}$ on $\mu$ in the last.
The integral in the second line is solved by Taylor expanding the term in square brackets around $u = 0$, integrating term by term and summing up the resulting series:
\begin{align}\label{eq:Taylor1}
\!\int_0^{\infty}\! \frac{{\df} u}{u}e^{- \frac{u}{\beta_{\mu_0}}}\! \biggl[\!
\biggl( \frac{\mu_0}{\cal Q} \biggr)^{\!\!2 u} - 1 \biggr] = & \,2\!\int_0^{\infty}\! {\df} u \,e^{-
\frac{u}{\beta_{\mu_0}}} \sum_{n = 0} \frac{(2u)^n}{(n + 1) !} \log^{n+1}\! \biggl(
\frac{\mu_0}{\cal Q} \biggr) \\
= &\sum_{n = 1} \frac{1}{n} \biggl[ 2\beta_{\mu_0} \log\! \biggl( \frac{\mu_0}{\cal Q}
\biggr)\! \biggr]^n = - \!\log\! \biggl[ 1 + 2 \beta_{\mu_0} \log\! \biggl(
\frac{\cal Q}{\mu_0} \biggr) \!\biggr] . \nonumber
\end{align}
Introducing the results in Eq.~\eqref{eq:mu0workout} into Eq.~\eqref{eq:no-cusp-sumup} and defining $\beta_{\cal Q}\equiv \beta(\cal Q)$ one finds
\begin{equation}\label{eq:LamSumNocusp}
\beta_0\,\delta\!A(\mu) = \!\!\int_0^{\infty}\! \frac{{\df} u}{u} \biggl(\frac{\Lambda_{\rm QCD}}{\cal Q} \biggr)^{\!\!2 u}\!\bigl[F(0, u) - F (0, 0)\bigr] \!
+\!\!\int_{- \beta}^0\!\frac{{\df} \varepsilon}{\varepsilon}
\bigl[F (\varepsilon, 0) - F (0, 0)\bigr] \!- \frac{\gamma_0}{2} \log\!\biggl(\frac{\beta}{\beta_{\cal Q}}\biggr) .
\end{equation}
The same answer can be obtained using Eq.~\eqref{eq:mu0workout} with $\mu_0=\mu$ directly in Eq.~\eqref{eq:FOnocusp}.
The term with potential ambiguities depends only on the value of $\Lambda_{\rm QCD}$, and once this parameter has been fixed
it can be computed once and used for any value of $\mu$.\footnote{For a series without anomalous dimension one defines the function
$H(\varepsilon, u) \equiv
F(\varepsilon, u)/u$ which is finite as $u\to 0$.
Taking into account that $F(\varepsilon,0)=0$ one has that the anomalous dimension is zero, and from Eq.~\eqref{eq:LamSumNocusp} the closed from for the series
can be readily inferred:
\begin{equation}
A = 1 + \frac{1}{\beta_0} \!\int_0^{\infty} \!\mathd u \biggl(\frac{\Lambda_{\tmop{QCD}}}{\cal Q} \biggr)^{\!\!2 u} H (0, u) \,.
\end{equation}
Since $H (0, u) = e^{5/3}b(0,u)$ there is full agreement with Eq.~\eqref{eq:Anogam}.}

We finish this section computing
the dependence of the fixed-order coefficients on \mbox{$\ell\equiv\log(\mu/{\cal Q})$}. Making this logarithmic dependence explicit the series can be written as
\begin{equation}\label{eq:logexplicit}
\beta_0\, \delta\!A(\mu) = \sum_{i = 1}\beta^i a_{i,i-1}^\mu = \sum_{i = 1}\beta^i\sum_{j=0}^i a_{i,i-1,j}\,\ell^j\,,
\end{equation}
with $a_{i,i-1} \equiv a_{i,i-1,0}$. Applying a derivative with respect to $\log(\mu)$ (which acts both on $\beta$ and $\ell$) and imposing that the series has the
right anomalous dimension we obtain
\begin{align}\label{eq:recGA}
\sum_{i = 1} \hat\gamma_A^{i - 1} \beta^i = &\,
\sum_{i = 1} \beta^i \sum_{j = 0}^{i - 1} (j + 1) a_{i, i-1, j + 1} \ell^j - 2
\sum_{i = 1} (i - 1) \beta^i \sum_{j = 0}^{i - 1} a_{i - 1, i - 2, j} \ell^j\,,\\
a_{i, i - 1,1} = &\, \hat\gamma_A^{i - 1} + 2(i - 1) a_{i - 1, i - 2}=2(i - 1)!\, F_{0,i-1} \,,\nonumber\\
a_{i, i-1, j>1} =&\, \frac{2 (i - 1)}{j} a_{i - 1, i-2, j - 1} = \frac{2^j(i-1)!}{j!}F_{0,i-j} \,,\nonumber
\end{align}
where in the first and second terms of the top line we have shifted $j\to j+1$ and $i\to i-1$, respectively, as compared to the summation in Eq.~\eqref{eq:logexplicit}, to have the
sum expressed in terms of $\beta^i$ and $\ell^j$. To obtain the second ($j=0$) and third ($j\geq 1$) lines we simply equate powers of $\beta$ and $\ell$ on both sides of the first
line. In the last equality of the second line we have used Eq.~\eqref{eq:nologFOnocusp} to write the result in terms of $F_{0,i-1}$, making it compatible with the closed form show
in the last line. This final form is obtained using the first equality in the last line recursively $j$ times, until $j=1$ is reached. The same
result can be obtained by a direct computation, expanding in Eq.~\eqref{eq:A0sum} the function $F^\mu(0, u)$ in powers of $u$:
\begin{align}\label{eq:l-explicit}
e^{2u\ell}F(0, u) &\,= \sum_{j=0}\frac{(2\ell)^j}{j!}\sum_{i=0}F_{0,i}\, u^{i+j} = \nonumber
\sum_{j=0}\frac{(2\ell)^j}{j!}\sum_{i=j}F_{0,i-j}\, u^i = \sum_{i=0} u^i \sum_{j=0}^i \frac{2^j}{j!}F_{0,i-j}\,\ell^j\,,\\
a_{i, i - 1,j} &\,= \frac{2^j(i-1)!}{j!}F_{0,i-j} - \delta_{j0}\frac{(-1)^i F_{i,0}}{i} \,.
\end{align}

\section{\boldmath Series with cusp anomalous dimension}
\label{sec:SCETren}
In this section we present a novel computation, adapting the derivation presented in Sec.~\ref{sec:noSCETren} to series with cusp anomalous
dimension. We will use our result to quantities relevant for effective field theories applied to jets initiated by massless and massive quarks.
\subsection{Derivation}
As already anticipated, the function $F(\varepsilon,u)$ defined in the second line of Eq.~\eqref{eq:sumF} is now singular when $u\to 0$.
This is related to $1/\varepsilon^2$ poles in one-loop amplitudes, which signals the presence of Sudakov (double) logarithms at this order (the function is however
regular at $\varepsilon=0$). Therefore we define
\begin{equation}
G^\mu(\varepsilon, u) \equiv u F^\mu(\varepsilon, u) = \sum_{m = 0} u^m \sum_{j = 0} \varepsilon^j G^\mu_{j, m}\,.
\end{equation}
The bare series $A_0=1+\delta A_0$, after performing manipulations completely analogous to those used in
the derivation of Eq.~\eqref{eq:nocuspDer}, takes the following form:
\begin{align}
\beta_0\,\delta\!A_0 \equiv\, & \beta_0\Biggl[ \delta A(\mu) + \sum_{j = 1}^\infty \frac{\delta A^{1/\varepsilon^j}}{\varepsilon^j}\Biggr]\!=
\sum_{n = 1} \frac{G^\mu(\varepsilon, n\varepsilon)}{n^2 \varepsilon} \sum_{i = 0} \frac{(- 1)^i (n)_i}{i!} \biggl(
\frac{\beta}{\varepsilon} \biggr)^{i + n}\\
= \,& \sum_{j = - \infty}^0 \varepsilon^j \sum_{i = \max (1,
- j - 1)} (- \beta)^i \sum_{m = 0}^{j + i + 1} G^\mu_{i + j + 1 - m, m} \sum_{n =
1}^i (- 1)^n \binom{i - 1}{n - 1} n^{m - 2} + \mathcal{O}(\varepsilon)\,,\nonumber
\end{align}
where again the sum in $j$ has been restricted to non-positive values, since those are the only ones necessary for renormalizing the series at $\mathcal{O}(1/\beta_0)$.
As already discussed, the sum over $n$ is zero unless $m = 0, 1$ or $m \geq i + 1$. The result for $m=1$ has been already presented in Eq.~\eqref{eq:sum2},
while for $m=0$ the sum is worked out as follows
\begin{equation}\label{eq:sum3}
\sum_{n = 1}^i \frac{(- 1)^n}{n^2} \binom{i - 1}{n - 1} = \sum_{n = 1}^i \frac{(- 1)^n}{i n} \frac{i!}{n! (i - n) !} = \frac{1}{i} \sum_{n = 1}^i
\frac{(- 1)^n}{n} \binom{i}{n} = - \frac{H_i}{i}\,,
\end{equation}
where we have expressed the binomial in terms of factorials and have used $(i-1)! = i!/i$, $n^2(n-1)! = n\times n!$ in the second equality, such that a binomial can be written
in the third. The last form is obtained with Eq.~\eqref{eq:HnSum}, derived in Appendix~\ref{sec:AppSum}. All in all we find
\begin{align}\label{eq:A0cusp}
\beta_0 \,\delta\!A_0 =\, & \sum_{j = - \infty}^0 \varepsilon^j
\sum_{i = \max (1, - j - 1)} \!\! (- \beta)^i \Biggl[ \,\sum_{m = 0}^j G^\mu_{j -
m, m + i + 1} \sum_{n = 1}^i (- 1)^n \binom{i - 1}{n - 1} n^{m + i - 1}
\\
& - \frac{H_i}{i} G_{i + j + 1, 0} \Biggr] - \sum_{i =
\max (1, - j)} (- \beta)^i \frac{G^\mu_{i + j, 1}}{i} \,, \nonumber
\end{align}
with $G^\mu_{i,0}=G_{i,0}$.
If $j < 0$ the sum over $m$ vanishes and only the terms in the second line contribute. Given the ``max'' condition in the lower limit of the
sum over $i$ we treat $j=-1$ separately. On the other hand, $j = 0$ implies $m = 0$, leaving a single term in the sum:
\begin{align}\label{eq:Aep}
\beta_0 A_0^{1 / \varepsilon} =\, & - \sum_{i = 1} \frac{(- \beta)^i}{i} \bigl(H_i G_{i, 0} + G^\mu_{i - 1, 1}\bigr)\,, \\
\beta_0 A_0^{1 / \varepsilon^j} =\, & \sum_{i = j} (- \beta)^i \biggl[ \frac{H_{i - 1}}{\beta (i - 1)} G_{i
- j, 0} - \frac{1}{i} G^\mu_{i - j, 1} \biggr], \nonumber\\
\beta_0 \delta A(\mu) =\,& \sum_{i = 1} \beta^i \Bigl[
G^\mu_{0, i + 1} (i - 1) ! - \frac{(- 1)^i}{i} \bigl(G^\mu_{i, 1} + H_i G_{i + 1, 0} \, \bigr)\! \Bigr] .\nonumber
\end{align}
From this result we obtain the following expression for the non-logarithmic term of the renormalized fixed-order coefficients:
\begin{equation}\label{eq:nologcusp}
a_{i,i-1} =G_{0, i + 1} (i - 1) ! - \frac{(- 1)^i}{i} \bigl( G_{i, 1} + H_i G_{i + 1, 0}\bigr)
= G_{0, i + 1} (i - 1) ! - \frac{\hat\gamma_A^i}{2} - \frac{\hat\Gamma_{\!\!A}^{i+1}}{4(i+1)}\,,
\end{equation}
where we have anticipated the results for the anomalous dimensions to be found in Eqs.~\eqref{eq:cuspPart} and \eqref{eq:nocuspPart}.
We derive next closed expressions in terms of $G(\varepsilon,u)$ for the various sums. We start with the divergent terms, and work out the following
sums which are valid for any value of $j$:
\begin{equation}
\sum_{i = j} (- \beta)^i \frac{G^\mu_{i - j, 1}}{i} = \sum_{i = 0} (-
\beta)^{i + j} \frac{G^\mu_{i, 1}}{i + j} = -\!\! \int_{- \beta}^0\! {\df}
\varepsilon \,\varepsilon^{j - 1} \sum_{i = 0} \varepsilon^i G^\mu_{i, 1}
= -\!\! \int_{- \beta}^0 \!{\df} \varepsilon\, \varepsilon^{j - 1}
\frac{{\df}G^\mu (\varepsilon, u)}{{\df} u} \biggr|_{u = 0} .
\end{equation}
This time, the coefficients $A^{1/\varepsilon^j}$
exhibit explicit $\mu$ dependence, which however happens to be linear in $\log\mu$.
Using $G^\mu(\varepsilon, u) = (\mu/{\cal Q})^{2u} G(\varepsilon, u)$ we find $G^\mu_{i,1} = G_{i,1} + 2 \log(\mu/{\cal Q}) G_{i,0}$ and the related result
\begin{equation}
\frac{{\df}G^\mu (\varepsilon, u)}{{\df} u} \biggr|_{u = 0} = \frac{{\df}G (\varepsilon, u)}{{\df} u} \biggr|_{u = 0}
+ \,2 \log\!\biggl(\frac{\mu}{\cal Q}\biggr)G (\varepsilon, 0)\,.
\end{equation}
For series which involve harmonic numbers we use identity
\begin{equation}\label{eq:Harm-sum-id}
\frac{(- \beta)^i}{i} H_i =\int_{- \beta}^0\! {\df} \varepsilon \log \biggl( 1 + \frac{\varepsilon}{\beta}\biggr) \varepsilon^{i - 1}\,,
\end{equation}
for which a proof is provided in Eq.~\eqref{eq:Hn2Proof} of Appendix~\ref{sec:AppSum}. With this result we obtain a closed form necessary
to find a resumed expression for $A^{1/ \varepsilon}$, while for higher-power divergences we obtain integrals which do not involve UV subtractions
at $\varepsilon=0$:
\begin{align}
&- \!\sum_{i = 1} \frac{(- \beta)^i}{i} H_i G_{i, 0} =\!\! \int_{- \beta}^0\!
\frac{ {\df} \varepsilon}{\varepsilon} \log \!\biggl( \frac{\beta}{\beta +
\varepsilon} \biggr) \!\sum_{i = 1} \varepsilon^i G_{i, 0}
= \!\!\int_{- \beta}^0\!
\frac{ {\df} \varepsilon}{\varepsilon} \log\! \biggl( \frac{\beta}{\beta +
\varepsilon} \biggr)\! \bigl[G (\varepsilon, 0) - G (0, 0)\bigr],\nonumber\\
& \sum_{i = 0} (- \beta)^{i - j - 1} \frac{H_{i - j - 1}\,G_{i,0}}{i - j - 1} = \!\!\int_{- \beta}^0 \frac{{\df} \varepsilon}{\varepsilon^{1 + j}}\log \Bigl( 1 +
\frac{\varepsilon}{\beta} \Bigr) \! \sum_{i = 0}
\varepsilon^i G_{i, 0}
= \!\!\int_{- \beta}^0 \frac{{\df}
\varepsilon}{\varepsilon^{1 + j}} \log \Bigl( 1 + \frac{\varepsilon}{\beta}
\Bigr) G (\varepsilon, 0) \,.
\end{align}
Finally we need the following closed forms for sums related to the renormalized series:
\begin{align}\label{eq:cusp-sums}
\sum_{i = 1} (- \beta)^i \frac{G^\mu_{i, 1}}{i} = & - \!\!\int_{- \beta}^0 \!\frac{{\df}\varepsilon}{\varepsilon} \sum_{i = 1} \varepsilon^i G^\mu_{i, 1} = -
\!\! \int_{- \beta}^0 \frac{{\df} \varepsilon}{\varepsilon} \frac{{\df}}{{\df} u} \!\bigl[G^\mu (\varepsilon, u) - G^\mu(0, u)\bigr]_{\!u = 0}\,, \\
\sum_{i = 1} \beta^i G^\mu_{0, i + 1} \Gamma(i) = & \int_0^{\infty}\!\frac{ {\df} u}{u^2}\,
u^{- \frac{u}{\beta}} \sum_{i = 2} u^i G^\mu_{0, i} \nonumber\\
= & \int_0^{\infty} \frac{{\df} u}{u^2} \,u^{- \frac{u}{\beta}} \!\biggl[ G^\mu(0, u) - G (0, 0)
- u \frac{{\df}G^\mu(0, \tau)}{{\df} \tau} \biggr|_{\tau = 0} \, \biggr] , \nonumber\\
\sum_{i = 1} \frac{(- \beta)^i}{i} G_{i + 1, 0} H_i =\,&
\int_{- \beta}^0\! \frac{{\df} \varepsilon}{\varepsilon^2} \log \!\biggl( 1 + \frac{\varepsilon}{\beta} \biggr)\! \sum_{i = 2} \varepsilon^iG_{i, 0} \nonumber\\
=& \int_{- \beta}^0 \! \frac{{\df} \varepsilon}{\varepsilon^2} \log \!\biggl( 1 + \frac{\varepsilon}{\beta} \biggr)\!
\biggl[ G (\varepsilon, 0) - G (0, 0) - \varepsilon \frac{{\df} G(\tau, 0)}{{\df} \tau} \biggr|_{\tau = 0} \,\biggr].\nonumber
\end{align}
The inverse Borel transform integral involves now two subtractions around $u=0$, which are necessary to render the integrand finite at this value given
the overall $1/u^2$ factor. The contribution in the first line involves a $u$-derivative of the once subtracted $G(\varepsilon,u)$ function
around $\varepsilon=0$ that ensures a smooth behavior towards the upper integration limit. On the other hand, the $u$ derivative makes
this term dependent on $\mu$. We can also explicitly display the $\mu$ dependence of the derivative in the first line of Eq.~\eqref{eq:cusp-sums}
and the subtraction in the inverse Borel transform (last line)
\begin{align} \frac{{\df}}{{\df} u} \bigl[G^\mu(\varepsilon, u) - G^\mu(0, u)\bigr]_{u = 0} =\,&
2 \log\!\biggl(\frac{\mu}{\cal Q}\biggr)\!\bigl[G (\varepsilon, 0) - G (0, 0)\bigr] + \frac{{\df}}{{\df} u} \!\bigl[G(\varepsilon, u) - G(0, u)\bigr]_{\!u = 0}\,,\nonumber\\
\frac{{\df}G^\mu(0, u)}{{\df} u} \biggr|_{u = 0} =\,& \frac{{\df}G(0, u)}{{\df} u} \biggr|_{u = 0} + \,2 \log\!\biggl(\frac{\mu}{\cal Q}\biggr)G (0, 0)\,.
\end{align}
We can combine the various results to write down a closed form for the bare series, which we split as $\delta\!A_0 = \delta Z+ \delta\!A(\mu)$, with renormalized series
$A(\mu) = 1 + \beta_0\delta\!A(\mu)$ and renormalization factor $\delta Z = Z - 1 = \delta Z_{\rm nc} + 2\log(\mu/{\cal Q}) \delta Z_{\rm cusp}$:\footnote{If $G(\varepsilon,0)=0$ one has $\delta Z_{\rm cusp}=0$ and
$\delta Z_{\rm nc} $ becomes equal to the result in Eq.~\eqref{eq:ZNC}.}
\begin{align}\label{eq:Z-res}
\beta_0\delta Z_{\rm cusp} =&\, \sum_{j = 1}^{\infty} \frac{1}{\varepsilon^j} \!\int_{- \beta}^0\! {\df} h\,h^{j-1}\,G(h,0)\,,\\
\beta_0\delta Z_{\rm nc} =&\, \frac{1}{\varepsilon}\! \int_{- \beta}^0 \! {\df} h\, \biggl\{\frac{{\df}G (h, u)}{{\df} u} \biggr|_{u = 0} - \frac{1}{h} \log\! \biggl(1 + \frac{h}{\beta} \biggr) [G (h, 0) - G (0, 0)] \biggr\}\nonumber \\
& +\sum_{j = 2}^{\infty} \frac{1}{\varepsilon^j} \!\int_{- \beta}^0\! {\df} h\,h^{j-1} \biggl[ \frac{{\df} G (h, u)}{{\df} u} \biggr|_{u = 0} - \log\! \biggl( 1 +
\frac{h}{\beta} \biggr) G (h, 0) \biggr],\nonumber\\
\beta_0\,\delta\!A(\mu) =&\, \int_0^{\infty} \! \frac{{\df} u \,e^{- \frac{u}{\beta}}}{u^2} \biggl\{\!\!\biggl(\frac{\mu}{\cal Q}\biggr)^{\!\!2u} \! G (0,
u) -\! \biggl[1 + 2u \log\!\biggl(\frac{\mu}{\cal Q}\biggr)\!\biggr] G (0, 0)- u \frac{{\df}G (0, \tau)}{{\df} \tau} \biggr|_{\tau = 0} \biggr\} \nonumber\\
& +\! \int_{- \beta}^0 \! \frac{{\df} h}{h} \biggl\{ \frac{{\df}}{{\df} u}\! \bigl[G (h, u) - G (0, u)\bigr]_{\!u = 0}
+ 2 \log\!\biggl(\frac{\mu}{\cal Q}\biggr) \bigl[G (h, 0) - G (0, 0)\bigr]\nonumber\\
&- \!\frac{1}{h} \log\! \biggl(1 + \frac{h}{\beta} \biggr) \!\biggl[G (h, 0) - G (0, 0) - h \frac{{\df} G(\tau, 0)}{{\df} \tau}
\biggr|_{\tau = 0} \,\biggr] \!\biggr\}.\nonumber
\end{align}
Ambiguities are again confined to the inverse Borel transform integral, and it will be shown in Sec.~\ref{sec:RGEcusp} that they do not
depend on the renormalization scale.

We finish this section deriving a closed expression for the cusp ($\Gamma_{\!\!A}$) and non-cusp ($\gamma_A$) anomalous dimensions, which we define as
${\df}\!\log\! A/{\df}\! \log\!\mu = \gamma_A + \log(\mu/{\cal Q}) \Gamma_{\!\!A}$. Following the same steps employed to find the
anomalous dimension in Sec.~\ref{sec:noSCETren} we find for $\Gamma_{\!\!A}$ the following closed form and fixed-order expansion:
\begin{align}\label{eq:cuspPart}
\Gamma_{\!\!A}(\beta) = \,& \frac{1}{\beta_0}\sum_{i=0}\hat \Gamma_{\!\!A}^i \,\beta^{i+1} =
4 (\varepsilon + \beta) \beta \frac{\mathd}{\mathd \beta} \delta Z_{\rm cusp}
= \frac{4 \beta}{\beta_0} G(- \beta, 0)
\,,\\
\hat \Gamma_{\!\!A}^i = \,& 4(-1)^i G_{i, 0}\,,\qquad
\Gamma_{\!\!A}^i = \hat \Gamma_{\!\!A}^i \,\beta_0^{i}
\,,\nonumber
\end{align}
making clear that the cusp anomalous dimension is not afflicted by renormalons. For a series without cusp $G(\varepsilon, 0)=0$ and therefore $\Gamma_{\!\!A}=0$.
Likewise, the non-cusp anomalous dimension is derived as follows
\begin{align}
&\frac{\beta_0}{2} \gamma_A(\beta) = (\varepsilon + \beta) \beta \frac{\mathd}{\mathd\beta} \delta Z_{\rm nc} - \delta Z_{\tmop{cusp}} \\
& \,=\!-\!\sum_{i = 1} (- \beta)^i \bigl(G_{i - 1, 1} + H_i G_{i, 0}\bigr)\! +\!\!\! \sum_{j = - \infty}^{-2} \!\!\varepsilon^j \!\sum_{i = - j}\! (-\beta)^i \frac{G_{i + j, 0}}{i}
+ \frac{1}{\varepsilon} \!\sum_{i = 2} (- \beta)^i \bigl(H_{i - 1} G_{i - 1, 0} +\! G_{i - 2, 1}\bigr)\nonumber \\
&\quad\! + \!\!\! \sum_{j = - \infty}^{-2} \! \! \varepsilon^j \sum_{i = 1 - j} (- \beta)^i \biggl[ \frac{H_{i - 1} G_{i + j - 1, 0}}{\beta} - G_{i + j - 1, 1} \biggr]\!
+ \!\frac{1}{\varepsilon}\!\sum_{i = 2} (- \beta)^i \biggl[ \frac{H_{i - 1} G_{i-2, 0}}{\beta} - G_{i - 2, 1} \biggr] \nonumber\\
&\quad\!- \! \! \!\sum_{j = - \infty}^{- 2} \! \! \varepsilon^j \sum_{i = 1 - j} (- \beta)^i \biggl[ \frac{H_{i - 2} G_{i + j - 1, 0}}{\beta} - G_{i + j - 1,
1} \biggr] \! + \!\frac{1}{\varepsilon} \!\sum_{i = 1} (-\beta)^i \frac{G_{i - 1, 0}}{i}\,, \nonumber
\end{align}
where we have pulled out of each sum the $1/\varepsilon$ term and have shifted $i$ and $j$ by one unit wherever necessary to write all sums in terms of
$\varepsilon^j$ and $(-\beta)^i$ (except for some terms involving harmonic numbers which have an additional explicit $1/\beta$ factor). It is not hard to realize that all terms cancel
out except for the very first sum, from which the fixed-order coefficients can be read off. The sum can be expressed in a closed form:
\begin{align}\label{eq:nocuspPart}
\gamma_A(\beta) = \,& \frac{1}{\beta_0}\sum_{i=0}\hat \gamma_{A}^i \,\beta^{i+1}
=\frac{2}{\beta_0}\! \biggl[ \beta \sum_{i = 0} (- \beta)^i G_{i, 1} - \sum_{i =1} (- \beta)^i H_i G_{i, 0} \biggr]\\
=\, & \frac{2}{\beta_0} \biggl\{ \beta \frac{{\df}G(-\beta,u)}{{\df} u}
\biggr|_{u = 0} +\! \int_0^1\!\frac{{\df} x}{1 - x} \bigl[ G (-\beta x, 0) - G (- \beta, 0) \bigr] \!\biggr\},\nonumber\\
\hat \gamma_{A}^i = \,& 2(-1)^i \bigl(G_{i,1} + H_{i+1}G_{i+1,0}\bigr)
= 2(-1)^i G_{i,1} - \frac{H_{i+1}}{2} \hat \Gamma_{\!\!A}^{i+1} \,.\nonumber
\end{align}
Since, as we shall see, the function $G(\beta, 0)$ has singularities on the positive real axis only, no pole is crossed when carrying out the integration in the second line,
making thus clear that the non-cusp anomalous dimension has no ambiguities. Furthermore, this integral has no divergence at $x\to 1$
because the term in squared brackets tends to zero in this limit.
To obtain the last term in the second line we have used\footnote{One can obtain the same result applying the derivative directly into the integral form of
$\delta Z_{\rm nc}^{1 / \varepsilon}$. To that end we change variables $h= -\beta x$ in the second term of Eq.~\eqref{eq:Z-res}. To tame the divergence occurring at
$x=1$ we add a regulator $\delta$ to the argument of the logarithm. Defining $G'\!(y,0) \equiv {\df}G(y, 0)/{\df}y$ one has
\begin{align}
&\beta\frac{\df}{{\df} \beta}\!\!\int_0^1 \!\frac{{\df} x}{x}\log(1-x+\delta) \bigl[G(-x\beta,0) -G(0,0)\bigr]\!=
-\beta\!\!\int_0^1\!{\df} x \log(1-x+\delta) G'\!(-x\beta, 0)\\
&\qquad\qquad\qquad\qquad\qquad\qquad=\!\int_0^1\!{\df} x\,\frac{G(-x\beta,0)}{1-x+\delta} + G(-\beta,0)\log(\delta) - \log(1+\delta) G(0,0) \nonumber\\
&\qquad\qquad\qquad\qquad\qquad\qquad= \!\int_0^1\!\frac{{\df} x}{1-x+\delta}\bigl[ G(-x\beta,0) - G(-\beta,0)\bigr] \!+ \!\bigl[G(-\beta,0) - G(0,0)\bigr] \log(1+\delta)\,,\nonumber
\end{align}
where integration by parts has been used to obtain the second line, while to get the third one writes \mbox{$\log(\delta)=\log(1-\delta)-\int_0^1 {\df} x/(1-x)$}.
When letting $\delta\to0$ the last term in the third line vanishes due to the global $\log(1+\delta)$ factor, reproducing the result in second line of Eq.~\eqref{eq:nocuspPart}.}
\begin{equation}
\int_{- \beta}^0 \frac{{\df} \varepsilon}{\beta + \varepsilon} \Bigl[\varepsilon^i - (- \beta)^i\Bigr] = \int_{- \beta}^0 {\df} \varepsilon \sum_{j
= 0}^{i - 1} \varepsilon^j (- \beta)^{i - j - 1} = - (- \beta)^i \sum_{j =1}^i \frac{1}{j} = - (- \beta)^i H_i \,,
\end{equation}
where in the first step the identity $a^i - b^i= (a-b)\sum_{j=0}^{i-1} a^j b^{i-1-j}$ is applied, in the second we integrate term by term, and identifying the harmonic
number's definition the last expression is obtained. In Eq.~\eqref{eq:nocuspPart} we switch variables $\varepsilon = -\beta x$ to map the integration to the unit range $[0,1]$.
If the series has no cusp anomalous dimension $G(\varepsilon,0)=0$ and the integral in the second line of Eq.~\eqref{eq:nocuspPart} vanishes. In this case
$[{\df}G(-\beta,u)/{\df}u]_{u=0}= F(-\beta,0)$ and the result in the last line of Eq.~\eqref{eq:gammaNC} is recovered.

\subsection{Exact solution to the RGE}\label{sec:RGEcusp}
Using the chain rule we can express the $\mu$ derivative in terms of a $\beta$ derivative. Furthermore, we can write the factor $\log(\mu/{\cal Q})$
multiplying $\Gamma_{\!\!A}$ as the difference of inverse $\beta$ couplings:
\begin{align}
- \frac{2 \beta^2}{A} \frac{\mathrm{\text{d}} A}{\mathrm{\text{d}} \beta}
= &\, \gamma_A (\beta) + \Gamma_{\!\!A} (\beta) \log \!\biggl( \frac{\mu}{\cal Q} \biggr) =
\gamma_A (\beta) + \frac{\Gamma_{\!\!A} (\beta)}{2} \biggl( \frac{1}{\beta} -
\frac{1}{\beta_{\cal Q}} \biggr), \\
\log\! \biggl[ \frac{A (\mu)}{A (\mu_0)} \biggr] = & -\! \frac{1}{2}
\int_{\beta_{\mu_0}}^{\beta_{\mu}} \frac{{\df} \beta}{\beta^2} \biggl[
\gamma_A (\beta) + \frac{\Gamma_{\!\!A} (\beta)}{2} \biggl( \frac{1}{\beta} -
\frac{1}{\beta_{\cal Q}} \biggr)\! \biggr] \nonumber\\
=& - \!\frac{1}{2\beta_0} \biggl( \hat\gamma_A^0 + \frac{\hat\Gamma^1_{\!\!A}}{2} -
\frac{\hat\Gamma^0_{\!\!A}}{2 \beta_{\cal Q}} \biggr) \!\log\! \biggl(
\frac{\beta_{\mu}}{\beta_{\mu_0}} \biggr) \!+ \frac{\hat\Gamma^0_{\!\!A}}{4\beta_0}\! \biggl(
\frac{1}{\beta_{\mu}} - \frac{1}{\beta_{\mu_0}} \biggr) + \omega ({\cal Q}, \mu, \mu_0)\,, \nonumber
\end{align}
where in the last line we have pulled out those terms that after integration are not regular if either $\beta_{\mu}$ or $\beta_{\mu_0}$ tend to zero.
All regular contributions are kept in a running kernel which is defined as the difference of two terms, each of them depending on a single renormalization scale,
$\omega({\cal Q},\mu,\mu_0)\equiv\omega({\cal Q},\mu)-\omega({\cal Q},\mu_0)$ with
\begin{equation}\label{eq:wTilde}
\omega ({\cal Q}, \mu) = - \frac{1}{2} \!\int_0^{\beta_{\mu}} \!\frac{{\df}
\beta}{\beta^2} \biggl[\gamma_A (\beta) - \beta \gamma'_A (0) -
\frac{\Gamma_{\!\!A}(\beta) - \beta\, \Gamma'_{\!\!A} (0)}{2 \beta_{\cal Q}}
+ \frac{\Gamma_{\!\!A}
(\beta) - \beta\, \Gamma'_{\!\!A} (0) - \beta^2 \Gamma_{\!\!A}' (0) / 2}{2 \beta}
\biggr] .
\end{equation}
Expanding strictly to leading order in large-$\beta_0$ one can express the renormalized series as:
\begin{equation}\label{eq:cusp-run-match}
\delta\!A (\mu) = \delta\!A (\mu_0) + \frac{1}{2\beta_0} \!\biggl( \hat\gamma_A^0 + \frac{\hat\Gamma^1_{\!\!A}}{2} -
\frac{\hat\Gamma^0_{\!\!A}}{2 \beta_{\mu_0}} \biggr)\! \log \!\biggl( \frac{\beta_{\mu
0}}{\beta_{\mu}} \biggr) + \frac{\hat\Gamma^0_{\!\!A}}{2\beta_0} \log\!\biggl( \frac{\mu}{\mu_0} \biggr)\! + \omega ({\cal Q}, \mu, \mu_0)\, ,
\end{equation}
where we have used $1/\beta_\mu - 1/\beta_{\mu_0} = 2\log(\mu/\mu_0)$, relation that stems from the solution to the RGE equation for the $\beta$ coupling in
Eq.~\eqref{eq:betaRun}. Let us work out each term appearing in Eq.~\eqref{eq:wTilde}, starting with the ones involving $\Gamma_{\!\!A}$.
Using the closed form in Eq.~\eqref{eq:cuspPart}, noting that $\beta_0\Gamma_{\!\!A}'(0)=4G(0,0)$ and $\beta_0\Gamma_{\!\!A}''(0)=8[{\df}G(\beta,0)/{\df}\beta]_{\beta=0}$, we get
\begin{align}\label{eq:Cpiece}
- \frac{\beta_0}{2}\! \int_0^{\beta_{\mu}}\! \frac{{\df} \beta}{\beta^2}
\frac{\Gamma_{\!\!A} (\beta) - \beta \,\Gamma'_{\!\!A} (0)}{2 \beta_{\cal Q}}
= &\, \frac{1}{\beta_{\cal Q}} \!\int_{- \beta_{\mu}}^0 \!\frac{{\df} \varepsilon}{\varepsilon}
\bigl[G (\varepsilon, 0) - G (0, 0)\bigr]\,, \\
\frac{\beta_0}{4}\! \int_0^{\beta_{\mu}}\! \frac{{\df} \beta}{\beta^3}
\biggl[\Gamma_{\!\!A} (\beta) - \beta\, \Gamma'_{\!\!A} (0) - \frac{\beta^2}{2} \Gamma_{\!\!A}'' (0)\biggr]
= & \!\int_{- \beta_{\mu}}^0 \!\frac{{\df}
\varepsilon}{\varepsilon^2} \biggl[ G (\varepsilon, 0) - G (0, 0) -
\varepsilon \frac{{\df} G (\tau, 0)}{{\df} \tau}
\biggr|_{\tau = 0}\, \biggr] . \nonumber
\end{align}
To work out closed expressions for the term involving the non-cusp anomalous dimension we use the expression for $\gamma_A$ as given in
Eq.~\eqref{eq:nocuspPart}, taking into account that
\begin{align}\label{eq:gamader}
\frac{\beta_0}{2} \gamma'_A (0) = \,& \frac{{\df}}{{\df} u} G (0,u) \biggr|_{u = 0} + \!\int_0^1 \frac{{\df} x}{1 - x}
\frac{{\df}}{{\df} \beta} \!\bigl[G (- \beta x, 0) - G (- \beta, 0)\bigr]_{\!\beta =0} \\
= \,& \frac{{\df}}{{\df} u} G (0,u) \biggr|_{u = 0} +\! \int_0^1{\df} x \frac{{\df} G (\tau, 0)}{{\df} \tau}\biggr|_{\tau = 0} \,,\nonumber
\end{align}
where to arrive at the second line the $\beta$ derivative has been computed applying the chain rule. Changing the integration variable $\varepsilon= -\beta x $ in
Eqs.~\eqref{eq:nocuspPart} and \eqref{eq:gamader} one finds
\begin{align}\label{eq:noCpiece}
& \frac{\beta_0}{2}\! \int_0^{\beta_{\mu}} \!\frac{{\df} \beta}{\beta^2}
\bigl[\gamma_A (\beta) - \beta \gamma'_A (0)\bigr] = \!
\int_0^{\beta_{\mu}}\! \frac{{\df} \beta}{\beta} \frac{{\df}}{{\df} u}\! \bigl[G
(- \beta, u) - G (0, u)\bigr]_{\!u = 0} \\
&\qquad\qquad\quad + \! \int_0^{\beta_{\mu}} \!\frac{{\df} \beta}{\beta^2} \!
\int_{- \beta}^0\! \frac{{\df} \varepsilon}{\beta + \varepsilon}
\biggl[ G (\varepsilon, 0) - G (- \beta, 0) - (\beta + \varepsilon)
\frac{{\df} G (\tau, 0)}{{\df} \tau} \biggr|_{\tau =
0}\, \biggr] \nonumber\\
& \qquad\qquad= \!\int_0^{\beta_{\mu}}\! \frac{{\df} \beta}{\beta}
\frac{{\df}}{{\df} u}\! \bigl[G (- \beta, u) - G (0, u)\bigr]_{\!u = 0} \nonumber\\
& \qquad\qquad\quad+ \!\int_{- \beta_{\mu}}^0 \!\frac{{\df}
\varepsilon}{\varepsilon^2} \biggl[ G (\varepsilon, 0) - G (0, 0) -
\varepsilon \frac{{\df} G (\tau, 0)}{{\df} \tau}
\biggr|_{\tau = 0}\, \biggl]\! \biggl( \int_{- \varepsilon}^{\beta_{\mu}}\!
\frac{{\df} \beta}{\beta + \varepsilon} \frac{\varepsilon^2}{\beta^2} -\!
\int_{- \varepsilon}^0 \frac{{\df} \beta}{\varepsilon + \beta} \biggr)
\nonumber\\
& \qquad\qquad=- \!\int_{- \beta_{\mu}}^0\! \frac{{\df}
\varepsilon}{\varepsilon} \frac{{\df}}{{\df} u}\! \bigl[G (\varepsilon, u) - G(0, u)\bigr]_{\!u = 0} \nonumber\\
& \qquad\qquad\quad+\! \int_{- \beta_{\mu}}^0\! \frac{{\df}
\varepsilon}{\varepsilon^2} \biggl[ G (\varepsilon, 0) - G (0, 0) -
\varepsilon \frac{{\df} G (\tau, 0)}{{\df} \tau}
\biggr|_{\tau = 0} \,\biggr] \!\biggl[ \log\! \biggl( 1 +
\frac{\varepsilon}{\beta_{\mu}} \biggr) - 1 -
\frac{\varepsilon}{\beta_{\mu}} \biggr] , \nonumber
\end{align}
where to obtain the second equality the following manipulations have been performed to the integral in the second line
\begin{align}\label{eq:manipulation}
& \int_0^{\beta_{\mu}}\! \frac{{\df} \beta}{\beta^2} \!\int_{- \beta}^0\!
\frac{{\df} \varepsilon}{\beta + \varepsilon} \biggl[ G (\varepsilon, 0) - G(- \beta, 0) - (\beta + \varepsilon) \frac{{\df} G (\tau,
0)}{{\df} \tau} \biggr|_{\tau = 0}\, \biggr] \\
& \quad\qquad\qquad= \!\int_0^{\beta_{\mu}} \!\frac{{\df} \beta}{\beta^2} \! \int_{- \beta}^0
\frac{{\df} \varepsilon}{\beta + \varepsilon} \biggl[ G (\varepsilon, 0) - G
(0, 0) - \varepsilon \frac{{\df} G (\tau, 0)}{{\df}
\tau} \biggr|_{\tau = 0}\, \biggr] \nonumber\\
&\quad\qquad\qquad\quad -\! \int_0^{\beta_{\mu}} \!\frac{{\df} \varepsilon}{\varepsilon^2}
\int_{- \varepsilon}^0 \frac{{\df} \beta}{\beta + \varepsilon} \biggl[ G (-
\varepsilon, 0) - G (0, 0) + \varepsilon \frac{{\df} G (\tau,
0)}{{\df} \tau} \biggr|_{\tau = 0} \,\biggr] \nonumber\\
& \quad\qquad\qquad=\, \int_{- \beta_{\mu}}^0\! \frac{{\df} \varepsilon}{\varepsilon^2}
\biggl[ G (\varepsilon, 0) - G (0, 0) - \varepsilon \frac{{\df} G
(\tau, 0)}{{\df} \tau} \biggr|_{\tau = 0}\, \biggr]\!
\biggl( \int_{- \varepsilon}^{\beta_{\mu}}\! \frac{{\df} \beta}{\beta +
\varepsilon} \frac{\varepsilon^2}{\beta^2} + \!\int^{- \varepsilon}_0\!
\frac{{\df} \beta}{\beta + \varepsilon} \biggr). \nonumber
\end{align}
To obtain the second line we add and subtract $G(0, 0)$ within the square brackets in the first line, collect terms that depend on $\varepsilon$ and $\beta$ separately [\,such that
$G (0, 0)$ is subtracted in each one\,], followed by renaming $\varepsilon \leftrightarrow \beta$ in the term that only depends on $\beta$ (which appears in the third line); to get the
last expression we switch the order of integration in the second line integral and flip the sign of the two integration variables $\varepsilon \rightarrow - \varepsilon$,
$\beta \rightarrow - \beta$ in the third. Since $0\leq -\varepsilon \leq \beta_\mu$, the two integrands in the last line become equal in the vicinity of the pole that happens at
$\beta=-\varepsilon$, and since the singularities are approached from opposites sides they cancel when the two terms are added up if the integration is performed symmetrically.
Equivalently, to obtain the last line of Eq.~\eqref{eq:noCpiece} one can use a regulator $\delta\geq 0$, and keeping track of the changes of variables carried out one
obtains the following regularized result:

\begin{align}
& \!\!\int_{-\varepsilon}^{\beta_{\mu}} \!\!\frac{\mathd \beta}{\beta + \varepsilon + \delta}\frac{1}{\beta^2}+
\frac{1}{\varepsilon^2}\! \!\int^{-\varepsilon}_0\! \!\frac{\mathd \beta}{\beta + \varepsilon-\delta} =
\frac{1}{(\delta + \varepsilon)^2} \log \biggl(\! -\frac{\varepsilon}{\delta}
\frac{\beta_{\mu} + \delta + \varepsilon}{\beta_{\mu}} \biggr) \\
&\qquad\qquad\qquad\quad-\!\biggl( 1 + \frac{\varepsilon}{\beta_{\mu}} \biggr) \frac{1}{\varepsilon (\delta + \varepsilon)}
+ \frac{1}{\varepsilon^2} \!\log \biggl( \frac{\delta}{\delta - \varepsilon} \biggr)\nonumber
\xrightarrow[\delta \rightarrow 0]{} \frac{1}{\varepsilon^2}\! \biggl[ \log \biggl( 1 + \frac{\varepsilon}{\beta_{\mu}}\biggr)
- 1 - \frac{\varepsilon}{\beta_{\mu}} \biggr],\nonumber
\end{align}
where in the last expression we take the $\delta\to 0$ limit, now smooth.
Combining the results in Eqs.~\eqref{eq:Cpiece} and \eqref{eq:noCpiece} and using the 2-loop cusp anomalous dimension coefficient
\mbox{$G_{1,0}=[{\df}G(\varepsilon,0)/{\df}\varepsilon]_{\varepsilon=0}=-\hat\Gamma_{\!\!A}^1/4$} we get
\begin{align}\label{eq:cuspRGE}
\beta_0\,\omega ({\cal Q}, \mu) = & \!\int_{- \beta_{\mu}}^0\!
\frac{{\df} \varepsilon}{\varepsilon} \frac{{\df}}{{\df} u} \!\bigl[G (\varepsilon, u) - G (0, u)\bigr]_{\!u = 0} - 2\log\!\biggl(\frac{\cal Q}{\mu}\biggr)\!
\!\int_{- \beta_{\mu}}^0 \!\frac{{\df} \varepsilon}{\varepsilon} \bigl[G (\varepsilon, 0) - G (0,0)\bigr]\nonumber\\
& - \!\int_{- \beta_{\mu}}^0 \!\frac{{\df} \varepsilon}{\varepsilon^2}
\biggl[ G (\varepsilon, 0) - G (0, 0) - \varepsilon \frac{{\df} G
(\tau, 0)}{{\df} \tau} \biggr|_{\tau = 0} \,\biggr] \!\log
\biggl( 1 + \frac{\varepsilon}{\beta_{\mu}} \biggr)+ \frac{\hat\Gamma_{\!\!A}^1}{4}\,,
\end{align}
where the logarithm in the first line appears from the solution to the RGE equation for the strong coupling $1/\beta_{\cal Q} - 1/\beta_\mu = 2\log({\cal Q}/\mu)$.
The last term in the second line is constant and therefore vanishes when computing $\omega ({\cal Q}, \mu_1,\mu_2)$. In Appendix~\ref{sec:AppManipulo}
we give an alternative derivation of this result based on series expansions.

Introducing Eq.~\eqref{eq:cuspRGE} and the form shown in Eq.~\eqref{eq:Z-res} for $A(\mu_0)$
in Eq.~\eqref{eq:cusp-run-match} we obtain a resummed expression for the renormalized series $A(\mu)$, with $\beta\equiv \beta(\mu)$:
\begin{align}\label{eq:cusp-resum}
\beta_0\,\delta\! A (\mu) =\, & \frac{1}{2}\biggl(\!
\hat\gamma_A^0 + \frac{\hat\Gamma^1_{\!\!A}}{2} - \frac{\hat\Gamma^0_{\!\!A}}{2
\beta_{\cal Q}} \biggr)\! \log\! \biggl( \frac{\beta_{\mu_0}}{\beta_{\mu}} \biggr)\! +
\frac{\hat\Gamma^0_{\!\!A}}{2} \log \!\biggl( \frac{\mu}{\mu_0} \biggr)\\
& + \!\int_0^{\infty} \!\frac{{\df} u}{u^2}e^{- \frac{u}{\beta_{\mu_0}}} \biggl\{\!
G (0, u)\! \biggl( \frac{\mu_0}{\cal Q} \biggr)^{\!\!2u} - G (0, 0) \!\biggl[ 1 + 2u
\log\! \biggl( \frac{\mu_0}{\cal Q} \biggr) \!\biggr] - u
\frac{{\df}G (0, \tau)}{{\df} \tau} \biggr|_{\tau = 0} \biggr\} \nonumber\\
& + 2\log\!
\biggl( \frac{\mu}{\cal Q} \biggr) \!\! \! \int_{- \beta_{\mu}}^0\!\!\! \frac{{\df}
\varepsilon}{\varepsilon} \bigl[G (\varepsilon, 0) - G (0, 0)\bigr] +\! \int_{- \beta_{\mu}}^0 \!\!\frac{{\df} \varepsilon}{\varepsilon}
\biggl\{ \!\frac{{\df}}{{\df} u}\! \bigl[G (\varepsilon, u) - G (0, u)\bigr]_{\!u = 0}\nonumber \\
&\,-\! \frac{1}{\varepsilon} \log\! \biggl( 1 + \frac{\varepsilon}{\beta_{\mu}} \biggr) \! \biggl[ G (
\varepsilon, 0) - G (0, 0) - \varepsilon \frac{{\df} G (\tau,0)}{{\df} \tau} \biggr|_{\tau = 0} \,\biggr] \!\biggr\} .\nonumber
\end{align}
We have displayed all dependence on $\mu$ and $\mu_0$ explicitly, and shown that the ambiguous term in the second line is independent of $\mu$.
It can be shown that it is also $\mu_0$ independent, and to that end we write the expression between curly brackets in the second
line as follows:
\begin{align}\label{eq:fudgeCusp}
& \biggl( \frac{\mu_0}{\cal Q} \biggr)^{\!\! 2u}\biggl[G (0, u) - G (0, 0)- u \frac{{\df}G (0, \tau)}{{\df} \tau} \biggr|_{\tau = 0}\,\biggr]
+ u \biggl[\! \biggl( \frac{\mu_0}{\cal Q} \biggr)^{\!\! 2u} - 1 \biggr] \! \biggl(\frac{\hat \gamma_A^0}{2}+\frac{\hat \Gamma_{\!\!A}^1}{4}\biggr)\\
& \qquad+ \frac{\hat\Gamma_{\!\!A}^0}{4} \biggl[\! \biggl( \frac{\mu_0}{\cal Q} \biggr)^{\!\! 2u} - 1 - 2u \log\biggl( \frac{\mu_0}{\cal Q} \biggr) \!\biggr],
\nonumber
\end{align}
where we have used $G_{0,0}=\hat \Gamma_{\!\!A}^0/4$ and $G_{0,1} = \hat \gamma_A^0/2+\hat\Gamma_{\!\!A}^1/4$.
The twice-subtracted integral resulting from the last line can be solved by Taylor expanding around $u = 0$, integrating each term
and summing up the series:
\begin{align}\label{eq:Taylor2}
&\!\int_0^{\infty}\! \frac{{\df} u}{u^2}e^{- \frac{u}{\beta_{\mu_0}}}\! \biggl[\!
\biggl( \frac{\mu_0}{\cal Q} \biggr)^{\!\!2 u} \!\!- 1 -2u \log\biggl( \frac{\mu_0}{\cal Q} \biggr)\!\biggr] =4\! \!\int_0^{\infty}\! {\df} u \,e^{-
\frac{u}{\beta_{\mu_0}}}\! \sum_{n = 0} \frac{(2u)^n}{(n +2) !} \log^{n+2}\! \biggl(
\frac{\mu_0}{\cal Q} \biggr) \\
&\qquad \qquad \qquad \qquad \qquad \qquad= 2\log \biggl( \frac{\mu_0}{\cal Q}\biggr)\sum_{n = 1} \biggl(\frac{1}{n}-\frac{1}{n+1}\biggr)\biggl[ 2\beta_{\mu_0} \log \biggl( \frac{\mu_0}{\cal Q}
\biggr)\! \biggr]^n\nonumber \\
&\qquad\qquad \qquad \qquad \qquad \qquad= 2\log \biggl( \frac{\mu_0}{\cal Q}\biggr) \!+\! \biggl[2\log \biggl( \frac{\mu_0}{\cal Q}\biggr) -
\frac{1}{\beta_{\mu_0}}\biggr]\sum_{n = 1}\frac{1}{n}\biggl[ 2\beta_{\mu_0} \log \biggl( \frac{\mu_0}{\cal Q} \biggr)\! \biggr]^n \nonumber\\
&\qquad\qquad \qquad \qquad \qquad \qquad = 2\log \biggl( \frac{\mu_0}{\cal Q}\biggr)\!
-\frac{1}{\beta_{\cal Q}} \!\log \biggl(\frac{\beta_{\cal Q}}{\beta_{\mu_0}} \biggr),\nonumber
\end{align}
where in the last step we have used Eq.~\eqref{eq:betaRun} to re-write the sum's prefactor. Using
Eqs.~\eqref{eq:fudgeCusp}, \eqref{eq:Taylor1} and \eqref{eq:Taylor2} in Eq.~\eqref{eq:cusp-resum} one finds
\begin{align}\label{eq:cuspLambda}
\beta_0\,\delta\! A (\mu) = & \!\int_0^{\infty} \!\frac{{\df} u}{u^2} \biggl(\frac{\Lambda_{\rm QCD}}{\cal Q}\biggr)^{\!\!2u}\biggl[
G (0, u) - G (0, 0) - u \frac{{\df}G (0, \tau)}{{\df} \tau} \biggr|_{\tau = 0} \,\biggr]\\
& + 2\log\! \biggl( \frac{\mu}{\cal Q} \biggr) \!\! \! \int_{- \beta_{\mu}}^0\!\!\! \frac{{\df}
\varepsilon}{\varepsilon} \bigl[G (\varepsilon, 0) - G (0, 0)\bigr] +\! \int_{- \beta_{\mu}}^0 \!\!\frac{{\df} \varepsilon}{\varepsilon}
\biggl\{ \!\frac{{\df}}{{\df} u}\! \bigl[G (\varepsilon, u) - G (0, u)\bigr]_{\!u = 0}\nonumber \\
&\,-\! \frac{1}{\varepsilon} \log\! \biggl( 1 + \frac{\varepsilon}{\beta_{\mu}} \biggr) \! \biggl[ G (
\varepsilon, 0) - G (0, 0) - \varepsilon \frac{{\df} G (\tau,0)}{{\df} \tau} \biggr|_{\tau = 0} \,\biggr] \!\biggr\} \nonumber\\
& + \frac{1}{2}\biggl(\!
\hat\gamma_A^0 + \frac{\hat\Gamma^1_{\!\!A}}{2} - \frac{\hat\Gamma^0_{\!\!A}}{2
\beta_{\cal Q}} \biggr)\! \log\! \biggl( \frac{\beta_{\cal Q}}{\beta_{\mu}} \biggr)\! +
\frac{\hat\Gamma^0_{\!\!A}}{2} \log \!\biggl( \frac{\mu}{\cal Q} \biggr) \nonumber.
\end{align}
The same result can be found using Eqs.~\eqref{eq:fudgeCusp}, \eqref{eq:Taylor1} and \eqref{eq:Taylor2} with $\mu_0=\mu$ directly in Eq.~\eqref{eq:Z-res}.
In this closed form ambiguities are manifestly
$\mu_0$ independent, but explicitly depend on $\Lambda_{\rm QCD}$, hence making the connection to non-perturbative physics manifest.
The dependence on $\mu$ is fully contained in terms related to the anomalous dimensions. If $G(\varepsilon,0)=0$, the first term in the second
line vanishes, along with the third and $\hat\Gamma_{\!\!A}^{0,1}$, thus recovering the result in Eq.~\eqref{eq:LamSumNocusp}.

For the sake of completeness, we present the form of the evolution kernel between two renormalization scales in closed and expanded
forms. For the former, defining $\beta_i\equiv \beta(\mu_i)$, we find:
\begin{align}
\beta_0[A (\mu_2) - A (\mu_1)] = &\, \frac{1}{2}\! \biggl(\hat\gamma_A^0 + \frac{\hat\Gamma^1_{\!\!A}}{2} - \frac{\hat\Gamma^0_{\!\!A}}{2
\beta_{\cal Q}} \biggr)\! \log\! \biggl( \frac{\beta_1}{\beta_2} \biggr)+ \!\int_{- \beta_{2}}^{- \beta_{1}} \!\frac{{\df}
\varepsilon}{\varepsilon} \frac{{\df}}{{\df} u}\! \bigl[G (\varepsilon, u) - G(0, u)\bigr]_{\!u = 0} \nonumber\\
& -\! \int_{- \beta_{2}}^0\! \frac{{\df} \varepsilon}{\varepsilon^2}
\log \biggl( 1 + \frac{\varepsilon}{\beta_{2}} \biggr)\! \biggl[ G (
\varepsilon, 0) - G (0, 0) - \varepsilon \frac{{\df} G (\tau,
0)}{{\df} \tau} \biggl|_{\tau = 0} \,\biggr] \nonumber\\
& + \!\int_{- \beta_{1}}^0 \!\frac{{\df} \varepsilon}{\varepsilon^2}
\log \biggl(1 + \frac{\varepsilon}{\beta_{1}} \biggr)
\! \biggl[ G (\varepsilon, 0) - G (0, 0) - \varepsilon \frac{{\df}
G (\tau, 0)}{{\df} \tau} \biggr|_{\tau = 0} \,\biggr]
\nonumber\\
& + \frac{\hat\Gamma^0_{\!\!A}}{4} \biggl( \frac{1}{\beta_{2}}
- \frac{1}{\beta_{1}} \biggr)
+ \biggl( \frac{1}{\beta_{2}} - \frac{1}{\beta_{\cal Q}} \biggr)\! \int_{-
\beta_{2}}^0 \!\frac{{\df} \varepsilon}{\varepsilon} \bigl[G (\varepsilon, 0)- G (0, 0)\bigr] \nonumber\\
& - \biggl( \frac{1}{\beta_{\mu 1}} - \frac{1}{\beta_{\cal Q}}
\biggr)\! \int_{- \beta_{\mu 1}}^0\! \frac{{\df} \varepsilon}{\varepsilon} \bigl[G (\varepsilon, 0) - G (0, 0)\bigr] ,
\end{align}
where we have intentionally written all dependence on $\mu_i$ through $\beta_i$. Expanding $G(\varepsilon,0)$ and integrating each term one
arrives at the following perturbative expansion for the evolution kernel:
\begin{align}
\beta_0[A (\mu_2)\! -\! A (\mu_1)] = &\,
\frac{\hat\Gamma^0_{\!\!A}}{2} \! \Biggl[ \frac{1}{\beta_{\cal Q}}\!
\log\! \biggl( \frac{\beta_2}{\beta_1} \biggr)\! +\!
\frac{1}{\beta_{2}} \!-\! \frac{1}{\beta_1} \Biggr]_{\rm \!LL}\!\!
+\! \Biggl[\frac{\hat\Gamma^1_{\!\!A}}{2} \frac{\beta_{2} - \beta_{1}}{\beta_{\cal Q}}
- \!\biggl( \hat\gamma_{\!A}^0 +\! \frac{\hat\Gamma^1_{\!\!A}}{2} \biggr) \!\log\! \biggl( \frac{\beta_{2}}{\beta_{1}} \biggr)\!\Biggr]_{\rm \!NLL} \nonumber
\\
& - \sum_{i = 1} \Biggl[ \frac{\beta_{2}^i - \beta_{1}^i}{i} \biggl( \hat\gamma^i_{A} + \frac{\hat\Gamma^{i + 1}_{\!\!A}}{2
\beta_0} \biggr) - \frac{\hat\Gamma^{i + 1}_{\!\!A}}{2\beta_{\cal Q}} \frac{\beta_{2}^{i + 1} - \beta_{1}^{i +
1}}{i + 1} \Biggr]_{\rm \!N^{(i+1)}LL}\, ,
\end{align}
where the first and second terms in square brackets in the first line are the LL and NLL contributions, respectively, while to obtain an N$^n$LL-accurate result one has
to add the first $n-1$ terms of the sum in the second line.

We finish this section by presenting an expression for the fixed-order coefficients multiplying powers of $\ell = \log(\mu/{\cal Q})$. In this case, due to the cusp anomalous
dimension, one has an additional power of $\ell$ at each order as compared to series without cusp:
\begin{equation}\label{eq:logexplicitcusp}
\beta_0\, \delta\!A(\mu) = \sum_{i = 1}\beta^i a^\mu_{i,i-1} = \sum_{i = 1}\beta^i\sum_{j=0}^{i+1} a_{i,i-1,j}\,\ell^j\,.
\end{equation}
Imposing that the series reproduces both types of anomalous dimensions when acting with a derivative with respect to $\log\mu$ one finds
\begin{align}
\sum_{i = 1} \beta^i \bigl( \hat\gamma_A^{i - 1} + \hat\Gamma_{\!\!A}^{i - 1} \ell\bigr) &=
\sum_{i = 1} \beta^i\! \sum^i_{j = 0} (j + 1) a_{i,i-1, j + 1} \ell^j - 2\! \sum_{i= 2} (i - 1) \beta^i \! \sum^i_{j = 0} a_{i - 1, i-2,j} \ell^j\,,\\
a_{i,i-1,1} &= \hat\gamma_{\!A}^{i - 1} + 2 (i - 1) a_{i - 1, i-2} =2 G_{0, i} (i - 1) ! - \frac{2(- 1)^i}{i} G_{i, 0}\,,\nonumber\\
a_{i,i-1,2} & = \frac{1}{2} \hat\Gamma_{\!\!A}^{i - 1} + (i - 1) a_{i - 1, i-2, 1} = 2 G_{0, i-1} (i - 1) !\,,\nonumber\\
a_{i,i-1,j} & = \frac{2 (i - 1)}{j} a_{i - 1, i-2, j - 1} = \frac{2^{j} (i - 1)!}{j!} G_{0, i+1-j}\,,\nonumber
\end{align}
where we have again equated equal powers of $\beta$ and $\ell$ in the first line to obtain the relations between coefficients shown below, treating separately
the cases $j=0,1$ in the second and third lines, respectively.
To obtain the last equality in the second line we have
used the expressions for $\hat\gamma_A^i$ and $a_{i, i-1}$ in Eq.~\eqref{eq:nocuspPart} and \eqref{eq:nologcusp}, respectively. Likewise,
the expression for $\hat\Gamma_{\!\!A}^i$ in Eqs.~\eqref{eq:cuspPart} and the result for $a_{i, i-1,1}$ in the second line has been used to obtain the first equality
in the third. To find the last expression
in the third line we apply the first relation recursively $(j-1)$
times until we reach $j=2$, such that the previous line can be used (therefore, this last expression is valid for $j\geq 2$). One can obtain identical results
expanding the logarithms in the last line of Eq.~\eqref{eq:Aep}, complementing the derivation carried out in Eq.~\eqref{eq:l-explicit}
with a direct computation of $G^\mu_{i,1}$:
\begin{align}
\frac{\df}{{\df}u}\!\biggl[e^{2u\ell}G(\varepsilon,u)\biggr]_{\!u=0} = &\,
2\ell\, G(\varepsilon,0) + \frac{{\df} G(\varepsilon,u)}{{\df}u}\biggr|_{u=0}
= \sum_{i=0} \bigl(G_{i,1} + 2\ell\, G_{i,0}\bigr)\,,\\
a_{i,i-1,j} = &\, \frac{2^j(i - 1) ! }{j!}G_{0,i+1-j}- \frac{(- 1)^i}{i} \Bigl[2 \delta_{j1} G_{i,0}+ \delta_{j0} \bigl(H_i G_{i + 1, 0} + G_{i,1}\bigr)\!\Bigr].\nonumber
\end{align}

\section{Principal value prescription}\label{sec:ppv}
As already discussed, the inverse Borel transform presented in Eqs.~\eqref{eq:LamSumNocusp} and \eqref{eq:cuspLambda} is ambiguous (that is, there are poles
in the integration path) and one needs a prescription in order to assign a finite value, which in turn defines the so called Borel sum. In this section we present the principal
value (PV) prescription, which will be used in the applications discussed throughout the following sections.

Let us consider a function $f(u)$, denoting the set of its poles of order $m$ in the real positive axis by $\bigl\{ u^{(m)}_{1\leq n\leq n_p} \bigr\}$, where $u^{(a)}_n > u^{(b)}_m$
if $n > m$. In general one can have $n_p=\infty$. A pole of order $m$ usually implies lower order poles at the same position, so $u_n^{(m)}$ stands for all poles
of order $1\leq i\leq m$ at the same position. The asymptotic expansion for $f(u)$ therefore reads:
\begin{equation}\label{eq:fAsy}
f(u) \asymp \sum_{n=1}^{n_p}\sum_{i=1}^m\frac{f_n^i}{\bigr[u-u_n^{(m)}\bigr]^i}\,.
\end{equation}
To compute the integral's principal value along the real axis we split the integration in several pieces:
\begin{align}\label{eq:IntegralPoles}
&{\rm PV}\!\Biggl[\!\int_0^\infty\! \mathd u f (u)\Biggl] = \!\int_{0}^{u_1 - \delta_1} \!\mathd u f(u)
+\! \sum_{n = 1}^{n_p-1}\!\int_{u_n + \delta_n}^{u_{n+1} - \delta_{n+1}} \!\mathd u f(u)
+\! \int_{u_n + \delta_n}^{\infty} \!\mathd u f(u) \\
&\quad+\!\sum_{n=1}^{n_p}\sum_{i=1}^m
\Biggl\{\!\int_{u_n - \delta_n}^{u_n + \delta_n} \mathd u\Biggl[f(u)-\frac{f_n^i}{\bigr[u-u_n^{(m)}\bigr]^i}\Biggr]\!
+ f_n^i\,{\rm PV}\!\Biggl[\int_{u_n - \delta_n}^{u_n + \delta_n} \frac{\mathd u}{\bigr[u-u_n^{(m)}\bigr]^i} \Biggr]\!\Biggr\},\nonumber
\end{align}
where $\delta_i$ are finite positive numbers such that $u_i + \delta_i \leq u_{i+1} - \delta_{i+1}$ for $1\leq i < n_p-1$.\footnote{Strictly speaking it is enough to require $u_i + \delta_i \leq u_{i+1}$, but we implement the most restrictive condition to avoid overlapping between consecutive patches.} In case there are infinitely many poles
the last integral in the first line is zero and the first sum in the second line extends to infinity. The result in Eq.~\eqref{eq:IntegralPoles}
is independent of the specific values of $\delta_i$ as long as they verify the condition just stated.
Thanks to the pole subtractions, only the last term within the sum in the second line of Eq.~\eqref{eq:IntegralPoles} needs regularization.
The PV prescription for an order-$i$ pole, with $i$ a natural number, is defined as\footnote{In full QCD poles turn to branch cuts, and the
PV prescription has to be adapted accordingly.}
\begin{equation}\label{eq:PV}
{\rm PV}_{\!\pm}\! \Biggl[\int_{u_n - \delta}^{u_n + \delta}\! \frac{ \mathd u}{(u - u_n)^i} \Biggr] \! \equiv \lim_{\ep\to 0}
\Biggl[ \!\int_{u_n - \delta}^{u_n - \ep} \!\frac{\mathd u}{(u -
u_n)^i} + \!\int_{C^{\pm}_{\ep, u_n}}\!\! \frac{\mathd u}{(u - u_n)^i}
+ \!\int_{u_n + \ep}^{u_n + \delta}\!\! \frac{\mathd u}{(u - u_n)^i} \Biggr] ,
\end{equation}
where $\delta >\ep > 0$ and $C^{+(-)}_{\ep, u_n}$ denotes an upper (lower) semi-circumference of radius $\ep$ centered at $u_n$ in the (anti-)clockwise
direction. In our prescription we want to preserve two important properties of ordinary integration: a) integrating an odd function $f(u)$ with respect to the point $u_0$,
that is $f(u-u_0) = -f(u_0-u)$, in the interval $\{u_0-\delta,u_0+\delta\}$ results in zero; b) if the integrand is real, so is the integral. Therefore we will take the average
of ${\rm PV}_{+}$ and ${\rm PV}_{-}$ as our definition:
\begin{align}
{\rm PV}\!\Biggl[\int {\rm d}u f(u)\Biggr] =\,& \frac{1}{2}\Biggl\{{\rm PV}_+\!\Biggl[\int {\rm d}u f(u)\Biggr] + {\rm PV}_-\!\Biggl[\int {\rm d}u f(u)\Biggr]\Biggr\},\\
\delta_{\Lambda }\Biggl[\int {\rm d}u f(u)\Biggr] =\,&\frac{1}{2\pi i}\Biggl\{{\rm PV}_+\!\Biggl[\int {\rm d}u f(u)\Biggr] - {\rm PV}_-\!\Biggl[\int {\rm d}u f(u)\Biggr]\Biggr\},\nonumber
\end{align}
where in the second line we give the standard prescription to assign an ambiguity to the Borel sum, which turns out to be a circular path enclosing the pole, such that
the residue theorem can be used. Therefore, if the integrand is real, so is the ambiguity. Moreover, it can be shown that the numerical value obtained in this way
is close to the smallest term of the series (in absolute value), which is normally taken as the intrinsic ambiguity of an asymptotic series \cite{Beneke:1998ui}.
Needless to say, only simple poles contribute to the total ambiguity, which is simply the sum of their residues:
\begin{equation}
\delta_{\Lambda }\!\Biggl[\!\int_0^\infty\! \mathd u f (u)\Biggl] = \sum_{n=1}^{n_p} f_n^1\,.
\end{equation}
Furthermore, since the integration path in Eq.~\eqref{eq:PV} is symmetric around $u=u_0$, with our prescription the principal value is zero if $i=2k+1$ for $k\geq 0$.
On the other hand if $i=2k$ one has
\begin{align}
\int_{u_n - \delta}^{u_n - \ep} \!\frac{ \mathd u}{(u - u_n)^{2k}} + \int_{u_n + \ep}^{u_n + \delta} \!\frac{ \mathd u}{(u - u_n)^{2k}} = \,&
\frac{2}{2k - 1} \biggl[ \frac{1}{\varepsilon^{2k - 1}} - \frac{1}{\delta^{2k -1}} \biggr],\\
\frac{1}{2} \Biggl[\int_{C^{+}_{\ep, u_n}} \frac{ \mathd u}{(u - u_n)^{2k}} + \int_{C^{-}_{\ep, u_n}} \frac{ \mathd u}{(u - u_n)^{2k}}\Biggr] = \,&
\frac{-2}{2k - 1} \frac{1}{\varepsilon^{2k - 1}} \,,\nonumber
\end{align}
such that the divergence $1/\varepsilon^{2k - 1}$ cancels out when adding the two terms, leaving a finite remainder. We can write a general formula
for the PV integral in Eq.~\eqref{eq:PV} when the symmetric prescription is used using the modulo operation:
\begin{equation}\label{eq:asyPole}
{\rm PV}\! \Biggl[\int_{u_n - \delta}^{u_n + \delta}\! \frac{ \mathd u}{(u - u_n)^{i}} \Biggr] =
-\frac{2}{i - 1}\frac{\delta_{0,{\rm mod}(i,2)}}{\delta^{i -1}} \,,
\end{equation}
with $\delta_{0,{\rm mod}(i,2)}$ a Kronecker delta symbol which is non-zero only if $i$ is an even number.

One can figure out the contribution of each
singularity to the asymptotic behavior of the fixed-order coefficients by expanding the pole in powers of $u$ and integrating
term by term, which is equivalent to the replacement $u^n\to \beta^{1+n}\Gamma(n+1)$:
\begin{equation}\label{eq:poleA}
\int_0^{\infty}\! \frac{{\df} u}{(u-u_0)^m}\,e^{-\frac{u}{\beta}}=
\frac{(-1)^m}{u_0^{m-1}}\sum_{n=1} (m)_{n-1}
\biggl(\frac{\beta}{u_0}\biggr)^{\!\!n}\,.
\end{equation}
Therefore one has that for larger (smaller) pole's order ($u_0$) the asymptotic behavior becomes worse. Using the asymptotic expansion
shown in Eq.~\eqref{eq:fAsy}, the contribution to the fixed-order coefficients of a series
coming from the poles of its Borel transform takes the following form:
\begin{equation}\label{eq:aAsy}
\sum a_n\beta_{\cal Q}^n =\int_0^\infty\! \mathd u\,e^{-\frac{u}{\beta_{\cal Q}}} f (u)\,, \qquad\quad
a_{n}^{\rm asy} = \sum_{k=1}^{n_p} \sum_{i=1}^m \frac{(-1)^i (i)_{n-1}}{(u_k^i)^{n+i-1}}f_n^i\,.
\end{equation}
The exact coefficients are dominated by the
poles closest to the origin of the Borel plane. We can also work out the case $\mu \neq {\cal Q}$ taking into account that
\mbox{$\beta_{\cal Q} = \beta_\mu/(1 - 2 \beta_\mu \ell)$} with $\ell = \log(\mu/{\cal Q})$, which upon re-expansion in terms of powers of $\beta_\mu$ results in
\begin{equation}\label{eq:expAlpha}
\sum_{j=0} \sum_{n=1} a_n\frac{(n)_j}{j!} (2\ell)^j\beta_{\mu}^{j+n} =
\sum_{n=1} \beta_{\mu}^{n} \sum_{j=0}^{n-1} \binom{n-1}{j} (2\ell)^j a_{n-j}\,,
\end{equation}
where we have shifted $n\to n-i$, reversed the order of the sums and expressed the Pochhammer symbol in terms of a binomial to obtain a canonical
expansion in powers of $\beta_\mu$ and $\ell$. Applying this result to the contribution coming from a pole of order $i$ as shown in Eq.~\eqref{eq:poleA},
for $n\gg 1$ we get
\begin{equation}\label{eq:aprox}
\!\!\sum_{j=0}^{n-1} \frac{\Gamma(n + i - j-1)\Gamma(n)(2 u_0 \ell)^j}{\Gamma(i)\Gamma(n-j)\Gamma(j+1)}
\xrightarrow[n \to \infty]{}(i)_{n-1}\!\sum_{j=0}^{\infty} \frac{(2 u_0 \ell)^j}{j!}
= (i)_{n-1} \!\biggl(\frac{\mu}{\cal Q}\biggr)^{\!\!2u_0}\,,
\end{equation}
where we have used that for large $n$ the sum is dominated by small values of $j$, therefore setting $n-j\to n$ (this approximation is more accurate for small
values of $i$).
The dependence on $\mu/{\cal Q}$ does not depend on $n$, and therefore is common to all powers of $\beta_\mu$ generated by a given pole.
With this result the asymptotic expression for the $\mu$-dependent fixed-order coefficients is obtained:
\begin{equation}\label{eq:aAsyMu}
a_{n}^{\rm asy}(\mu) = \sum_{k=1}^{n_p} \sum_{i=1}^m \biggl(\frac{\mu}{\cal Q}\biggr)^{\!\!2u_k^i}\frac{(-1)^if_n^i}{(u_k^i)^{n+i-1}}(i)_{n-1}\,.
\end{equation}
Since the approximation used in Eq.~\eqref{eq:aprox} is exact if $\ell=0$, the result above is in agreement with the second line of Eq.~\eqref{eq:aAsy} for $\mu = \cal Q$.
When applied to simple poles one can use the identity $(1)_{n-1}=\Gamma(n)$.

For the numerical computations in this article we have developed two independent numeric codes, written in
\texttt{Mathematica}~\cite{mathematica} and \texttt{Python}~\cite{Rossum:1995:PRM:869369}, that agree within $15$ decimal places.
While we use built-in native functions in \texttt{Mathematica}, for evaluating special functions and performing numerical integrals in
\texttt{Python}, the \texttt{SciPy} module~\cite{Virtanen:2019joe}, that builds on the \texttt{NumPy} package~\cite{oliphant2006guide}, is used.

\section{\boldmath Application to the \texorpdfstring{$\overline{\rm MS}$}{MSbar} and MSR masses}
\label{sec:MSbarPole}
\begin{figure}[t]\centering
\includegraphics[width=0.3\textwidth]{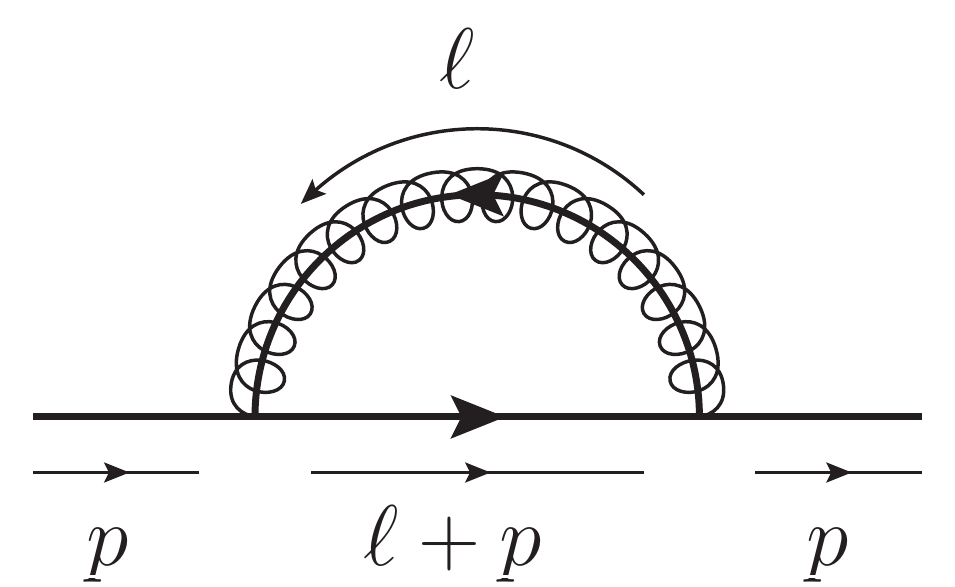}
\caption{One-loop quark self-energy diagram for a massive or massless quark, with $\ell$ the virtual loop momentum and
$p$ the quark off-shell momentum.\label{fig:selfenergy}}
\end{figure}
In this section we review the calculation of the relation between the $\overline{\rm MS}$ and pole quark masses. From this computation we will also
obtain the wave-function renormalization for a massive quark in the on-shell scheme, necessary for the SCET to bHQET matching
coefficient computation carried out in Sec.~\ref{sec:bHQET}, as well as the anomalous dimensions for the $\overline{\rm MS}$ and MSR masses.

\subsection[Massive quark self-energy and the $\overline{\rm MS}$ mass]{\boldmath Massive quark self-energy and the $\overline{\rm MS}$ mass}
The Feynman diagram shown in Fig.~\ref{fig:selfenergy} with a modified gluon propagator is computed in $d=4-2\varepsilon$ dimensions. We do not expand the result
for small $\varepsilon$ and split the self-energy in two Dirac structures in the usual way:\footnote{Our definition for the quark self-energy differs from others by a sign.}
\begin{align}
\Sigma_q(p,m_0) = &-\! i g_0^2 C_{\!F}\!\! \int \!\!\frac{{\df}^d \ell}{(2 \pi)^d} \frac{ \gamma^{\mu} [ ( \ell \!\! / + p \!
\!\! / ) + m_0 ] \gamma_{\mu}}{(- \ell^2)^{1 + u}[(\ell + p)^2 - m_0^2]}
\, = p \! \!\! / \,\Sigma_q^p (p^2, m_0) + m_0\Sigma_{q}^m (p^2, m_0)\,, \nonumber \\
\Sigma_q^m (p^2, m_0) =& - \!2C_{\!F} \frac{g_0^2}{(4 \pi)^{2-\varepsilon}} \frac{(2 - \varepsilon) \Gamma (h +
\varepsilon) \Gamma (1 - u - \varepsilon)}{ \Gamma (2 -
\varepsilon) m_0^{2 (u+\varepsilon)}} \,{}_2 F_1 \!\biggl(1+u, u + \varepsilon ; 2 - \varepsilon,
\frac{p^2}{m_0^2} \biggr),\nonumber\\
\Sigma_q^p (p^2, m_0) = &\,C_{\!F} \frac{g_0^2}{(4 \pi)^{2-\varepsilon}} \frac{\Gamma (1 - u - \varepsilon)
\Gamma (u +
\varepsilon)}{\Gamma (1 - \varepsilon) m_0^{2 (u+\varepsilon)}} \biggl[\! \biggl( 1 + \frac{m_0^2}{p^2}\biggr)
{}_2 F_1 \!\biggl( 1+u, u + \varepsilon ; 2 - \varepsilon,
\frac{p^2}{m_0^2} \biggr) \nonumber\\
& - \frac{m_0^2}{p^2} \,{}_2 F_1 \!\biggl( u, u - 1 + \varepsilon ; 2 -
\varepsilon, \frac{p^2}{m_0^2} \biggr)\! \biggr] ,
\end{align}
\renewcommand{\arraystretch}{1.15}
\begin{table}[t!]
\centering
\begin{tabular}{|c|cccc|}
\hline
$n$ & $\hat{\gamma}^n_m$ & $\hat{\gamma}^n_R$ & $\hat{\gamma}^{J,n}_R$& $\hat{\gamma}^{J',n}_R$\\
\hline
$0$ & $-4$ &$5.33333$ &$4.74953$ &$6.70269$\\
$1$ & $-3.33333$ &$14.3261$ &$15.304$ &$12.4616$\\
$2$ & $3.88889$ &$-5.98376$ &$1.86071$ &$-3.17481$\\
$3$ & $5.00534$ &$21.989$ &$-15.9745$ &$-12.8867$\\
$4$ & $0.442501$ &$-26.3831$ &$-13.2198$ &$0.563555$\\
$5$ & $-1.55019$ &$535.716$ &$17.2635$ &$7.85014$\\
$6$ & $-0.670009$ &$-1397.07$ &$16.6685$ &$6.55289$\\
$7$ & $0.0812939$ &$19031.9$ &$45.4207$ &$22.2101$\\
$8$ & $0.123507$ &$-100044$ &$359.12$ &$49.373$\\
$9$ & $0.022317$ &$1.23114\times 10^6$ &$897.07$ &$312.347$\\
\hline
\end{tabular}
\caption{Numeric coefficients for the $\MSBar$ mass and various $R$ anomalous dimensions. The hatted coefficients are defined as $f \equiv (1 / \beta_0)
\sum^{\infty}_{n = 0} \hat{f}_n \beta^{n + 1}$.\label{tab:gammaR}}
\end{table}
\!\!with ${}_2F_1$ the Gauss hypergeometric function, $g_0$ and $m_0$ the bare strong coupling and mass, respectively, and $C_{\!F}=(N_c^2-1)/(2N_c)$. The relation between
the scale dependent $\overline{\rm MS}$ mass $\overline m(\mu)$ and the scale-independent pole mass $m_p$ at one loop with a modified gluon propagator takes the
following form
\begin{align}
Z^{\overline{\rm MS}}_m\, \frac{\overline m(\mu)}{m_p} =&\, 1 + \Sigma_q^m (m_p^2, m_p) + \Sigma_q^p (m_p^2,m_p)\\
=&\, 1 - 2C_{\!F} \frac{g^2_0}{(4 \pi)^{2-\varepsilon}} \frac{\Gamma (u + \varepsilon)}{m_p^{2 (u+\varepsilon)}} \frac{\Gamma
(1 - 2 u - 2 \varepsilon)}{\Gamma (3 - u - 2 \varepsilon) } (3 - 2\varepsilon) (1 - u - \varepsilon)\,,\nonumber
\end{align}
where $Z^{\overline{\rm MS}}_m$ (which depends on $\mu$ such that the right-hand-side does not) contains only $1/\varepsilon^n$ poles and ensures
$\overline m(\mu)/m_p$ is finite. From this result we can identify the function $b(\varepsilon,u)$
in Eq.~\eqref{eq:1-loopbare}, with ${\cal Q}=m_p$, and compute $F(\varepsilon,u)$ as defined in Eq.~\eqref{eq:sumF}. To make later formulas as simple as possible
we define the function corresponding to the series
$\delta_{\overline{\rm MS}}( \overline{m}) \equiv m_p - \overline{m}$, which only differs by a sign and has ${\cal Q}=\overline{m}\equiv\overline{m}(\overline{m})$:
\begin{figure*}[t!]
\subfigure[]
{\includegraphics[width=0.48\textwidth]{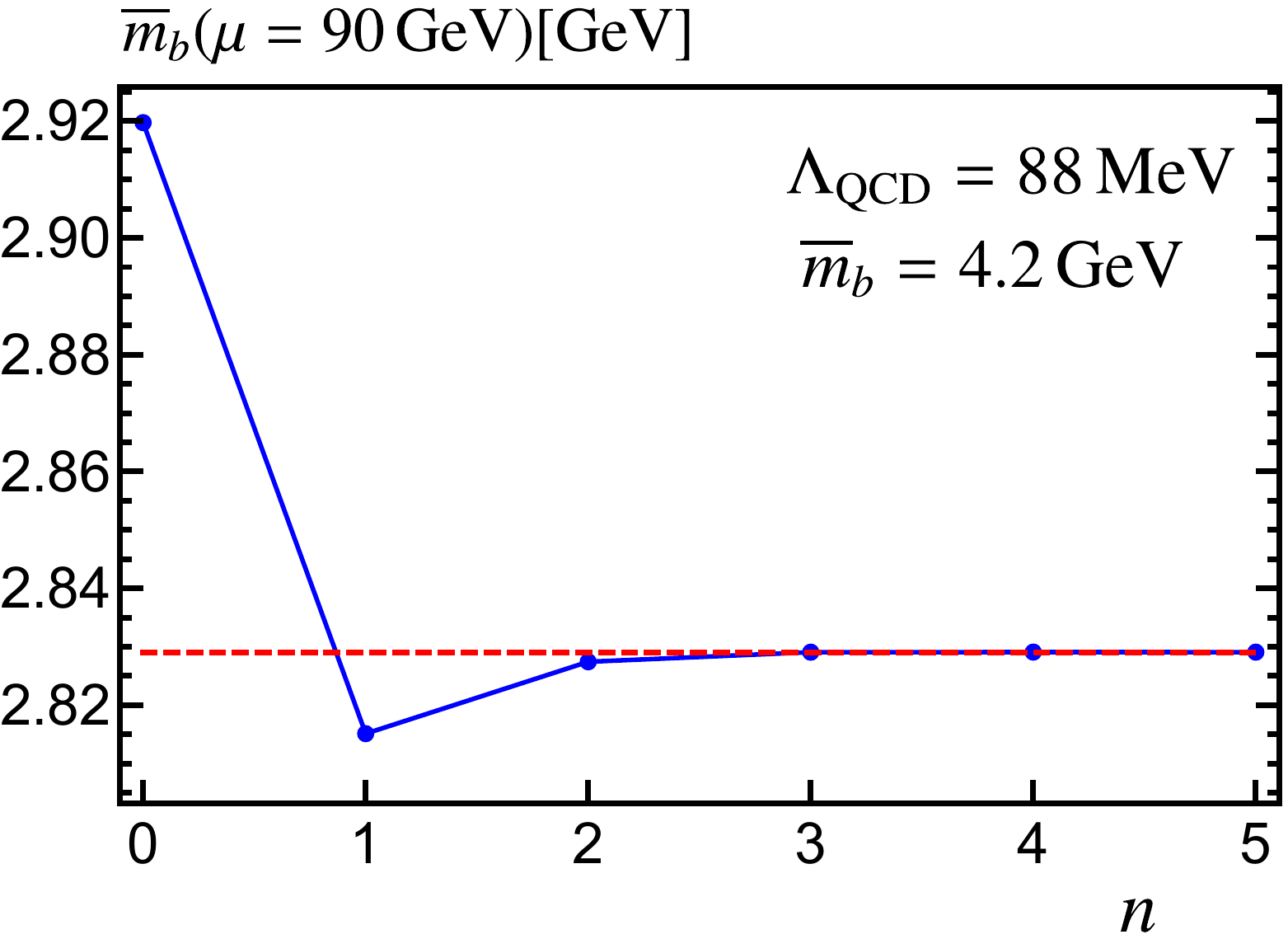}
\label{fig:MSbar}}~~
\subfigure[]{\includegraphics[width=0.46\textwidth]{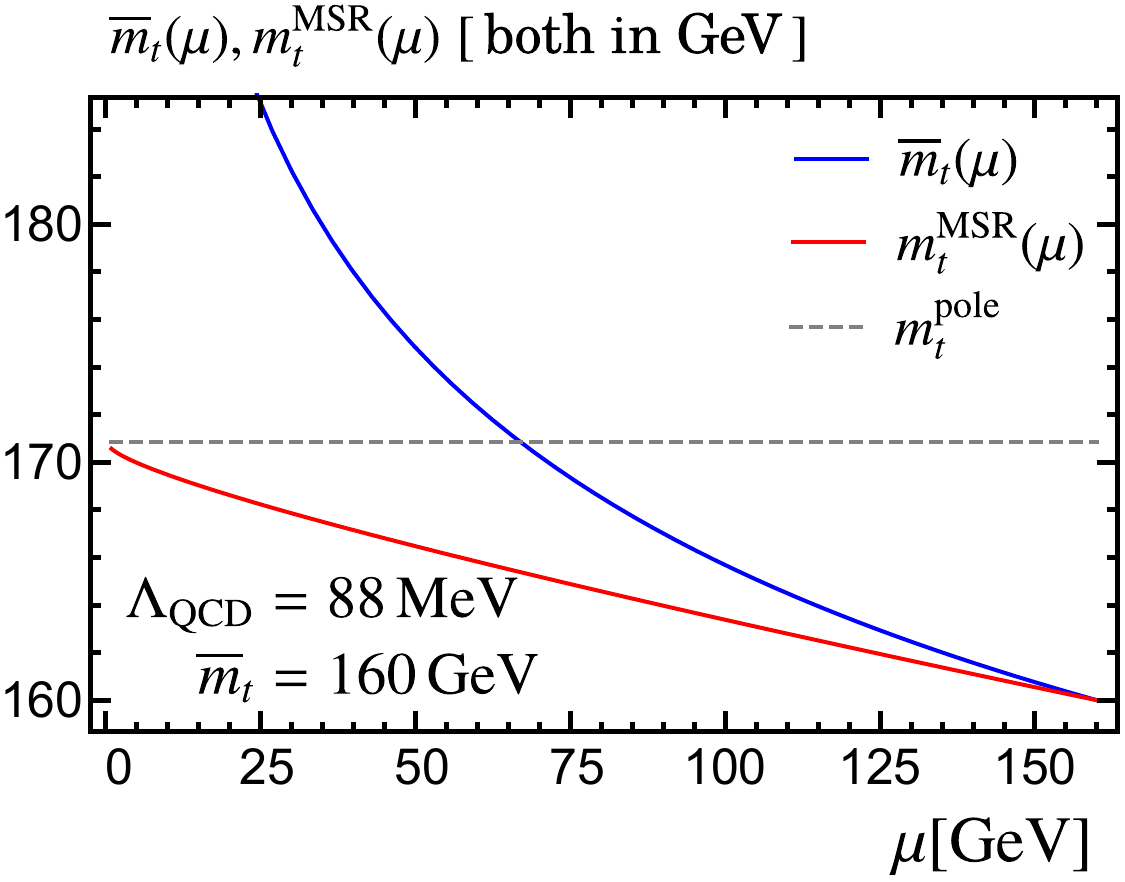}
\label{fig:MSR-MSbar-pole}}
\caption{Left panel: bottom quark mass in the $\overline{\rm MS}$ scheme at $\mu=90\,$GeV obtained from $\overline{m}_b$. Blue dots have
N$^n$LL accuracy, while the dashed red line uses the exact evolution kernel. Right panel: exact prediction for the top quark mass in the
$\overline{\rm MS}$ (blue), MSR (red) and pole (gray) schemes.
The pole and MSR mass uncertainties are too small to be visible in the plot.}
\label{fig:MSbar-MSR-pole}
\end{figure*}
\begin{equation}
F_{\rm \overline{MS}}(\varepsilon, u) = 2C_{\!F} D (\varepsilon)^{\frac{u}{\varepsilon} - 1}
\frac{(1 - u) (3 - 2 \varepsilon) \Gamma (1 + u) \Gamma (1 - 2u)e^{\varepsilon \gamma_E}}{\Gamma (3 - u - \varepsilon)}\,.
\end{equation}
From this result one can find a closed expression for the ${\rm \overline{MS}}$ mass anomalous dimension $\gamma_m$, first computed in
Ref.~\cite{PalanquesMestre:1983zy} (see also Ref.~\cite{Grozin:2003gf}):\footnote{Note that the ${\rm \overline{MS}}$ mass anomalous dimension is defined as
$\frac{\mu}{\overline{m}}\frac{{\df} \overline m}{{\df}\mu} = 2\gamma_m$, and therefore \mbox{$\gamma_m = -\gamma_{\rm \overline{MS}}/2$}.}
\begin{align}
&\beta_0\gamma_m (\beta) = -\beta F_{\rm \overline{MS}}(-\beta, 0)=- \frac{C_{\!F} \beta (3 + 2 \beta) \Gamma (4 + 2 \beta)}{ 3(2 +\beta) \Gamma (1 - \beta) \Gamma (2 + \beta)^3 }\\
&\qquad=- 3 C_{\!F} \beta \exp \!\biggl[ \frac{5}{6} \beta + \sum_{n = 1}
\frac{(- \beta)^n}{n} \Bigl\{ 2 + 2^{- n} - 2^n (1 + 2 \times 3^{- n}) + \zeta_n [2^n - 3 - (- 1)^n] \Bigr\} \!\biggr].\nonumber
\end{align}
In the second line we have written the expression in a way that facilitates its expansion in powers of $\beta$
by computer programs, such that the fixed-order coefficients can be easily obtained. To carry out the expansion of
the exponential of a series we use the following recursive relation, which is derived imposing $\exp[f(x)]'= f'(x) \exp[f(x)]$:
\begin{equation}\label{eq:exp-expand}
\exp\! \Biggl[ \,\sum_{i = 1} a_i \,x^i \Biggr] = 1 + \sum_{i = 1} f_i\, x^i
\equiv \sum_{i = 0} f_i \,x^i, \qquad
f_{i + 1} = \frac{1}{i + 1} \sum_{j = 0}^i (j + 1) f_{i - j} \,a_{j + 1}\,,
\end{equation}
Our result reproduces the full QCD leading flavor structure up to 5 loops \cite{Vermaseren:1997fq,Chetyrkin:1997dh,Baikov:2014qja,Luthe:2016xec}, collected in the first
column of Table~\ref{tab:gammaR}. Since the
pole closest to the origin in the function $F_{\rm \overline{MS}}(\beta, 0)$ is located at $\beta=-2.5$, the convergence radius of $\gamma_m (\beta)$ is $\Delta\beta = 5/2$.
Interestingly, since all poles in $F_{\rm \overline{MS}}(\beta, 0)$ lay on the positive real axis, $\gamma_m$ has no singularities in the physical region $\alpha_s>0$.
The series is thus not ambiguous and provided that $\beta<2.5$ the partial sum will eventually converge to the exact result.

To obtain the fixed-order relation between the pole and ${\rm \overline{MS}}$ masses we can apply Eq.~\eqref{eq:nologFOnocusp}, and for that we
need\footnote{An equivalent result was found in Refs.~\cite{Beneke:1994sw,Ball:1995ni}.}
\begin{align}
F_{\rm \overline{MS}}(0, u) \equiv\,& \frac{6 C_{\!F} e^{\frac{5 u}{3}} \Gamma (1 - 2 u) \Gamma (1 + u)}{(2 - u) \Gamma (1 -u)}\\
= \,&3 C_{\!F} \exp\! \biggl[ \frac{13 u}{6} + \sum_{n = 2} \frac{u^n}{n} \Bigl\{ 2^{- n} +
\zeta_n [2^n - 1 + (- 1)^n] \Bigr\}\! \biggr],\nonumber
\end{align}
\begin{figure*}[t!]
\subfigure[]
{\includegraphics[width=0.48\textwidth]{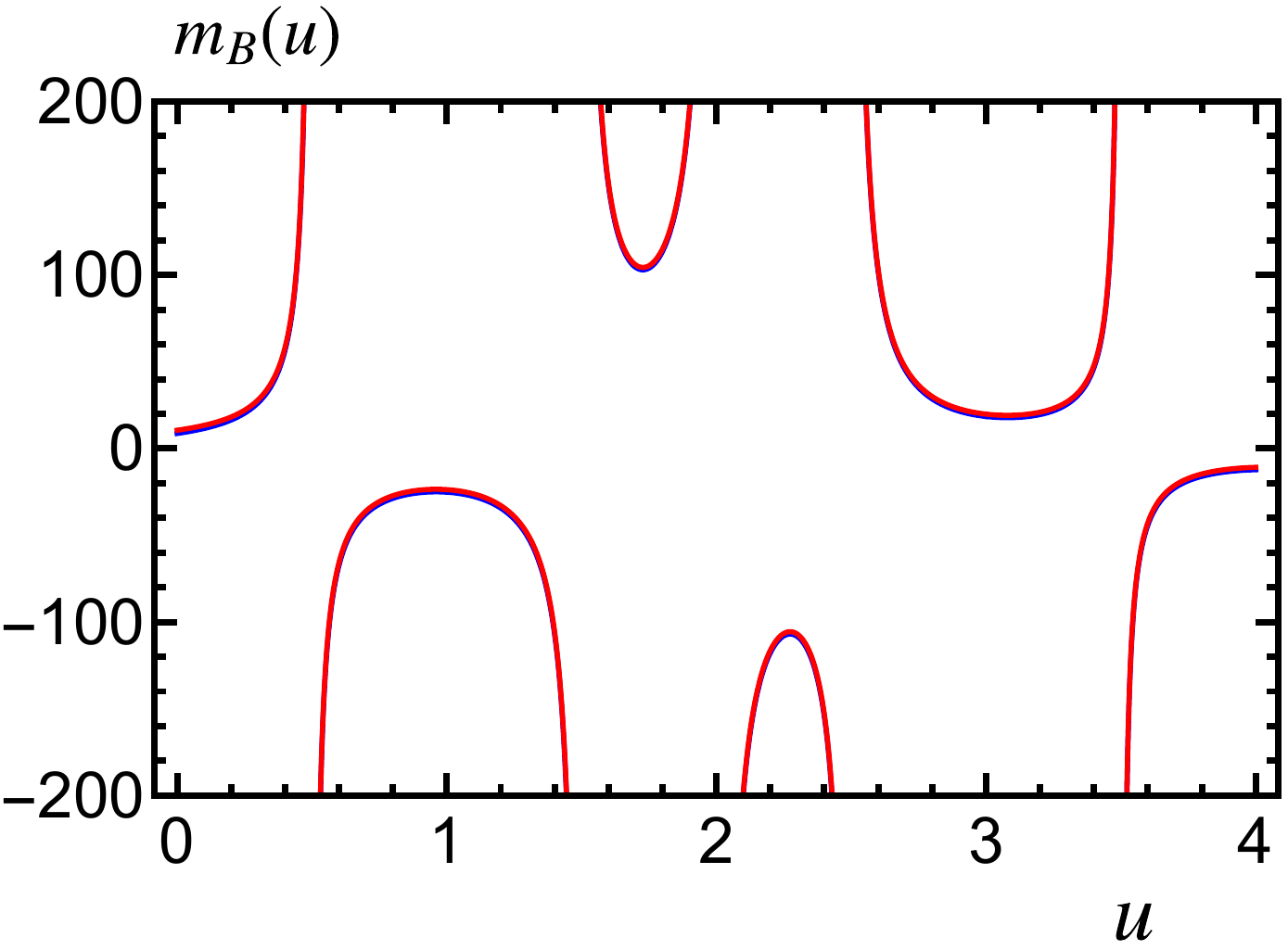}
\label{fig:FMS}}~~
\subfigure[]{\includegraphics[width=0.48\textwidth]{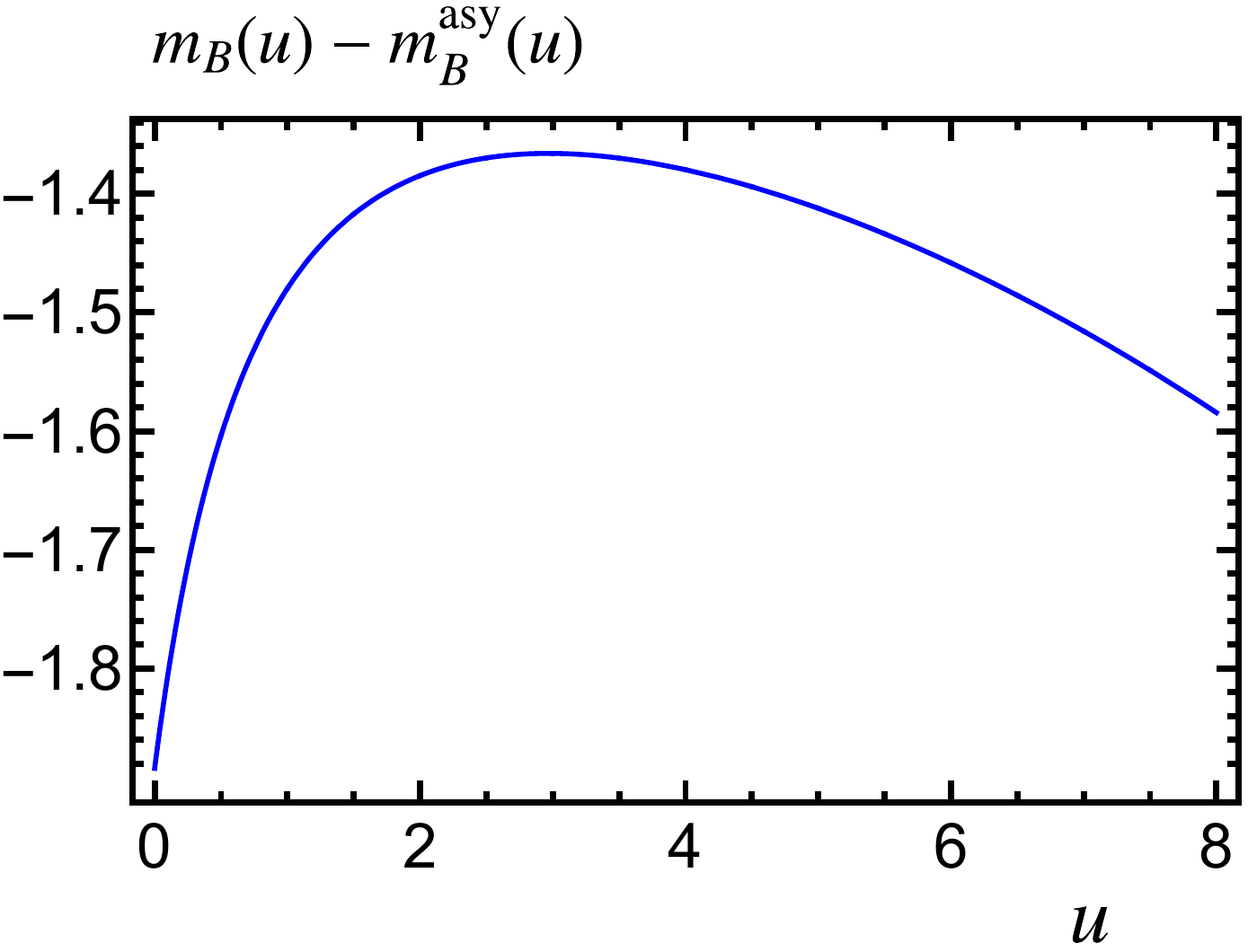}
\label{fig:FMS-diff}}
\caption{Pole structure of the inverse Borel function $m_B(u)$ for the relation between the pole and $\MSBar$ masses. In the left panel we show both
$m_B(u)$ (blue) and its asymptotic expansion (red) [\,see Eq.~\eqref{eq:mBasy}\,], although they are hardly distinguishable.
In the right panel the difference between $m_B(u)$ and its asymptotic expansion, free from poles, is shown.}
\label{fig:asy}
\end{figure*}
\!\!where in the second line the argument of the exponential has been expanded in powers of $u$. The fixed-order coefficients obtained using Eq.~\eqref{eq:exp-expand} reproduce full QCD
leading flavor results up to $\mathcal{O}(\alpha_s^4)$~\cite{Tarrach:1980up,Gray:1990yh,Melnikov:2000qh,Marquard:2016dcn}, collected in the first column of Table~\ref{tab:masses}.
In Fig.~\ref{fig:MSbar} we show
how the perturbative $\overline{\rm MS}$ evolution for the bottom quark mass is quickly convergent and already at NLL is very close to the exact value.\footnote{Unless indicated
otherwise, for the analyses carried out in this article we use $n_f=5$ active flavors and the following numerical values: $\overline{m}_t=160\,$GeV,
$\overline{m}_b=4.2\,$GeV, $\overline m_c =1.3\,$GeV and $\Lambda^{n_f=5}_{\rm QCD}=88.3\,$GeV.} In
Fig.~\ref{fig:MSR-MSbar-pole} we compare the $\overline{\rm MS}$ and MSR (see Sec.~\ref{sec:MSR}) evolution for the top quark mass for values of the renormalization
scale smaller than $\overline{m}_t$. Both schemes coincide for $\mu=\overline{m}_t$ and, even though the $\overline{\rm MS}$ evolution is unphysical for those
scales, growing out of control and becoming even larger than the pole mass, we observe that $\overline{m}_t( \overline{m}_t/2)\simeq m_t^{\rm pole}$.
The numerical value of the MSR mass monotonically and smoothly grows as $R\to 0$ to reach $m_t^{\rm pole}$ precisely
at this limiting value. R-evolution for $R<\overline{m}_t$ is physical and should be used
for processes that probe scales smaller than the quark mass itself.

The modified one-loop massive quark wave-function renormalization in the on-shell scheme $Z_\xi^{\rm OS}$ has the form:
\begin{align}\label{eq:wave}
Z_\xi^{\rm OS} =&\,1 - \Sigma_q^p (m_p^2, m_p) - 2 m_p^2 \frac{\df}{{\df} p^2}\!
\biggl[\Sigma_q^p (p^2, m_p) + \Sigma_q^m (p^2, m_p)\biggr]_{\!p^2= m_p^2} \\
=&\, 1 - \frac{2C_{\!F}\,g_0^2}{(4 \pi)^{2-\varepsilon}} \frac{\Gamma (u + \varepsilon) \Gamma (1 - 2 u- 2 \varepsilon)}
{\Gamma (3 - u - 2 \varepsilon) m^{2 (u+\varepsilon)}} (3 - 2 \varepsilon) (1+u) (1 - u - \varepsilon) \,.\nonumber
\end{align}
In this approximation one has $Z_\xi^{\rm OS}= (1+u)Z_m^{\overline{\rm MS}}\,\overline{m}(\mu)/m_p$,
and therefore $Z_m^{\overline{\rm MS}}\,\overline{m}(\mu)/m_p$ and $Z_\xi^{\rm OS}$ agree in full QCD at one loop.
\begin{figure*}[t!]
\subfigure[]
{\includegraphics[width=0.49\textwidth]{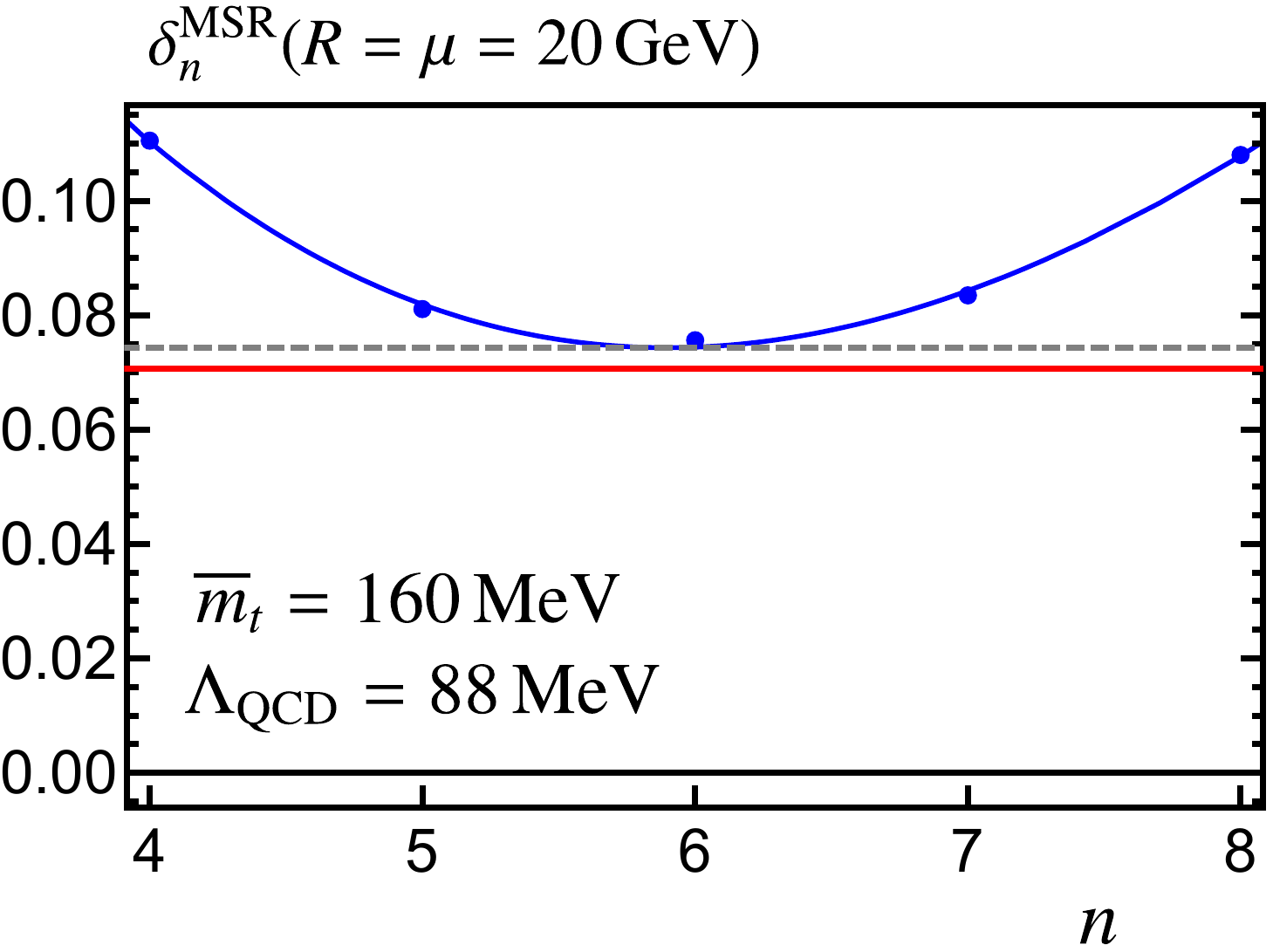}
\label{fig:AmbMin}}~~
\subfigure[]{\includegraphics[width=0.465\textwidth]{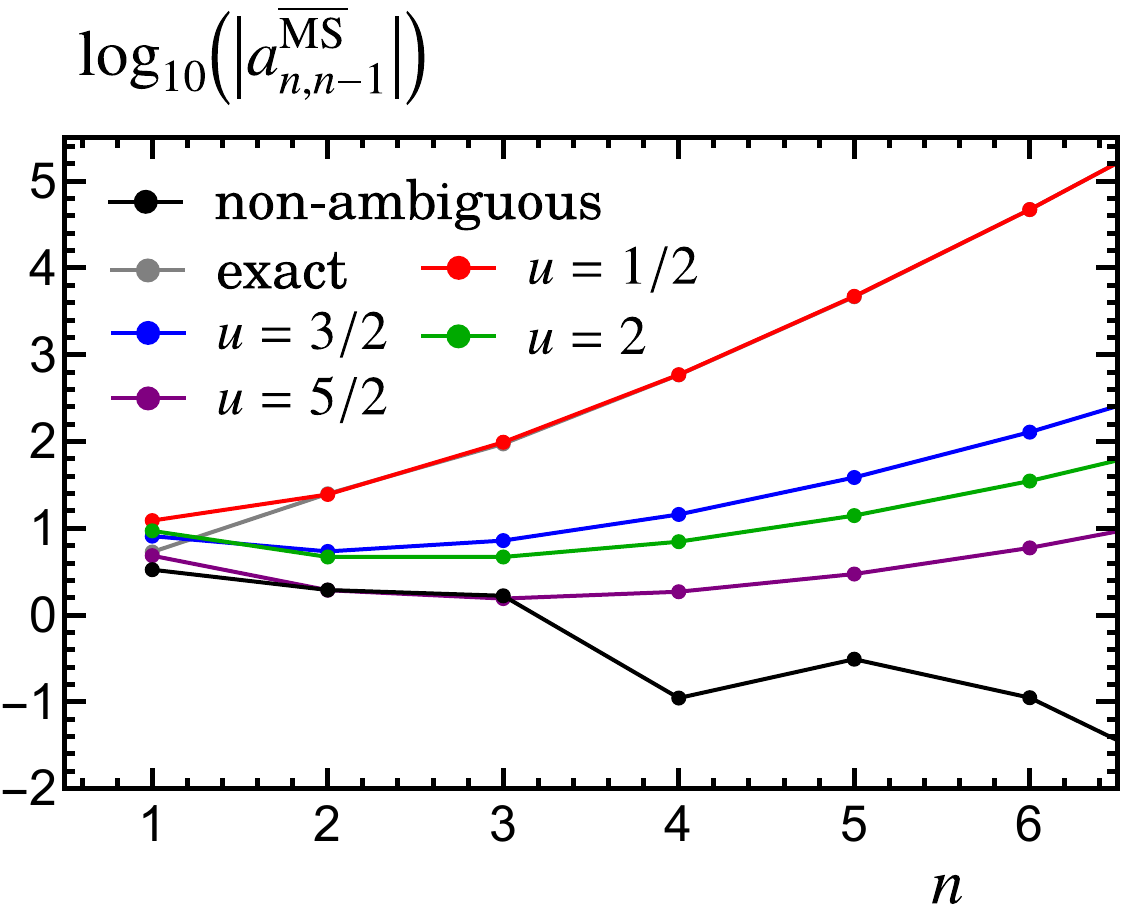}
\label{fig:poles}}
\caption{Left panel: Fixed-order corrections $\delta_n^{\rm MSR}$ for the series $\delta_{\rm MSR}(R)$
for
\mbox{$R=\mu=20\,$GeV}. The five smallest terms (blue dots) are shown together with a fit to a cubic polynomial (blue line). The ambiguity
is represented as a red line, while the minimal term cubic interpolation appears as a gray dashed line.
Right panel: Different contributions to the series coefficients $a_{n,n-1}^{\MSBar}$ that relate the pole and $\MSBar$ masses. Gray dots show the exact coefficients,
while red, green, blue and purple correspond to the contributions from the first four poles in the Borel transform $m_B(u)$. The black dots depict the non-ambiguous
contribution, that is, the second term in Eq.~\eqref{eq:nologFOnocusp}.}
\label{fig:ambPoles}
\end{figure*}

\subsection{Renormalon pole normalization and the RS mass}
\begin{figure*}[t!]
\subfigure[]
{\includegraphics[width=0.46\textwidth]{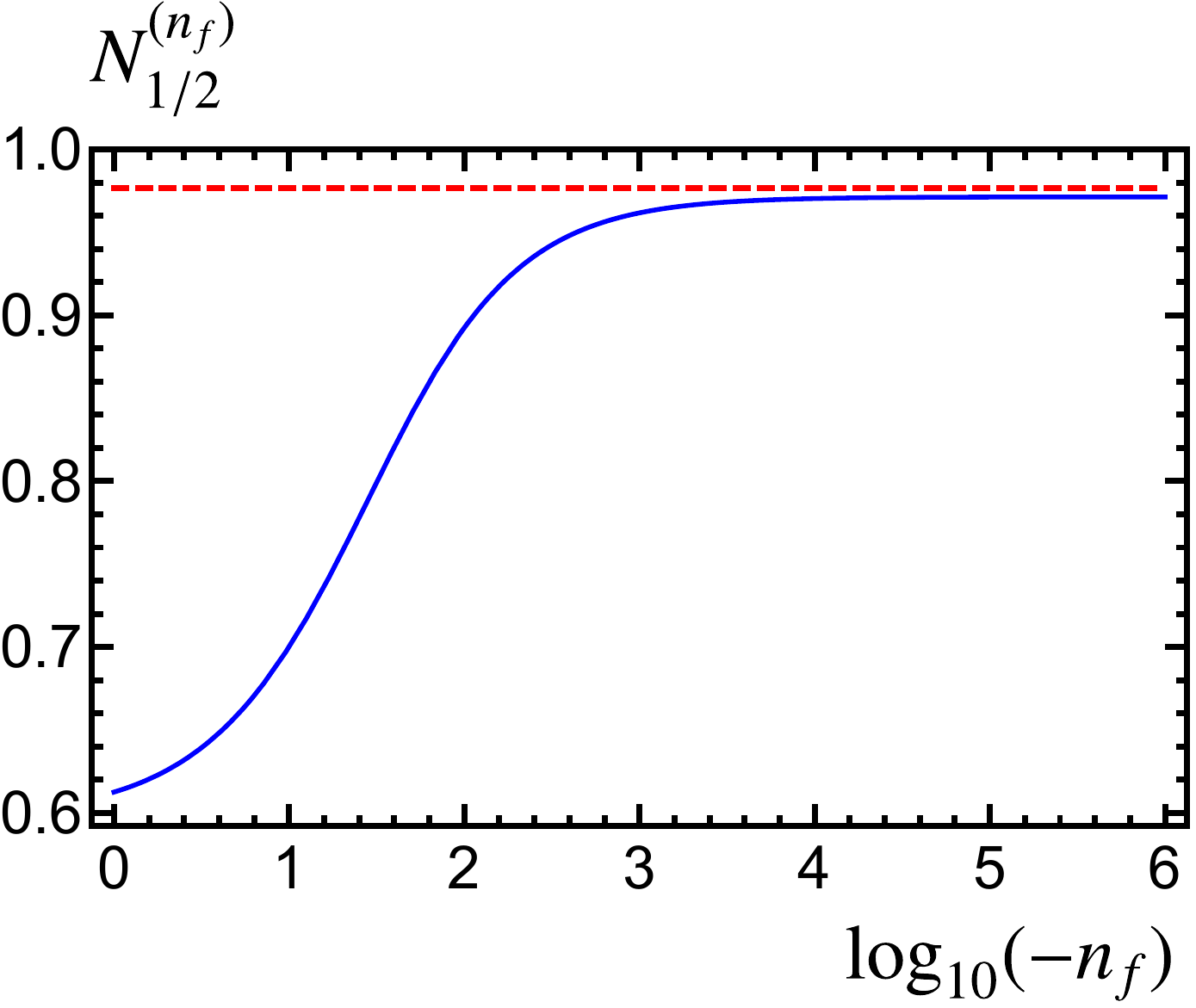}
\label{fig:N12}}~~
\subfigure[]{\raisebox{3mm}[0pt][0pt]{\includegraphics[width=0.48\textwidth]{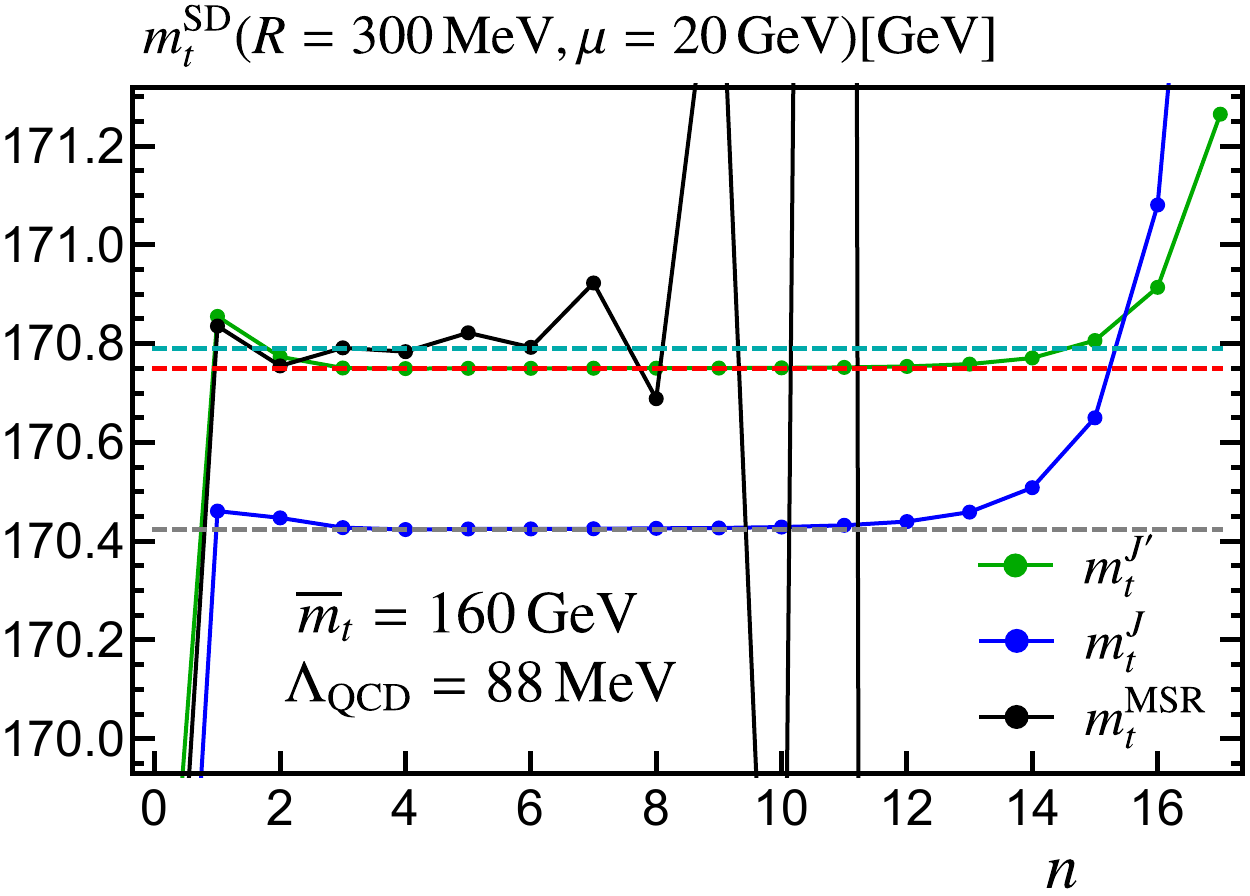}}
\label{fig:MSR}}
\caption{Left panel: Normalization of the leading pole renormalon $N^{(n_f)}_{1 / 2}$ for the series $m_p - \overline{m}$
as a function of the active number of flavors $n_f$, shown as a blue line,
and its large-$\beta_0$ limit $N^{\beta_0}_{1 / 2}$ appearing as a horizontal dashed red line. Right panel: Comparison of the MSR top quark
mass $m_t^{\rm MSR}(R)$ (black dots), jet mass in the derivative $m_t^J(R,\mu)$ (blue
dots) and non-derivative $m_t^{J'}(R)$ (green dots) schemes for $R=300\,$MeV and $\mu=20\,$GeV, as computed with a partial sum including $n+1$ terms or in an
exact form (cyan, gray and red dashed lines, respectively).}
\label{fig:RenSumRul}
\end{figure*}

To study the asymptotic behavior of the series $m_p - \overline{m}$ we look at the poles of its Borel transform. To that end it is enough to analyze
$m_B(u) =\! \bigl[F_{\rm \overline{MS}}(0, u) - F_{\rm \overline{MS}}(0, 0)\bigr]/u$, a function that can be seen in Fig.~\ref{fig:FMS}. The subtraction ensures
there is no singularity at $u=0$, but plays no role for poles with $u>0$. We find the following asymptotic expansion:
\begin{equation}\label{eq:mBasy}
m_B(u) \asymp - C_{\!F} \!\Biggl[ \frac{e^{10 / 3}}{2 (u - 2)} + 6 \sum_{k = 0}
\frac{e^{\frac{5 k}{3} + \frac{5}{6}} \Gamma \bigl(\frac{1}{2}+k\bigr)}{(3-2 k)\, (2k)! \,
\Gamma\bigl(\frac{1}{2}-k\bigr)\bigl(u-\frac{2k+1}{2}\bigr)} \Biggr].
\end{equation}
The difference between $m_B(u)$ and its asymptotic expansion is shown in Fig.~\ref{fig:FMS-diff} (in the range $0<u<8$
the largest deviation is only $20\%$). We can apply Eq.~\eqref{eq:asyPole} to the various terms in Eq.~\eqref{eq:mBasy} to obtain the contribution of each
pole to the series coefficients $a_{n,n-1}^{\overline {\rm MS}}$ defined in Eq.~\eqref{eq:MSR}.
This can be seen in Fig.~\ref{fig:poles}, where the exact value is shown in gray, while the contribution of the first four poles appear
in colors other than black. The picture clearly shows how the $u=1/2$ pole overly dominates already at low values of $n$. The black dots represent the non-ambiguous contribution coming from the second term in Eq.~\eqref{eq:pole-closed},
which does not diverge as $n\to\infty$ and quickly becomes completely negligible.
The leading pole is located at $u=1/2$, and its residue fixes the norm of the pole-mass renormalon $N^{\beta_0}_{1/2}$
\begin{align}\label{eq:leadingPole}
m_B(u) = N^{\beta_0}_{1 / 2} \biggl(\frac{4\pi}{\beta_0} \frac{1}{1-2u} \biggr)\! + \mathcal{O}\bigl[(1-2u)^0\bigr]\,,
\end{align}
to the $\beta_0$-independent value $N^{\beta_0}_{1 / 2}=C_{\!F} e^{5/6}/\pi\simeq 0.976564$. In Ref.~\cite{Hoang:2017suc} a sum rule to compute the
value of $N^{(n_f)}_{1 / 2}$ was derived using R-evolution. In Fig.~\ref{fig:N12} it is shown how the value of $N^{(n_f)}_{1 / 2}$ computed in full QCD using the fixed-order
relation between the pole and $\overline{\rm MS}$ masses at four loops (blue line) approaches $N^{\beta_0}_{1 / 2}$ (red dashed line)
when $n_f$ tends to $-\infty$. The ambiguity of the series can be computed in a compact form, and its leading term is independent of $\overline m$
\begin{table}[t!]
\centering
\begin{tabular}{|c|cccc|}
\hline
$n$ & $m^{\rm MSR}$ & $m^{\rm RS}$ & $m^J$ & $m^{J'}$\\
\hline
$1$ & $5.33333$ &$12.2719$ &$4.74953$ &$6.70269$\\
$2$ & $24.9928$ &$24.5437$ &$24.8031$ &$25.867$\\
$3$ & $93.9875$ &$98.175$ &$101.073$ &$100.293$\\
$4$ & $585.914$ &$589.05$ &$590.464$ &$588.872$\\
$5$ & $4660.93$ &$4712.4$ &$4710.49$ &$4711.54$\\
$6$ & $47145$ &$47124$ &$47122.2$ &$47123.3$\\
$7$ & $564343$ &$565488$ &$565483$ &$565486$\\
$8$ & $7.91983\times 10^6$ &$7.91683\times 10^6$ &$7.9168\times 10^6$ &$7.91683\times 10^6$\\
$9$ & $1.26617\times 10^8$ &$1.26669\times 10^8$ &$1.26669\times 10^8$ &$1.26669\times 10^8$\\
$10$ & $2.28034\times 10^9$ &$2.28005\times 10^9$ &$2.28005\times 10^9$ &$2.28005\times 10^9$\\
\hline
\end{tabular}
\caption{Numeric values for the non-logarithmic coefficients for various short-distance masses $a_{n,n-1}$.\label{tab:masses}}
\end{table}
\begin{align}\label{eq:MS-amb}
\delta_{\Lambda} m_p =\,& \frac{\overline m}{2\beta_0}\Biggl\{ \frac{e^{\frac{5}{6}} \Lambda_{\rm QCD}}{\overline m}
\Biggl[\! \Biggl(\frac{e^{\frac{5}{6}}\Lambda_{\rm QCD}}{\overline m}\Biggr)^{\!\!2}-2\Biggr]\!
\sqrt{4+\biggl(\frac{e^{\frac{5}{6}}\Lambda_{\rm QCD}}{\overline m}\biggr)^{\!\!2}}-\biggl(\frac{e^{\frac{5}{6}}\Lambda_{\rm QCD}}{\overline m}\biggr)^{\!\!4}\Biggr\}\\
=\,&-\!\frac{2 e^\frac{5}{6}C_F}{\beta_0}\Lambda_{\rm QCD} + \mathcal{O}\bigl(\Lambda_{\rm QCD}^3\bigr)\,,\nonumber
\end{align}
where here and in what follows we keep minus signs in ambiguities to track down potential ambiguity cancellations. Using
\mbox{$\alpha_s^{(n_f=5)}(m_Z)=0.1181$}
and continuous matching for the strong coupling at the scales
$\overline{m}_b$ and $\overline{m}_c$ we obtain \mbox{$\Lambda^{n_f=\{3,4,5\}}_{\rm{QCD}}=\{88,120,144\}\,$MeV}, which are roughly a
factor of $2$ smaller than the full QCD values quoted in Eq.~(2.25) of
Ref.~\cite{Hoang:2017btd}. The leading ambiguities in the large-$\beta_0$ limit for heavy quarks with $n_f=3,4$ and $5$ active flavors are $71\,$MeV, $89\,$MeV and $98\,$MeV,
respectively (to be compared with their full QCD counterparts $180\,$MeV, $215\,$MeV and $250\,$MeV found in \cite{Hoang:2017btd}, which are a factor of $2.5$ larger).
In Fig.~\ref{fig:AmbMin} we show with a red line the top quark mass ambiguity
computed with Eq.~\eqref{eq:MS-amb}, which is compared to the estimate obtained by the minimal correction term (gray dashed line), where we have defined
$\delta_n^{\rm MSR}(R,\mu)\equiv R \beta_\mu^n \sum_{i=0}^{n-1} a_{n,n-1,i}^{\overline {\rm MS}} \log^i (\mu/R)$.
The two estimates are in reasonable agreement.

Equation.~\eqref{eq:leadingPole} can be used to write down the relation between the pole and RS masses~\cite{Pineda:2001zq} at leading order in $1/\beta_0$
(see coefficients in the second column of Table~\ref{tab:masses}):\footnote{For simplicity we consider only the ``non-primed'' version of the RS mass.}
\begin{align}\label{eq:RSmass}
\delta_{\rm RS}(R) = m_p - m^{\rm RS} (R) \equiv &\, \frac{4RC_{\!F}\,e^{\frac{5}{6}}}{\beta_0} \!\!\int_0^{\infty} \!\frac{{\df} u}{1-2u} \,e^{- \frac{u}{\beta_R}}
= \frac{4RC_{\!F}\,e^{\frac{5}{6}}}{\beta_0} \!\!\int_0^{\infty} \!\frac{{\df} u}{1-2u} \biggl(\frac{\Lambda_{\rm QCD}}{R} \biggr)^{\!\!2 u} \nonumber \\
=&\,\frac{2RC_{\!F}e^{\frac{5}{6}}}{\beta_0} \sum_{n=1}2^n\Gamma(n) \beta^{n}
\equiv \frac{R}{\beta_0}\!\sum_{n=1} a_{n,n-1}^{\rm RS}\beta_R^n \,,
\end{align}
where in the second line we show the fixed-order relation between $m_p$ and $m^{\rm RS}$, obtained using Eq.~\eqref{eq:poleA} with $m=1$ and $u_0=1/2$. The
series has the same leading renormalon at $u=1/2$ as the $\overline{\rm MS}$ or MSR masses (see Sec.~\ref{sec:MSR}). Therefore, as can be seen in
Fig.~\ref{fig:poles}, the RS fixed-order coefficients $a_n^{\rm RS}$ (red dots) tend to $a_n^{\overline{\rm MS}}$ (in gray) defined in Eq.~\eqref{eq:MSR} for
large $n$. The difference $m^{\rm MSR} (R_1)-m^{\rm MSR} (R_2)$ is free from any renormalon ambiguity and the integration converges for $u\to\infty$ provided
that $\max(R_1,R_2) > \Lambda_{\rm QCD}$. The closed form coincides with ${\rm PV}[\delta_{\rm RS}(R_2)]-{\rm PV}[\delta_{\rm RS}(R_1)]$:
\begin{align}\label{eq:RSdiff}
m^{\rm RS} (R_1)-m^{\rm RS} (R_2) = \frac{2C_{\!F}\,e^{\frac{5}{6}}\Lambda_{\rm QCD}}{\beta_0}
\biggl[{\rm Ei} \!\biggl( \frac{1}{2 \beta_2} \biggr) - {\rm Ei}\! \biggl( \frac{1}{2
\beta_1} \biggr)\!\biggr] ,
\end{align}
with ${\rm Ei}(x)$ the exponential integral function. One can compute the R-anomalous dimension for the RS mass in the way explained in Sec.~\ref{sec:MSR},
and the result is very simple, \mbox{$\beta_0\gamma_{\rm RS}(\beta)=4C_{\!F} e^{5/6}\beta$}, with only the one-loop term $\hat\gamma_{\rm RS}^0$
non-zero. This implies that Eq.~\eqref{eq:RSdiff} is fully compatible with the result shown in Eq.~\eqref{eq:Revol-FO} truncated at leading-log, such that for the RS
mass exact and fixed-order evolution exactly coincide. To relate the RS and $\overline{\rm MS}$ masses one
can take the difference $\delta_{\overline{\rm MS}}( \overline{m})-\delta_{\rm RS}(\overline{m})$:
\begin{align}\label{eq:pole-closed}
m^{\rm RS}(\overline{m}) - \overline{m} =\,& \overline{m}\,\Bigg\{\!\int_0^{\infty} \!{\df} u \,e^{- \frac{u}{\beta_{\overline{m}}}}\!
\Biggr[\frac{F_{\overline {\rm MS}}(0, u) - F_{\overline {\rm MS}}(0, 0)}{u}
-\frac{4C_{\!F}\,e^{\frac{5}{6}}}{1-2u}\Biggl] \\
&+ \!\!\int_{- \beta_{\overline{m}}}^0 \!\!\frac{{\df} \varepsilon}{\varepsilon}\,
[F_{\overline {\rm MS}}(\varepsilon, 0) - F_{\overline {\rm MS}}(0, 0)]\!\Biggr\}\, ,\nonumber
\end{align}
with $\beta_{\overline{m}} = \beta(\overline{m})$. The matching relation above is free from the $u=1/2$ renormalon, but has higher order singularities.
After determining $m^{\rm RS}(\overline{m})$ from Eq.~\eqref{eq:pole-closed} the RS mass can be computed at any other scale $R$ using Eq.~\eqref{eq:RSdiff}.

\subsection{The MSR mass its and R-evolution}\label{sec:MSR}
The MSR scheme~\cite{Hoang:2008yj,Hoang:2017suc} is defined from the series relating the pole and
${\rm \overline{MS}}$ masses $\delta_{\overline{\rm MS}}( \overline{m})$:
\begin{align}\label{eq:MSR}
\!\delta_{\rm MSR}(R)=\,&m_p - m^{\rm MSR} (R) =
\frac{R}{\beta_0} \!\sum_{n=1} a_{n,n-1}^{\overline {\rm MS}} \beta_R^n
= R \!\sum_{n=1} \beta_\mu^n
\sum_{i=0}^{n-1} a_{n,n-1,i}^{\overline {\rm MS}} \log^i\!\Bigl(\frac{\mu}{R}\Bigr)\\[-0.1cm]
=\,& \frac{R}{\beta_0}\biggl\{\int_0^{\infty} \!\frac{{\df} u}{u} \,e^{- \frac{u}{\beta_R}}
[F_{\overline {\rm MS}}(0, u) - F_{\overline {\rm MS}}(0, 0)] + \!\!\int_{- \beta_R}^0 \!\!\!\frac{{\df} \varepsilon}{\varepsilon}\,
[F_{\overline {\rm MS}}(\varepsilon, 0) - F_{\overline {\rm MS}}(0, 0)]\!\biggr\},\nonumber
\end{align}
where in the second line we have written an exact form in the large-$\beta_0$ limit with \mbox{$\beta_R=\beta(R)$}.
The fixed-order series us expanded in powers of $\beta_R$ or \mbox{$\beta_\mu=\beta(\mu)$} in the first line, where
$a_{n,n-1,i}^{\overline {\rm MS}}$ are the coefficients that appear in the fixed-order expansion of $\delta_{\overline{\rm MS}}$.
The definition above implies $m^{\rm MSR}(\overline{m})=\overline{m}$, such that it is convenient to take $R=\overline{m}$
as a boundary condition to determine the MSR mass at any other scale $R$. When taking the difference of MSR masses at different values of $R$,
\mbox{$\Delta^{\rm MSR} (R_2, R_1)\equiv m^{\rm MSR}(R_2)-m^{\rm MSR} (R_1)$}, the inverse-Borel term will not cancel. However, due to the $R$ prefactor in
Eq.~\eqref{eq:MSR}, the integrand's pole at $u=1/2$ becomes $R$-independent and thus vanishes when subtracting two series, making the difference less
ambiguous than each series taken by itself.
\begin{figure*}[t!]
\subfigure[]
{\includegraphics[width=0.465\textwidth]{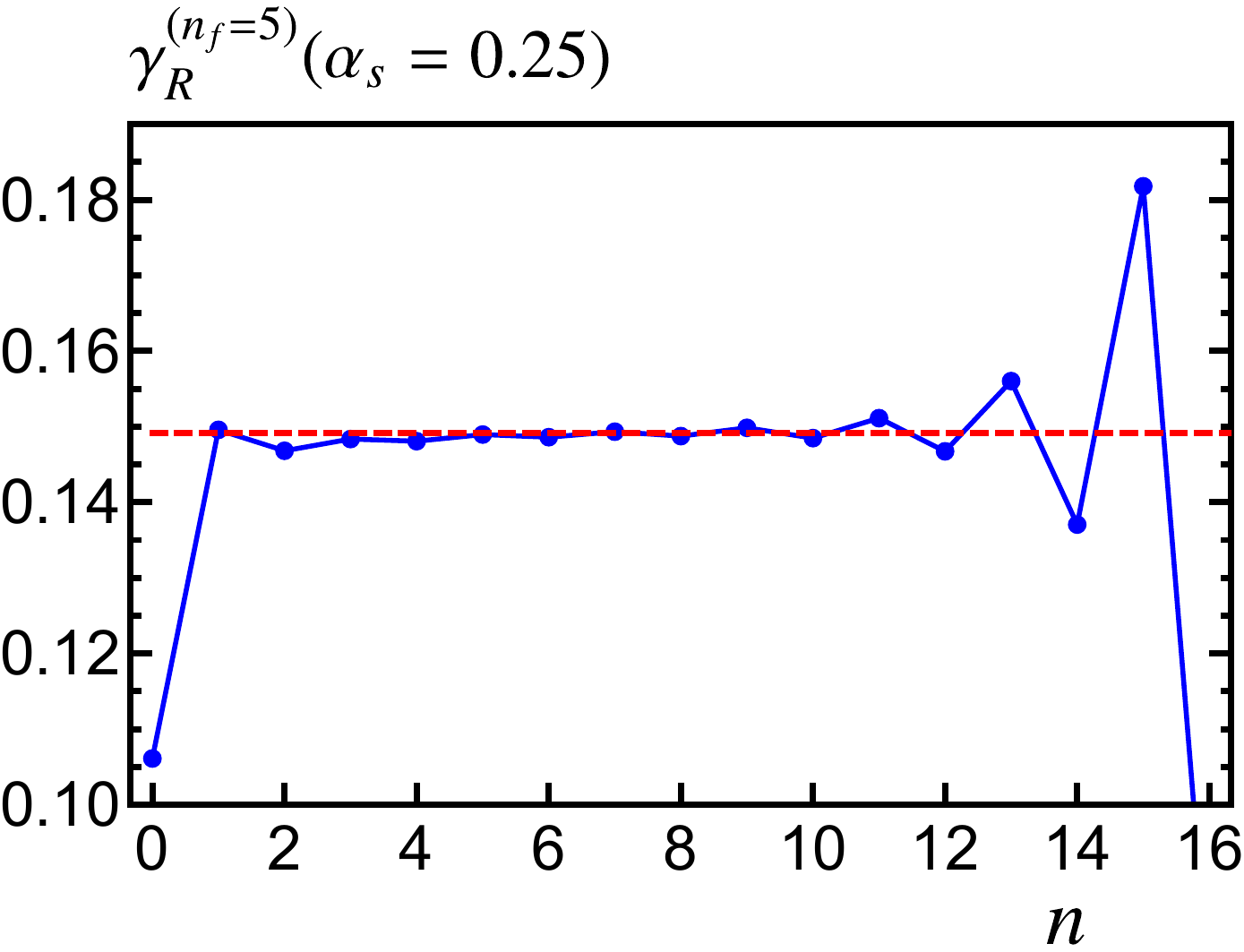}
\label{fig:MSRgamma}}~~~~
\subfigure[]{\includegraphics[width=0.46\textwidth]{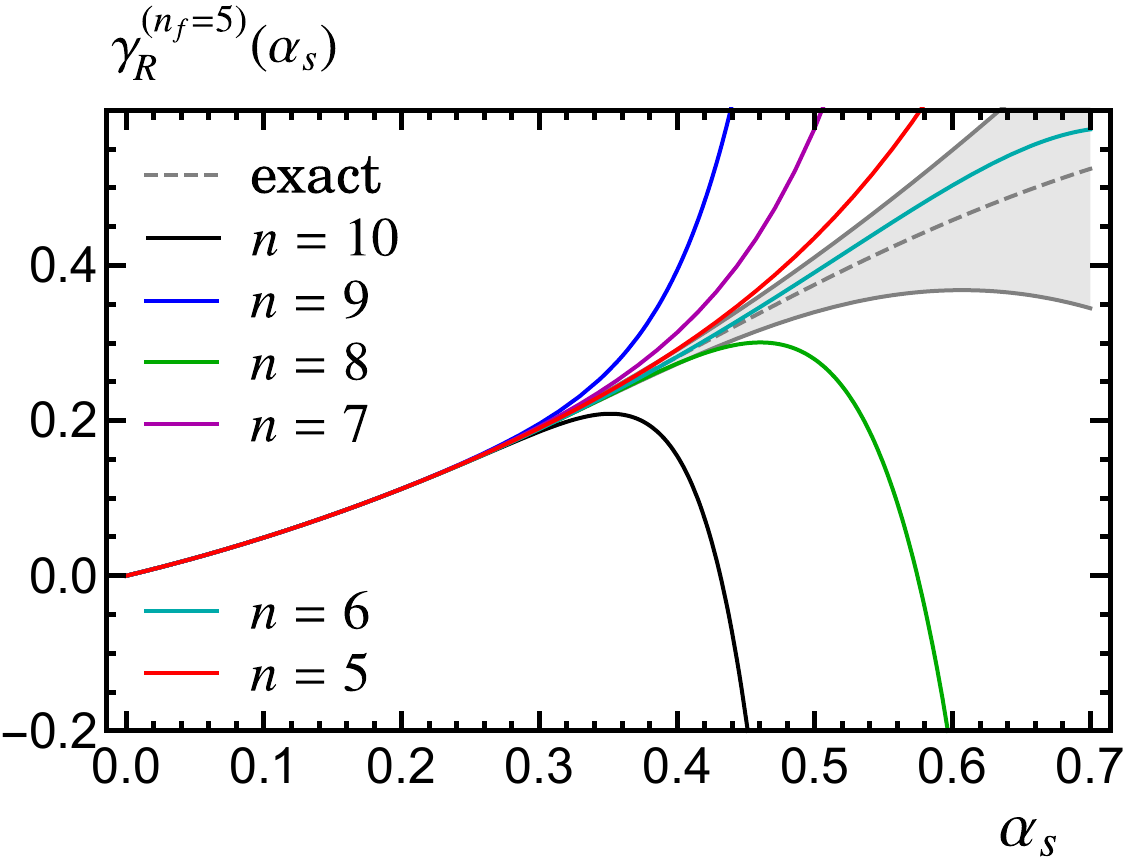}
\label{fig:gRAlpha}}
\caption{Left panel: comparison of the R-anomalous dimension $\gamma_R(\alpha_s=0.25)$ computed with a partial sum including $n+1$ terms
(blue dots) or through the closed integral form (red dashed line).
Right panel: Dependence of $\gamma_R$ with $\alpha_s$ in its exact form (gray band) and when including up to $\gamma_R^n$
with $n=5,6,7,8,9,10$ in red, cyan, magenta, green, blue and black, respectively.}
\label{fig:gammaR}
\end{figure*}

Taking into account that the pole mass is $R$-independent, one can compute a closed form for the MSR mass R-anomalous dimension using the
second line of Eq.~\eqref{eq:MSR}:
\begin{align}\label{eq:MSRgammaExact}
\gamma_R (\beta_R) =& -\! \frac{{\df} m^{\rm MSR} (R)}{{\df} R} = \frac{1}{\beta_0}\biggl\{\int_0^{\infty} \!\frac{{\df} u}{u}(1-2u) e^{- \frac{u}{\beta_R}}
\bigl[F_{\overline {\rm MS}}(0, u) - F_{\overline {\rm MS}}(0, 0)\bigr]\\[-0.1cm]
& + \!\!\int_{- \beta_R}^0 \!\!\!\frac{{\df} \varepsilon}{\varepsilon}
\bigl[F_{\overline {\rm MS}}(\varepsilon, 0) - F_{\overline {\rm MS}}(0, 0)\bigr] +
2 \beta_R \bigl[F_{\overline {\rm MS}}(-\beta_R, 0) - F_{\overline {\rm MS}}(0, 0)\bigr]\!\biggr\},\nonumber
\end{align}
where the pole at $u=1/2$ in the integrand of the first-line cancels, pushing the leading renormalon to $u=3/2$, thus making $\gamma_R$
less ambiguous than $\delta_{\rm MSR}$. The corresponding fixed-order coefficients $\gamma_R^n$, collected in the second
column of Table~\ref{tab:gammaR}, can be obtained
from to the perturbative expression for $\delta_{\rm MSR}$:
\begin{align}
\beta_0 \gamma_R (\beta) \equiv\,& \sum_{n=0}\hat \gamma_R^n\, \beta^{n+1} =
\sum_{n=0}\Bigl(a^{\rm \overline{MS}}_{n+1,n} -2 n \, a^{\rm \overline{MS}}_{n,n-1}\Bigr) \beta^{n+1}\,, \label{eq:gammaR} \\[-0.2cm]
\hat \gamma_R^{n} = &\, n! \Bigl[F^{\overline {\rm MS}}_{0,n+1} - 2(1-\delta_{n,0})F^{\overline {\rm MS}}_{0,n}\,\Bigr] +
(-1)^n\!\Biggl[\frac{ F^{\overline {\rm MS}}_{n+1,0}}{n+1} + 2 (1-\delta_{n,0}) F^{\overline {\rm MS}}_{n,0}\Bigg],\nonumber
\end{align}
reproducing the leading flavor structure of full QCD up to four loops~\cite{Hoang:2017suc}. In Fig~\ref{fig:MSRgamma} we compare the exact form for
$\gamma_R$ (red dashed line) in Eq.~\eqref{eq:MSRgammaExact} with the partial sum
of Eq.~\eqref{eq:gammaR} at various orders for a relatively large value of $\alpha_s$. We observe the expected behavior: the partial sum oscillates
around the exact value until $n\simeq 10$, and starts diverging after that order. In Fig.~\ref{fig:gRAlpha} we show the dependence of the closed form for $\gamma_R$
with $\alpha_s$ (gray band) with the partial sum up to $\gamma_R^n$ for $5\leq n\leq 10$.

One can integrate the R-anomalous RGE equation using $\gamma_R$ as given in Eq.~\eqref{eq:gammaR}. Switching the integration variable
from $R$ to $\beta$ through $\Lambda_{\rm{QCD}}$ using Eq.~\eqref{eq:betaRun} and reversing the integration order, the $\beta$
integrals can be carried out analytically. Following this procedure one of course finds
$\Delta^{\rm MSR} (R_2, R_1)=\delta_{\rm MSR}(R_1)- \delta_{\rm MSR}(R_2)$, with $\delta_{\rm MSR}$ defined in Eq.~\eqref{eq:MSR}.
This is an important cross check for our derivation.
For completeness we provide the perturbative R-evolution running kernel in the \mbox{large-$\beta_0$} limit, simply integrating each term
in the fixed-order expansion for $\gamma_R$:
\begin{align}\label{eq:Revol-FO}
&2\beta_0 \Delta^{\rm MSR} (R_2, R_1) = \Lambda_{\rm{QCD}} \sum_{n = 1} \hat\gamma^{n - 1}_R \!
\biggl[ \beta_2^{n - 1} E_n \!\biggl( - \frac{1}{2 \beta_2} \biggr) -
\beta_1^{n - 1} E_n \!\biggl( - \frac{1}{2 \beta_1} \biggr)\! \biggr] \\
& \qquad = \Lambda_{\rm{QCD}}\!\biggl[{\rm Ei} \!\biggl( \frac{1}{2 \beta_2} \biggr) - {\rm Ei}\! \biggl( \frac{1}{2
\beta_1} \biggr)\!\biggr] \!\sum_{n = 0} \frac{\hat\gamma^n_R}{2^{n} n!} + \sum_{i = 1} \Gamma(i)\!
\Bigl[ R_2 (2 \beta_2)^i - R_1 (2\beta_0)^i \Bigr]\! \sum_{n = i} \frac{\hat\gamma^n_R}{2^{n} n!}\,,\nonumber
\end{align}
with $E_n (x)=\int_1^{\infty} {\rm d} t\,e^{- xt}/t^n $. To obtain the result in the second line
integration by parts has been used to relate $I_n(\beta)\equiv \int \!{\df} \beta\,\beta^{n-1} \exp[-1/(2\beta)]$ for consecutive values
of $n$. Iterating this relation until $n=0$ each $E_n (-x)$ can be written in terms of ${\rm Ei}(x)$ and a finite sum.\footnote{This result is
completely analogous to the strategy followed by the \texttt{C++} public code REvolver~\cite{Hoang:2021fhn}.} Combining this last sum with the one over
$\hat\gamma_R^n$ one arrives at the final expression. If the sum includes up to $\hat\gamma_R^n$ the result is N$^n$LL accurate. In Fig~\ref{fig:MSR}
we compare the value $m_t^{\rm MSR}(R)$ computed through Eq.~\eqref{eq:MSR} (cyan dashed line)
with the R-evolved result obtained with Eq.~\eqref{eq:Revol-FO} at various orders (black joined dots) for a small value of $R$.

The pole mass value estimated from the fixed-order expansion of $\delta_{\rm MSR}(R)$ together with $m^{\rm MSR}(R)$ obtained with perturbative R-evolution
depends at any finite order on $R$.
This matching relation can be expressed in powers of $\beta_\mu$,
and as long as $\mu/R\sim\mathcal{O}(1)$ there are no large logarithms in the expansion. The numerical value of $m_p$ will also depend on $\mu$
if the series is truncated.
The coefficients $\hat a_{n,n-1,i}^{\overline {\rm MS}}$ can be easily obtained
demanding that the series is \mbox{$\mu$-independent}. This provides recursion relation that can be solved exactly:\footnote{See e.g.\ the last first equality in the last line of
Eq.~\eqref{eq:recGA}. The recursion relation in full QCD has been presented in Ref.~\cite{Mateu:2018zym}.}
\begin{equation}
a_{n,n-1,i}^{\overline {\rm MS}} = \frac{2 (n - 1)}{i} \, a_{n-1,n-2,i-1}^{\overline {\rm MS}} =
2^i\binom{n-1}{i} a_{n-i, n-i-1}^{\overline {\rm MS}} \,,
\end{equation}
where the first equality is used $i$ times until one reaches $i=0$ to obtain the second. This result is equivalent to Eq.~\eqref{eq:expAlpha}, derived
re-expanding $\beta_R$ in powers of $\beta_\mu$ and $\ell$. For the RS mass (or equivalently for the MSR mass if $n$ is large enough) this
sum can be carried out exactly, and using Eq.~\eqref{eq:aAsyMu} for a simple pole located at $u=1/2$ we find
\begin{equation}\label{eq:RSmu}
R\!\sum_{i=0}^{n-1} a_{n,n-1,i}^{\rm RS} \log^i\!\Bigl(\frac{\mu}{R}\Bigr)
\xrightarrow[n \to \infty]{} 2^{n+1}e^{\frac{5}{3}}C_F\mu\,\Gamma(n) \,.
\end{equation}
Therefore, at large orders $R$ gets effectively replaced by $\mu$
in Eq.~\eqref{eq:RSmass} if we choose to express the series in powers of $\beta_\mu$. This makes clear that when two series with the same leading renormalon
are subtracted, the ambiguity will cancel only if both are expressed in terms of $\alpha_s$ evaluated at the same renormalization scale $\mu$. At the sight of
Eq.~\eqref{eq:aAsyMu}, and given that at large orders perturbative coefficients are dominated by the lowest singularities of the Borel transform, the same
conclusion can be drawn from poles
located anywhere in the positive real axis and for series carrying an anomalous dimension.

\begin{figure*}[t!]
\subfigure[]
{\includegraphics[width=0.465\textwidth]{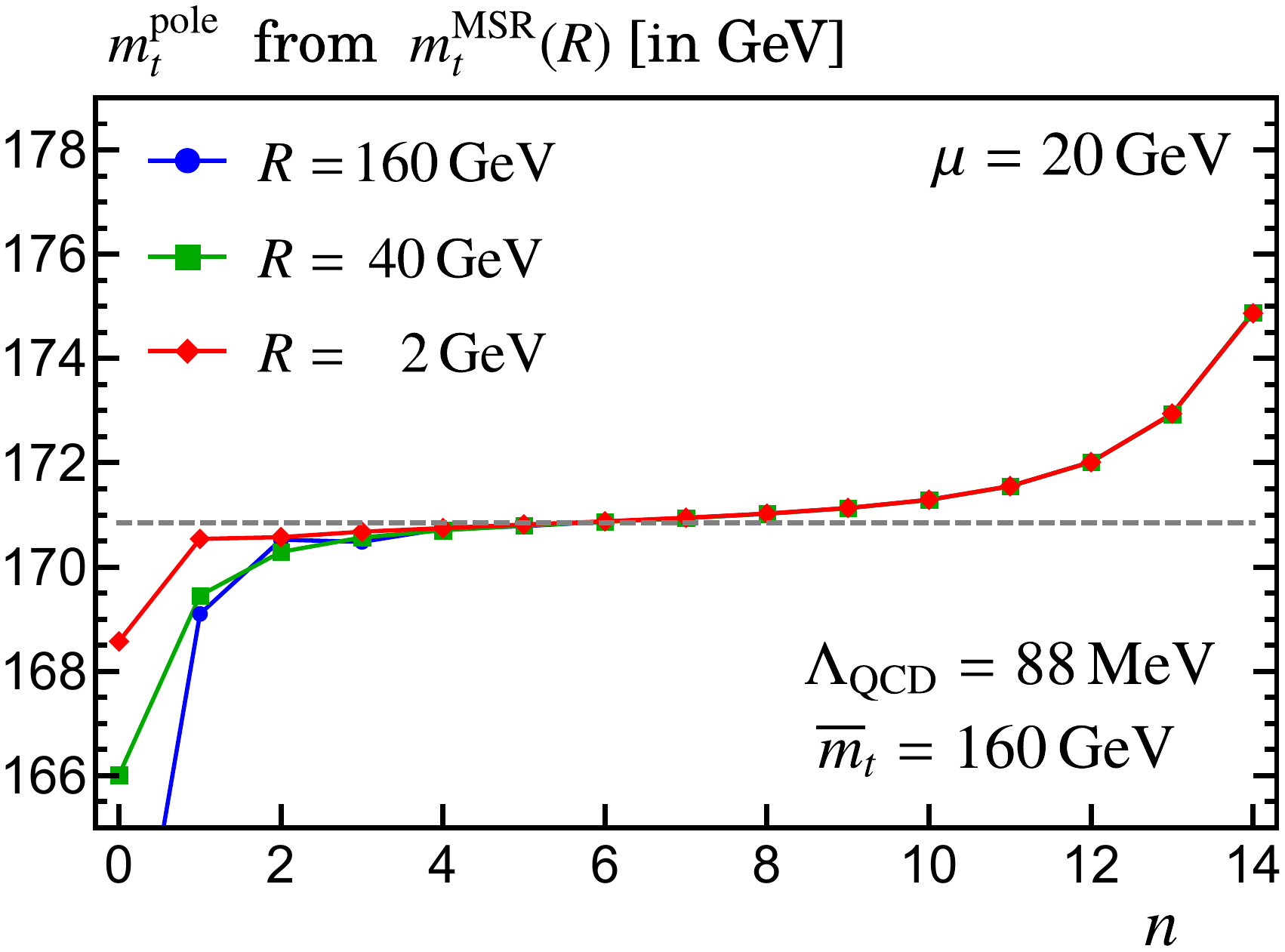}
\label{fig:poleR}}~~~~
\subfigure[]{\includegraphics[width=0.46\textwidth]{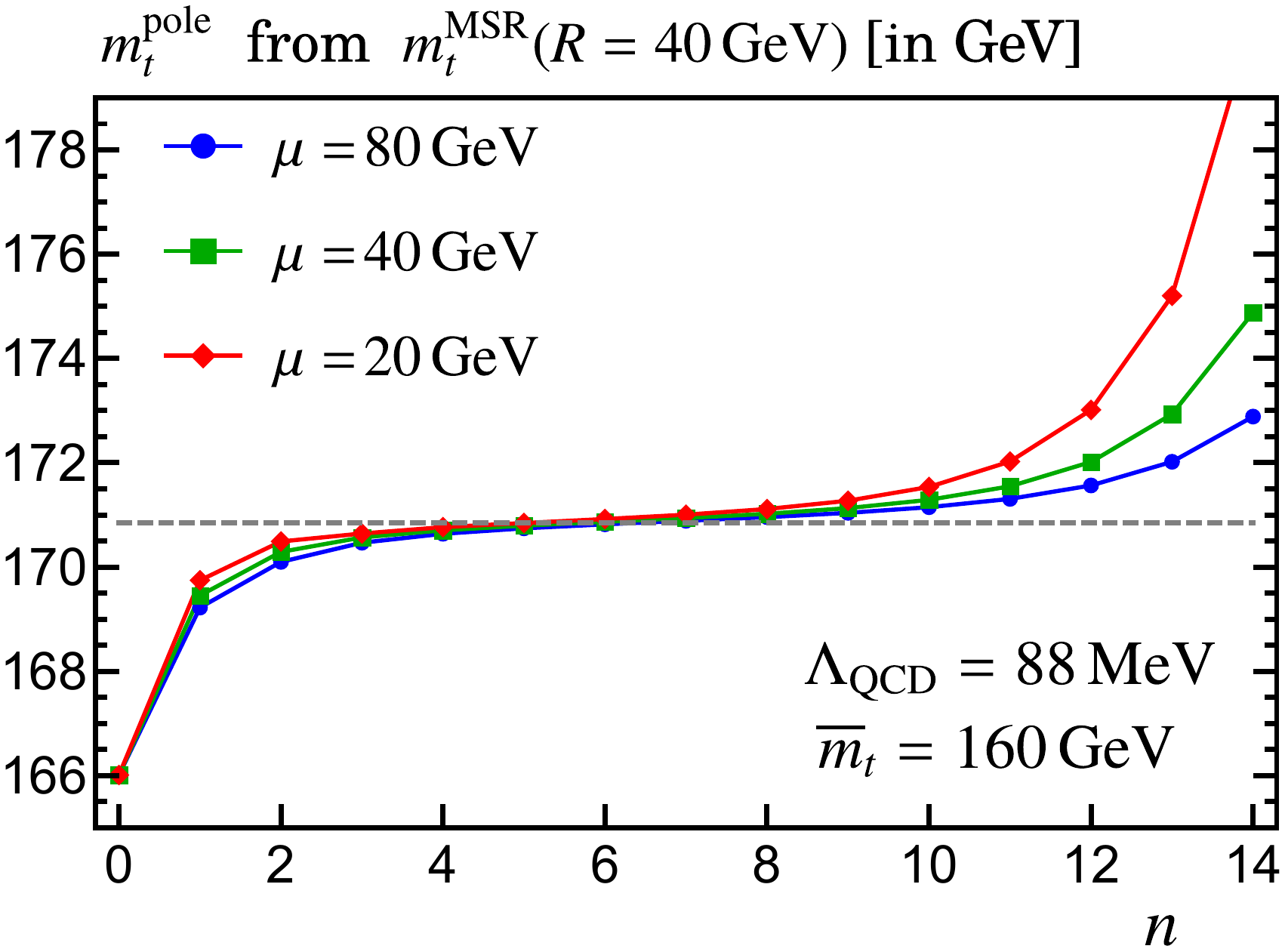}
\label{fig:poleMu}}
\caption{Determination of the top quark pole mass $m_t^{\rm pole}$
from its fixed order relation to the MSR mass, using as boundary condition $R=\overline{m}_t$.
In both panels the gray dashed line corresponds to the result computed as $\overline{m}_t+{\rm PV}[\delta_\MSBar(\overline{m}_t)]$.
The colored dots use $n$ terms in the fixed-order $\delta_{\rm MSR}$ series and R-evolution at N$^n$LL. In the left panel we keep $\mu=20\,$GeV and show $R=2\,$GeV,
$40\,$GeV and $160\,$GeV in red, green and blue, respectively. In the right panel we keep $R=40\,$GeV but use $\mu=20$\,GeV, $40\,$GeV and $80\,$GeV in
the same color ordering.} \label{fig:pole}
\end{figure*}
In Fig.~\ref{fig:pole} the pole mass is estimated from its perturbative relation to the MSR mass. The colored dots use N$^n$LL-accurate R-evolution
to compute $m_t^{\rm MSR}(R)$, and the fixed-order relation between the pole and MSR masses including $n$ terms. In the left panel we fix $\mu$ and pick three
values of $R$.
We observe that for small $R$ the exact value of the pole mass, computed as $\overline{m}_t+{\rm PV}[\delta_\MSBar(\overline{m}_t)]$,
is properly estimated \mbox{---\,within} perturbative uncertainties\,---
at low orders (since there are somewhat large logarithms for the choice $R=2\,$GeV, the partial sum does not exactly hit the exact value for low $n$).
For larger $R$ it takes more orders to adequately predict the pole mass.
The explanation for this behavior is simple: since the MSR mass at small values of $R$ is closer to the pole mass (formally both coincide if $R=0$) with a few matching
corrections one already predicts its value. On the other hand, the large-order behavior depends only on $\mu$.
In the right panel we keep $R$ fixed but change the value of $\mu$
(green dots are identical in both panels). The three series become flat more or less at the same time (since this depends only on $R$)
but for smaller $\mu$ the divergent behavior for $n>8$ is more pronounced.

\section{SCET computations}\label{sec:SCET}
Hadrons produced in $e^+ e^-$ collisions at high energies often adopt di-jet configurations. Jets are sprays of collimated particles traveling very fast in
nearly the same direction. In such configurations, final-state particles are either energetic and belong to a jet (collinear) or have small energies, populating
the phase-space region in between the jets (soft). Event shapes are kinematic variables constructed from the energy and momentum of the produced particles
that measure how di-jet an event looks from geometric considerations. In di-jet configurations one can use SCET
to derive factorization theorems that are valid at leading order in the power-expansion parameter of the theory. For thrust one has:
\begin{align}\label{eq:factSCET}
\frac{1}{\sigma_0} \frac{{\df} \sigma}{{\df} \tau} =&\, Q^2 H_Q (Q, \mu)\!\int \!{\df} \ell J_{\tau} (Q^2 \tau - Q \ell, \mu) S_{\tau} (\ell, \mu)\,,\\[-0.1cm]
J_{\tau} (s, \mu) \equiv& \int_0^s \!{\df} s' J_n (s - s', \mu) J_n (s', \mu)\,.\nonumber
\end{align}
In this section we compute the large-$\beta_0$ limit of two matrix elements that appear in Eq.~\eqref{eq:factSCET}: the hard matching coefficient $H_Q$
(common to all jet observables) and the hemisphere jet function $J_n$ (relevant for thrust, heavy-jet-mass and C-parameter). The computation corresponds
to one-loop diagrams with a modified gluon propagator, and since SCET matrix elements have cusp anomalous dimension, the technology developed in Sec.~\ref{sec:SCETren}
will be applied.

\subsection{Hard function}\label{sec:hard}
\begin{figure}[t]\centering
\includegraphics[width=0.15\textwidth]{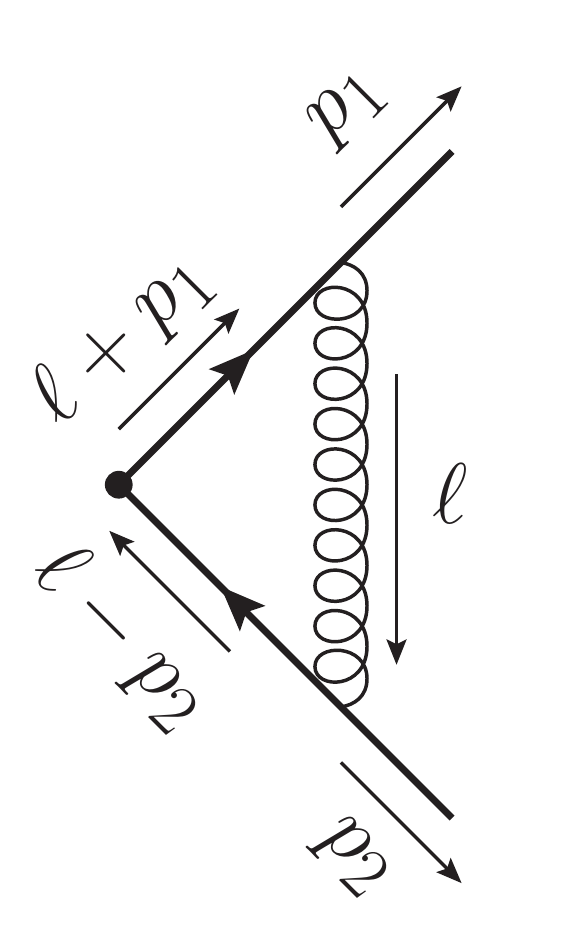}
\caption{Vector form factor for massless quarks with on-shell momenta $p_1$ and $p_2$ and virtual loop momentum $\ell$.\label{fig:hard}}
\end{figure}

The hard function $H_Q(Q,\mu)$ is the modulus squared of the matching coefficient $C_{\!H}(Q^2,\mu)$ between SCET and full QCD di-jet operators. For its computation
one needs to determine some matrix element (or Green function) of the vector current operator in both theories. Since $C_{\!H}$ does not depend
on the specific process used for its computation one usually chooses the simplest possibility, which in this case is the matrix element between
vacuum and a pair of on-shell quarks. This choice implies that all diagrams in SCET as well as self-energy diagrams in QCD are scaleless and
vanish. The QCD computation is often referred to as the quark vector form factor. Therefore, to obtain $H_Q$
at leading order in the large-$\beta_0$ expansion one only needs to compute the diagram shown in Fig.~\ref{fig:hard} with a modified gluon propagator. The virtual
correction $\delta C_H(Q)$ to the bare vector form factor $C_H=1+\delta C_H$ is then
\begin{align}\label{eq:CQ}
\delta C_H (Q^2) \gamma^{\mu} = &\, i g_0^2 C_{\!F} \!\!\int\!\! \frac{{\df}^d \ell}{(2 \pi)^d} \frac{\gamma^{\alpha}
( \ell \! \! / + p_1 \! \!\! \! \! /\,\,)
\gamma^{\mu} ( \ell \! \! / - p_2 \! \!\! \!\! /\,\, ) \gamma_{\alpha}}{(- \ell^2)^{1 + h} (\ell + p_1)^2
(\ell - p_2)^2} \,,\\[0.1cm]
\delta C_H (Q^2, \mu) = &\!
-\! \frac{2C_{\!F} g_0^2}{(4 \pi)^{2-\varepsilon}} \frac{\Gamma^2 (- u - \varepsilon) \Gamma (1 + u +
\varepsilon)}{\Gamma (3 - h - 2 \varepsilon) (- Q^2)^{h+\varepsilon}}\Bigl \{\! 2 - \varepsilon \bigr[3
+ h^2 - h (2 - 3 \varepsilon) - \varepsilon (3 - 2 \varepsilon)\bigr]\! \Bigr\}.\nonumber
\end{align}
From this result one can obtain the virtual correction to the bare hard matching coefficient defined as
$H_Q(Q) = |C_{\!H}(Q^2+i 0^+)|^2 \equiv 1 + \delta H_Q(Q) + \mathcal{O}(1/\beta_0^2)$.
For $e^+e^-$ even shapes one has $Q^2>0$ and keeping only terms linear in $1/\beta_0$ the correction $\delta H_Q$ equals twice the real part of $\delta C_H$:
\begin{equation}
\delta H_Q (Q) = 2 \delta C_H (-Q^2) \cos [\pi (h + \varepsilon)]\,.
\end{equation}
\!\!To obtain the cusp and non-cusp anomalous dimensions it is enough to consider $C_H (Q^2)$. Identifying $b(\varepsilon,h)$ from Eq.~\eqref{eq:CQ}
with ${\cal Q}^2 = -Q^2$ we obtain
\begin{equation}
G_Q(\varepsilon,u) = - 2 C_{\!F} D (\varepsilon)^{\frac{u}{\varepsilon} - 1}
\frac{\Gamma^2 (1 - u) \Gamma (1 + u) e^{\varepsilon \gamma_E}}{\Gamma (3 - u - \varepsilon)} \Bigl\{ 2 - \varepsilon \bigl[3 - \varepsilon + u (u + \varepsilon - 2) \bigr] \!\Bigr\},
\end{equation}
\begin{figure*}[t!]
\subfigure[]
{\includegraphics[width=0.46\textwidth]{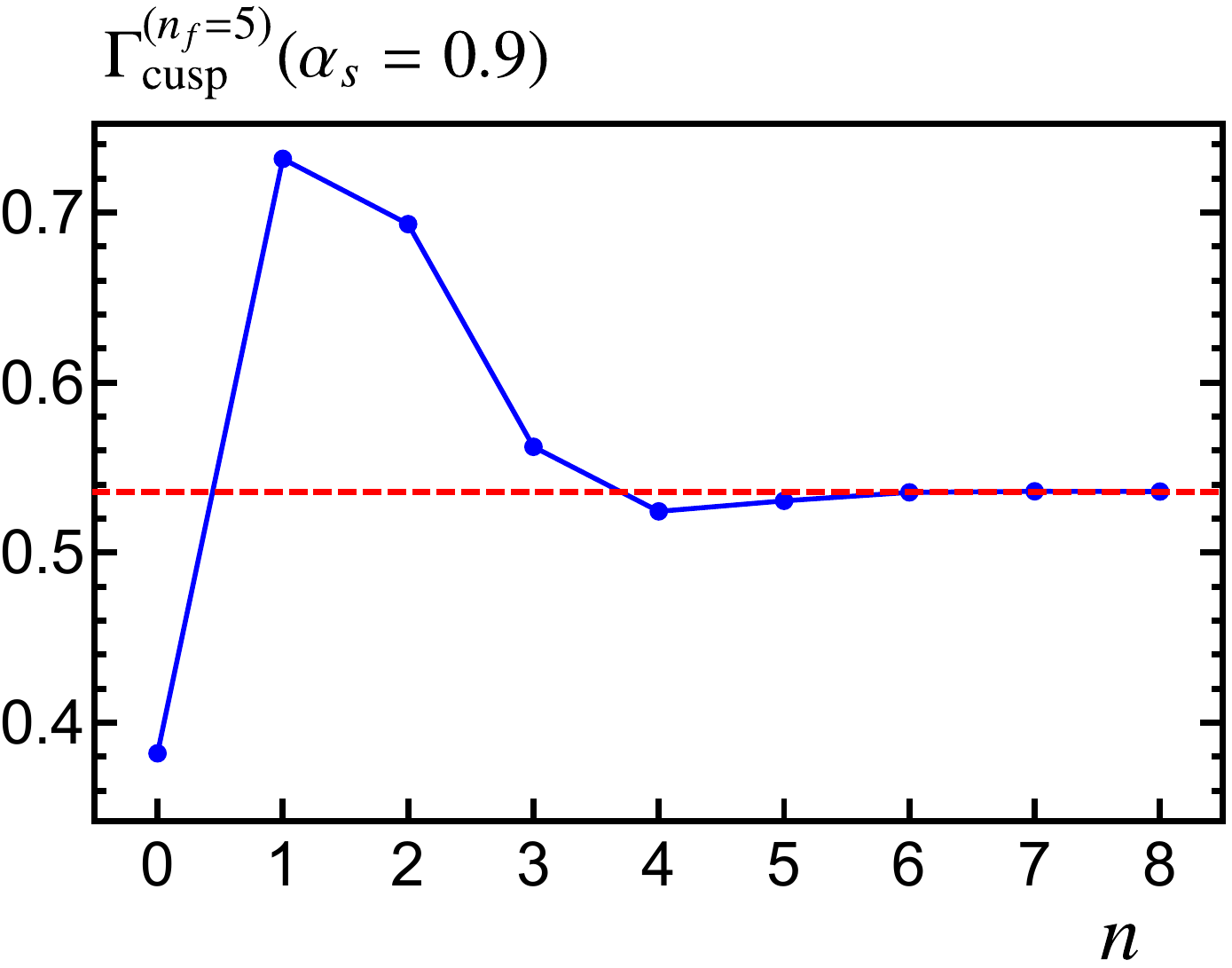}
\label{fig:cusp}}~~~~
\subfigure[]{\includegraphics[width=0.47\textwidth]{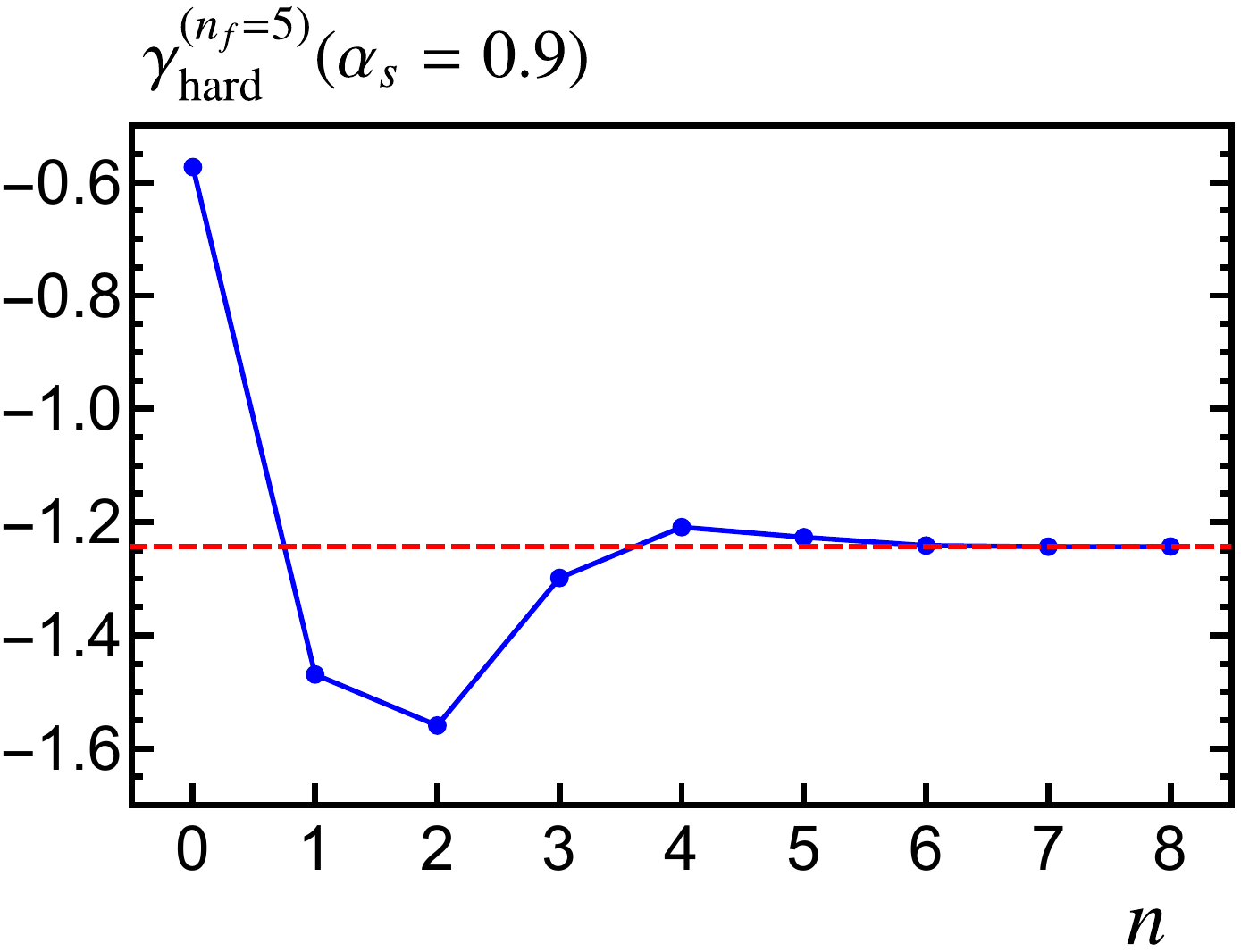}
\label{fig:gammaH}}\vspace*{-0.3cm}
\subfigure[]
{\includegraphics[width=0.46\textwidth]{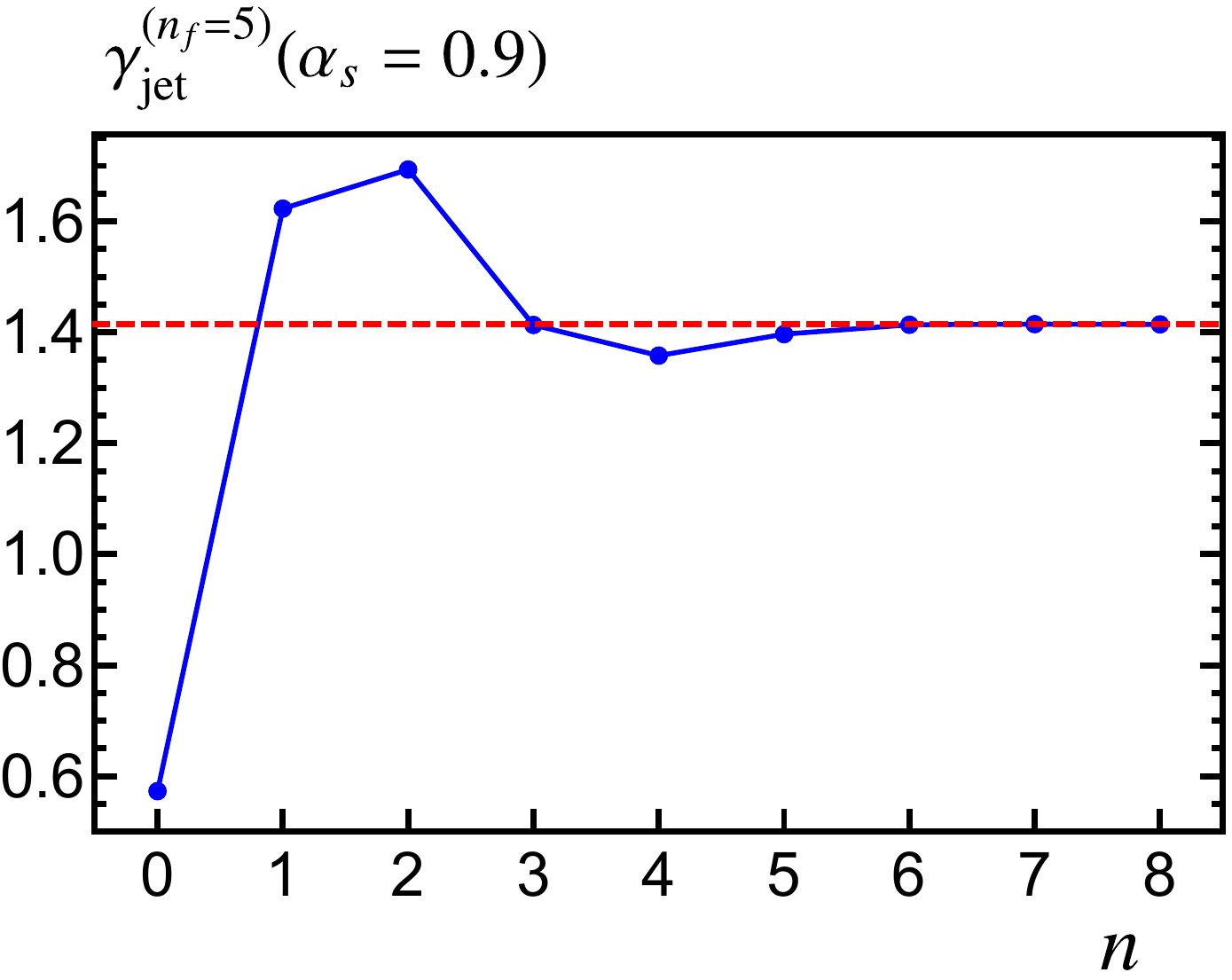}
\label{fig:gammaJ}}~~~~
\subfigure[]{\includegraphics[width=0.48\textwidth]{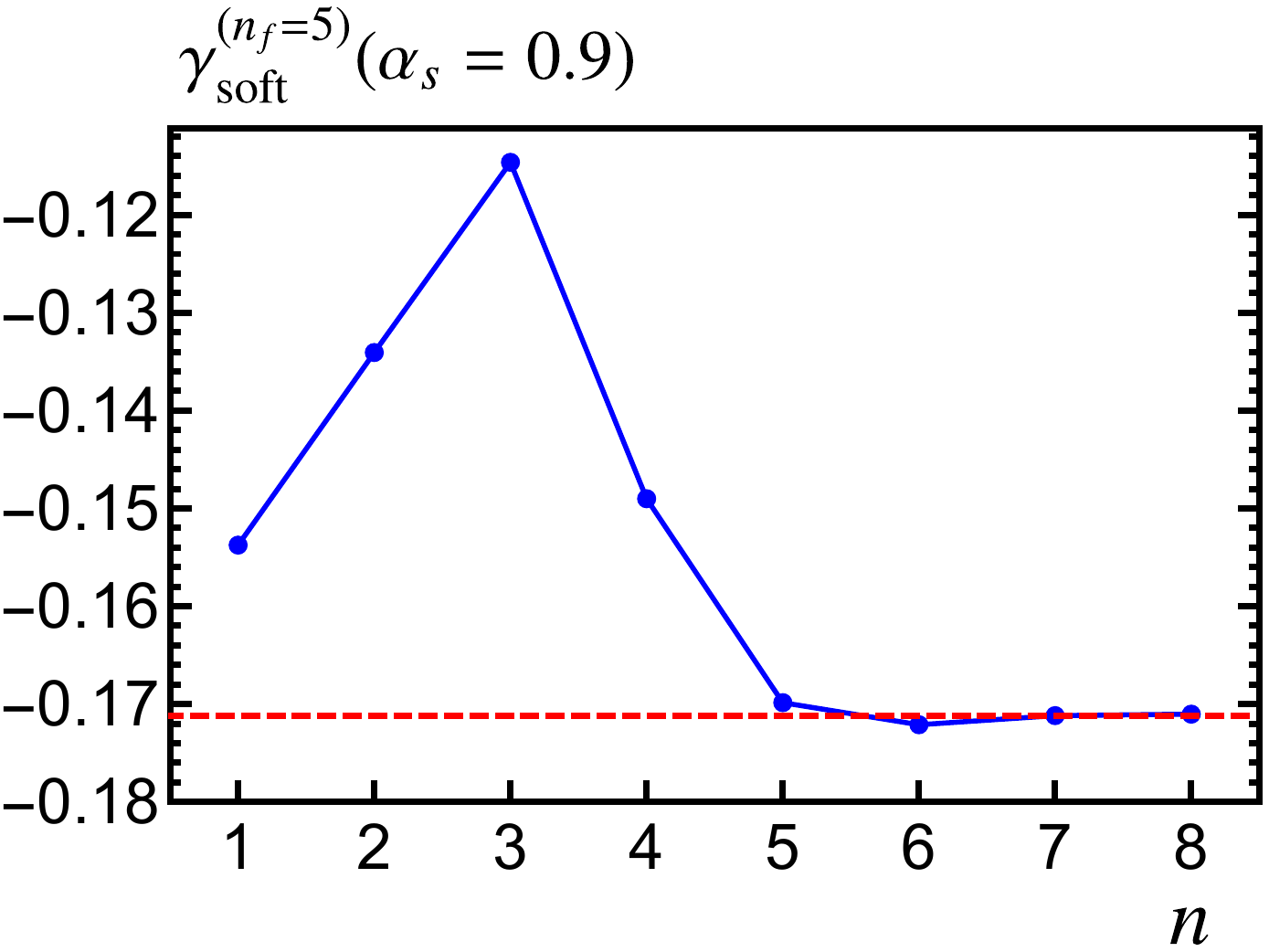}
\label{fig:gammaS}}
\caption{Comparison of the fixed-order partial sum with $n+1$ terms (blue dots) and exact results (red dashed line) for the cusp [\,panel (a)\,] and SCET non-cusp anomalous dimensions:
hard, jet, and soft [\,panels (b), (c), and (d), respectively\,], in the large-$\beta_0$ approximation for $\alpha_s=0.9$.} \label{fig:Hard}
\end{figure*}
and taking into account that $\Gamma_{\!C_Q} = -2 \Gamma_{\rm \!cusp}$ a closed form for the cusp anomalous dimension is obtained from the
equation above:
\begin{align}\label{eq:cusp-closed}
\Gamma_{\!\rm cusp}(\beta) = \,&-\!\frac{2 \beta}{\beta_0} G_Q(- \beta, 0) = \frac{2 C_{\!F}}{3 \pi} \frac{\sin (\pi
\beta) \Gamma (4 + 2 \beta)}{\beta_0 \Gamma (2 + \beta)^2}\\
= \,&\frac{4 \beta C_{\!F}}{\beta_0} \exp \!\biggl[\frac{5}{3} \beta + \sum_{n = 2}
\frac{(- \beta)^n}{n} \Bigl\{ 1 - 2^n (1 + 3^{- n}) + \zeta_n [2^n - 3 - (- 1)^n] \Bigr\} \!\biggr],\nonumber
\end{align}
where the second line has been partially expanded to easily compute the fixed-order coefficients recursively. Our result agrees with Ref.~\cite{Scimemi:2016ffw},
and the fixed-order coefficients
reproduce the leading flavor structure of full-QCD up to
$\mathcal{O}(\alpha_s^4)$~\cite{Korchemsky:1987wg,Moch:2004pa,Henn:2019swt,Huber:2019fxe}, collected in the first column of Table~\ref{tab:gamma}.
The convergence radius of $\Gamma_{\!\rm cusp}$
is set by the distance to the pole closets to the origin of the function $G_Q(- \beta, 0)$, which happens to be at $\beta = -2.5$. Furthermore, $\Gamma_{\!\rm cusp}(\beta)$ has no
singularities for positive $\beta$, and the series is, as expected, free from renormalons.
In Fig.~\ref{fig:cusp} we compare the exact result for the cusp anomalous dimension
to the partial sum of the fixed-order expansion up to $9$ loops. Even though we use an unphysically
large value for the strong coupling $\alpha_s=0.9$, the fixed-order expansion converges to the exact value.

\begin{figure*}[t]\centering
\subfigure[]
{\includegraphics[width=0.47\textwidth]{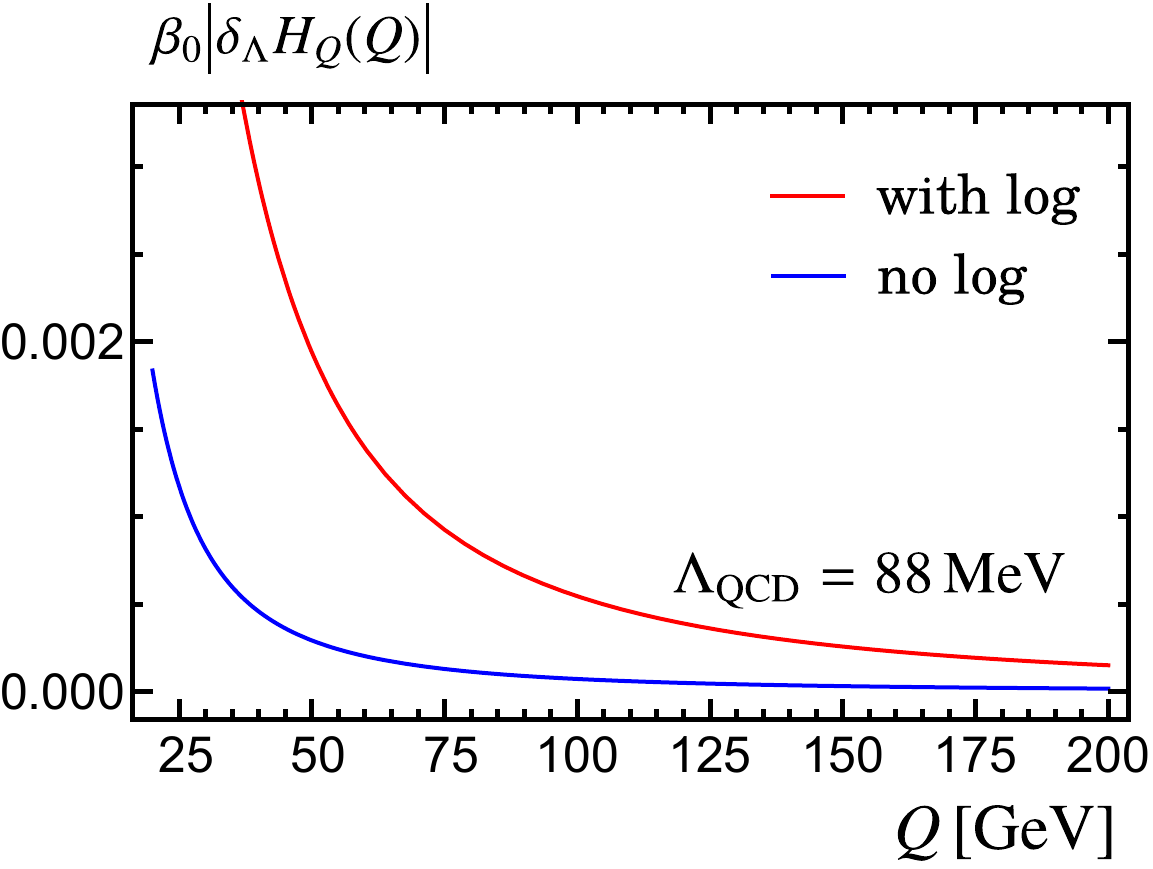}
\label{fig:hardAmb}}~~~~
\subfigure[]{\includegraphics[width=0.46\textwidth]{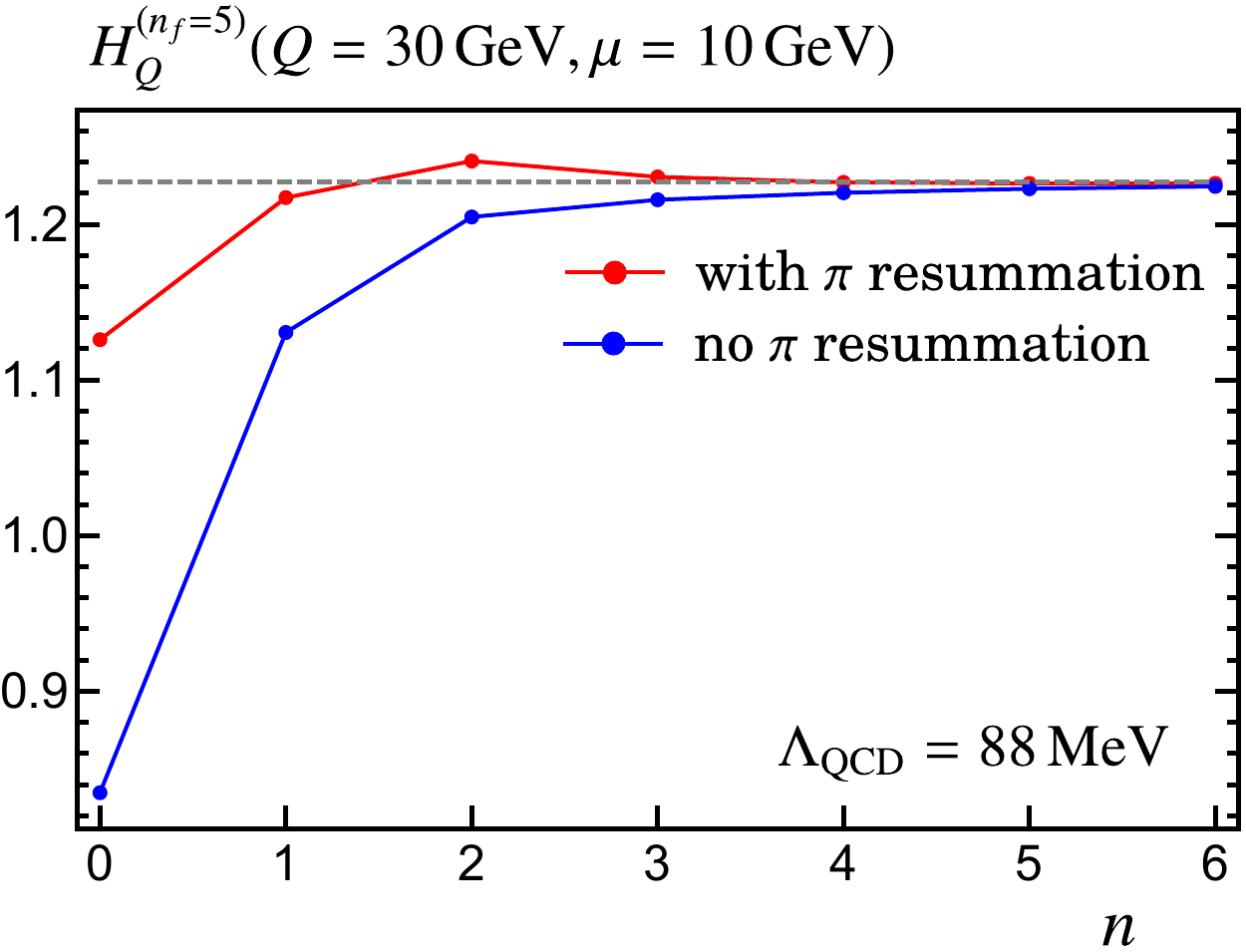}
\label{fig:piSum}}
\caption{Left panel: dependence of the SCET hard function
ambiguity (in absolute value) at leading order in $1/\beta_0$ with the center-of-mass energy $Q$.
The red line is the exact result, while in blue we neglect the logarithmic term [\,see Eq.~\eqref{eq:hardAmb}\,]. Right panel: comparison of the exact
result (gray dashed line) for $H_Q(Q = 30\,{\rm GeV}, \mu=10\,{\rm GeV})$
with the fixed-order expansion including $n+1$ terms in the
matching and resummation at N$^n$LL. Red dots include $\pi$-resummation (with $\mu_0=-iQ$), while blue dots
do not (using $\mu_0 = Q$). \label{fig:hardOrd}}
\end{figure*}
To determine the non-cusp hard anomalous dimension $\gamma_H(\beta)$ the following function is needed:
\begin{align}\label{eq:gH}
\frac{{\df}G_Q(-\beta,u)}{{\df} u}
\biggr|_{u = 0} =\,& \frac{C_F \Gamma (4 + 2 \beta)}{3 \beta\, \Gamma (2 + \beta)^3 \Gamma (1 - \beta)} \biggl\{\beta^2 - (1 + \beta) \log\! \biggl[
\frac{e^{2 \beta \gamma_E} \Gamma (4 + 2 \beta)}{6 \Gamma (2 +\beta)^2 \Gamma (1 - \beta)} \biggr] \\
&- \!\beta(1 + \beta) \psi^{(0)} (3 + \beta)\biggr\}\nonumber\\
=\,& C_F \exp\! \biggl[ \frac{2}{3} \beta + \sum_{n = 2}
\frac{(-\beta)^n}{n} \Bigl\{2 - 2^n (1 + 3^{- n}) + \zeta_n \bigl[2^n - 3 - (- 1)^n\bigr]\!\Bigr\}\!\biggr] \nonumber\\
&\times\! \Biggl\{ \sum_{n = 1} \biggl(-\frac{\beta}{2} \biggr)^{\!\!n} - \frac{4\beta}{3} - \frac{19}{3} +
2 (1 + \beta)\! \Biggl[ \, \sum_{n = 1} \frac{(-\beta)^n}{n + 1} \Bigr\{1 - 2^{n + 1} \nonumber \\
& \quad-\! \biggl(\frac{2}{3}\biggr)^{\!\!n+1}\!+ \zeta_{n + 1} \bigl[2^{n + 1} + (- 1)^n + n -1\bigr]\!\Bigr\} \!\Biggr] \!\Biggr\} ,\nonumber
\end{align}
where in the second line we have written a partially expanded expression which is the starting point of a convenient recursive algorithm to obtain the
fixed-order coefficients. This result can be used in Eq.~\eqref{eq:nocuspPart} together with the analytic form of
$G_{Q}(-\beta,0)$ provided in Eq.~\eqref{eq:cusp-closed} to obtain a closed form for $\gamma_H(\beta)\equiv\gamma_{C_H}\!(\beta) $, as well as its fixed-order coefficients. We find
full agreement for those when compared to the leading flavor structure of full QCD up to
$\mathcal{O}(\alpha_s^4)$~\cite{vanNeerven:1985xr,Matsuura:1988sm,Moch:2005id,Huber:2019fxe}, see second column of Table~\ref{tab:gamma}. To multiply out two series we use the
general result
\begin{equation}\label{eq:expProd}
\Biggl( \sum_{i = n} a_i x^i \!\Biggr) \!\Biggl( \,\sum_{j = m} b_j x^j\! \Biggr) =
\sum_{j = n + m} \! \! \! \! x^j \sum_{i = n}^{j - m} a_i b_{j - i} \,.
\end{equation}
\begin{table}[t!]
\centering
\begin{tabular}{|c|cccccc|}
\hline
$n$ & $\hat{\Gamma}^n_{\!\!\tmop{cusp}}$ & $\hat{\gamma}_H^n$ & $\hat{\gamma}_J^n$ & $\hat{\gamma}_S^n$
& $\hat{\gamma}_{H_m}^n$ & $\hat{\gamma}_B^n$\\
\hline
$1$ & $5.33333$ &$-16$ &$8$ &$0$ &$-10.6667$ &$5.33333$\\
$2$ & $8.88889$ &$-45.5782$ &$26.6989$ &$-3.90981$ &$-17.7778$ &$12.7987$\\
$3$ & $-1.77778$ &$-8.35288$ &$3.26488$ &$0.911559$ &$3.55556$ &$-2.68934$\\
$4$ & $-11.0442$ &$43.9815$ &$-23.6303$ &$1.6396$ &$22.0883$ &$-12.6838$\\
$5$ & $-5.83051$ &$27.5444$ &$-8.49109$ &$-5.2811$ &$11.661$ &$-0.549408$\\
$6$ & $1.73202$ &$-9.98024$ &$10.8212$ &$-5.83111$ &$-3.46405$ &$7.56314$\\
$7$ & $2.54404$ &$-14.9285$ &$8.61429$ &$-1.15006$ &$-5.08807$ &$3.6941$\\
$8$ & $0.56906$ &$-3.87091$ &$1.07053$ &$0.864928$ &$-1.13812$ &$-0.295868$\\
$9$ & $-0.259965$ &$1.00721$ &$-0.756306$ &$0.252702$ &$0.519929$ &$-0.512666$\\
$10$ & $-0.157657$ &$0.398943$ &$0.114127$ &$-0.313598$ &$0.315314$ &$0.155941$\\
\hline
\end{tabular}
\caption{Numeric coefficients for the cusp anomalous dimension and the
non-cusp anomalous dimensions of the hard and mass-scale matching
coefficients, the SCET jet function and the bHQET jet function. The hatted
coefficients are defined as $f \equiv (1 / \beta_0) \sum^{\infty}_{n = 0}
\hat{f}_n \beta^{n + 1}$. For the cusp anomalous dimensions we observe
$\Gamma_{\!\!H} = 2\Gamma_{\!\!S} = -2\Gamma_{\!\!J} = -2\Gamma_{\!\!B} = -4\Gamma_{\!\rm cusp} = 2 \Gamma_{\!\!H_m}$. \label{tab:gamma}}
\end{table}
In our case, one of the two series results from the expanding the exponential in Eq.~\eqref{eq:gH} using the recursive formula in Eq.~\eqref{eq:exp-expand}.
The convergence radius of the series is again $\Delta \beta = 2.5$.
In Fig.~\ref{fig:gammaH} we compare for $\alpha_s=0.9$
the exact and fixed-order results for $\gamma_H$,
finding again that the latter converges to the former already at $6$ loops.

The last piece of information necessary to determine $C_{\!H}$
at $\mathcal{O}(1/\beta_0)$ is
\begin{equation}
\! G_{\!Q}(0,u) = \!-\frac{4 \pi C_F e^{\frac{5 u}{3}} u \csc (\pi u)}{2 -3 u + u^2} =\! - 2 C_F \exp\! \biggl[
\frac{19u}{6} + \sum_{n = 2} \frac{u^n}{n} \Bigr\{ 1 + 2^{- n}
+ \zeta_n \bigl[1 + (- 1)^n\bigr]\! \Bigr\} \!\biggr] ,
\end{equation}
with ${\cal Q}= Q$. Using the Taylor expansion of this function in Eq.~\eqref{eq:nologcusp} we fully reproduce the leading flavor structure of
full QCD up to $\mathcal{O}(\alpha_s^3)$~\cite{Matsuura:1987wt,Matsuura:1988sm,Gehrmann:2005pd,Moch:2005id,Lee:2010cg,Baikov:2009bg,Gehrmann:2010ue},
see first column of Table~\ref{tab:matEl}.
To compute $\delta H_Q$ one considers instead $G_{\!H}(0,u) = 2 \cos(\pi u) G_{Q}(0,u)$, and to determine the fixed-order coefficients the Taylor expansions
for $\cos(\pi u)$ and $G_{Q}(0,u)$ must be used in Eq.~\eqref{eq:expProd}.

The
function $H_{\!B}(u)\equiv\! \bigl[G_{\!H}(0,u) - (G_{\!H})_{0,0} - u (G_{\!H})_{0,1}\bigr]/u^2$ has two double poles at $u=1,2$ and an infinite number of simple poles at integer values of $u\geq 3$:
\begin{equation}
H_{\!B}(u) \asymp 8C_F\! \Biggl[\frac{e^\frac{5}{3}}{(u-1)^2}+\frac{5e^\frac{5}{3}}{3(u-1)} - \frac{e^\frac{10}{3}}{2(u-2)^2} - \frac{e^\frac{10}{3}}{12(u-2)}
- \!\sum_{n = 3} \frac{e^{\frac{5n}{3}}}{(n - 2) (n - 1) n (u - n)}\Biggr].
\end{equation}
These double poles unambiguously signal the presence of anomalous dimensions with $n_f$ dependence at leading order for dimension-$2$ and -$4$ operators of the
associated OPE.\footnote{Operators corresponding to simple poles might also carry anomalous dimension, but at leading-order they should not depend on $n_f$. We
thank A.~Hoang for clarifying this point.} The total ambiguity
can be expressed in the following compact form:
\begin{align}\label{eq:hardAmb}
&\!\delta_{\Lambda} H_Q = \frac{C_F}{\beta_0}\Biggl\{\frac{8 }{3}\biggl(\frac{e^\frac{5}{6}\Lambda_{\rm QCD}}{Q}\biggr)^{\!\!2}
\biggl[6 \log\! \biggl(\frac{\Lambda_{\rm QCD}}{Q}\biggr)\!+5\biggr]\! - \frac{2}{3}\biggl(\frac{e^\frac{5}{6}\Lambda_{\rm QCD}}{Q}\biggr)^{\!\!4}
\biggl[12 \log\! \biggl(\frac{\Lambda_{\rm QCD}}{Q}\biggr)\!+1\biggr] \nonumber\\
& \;\quad+4 \biggl(\frac{e^\frac{5}{6}\Lambda_{\rm QCD}}{Q}\biggr)^{\!\!2}\!-6 \biggl(\frac{e^\frac{5}{6}\Lambda_{\rm QCD}}{Q}\biggr)^{\!\!4}\!
+4\! \Biggl[\!\biggl(\frac{e^\frac{5}{6}\Lambda_{\rm QCD}}{Q}\biggr)^{\!\!2}-1\Biggr]^{\!2}
\log\!\Biggl[1-\!\biggl(\frac{e^\frac{5}{6}\Lambda_{\rm QCD}}{Q}\biggr)^{\!\!2}\Biggr] \!\Biggr\},
\end{align}
\begin{table}[t!]
\centering
\begin{tabular}{|c|cccc|}
\hline
$n$ & $H_Q$ & $\tilde J_n$ & $H_m$ & $\tilde B_n$\\
\hline
$1$ & $9.3721$ &$0.560352$ &$15.0532$ &$7.52658$\\
$2$ & $15.0279$ &$10.7871$ &$59.9652$ &$29.0465$\\
$3$ & $117.452$ &$21.3687$ &$267.538$ &$112.621$\\
$4$ & $709.259$ &$50.6136$ &$1656.73$ &$661.256$\\
$5$ & $4302.01$ &$193.805$ &$13629.1$ &$5290.68$\\
$6$ & $28592.4$ &$961.578$ &$138593$ &$52915.6$\\
$7$ & $214413$ &$5740.14$ &$1.67874\times 10^6$ &$634995$\\
$8$ & $1.78994\times 10^6$ &$40068.6$ &$2.36208\times 10^7$ &$8.88995\times 10^6$\\
$9$ & $1.66615\times 10^7$ &$320083$ &$3.78939\times 10^8$ &$1.42239\times 10^8$\\
$10$ & $1.70403\times 10^8$ &$2.87865\times 10^6$ &$6.83033\times 10^9$ &$2.56031\times 10^9$\\
\hline
\end{tabular}
\caption{Numeric values for the non-logarithmic coefficients for the various SCET and bHQET coefficients $a_{n,n-1}$.\label{tab:matEl}}
\end{table}
where in the first line we have explicitly singled out the first two leading
contributions, while the second line starts at $\mathcal{O}(\Lambda_{\rm QCD}^6)$. The logarithmic enhancement of the two leading terms appears due to the double nature of the
poles and is associated to the anomalous dimension just discussed. This entails that $Q^2 \delta_{\Lambda} H_Q$ retains some logarithmic
dependence on $Q$ in Fig.~\ref{fig:hardAmb} the dependence of $\delta_{\Lambda} H_Q$ in $Q$ is shown in red, while the ambiguity when the
logarithm is set to zero appears in blue. Despite the logarithmic enhancement, it is negligibly small even for the smallest values of $Q$ for which
SCET is applicable.

We compare next the exact result for the hard function $H_Q(Q,\mu)$ obtained with Eq.~\eqref{eq:cuspLambda} with its fixed-order expansion. For the latter we compute
$H_Q(Q,\mu_0)$ and subsequently run to $H_Q(Q,\mu)$. A natural choice for the matching scale that avoids explicit large logarithms is $\mu_0=Q$, but in this way $H_Q(Q,Q)$
has analytical continuation $\pi^n$ terms that can be traced as coming from the real part of $\log^n(-\mu^2/Q^2+i\,0^+)$ with $Q^2$ and $\mu^2$ both positive when transitioning
from Euclidean to Minkowskian regions, which therefore can also be summed up using RG evolution. In the renormalon formalism these factors come from expanding in powers
of $u$ the factor $\cos(\pi u)$ appearing in $G_{\!H}(0,u)$. To carry out this procedure one performs resummation in $C_{\!H}(Q^2,\mu)$ by computing first $C_{\!H}(Q^2,\mu_0)$ with
$\mu_0=-iQ$ and then running to from $\mu_0$ to $\mu$. From this resummed result for $C_{\!H}(Q^2,\mu)$ one obtains $H_Q(Q,\mu)$ simply taking the modulus squared
expanded to $\mathcal{O}(1/\beta_0)$. We compare these two approaches in Fig.~\ref{fig:piSum} for a relatively small value of the center-of-mass energy.
The red dots, which include $\pi$-resummation, converge to the exact answer much faster than the blue, which do not. Eventually both approaches reproduce the exact result, and
start diverging for $n>30$. The ambiguity of $H_Q(Q,\mu)$ is too small to be visible in the plot. Since the hard factor mainly affects the norm of the event-shape distributions,
this behavior might explain why the total-hadronic cross section obtained via a direct integration of the differential event-shape cross section, was undershot at low orders in
the thrust and C-parameter distributions in Refs.~\cite{Abbate:2012jh,Hoang:2014wka}. Similarly, Ref.~\cite{Bachu:2020nqn} found poor order-by-order convergence in peak
differential cross sections for boosted tops unless they were self-normalized.

\subsection{Jet function}\label{sec:jet}
\begin{figure*}[t!]
\centering
\subfigure[]
{\includegraphics[width=0.35\textwidth]{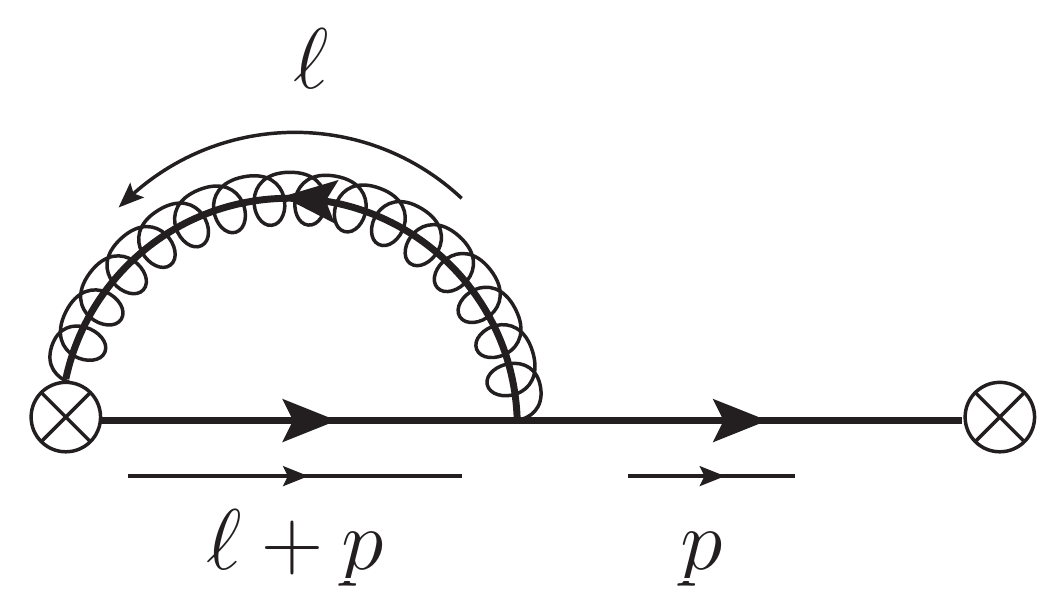}
\label{fig:vertex-correction}}~~~~
\subfigure[]{\includegraphics[width=0.35\textwidth]{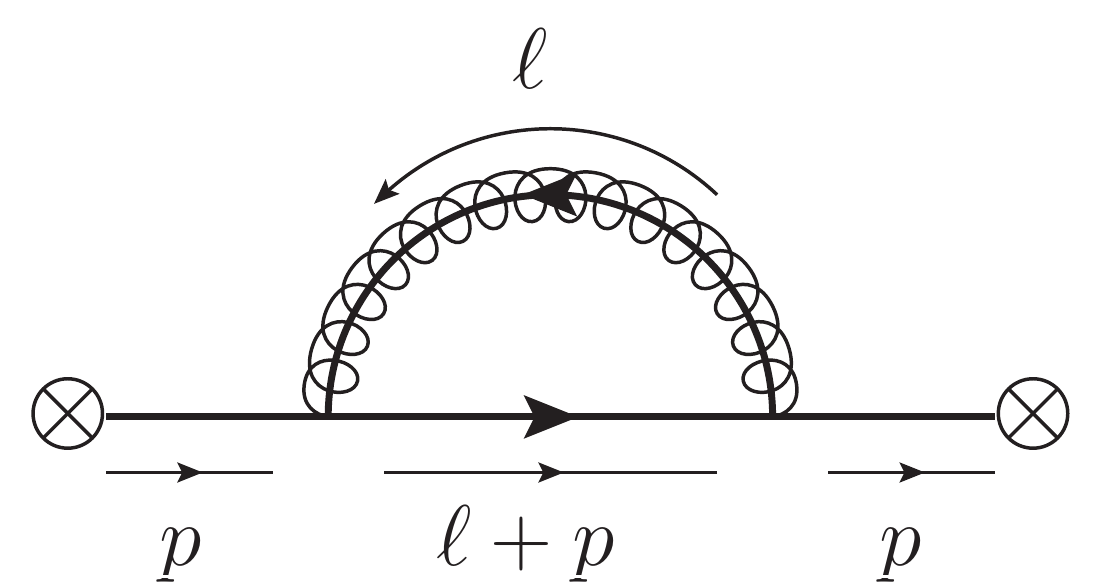}
\label{fig:quark-selfenergy-jet}}
\caption{Feynman diagrams for the one-loop jet function for massless quarks with off-shell momentum $p^2=s$ and virtual loop
momentum $\ell$. The cross represents a quark jet-field insertion. Diagram (a) has to be multiplied by a factor of two to account for the
symmetric contribution with the gluon line emitted from the right current. \label{fig:jet-function}}%which is not shown.} \label{fig:jet-function}
\end{figure*}
The SCET jet function describes energetic radiation branching from the initiating quark, and depends on the collinear measurement function, which is the same
for hemisphere masses, C-parameter and thrust. Furthermore, it is completely inclusive such that the jet function can be computed as the discontinuity of a
forward-scattering matrix element:
\begin{equation}
\mathcal{J}_n(s,\mu) =\frac{1}{8 \pi N_cQ} \!\int\! \mathd^d x\, e^{i p \cdot x}\, \tmop{Tr} \langle 0 | {\rm T}
\bigl\{ \chi_n (x) \overline{\chi}_{n, Q} (0) n \! \! \! / \bigr\} | 0 \rangle\,,
\end{equation}
with $d=4-2\varepsilon$, $\rm Tr$ a trace over spin and color, and $p^2=s$. We use the gauge-invariant jet field $\chi_n = W_n^\dagger \xi_n$, which is the product of a collinear
field $\xi_n$ and a path-ordered collinear Wilson line $W^\dagger_n$. Likewise, $\chi_{n, Q}$ is a jet field with total minus label momenta $\bar nP$ set to $Q$ with a Dirac delta
function. This overly simplifies the computation of the leading $1/\beta_0$ correction as one only needs to modify the gluon propagator of the one-loop diagrams
shown in Fig.~\ref{fig:jet-function}. The vertex diagram \ref{fig:vertex-correction} must be doubled to account for the symmetric contribution, which is omitted
along with the diagram in which the gluon is emitted and absorbed from both Wilson since it vanishes in the Feynman gauge.
Given that the jet functions for the three event shapes just mentioned are trivially related to one another, in this section we carry out the computation for the invariant mass of a single
hemisphere. The discontinuity of $\mathcal{J}_n(s,\mu)$ [\,that is, $\mathcal{J}_n(s+i0^+,\mu) - \mathcal{J}_n(s+i0^-,\mu)$\,]
yields the momentum-space jet function $J_n(s,\mu)$, with support for $s>0$. It can be converted to the position-space jet function
$\tilde J_n(y,\mu)$, with $y$ the variable conjugate to $s$, defined as:
\begin{equation}
\tilde{J}_n (y, \mu) = \int_0^{\infty} \text{d} s\, e^{- i s y} J_n (s, \mu)\, .
\end{equation}
While $\mathcal{J}_n(s,\mu)$ is a regular function, $J_n(y,\mu)$ for massless quarks contains Dirac delta and plus distributions. For its computation we will use the following identity:
\begin{equation}\label{eq:impart}
{\rm Im}[(- s - i 0^+)^{- 1 - \eta}\,] = \frac{\pi\,s^{- 1 - \eta}\,\theta(s)}{\Gamma (1 + \eta) \Gamma(- \eta)} \,.
\end{equation}
Both $\mathcal{J}_n(s,\mu)$ and $J_n(y,\mu)$ have dimensions of an inverse squared mass, while $s$ is a real-valued mass-dimension-two variable. On the contrary,
$\tilde J_n(y,\mu)$ is a dimensionless complex-valued regular function depending on the complex variable $y$, with mass dimension $-2$. At tree-level one has
\begin{equation}
\mathcal{J}^{\rm tree}_n(s)=-\frac{1}{2 \pi} \frac{1}{s + i 0^+}\,,\qquad J_n^{\rm tree}(s) = \delta(s)\,,\qquad \tilde J^{\rm tree}_n(y) = 1\,,
\end{equation}
and we define the corrections to the bare jet function in momentum and position space as $J_n(s,\mu) = \delta(s) + \delta J_n(s)$ and $\tilde J_n(y) = 1 + \delta \tilde J_n(y)$, respectively, which are computed
next.
The vertex contribution shown in Fig.~\ref{fig:vertex-correction}, and the self-energy correction of Fig.~\ref{fig:quark-selfenergy-jet} contribute to the forward-scattering matrix
element as follows
\begin{align}
\mathcal{J}_n^{\{a,b\}}(s) =\, & \frac{M^{\{a,b\}}(s)}{2 \pi} \frac{1}{s + i 0^+} \,,\\
M^a(s) =\, & i g_0^2 C_F\frac{1 + \varepsilon}{h + \varepsilon} \!
\int\!\! \frac{{\df}^d \ell}{(2 \pi)^d} \frac{1}{(- \ell^2)^{1 + h} (\ell + p)^2 } \nonumber\\
=\, &\frac{2 C_F}{(-s - i 0^+)^{h+\varepsilon}} \frac{g_0^2}{(4 \pi)^{2-\varepsilon}} \frac{\Gamma (h +
\varepsilon) \Gamma (2 - \varepsilon) \Gamma (- h - \varepsilon)}{\Gamma (h +1) \Gamma (2 - h - 2 \varepsilon)}\,,\nonumber\\
M^b(s) =\, & i g_0^2 C_F(1 - \varepsilon) \!\!
\int \!\!\frac{{\df}^d \ell}{(2 \pi)^d} \frac{1 - \frac{\ell^2}{p^2} }{(- \ell^2)^{1 + h} (\ell +p)^2}\nonumber\\
=\, & \frac{2C_F}{(-s-0^+)^{h+\varepsilon}} \frac{g_0^2}{(4 \pi)^{2-\varepsilon}} \frac{(1 - \varepsilon) \Gamma (2 - \varepsilon) \Gamma (1 - h -
\varepsilon) \Gamma (h + \varepsilon)}{\Gamma (1 + h) \Gamma (3 - h - 2\varepsilon)}\,.\nonumber
\end{align}
Adding the two contributions and using Eq.~\eqref{eq:impart} with $\eta=h+\varepsilon$
we obtain\footnote{The (dimensionless) cumulative jet function, defined as
$\Sigma_J(s_c,\mu) \equiv\! \int_0^{s_c}\!{\df} s \,J(s,\mu)$,
has the same cusp and non-cusp anomalous dimension as $\tilde J_n(s,\mu)$. Furthermore, the Borel transform of the series has no poles for $u>0$,
resulting in a finite convergence radius. Identifying ${\cal Q}^2=s_c$ one can write down a closed form for $\Sigma_J$ using
$G_{\Sigma_J}(0,u) = F_{\!\tilde J}(u)/\Gamma(1+u)$.}
\begin{align}
\delta J_n(s)=\,& \frac{2C_F g_0^2}{(4 \pi)^{2-\varepsilon}} \frac{s^{-1-h-\varepsilon}\,\Gamma (2 -
\varepsilon)}{\Gamma (1 + h) \Gamma (3 - h - 2 \varepsilon)} \biggl[ 5 - \varepsilon - \frac{2 (2 + h)}{h + \varepsilon} \biggr],\\
\delta \tilde J_n(x)=\,& \frac{2C_Fg_0^2}{(4 \pi)^{2-\varepsilon}} \frac{(i y)^{h+\varepsilon}\,\Gamma (2 - \varepsilon)
\Gamma (- h - \varepsilon)}{\Gamma (1 + h) \Gamma (3 - h - 2 \varepsilon)} \biggl[ 5 - \varepsilon - \frac{2 (2 + h)}{h +\varepsilon} \biggr].\nonumber
\end{align}
Let us compute the anomalous dimensions using the position-space jet function for which we can identify \mbox{${\cal Q}^2=-ie^{-\gamma_E}/y$}
and $b(\varepsilon,h)$, to find\footnote{The momentum-space jet function for $s>$ behaves as a series without cusp anomalous dimension [\,that is,
the associated $F(\varepsilon, u)$ function is regular at $u=0$\,]. Paradoxically, the (non-cusp) anomalous dimension of $J_n(s)$ for $s>0$ is precisely
$-4\Gamma_{\!\rm cusp}$. The series is not ambiguous, and identifying ${\cal Q}^2=s$ one can write down a closed form for $sJ_n(s,\mu)$ using
Eq.~\eqref{eq:LamSumNocusp} and noting that $F_J(0,u) = -G_{\Sigma_J}(0,u)$.}
\begin{equation}
G_{\!\tilde J}(\varepsilon,u) = D(\varepsilon)^{\frac{u}{\varepsilon} - 1}\frac{2 C_F e^{\gamma_E (\varepsilon-u) }
\Gamma (1-u) \Gamma(2-\varepsilon)}{\Gamma (3-u-\varepsilon)\Gamma (1+u-\varepsilon)}
\bigl[4 - 2 \varepsilon - u (3 - \varepsilon)\bigr]\,.
\end{equation}
Taking into account the relation $\Gamma_{\!\!\tilde J}=4\Gamma_{\!\rm cusp}$ we find that $\beta G_{\tilde J}(-\beta,0)/\beta_0$ reproduces the result found in
Eq.~\eqref{eq:cusp-closed}. To compute the jet non-cusp anomalous dimension we need also
\begin{align}
&\frac{{\df}G_{\!\tilde J}(-\beta,u)}{{\df} u}
\biggr|_{u = 0} = \frac{C_F \Gamma (4 + 2 \beta)}{3 (2 +\beta) \Gamma (2 +\beta)^3 \Gamma (1 -\beta)}\Biggl\{ 3 - \beta^2 \\
&\qquad\qquad\qquad\qquad\quad -\! \frac{2 (1 + \beta) (2 +\beta)}{\beta} \log\! \Biggl[\frac{6e^{-\gamma_E \beta} \Gamma(2 +\beta)^2
\Gamma (1 -\beta)}{\Gamma(4 + 2 \beta)} \Biggr] \!\Biggr\}\nonumber\\
&\qquad= C_F \exp\! \Biggl[ \frac{\beta}{6} + \sum_{n = 2}
\frac{(-\beta)^n}{n} \Bigl\{ 2 + 2^{- n} - 2^n (1 + 3^{- n}) + \zeta_n \bigl[2^n - 3
- (- 1)^n\bigr]\! \Bigl\}\! \Biggr]\! \Biggl\{\! 3 - \beta^2\nonumber\\
& ~\qquad +\! 2 (1 +\beta) (2 +\beta)\!
\Biggl[ \frac{5}{3}\! +\! \sum_{n = 1} \frac{(-\beta)^n}{n + 1} \Bigl\{ 2^{n + 1} (1
+ 3^{- n - 1}) - 1 + \zeta_{n + 1} \!\bigl[2 - 2^{n + 1} - (- 1)^n \bigr]\! \Bigr\} \!
\Biggr] \! \Biggr\}, \nonumber
\end{align}
where the second equality has partially expanded in powers of $\beta$. This expression can be used in Eq.~\eqref{eq:nocuspPart} together with
$G_{\!\tilde J}(-\beta,0) = -2G_{Q}(-\beta,0)$ to obtain a closed form for $\gamma_J(\beta)$, as well as its fixed-order coefficients. We find full agreement with the leading
flavor structure of the known full SCET results up to $\mathcal{O}(\alpha_s^3)$~\cite{Moch:2004pa}, see columns $3$ and $4$ in Table~\ref{tab:gamma}. The soft non-cusp
anomalous dimension can be computed from
a consistency condition and it reads $\gamma_S = -\gamma_H - \gamma_J$. In both cases the convergence radius is $\Delta \beta=5/2$.
In Figs.~\ref{fig:gammaJ} and \ref{fig:gammaS} we compare for $\alpha_s=0.9$
the exact and fixed-order results for the jet and soft non-cusp anomalous dimensions, respectively. The soft function has $\gamma_S^0=0$ and therefore the
partial sum starting from the first non-vanishing order is shown. We find good convergence in both cases.

To compute the non-logarithmic fixed-order coefficients $ a^J_{i, i-1,0}$ of the Fourier-space jet function, which are defined as
\begin{equation}
\tilde J_n = \sum_{i = 1} \beta^i \sum_{j = 0}^{i + 1} \hat 2^{-j} a^J_{i, i-1,j} \log^j (i y e^{\gamma_E} \mu^2)\,,
\end{equation}
where the factor $2^{-j}$ is necessary to compensate the $\mu^2$ that appears in the argument of the logarithm, we need one last function
\begin{equation}
G_{\!\tilde J}(0,u) = \frac{2 C_F e^{\left( \frac{5}{3} - \gamma_E \right) u} (4 - 3u)}{(1 - u) (2 - u) \Gamma (1 + u)} \\
=
4 C_F \exp \Biggl\{ \frac{29 u}{12} +\! \sum_{n = 2}
\frac{u^n}{n}\! \biggl[ 1 + 2^{- n} - \!\biggl(\frac{3}{4}\biggr)^{\!\!n} - (- 1)^n \zeta_n \biggr]\! \Biggr\}.
\end{equation}
Using its Taylor expansion in Eq.~\eqref{eq:nologcusp} we fully reproduce the leading flavor structure of the known coefficients in
full SCET up to $\mathcal{O}(\alpha_s^3)$~\cite{Lunghi:2002ju,Bauer:2003pi,Becher:2006qw,Bruser:2018rad,Banerjee:2018ozf},
see second column of Table~\ref{tab:matEl}.
\begin{figure*}[t!]
\subfigure[]
{\includegraphics[width=0.47\textwidth]{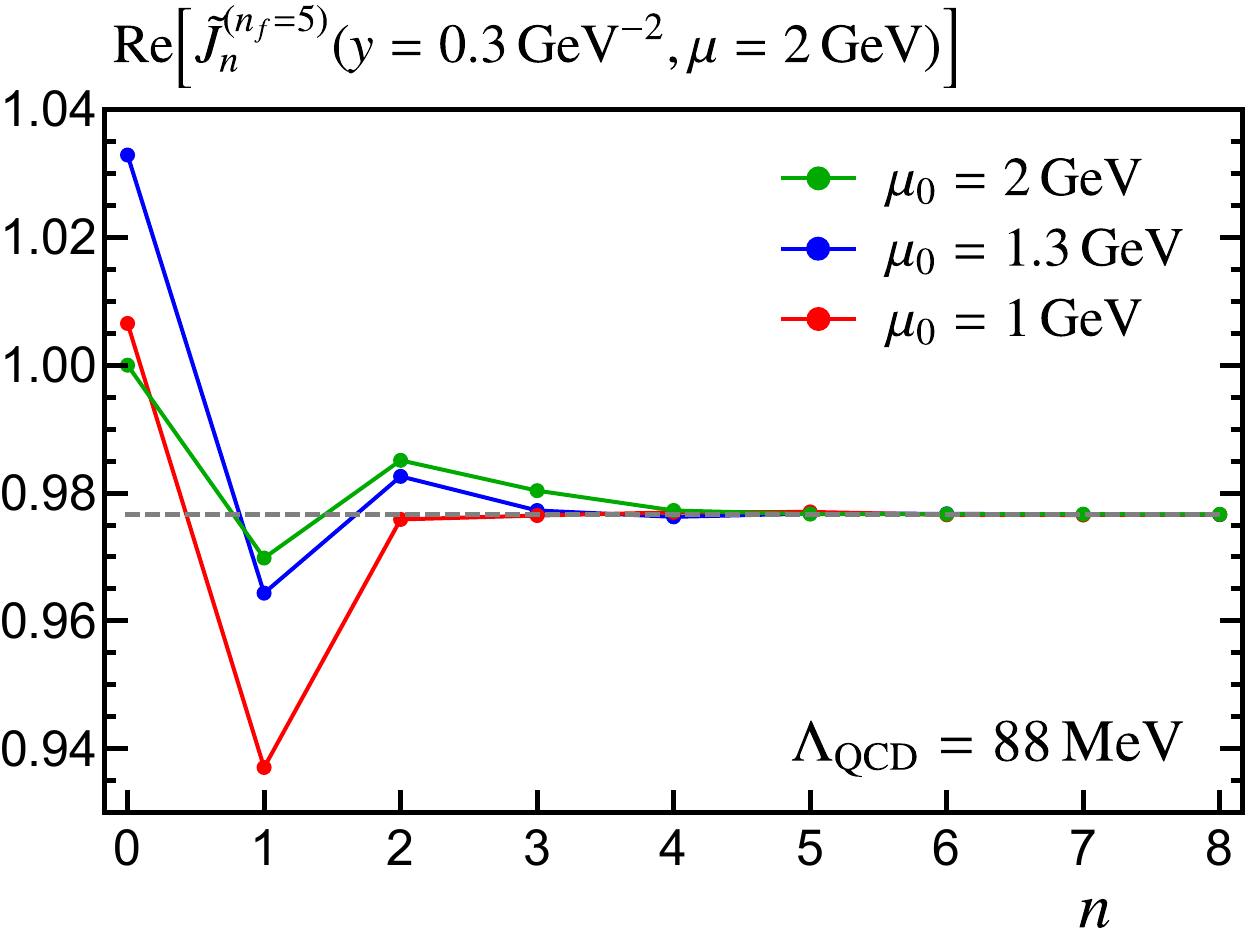}
\label{fig:jetRe}}~~~~
\subfigure[]{\includegraphics[width=0.465\textwidth]{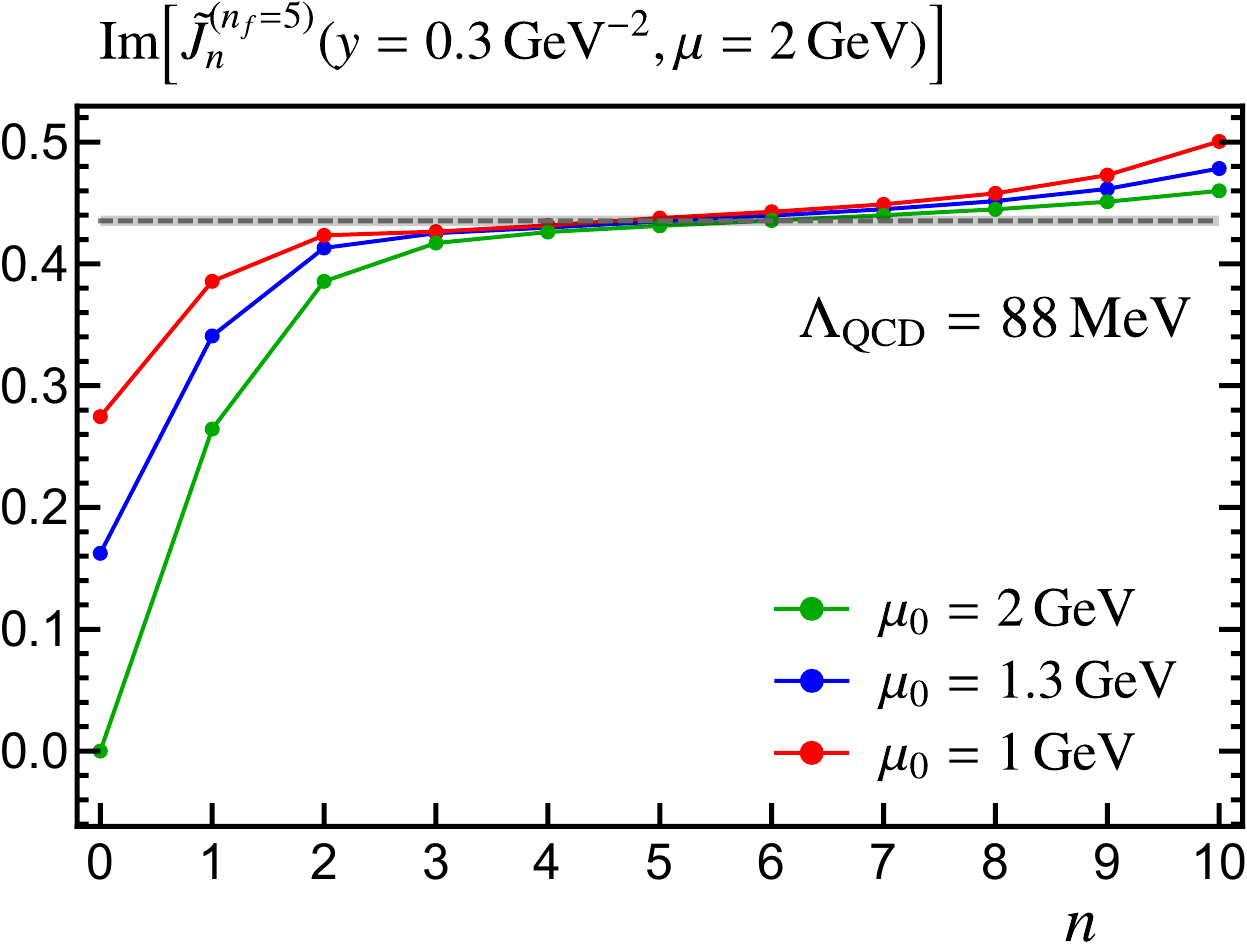}
\label{fig:jetIm}}
\caption{Comparison of the fixed-order partial sum (colored dots) and exact results (red dashed line) for the real (left panel) and imaginary (right panel) part
of the position-space SCET jet function $\tilde J_n(y,\mu)$, with $y=0.3\,{\rm GeV}^{-2}$ and $\mu=2\,$GeV. We show fixed-order results for three matching scales:
$\mu_0=2\,$GeV (green), $1.3\,$GeV (blue) and $1\,$GeV (red). \label{fig:Jet}}%We use $n_f=5$ active flavors and $\Lambda_{\rm QCD}=88\,$MeV.} \label{fig:Jet}
\end{figure*}

The function $\tilde J_{\!B}(u)\equiv \!\bigl[G_{\!\tilde J}(0,u) - (G_{\!\tilde J})_{0,0} - u(G_{\!\tilde J})_{0,1}\bigr]/u^2$ has only two simple poles at $u=1,2$, with the following asymptotic
expansion:
\begin{equation}
\tilde J_{\!B}(u) \asymp -C_F\!\Biggl[\frac{e^{\frac{10}{3}-2 \gamma_E}}{2 (u-2)}+\frac{2 e^{\frac{5}{3}-\gamma_E} }{u-1}\Biggr].
\end{equation}
Therefore the ambiguities in position and momentum space are
\begin{align}\label{eq:jetAmb}
\delta_{\Lambda} \tilde J_n &\,=-\frac{C_F}{\beta_0} \biggl[2\Bigl(i y e^{\frac{5}{3}} \Lambda_{\rm QCD}^2\Bigr)
+\frac{1}{2} \Bigl(i y e^{\frac{5}{3}} \Lambda_{\rm QCD}^2\Bigr)^{\!\!2}\,\biggr] ,\\
\delta_{\Lambda} J_n &\,=-\frac{2 e^{\frac{5}{3}} C_F}{\beta_0} \Lambda_{\rm QCD}^2\, \delta^{\prime\!}(s) + \mathcal{O}(\Lambda_{\rm QCD}^4)
\simeq \frac{2 e^{\frac{5}{3}} C_F}{\beta_0} \frac{\Lambda^2_{\rm QCD}}{s}\, \delta(s)\,,\nonumber
\end{align}
where in the second line the Dirac delta function's derivative is expressed as $\delta^{\prime\!}(s) \simeq -\delta(s)/s$, as obtained from the identity $[s\,\delta(s)]^\prime=0$.
For real values of $y$ the real (imaginary) part of the jet function is free from the $u=1(2)$ renormalon, but affected by the $u=2(1)$ ambiguity. For complex
$y$ both real and imaginary parts are affected by the two singularities. In the second line of Eq.~\eqref{eq:jetAmb} it is shown that when transforming
back to momentum space the leading ambiguity is proportional to $\delta^{\prime\!}(s)$.
Therefore, to find an ambiguous series in
momentum space one needs to consider a double cumulative (that is, the cumulative version of $\Sigma_J$)
or a weighted cumulative such as $\int_0^{s'} {\df} s \,s \,J_n(s,\mu)$.

In Fig.~\ref{fig:Jet} we compare the fixed-order expansion for the position-space jet function up to $(n+1)$-loops with the exact result, splitting in two separate
panels the real and imaginary parts, for a real value of the Fourier variable $y$. The ambiguity
is visible for the imaginary part only. We nevertheless observe nice convergence for different values of the matching scale.

\begin{figure*}[t!]
\centering
\subfigure[]
{\includegraphics[width=0.15\textwidth]{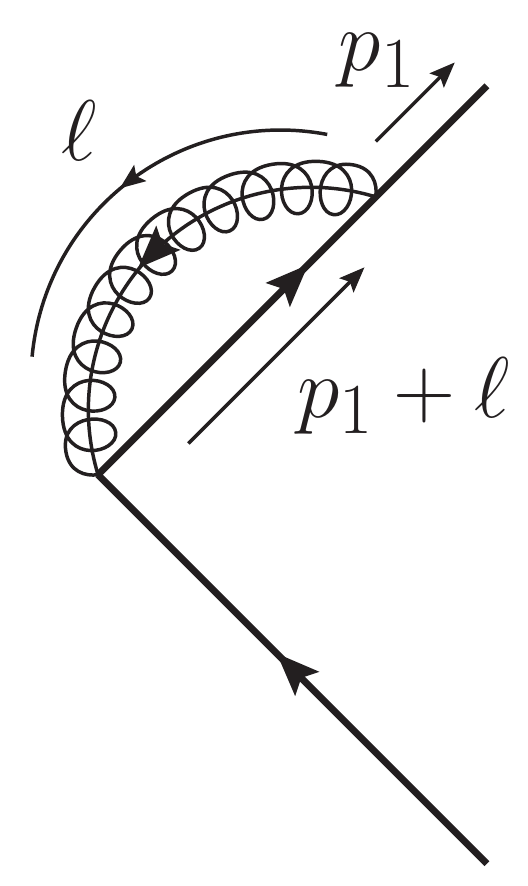}
\label{fig:collinear}}~~~~~~~~~~~~~~~~~~~~~~~~
\subfigure[]{\includegraphics[width=0.15\textwidth]{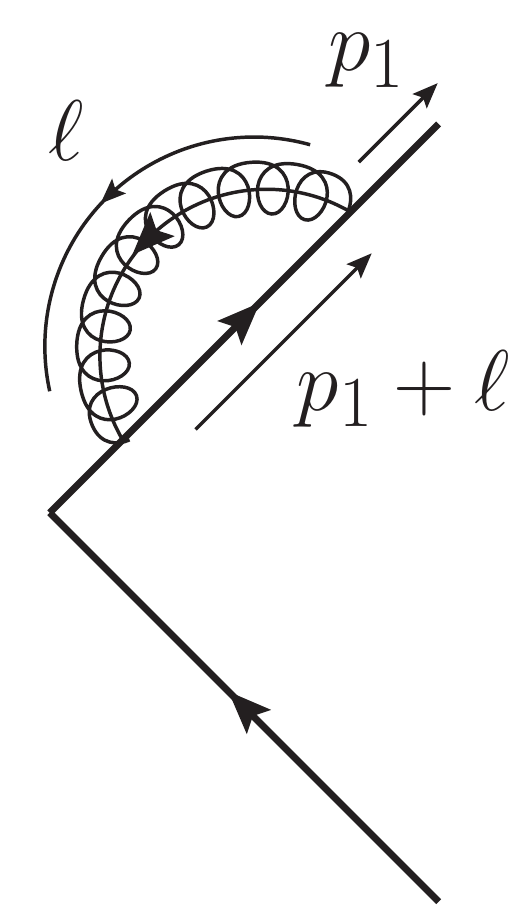}
\label{fig:collinear-self}}
\caption{SCET Feynman diagrams for the computation of the bHQET to SCET matching coefficient at one-loop for a massive on-shell quark with momentum
$p^2=m^2$ and virtual loop momentum $\ell$. We do not show the two symmetric diagrams, as they give identical results, or the contribution in which a
soft gluon is interchanged between the two massive collinear quarks, which vanishes, along with all bHQET diagrams.\label{fig:Cm}}
\end{figure*}

\section{bHQET computations}\label{sec:bHQET}
If the quarks primarily produced in the $e^+e^-$ collision at high energy are massive (e.g.\ the top quark) a new scale ---\,the quark mass $m$\,--- becomes relevant in
the problem. This mass changes the SCET jet function [\,secondary production of heavy quarks through gluon splitting affects also
the soft function starting at $\mathcal{O}(\alpha_s^2)$\,]. In the peak of the distribution
the jet invariant mass and $m$ are very similar, indicating that quarks
are very boosted. In this regime one can match SCET with massive quarks to bHQET, what allows to sum up a new class of large logarithms and treat
the top quark decay products in an inclusive way through the following factorization theorem~\cite{Fleming:2007qr,Fleming:2007xt}:
\begin{align}\label{eq:factbHQET}
\frac{1}{\sigma_0} \frac{{\df}\sigma_{\rm bHQET}}{\text{d} \tau} =&\, Q^2 H (Q,\mu_m) H_m\! \biggl( \!m, \frac{Q}{m}, \mu_m, \mu \!\biggr) \!\!\int \!\!{\df} \ell\,
B_{\tau}\! \biggl( \!\frac{Q^2 \tau - Q \ell}{m} - 2m, \mu \!\biggr) S_{\tau}(\ell, \mu)\,,\\
B_{\tau} (\hat{s}, \mu) =&\, m\! \int_0^{\hat{s}}\! \text{d} \hat{s}' B_n (\hat{s}- \hat{s}', \mu) B_n (\hat{s}', \mu)\,.\nonumber
\end{align}
The difference between Eqs.~\eqref{eq:factbHQET} and \eqref{eq:factSCET} is that the former has an additional (universal) hard factor $H_m$ that accounts for the
matching between SCET and bHQET, and the SCET jet function $J_n$ has been replaced by its bHQET counterpart $B_n$ (which takes different forms depending
on the observable considered, but we focus once more in the hemisphere mass). In this section we compute these two functions at leading order in
$1/\beta_0$.

\subsection{Hard massive factor}\label{sec:Cm}
To determine the hard massive function $H_m$ one needs to compute
the matching coefficient $C_m$ between the SCET and bHQET
massive di-jet operators. The optimal way of carrying out the computation is using the on-shell scheme and regularizing all infrared divergences within
dimensional regularization, making all bHQET diagrams vanish such that only SCET contributions need to be considered. The diagrams to take into
account are shown in Fig.~\ref{fig:Cm}, where both contributions need to be doubled
to account for the symmetric diagrams which are not explicitly shown. Furthermore, the contribution in which a soft gluon is exchanged between
the two massive quark lines also vanishes and has
been omitted. For diagram \ref{fig:collinear-self} we use our previous result on the wave-function renormalization $Z_\xi^{\rm OS}$ for a massive on-shell quark
in Eq.~\eqref{eq:wave}, since it is the same in SCET and full QCD.
\begin{figure*}[t!]
\subfigure[]
{\includegraphics[width=0.47\textwidth]{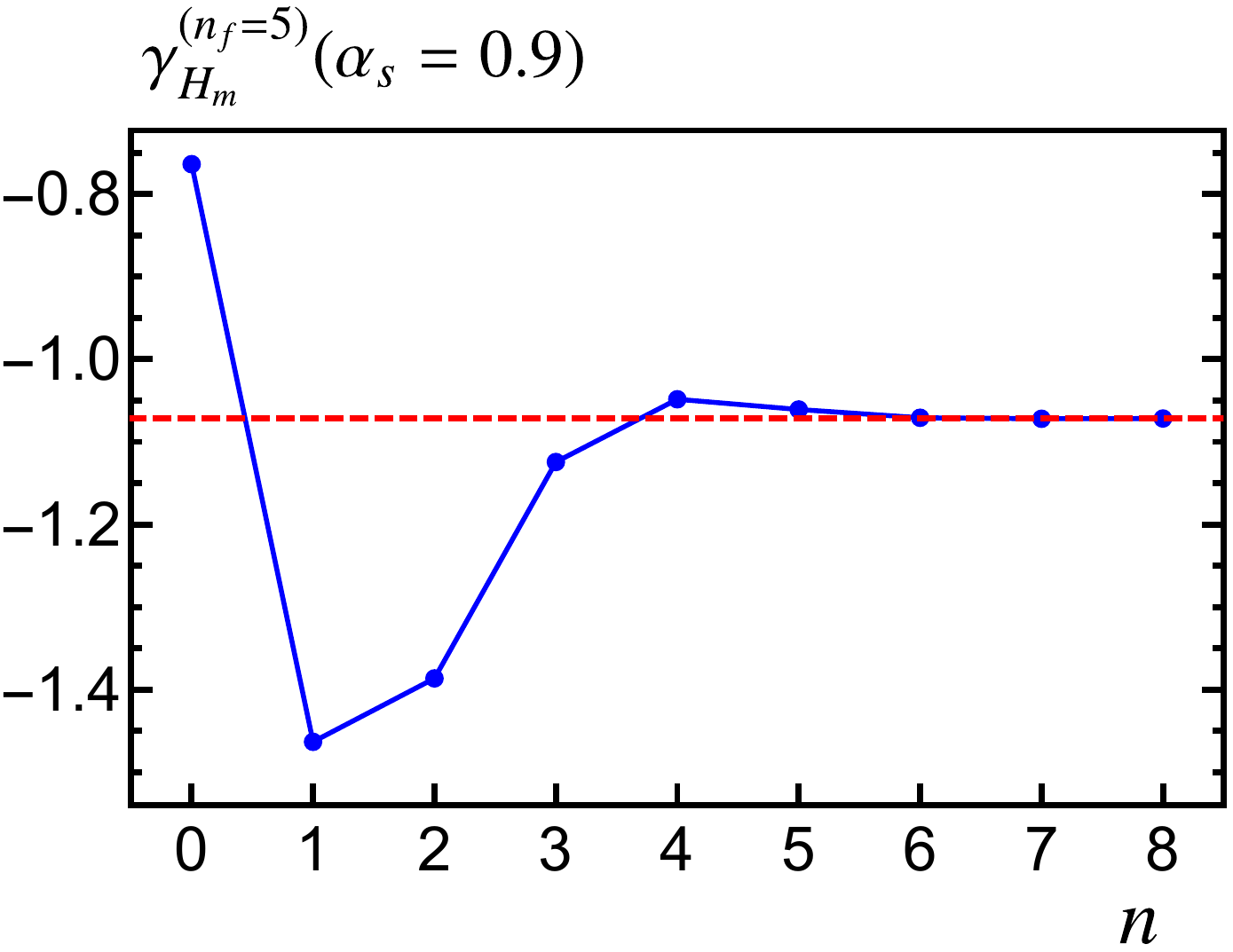}
\label{fig:gammaBHQET}}~~~~
\subfigure[]{\includegraphics[width=0.46\textwidth]{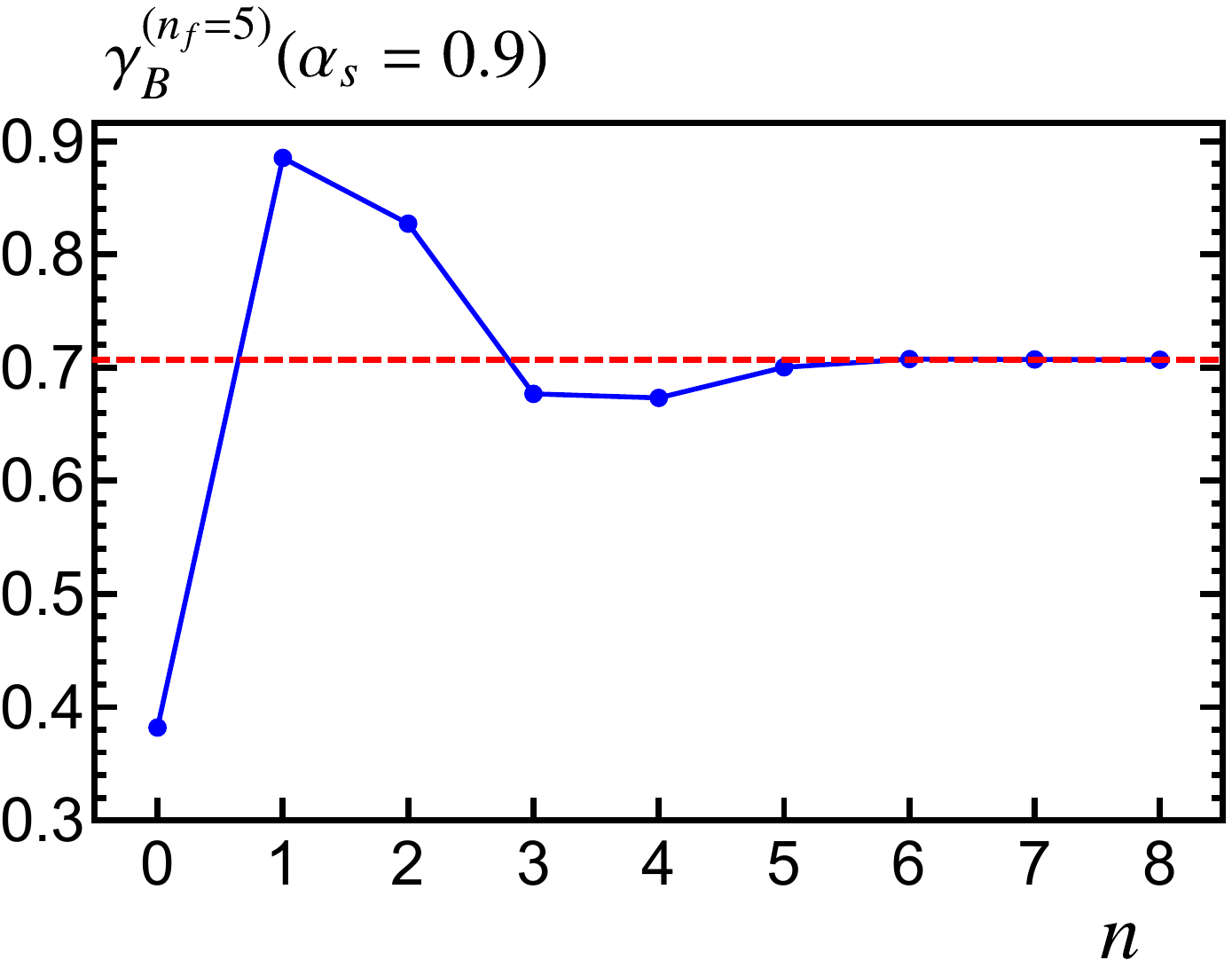}
\label{fig:gammaB}}
\caption{Comparison of the fixed-order expansion (blue dots) and exact results (red dashed line) for the non-cusp anomalous dimension of the bHQET matching coefficient
(left panel) and the bHQET jet function (right panel), in the large-$\beta_0$ approximation for $\alpha_s=0.9$.} \label{fig:BHQET}
\end{figure*}

At lowest one has $C_m = 1$. Defining the corrections to the bare matching coefficient as $C_m = 1 + \delta C_m(m)$, and using the massive SCET Feynman rules that can be found in
Ref.~\cite{Leibovich:2003jd}, the computation of diagram \ref{fig:collinear} using a modified gluon propagator yields
\begin{align}
\delta C_m^{(a)}(m) =\,& 2 i g_0^2 C_F\!\! \int \!\!\frac{\mathd^d \ell}{(2 \pi)^d}
\Biggl[ \frac{p_1^-}{(- \ell^2)^{1 + h} \ell^- [(\ell + p_1)^2 - m^2]} +
\frac{1}{(- \ell^2)^{1 + h} [(\ell + p_1)^2 - m^2]} \Biggr]\nonumber\\
=\,& \frac{g_0^2 C_F}{(4 \pi)^{2-\varepsilon}} \frac{(h + 1) \Gamma (h + \varepsilon) \Gamma (1
- 2 h - 2 \varepsilon)}{(h + \varepsilon) \Gamma (2 - h - 2 \varepsilon)m^{2(h+\varepsilon)}}\,.
\end{align}
Doubling the result above and including the wave-function renormalization we obtain the modified 1-loop correction
necessary to determine the leading $1/\beta_0$ term:
\begin{align}
\delta C_m(m) =\, & \frac{2 g_0^2 C_F}{(4 \pi)^{2-\varepsilon}} \frac{(1+h)
\Gamma (1 + h + \varepsilon) \Gamma (1 - 2 h - 2 \varepsilon)}{(h + \varepsilon)^2\,\Gamma (3 - h
- 2 \varepsilon)m^{2 (h+\varepsilon)}} \\
& \times \!\Bigl\{ 2 - 4 h (1 - \varepsilon)^2 + h^2 (3 - 2 \varepsilon) - \varepsilon \bigl[5 - (5 - 2 \varepsilon) \varepsilon\bigr]\! \Bigr\} .
\nonumber
\end{align}
Identifying ${\cal Q}=m$ we obtain the following form for the renormalon master function $G(\varepsilon,u)$:
\begin{align}\label{eq:GCM}
G_{C_m}\!(\varepsilon,u) =\,& 2 C_F D (\varepsilon)^{\frac{u}{\varepsilon} - 1} e^{\varepsilon \gamma_E} \frac{\Gamma (1 - 2 u) \Gamma (1+u)}{\Gamma (3 - u - \varepsilon)}\\
&\times (1 + u - \varepsilon) \Bigl\{2 - \varepsilon- u \bigl[4 + 2 (u - 1) \varepsilon - 3 u\bigr]\! \Bigr\}.\nonumber
\end{align}
For the $H_m$ factor one gets an additional factor of two: $G_{\!H_m}=2G_{C_m}$.
Equation~\eqref{eq:GCM} reproduces the cusp anomalous dimension noting that $\Gamma_{\!C_m} = 2\Gamma_{\!\rm cusp}$.

To determine the $C_m$ matching coefficient, on top of computing the relevant Feynman diagrams and
accounting for the wave-function renormalization, one needs to renormalize the SCET current multiplying the bare result with
$Z_{C_H}$ as computed in Sec.~\ref{sec:hard}. At leading order in $1/\beta_0$ this amounts to adding $\delta Z_{C_H}$ and
therefore the renormalized $C_m$ at this order is obtained simply by dropping the divergent terms, denoted by
$\delta Z_{C_m}^{(0)}$, from the bare computation. The renormalization factor for $C_m$
is then $\delta Z_{C_m}=\delta Z_{C_m}^{(0)} + \delta Z_{C_H}$ and accordingly, the $H_m$ anomalous dimension is
$\gamma_{H_m}=\gamma_H + 2\gamma_{C_m}^{(0)}$, where the last term is computed from Eq.~\eqref{eq:nocuspPart} with
the result $G_{\!C_m}\!(-\beta,0) = -G_{Q}(-\beta,0)$ using\footnote{The SCET current renormalization does not affect the direct determination of the
cusp anomalous dimension from Eq.~\eqref{eq:GCM}.}
\begin{align}
&\frac{{\df}G_{\!C_m}(-\beta,u)}{{\df} u} \biggr|_{u = 0} = \frac{C_F
\Gamma(4 + 2 \beta)}{3\Gamma(2 + \beta)^3 \Gamma (1 -\beta)} \Biggl\{ \!(1 + \beta) \psi^{(0)} (1 + \beta) - 2 \beta + \frac{1 + \beta}{2 + \beta} \\
&\qquad\qquad\qquad\quad\qquad - \frac{1 + \beta}{\beta} \log\! \Biggl[ \frac{6e^{-2 \beta\gamma_E} \Gamma(2 + \beta)^2
\Gamma (1 - \beta)}{\Gamma (4 + 2 \beta)} \Biggr]\! \Biggr\}\nonumber\\
& \qquad = 2 C_F \exp\! \Biggl[ \frac{2 \beta}{3} + \sum_{n = 2}
\frac{(-\beta)^n}{n} \Bigl\{ 2 - 2^n (1 + 3^{- n}) + \zeta_n \bigl[2^n - 3 - (- 1)^n \bigr]\!
\Bigr\}\! \Biggr]\! \Biggl\{\!- 2 \beta + (1 + \beta) \nonumber\\
& \qquad\quad\times \!\!\Biggl(\! \frac{13}{6} +\! \sum_{n = 1} \frac{(-\beta)^n}{n + 1}
\bigg\{2^{n + 1} +\! \biggl(\frac{2}{3}\biggr)^{\!\!n + 1}\! + \frac{n + 1}{2^{n+1}} - 1 + \zeta_{n + 1} \!\Bigr[1 - n
- (- 1)^n - 2^{n + 1}\Bigl]\!\biggr\} \!\!\Biggr) \!\!\Biggr\}.\nonumber
\end{align}
The second equality is expressed as a Taylor polynomial times the exponential of another series, in a form amenable to be re-expanded by a computer program with the algorithms
already explained. This result can be used to obtain a closed form for $\gamma_{H_m}$ and its fixed-order coefficients, whose series has convergence radius $\Delta \beta=5/2$. We
find full agreement with the leading flavor structure of the known full bHQET results up to $\mathcal{O}(\alpha_s^3)$~\cite{Fleming:2007xt,Jain:2008gb,Hoang:2015vua}, collected in the
fifth column of Table~\ref{tab:gamma}. In Fig.~\ref{fig:gammaBHQET} we compare the exact form of the bHQET non-cusp anomalous dimension for
$\alpha_s=0.9$ with its fixed-order partial sum including up to $\gamma_{H_m}^n$. Even for this enormous value of the strong coupling, the expanded result converges to the
exact answer.

\begin{figure*}[t!]
\subfigure[]
{\includegraphics[width=0.46\textwidth]{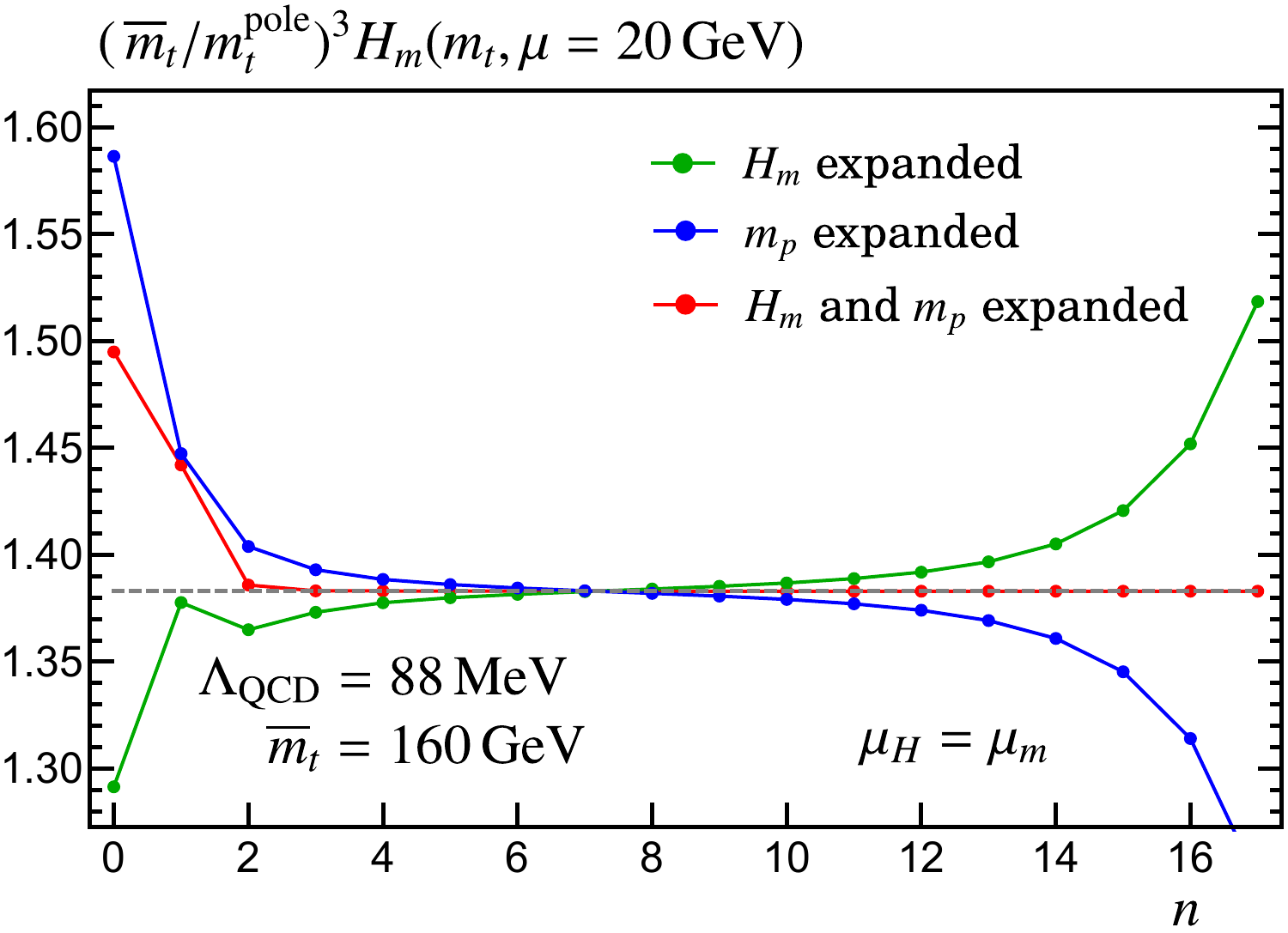}
\label{fig:HmNpExp}}~~~~
\subfigure[]{\includegraphics[width=0.46\textwidth]{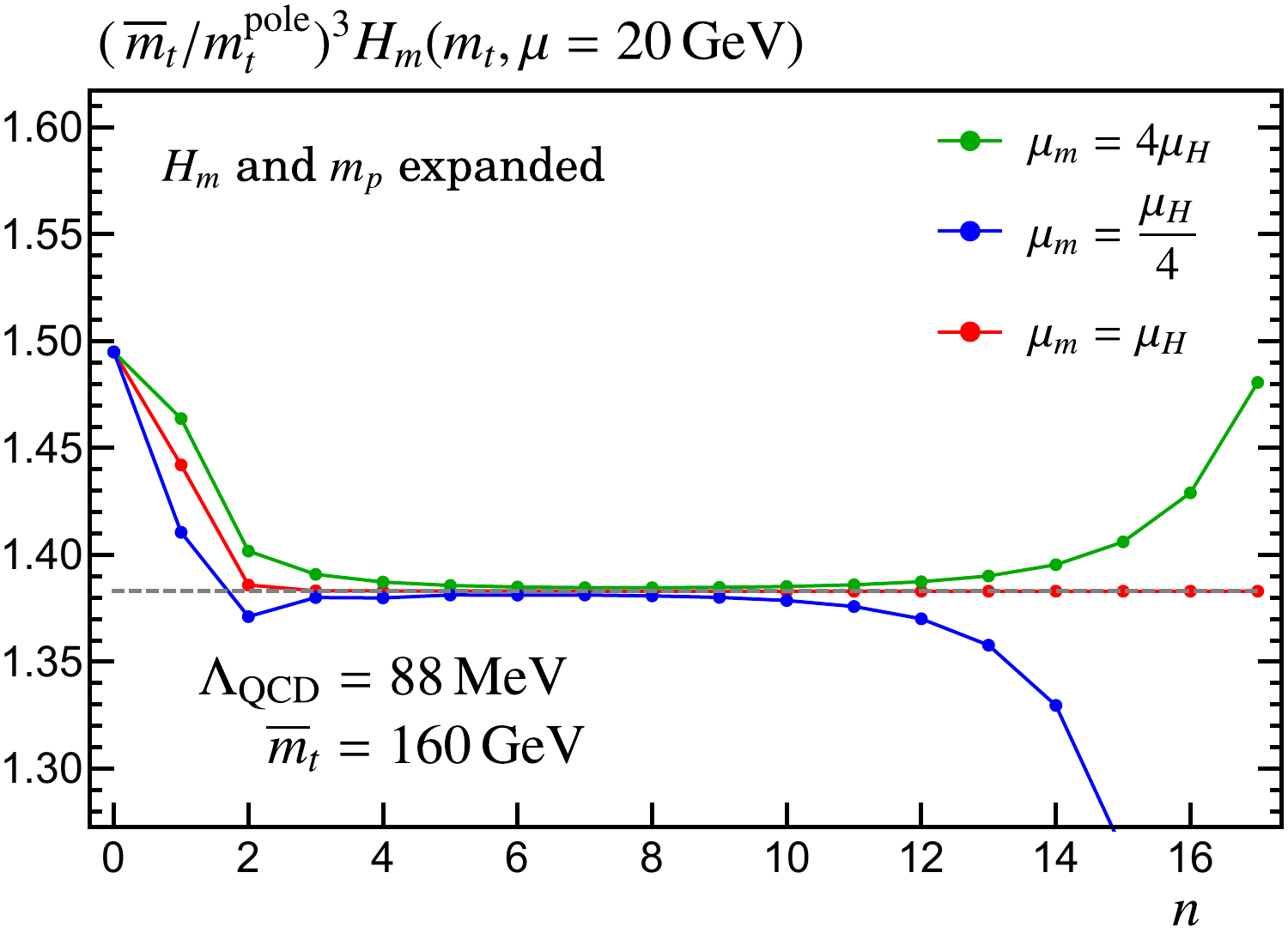}
\label{fig:HmExp}}
\caption{Comparison of the fixed-order partial sum (colored dots) and exact results (red dashed line) for the dimensionless and renormalon-free quantity
$(\overline m_t/m_t^{\rm pole})^3H_m(m_t,\mu)$,
with $\mu=20\,$GeV.
In panel (a) we expand only $H_m$ (green dots), only $m_p$ (blue dots), or both (red dots), using $\alpha_s({\overline m})$ as the expansion parameter in all cases.
In panel (b) we expand $H_m$ in terms of $\alpha_s({\overline m})$ and $m_p$ in powers of $\alpha_s(\mu_m)$ with
$\mu_m=4\overline m_t$ (green dots), $\mu_m=\overline m_t$ (red dots) and $\mu_m=\overline m_t/4$ (blue dots).} \label{fig:Hm}
\end{figure*}

To obtain the $C_m$ matching coefficient at leading order in $1/\beta_0$ we need the following function, which enters the inverse-Borel transform integral:
\begin{align}
G_{\!C_m}(0,u) = \,& 2 C_F(2 -4 u + 3 u^2)\frac{ e^{\frac{5 u}{3}} \Gamma (1 - 2 u) \Gamma (2 + u)}{\Gamma (3 - u)} \\
= \,& C_F (2 - 4 u + 3 u^2) \exp \!\Biggl[ \frac{25 u}{6} + \!\!\sum_{n = 2} \frac{u^n}{n}
\Bigl\{ 1 - (- 1)^n + 2^{- n} \!+ \zeta_n \bigl[2^n \!- 1 + (- 1)^n \bigr]\! \Bigr\}\! \Biggr] .\nonumber
\end{align}
where again we provide a partially expanded expression. With the result above we reproduce the leading flavor structure of the full bHQET coefficients up to
two loops~\cite{Fleming:2007xt,Hoang:2015vua}, see column $3$ in Table~\ref{tab:matEl}. The Borel transform
$C^m_{\!B}(u)\equiv \!\bigl[G_{C_m}(0,u) - (G_{C_m})_{0,0} - u(G_{C_m})_{0,1}\bigr]/u^2$
has only simple poles at half-integer values of $u$, including an $\mathcal{O}(\Lambda_{\rm QCD})$ renormalon.
Its asymptotic expansion can be written in terms of a finite and an infinite sum
\begin{align}
C^m_{\!B}(u) \asymp \,& C_F\Biggl\{ \,\sum_{k = 1}^5 \frac{(- 1)^k e^{\frac{5 k}{6}}\, [8 + k (3 k - 8)] \,\Gamma \bigl(
\frac{k}{2} + 2 \bigr)}{k \,\Gamma \bigl( 3 - \frac{k}{2} \bigr) k!}\frac{1}{u - \frac{k}{2}}\\
& -\sum_{k = 3} \frac{e^{\frac{5 k}{3} + \frac{5}{6}} \, [3 + 4 k (3 k - 1)]\,
\Gamma \bigl( k + \frac{5}{2} \bigr)}{(2 k + 1) \Gamma \bigl( \frac{5}{2} - k
\bigr) (2 k + 1) !} \frac{1}{u - \frac{2 k + 1}{2}}\Biggr\}.\nonumber
\end{align}
The leading ambiguity for $H_m$, which includes a factor of two, is three times as large as the pole mass ambiguity with the same sign:
\begin{equation}
\delta_{\Lambda} H_m = -\frac{6e^{\frac{5}{6}}C_F}{\beta_0}\frac{\Lambda_{\rm QCD}}{m}\,.
\end{equation}
Therefore, the combination $H_m/m_p^3$ is free from the leading ambiguity. To illustrate how this renormalon cancellation works in practice we refer to
Fig.~\ref{fig:Hm}, where the dimensionless quantity $(\overline m_t/m_t^{\rm pole})^3H_m(m_t,\mu)$ is considered, showing its exact value as
a gray dashed line in both panels.
In these plots we highlight some important aspects of consistently expanding perturbative series to achieve an effective renormalon cancellation. For the perturbative expansions
we use $\mu_H=\overline m_t$ as the matching scale, and therefore for a correct cancellation one needs to expand $(\overline m_t/m_t^{\rm pole})^3$ in powers of
$\alpha_s(\mu_H)$. In Fig.~\ref{fig:HmNpExp} we either keep $m_p$ exact expanding in powers of $\alpha_s(\mu_H)$ only $H_m(m_t,\mu_H)$ (green dots);
keep $H_m(m_t,\mu)$ exact expanding only $(\overline m_t/m_t^{\rm pole})^3$ (blue dots); or expand both (red dots), and only in the last case we see
a better-behaved perturbative expansion (keeping both factors exact generates the gray line). If Fig.~\ref{fig:HmExp} we expand
$H_m$ and $(\overline m_t/m_t^{\rm pole})^3$ in powers of $\alpha_s(\mu_H)$ and $\alpha_s(\mu_m)$, respectively, with $\mu_m=\xi \times\mu_H$ and $\xi = 4,1/4,1$
shown as green, blue, and red dots, respectively. As expected, only when $\xi=1$ the asymptotic behavior is removed. In both panels we observe that the removal
of the leading renormalon ambiguity makes the series approach the exact result after fewer orders have been included in the partial sum. The $H_m$ hard factor
mainly affects the norm of the distribution. In Ref.~\cite{Bachu:2020nqn} this renormalon was not properly accounted for, and this fact, together with the lack of $\pi$ summation
in $H_Q$, might explain the poor convergence found in peak cross sections for boosted top pairs unless the curves were self-normalized.

\subsection{Jet function}\label{sec:bJet}
The last matrix element we consider in this article is the bHQET jet function for hemisphere masses. The relevant Feynman diagrams coincide with those of the SCET
jet function shown in Fig.~\ref{fig:jet-function}, but the actual computation uses bHQET Feynman rules in which the heavy-quark momentum $p=mv+k$ with $v^2=1$ has a
residual component $k$. We follow the same computational strategy as for the determination of the SCET jet function in Sec.~\ref{sec:jet} based on the following
forward-scattering matrix element:
\begin{equation}
m\, \mathcal{B}_n(\hat s,\mu) = \frac{1}{8\pi N_c} \!\int\! \mathd^d x \,e^{i k \cdot x}\,
\tmop{Tr} \langle 0 | {\rm T}\bigl \{ h_v (x) \overline{h}_v (0) W_n (0) W_n^{\dag}(x) \bigr\} | 0 \rangle\,,
\end{equation}
with $h_v$ an HQET massive quark field, $\hat s=2v\cdot k$ and $W^{(\dagger)}_n$ the collinear Wilson line already introduced in Sec.~\ref{sec:jet}. Its discontinuity along the
branch cut $\hat s > 0$ defines the momentum-space jet function $B_n(\hat s, \mu)$, with support for $\hat s>0$. Both functions dimensions of an inverse squared mass,
although the kinematic variable
$\hat s\equiv(s-m^2)/m$
they depend on has dimensions of energy. Since the fixed-order expansion of $B_n(\hat s, \mu)$ contains Dirac delta and plus distributions [\,even though
$\mathcal{B}_n(\hat s,\mu)$ is a regular function\,], it is convenient to consider its Fourier transform
\begin{equation}
\tilde{B}_n (x, \mu) \equiv m\!\! \int_0^{\infty}\!\! \text{d} \hat{s}\, e^{- i \hat{s} x} B_n (\hat{s}, \mu)\,,
\end{equation}
which due to the prefactor $m$ is dimensionless and depends on $x$, the variable conjugate to $\hat s$ with dimensions of an inverse energy. At lowest order one
has
\begin{equation}
m\,\mathcal{B}^{\tmop{tree}}_n (\hat{s}) =
\frac{1}{2 \pi} \frac{i}{\hat s + i 0^+} \,,
\qquad m B_n^{\tmop{tree}} (\hat{s})= \delta (\hat{s}) \,,
\qquad {\tilde B}_n^{\tmop{tree}}(y) = 1 \,.
\end{equation}
Quantum corrections to the bare jet function, defined as $B_n(\hat s) = \delta(\hat s) + \delta B_n (\hat s)$ and \mbox{$\tilde B_n(x) = 1 + \delta \tilde B_n(x)$}, are computed next
at one-loop with a modified gluon propagator.
The diagram shown in Fig.~\ref{fig:vertex-correction} can be expressed as the vertex correction times an HQET propagator:
\begin{align}
m \mathcal{B}^a_n(\hat s)=\,& \frac{1}{2 \pi} \frac{i}{\hat{s} + i 0^+} \,\mathcal{M}_a(\hat s)\,,\\
\mathcal{M}_a(\hat s) =\,& i C_F g^2_0 (v \cdot \overline{n})\!\! \int \!\frac{\mathd^d \ell}{(2 \pi)^d}
\frac{1}{(- \ell^2)^{1 + h} [v \cdot (\ell + k)] \, \overline{n} \cdot \ell} \nonumber\\
=\, &\! - \! \frac{2C_Fg_0^2}{(4 \pi)^{2-\varepsilon}} (- \hat s - i 0^+)^{- 2 h-2\varepsilon}
\frac{\Gamma (- h - \varepsilon) \Gamma (2 h + 2 \varepsilon)}{\Gamma (1 +
h)}\,. \nonumber
\end{align}
On the other hand, the self-energy diagram in Fig.~\ref{fig:quark-selfenergy-jet} can be written in terms of the bHQET self-energy $\mathcal{M}_b$ and a squared
HQET propagator
\begin{align}
m \mathcal{B}^b_n(\hat s)=\,&\!-\! \frac{i}{\pi} \biggl( \frac{1}{- \hat s - i 0^+} \biggr)^{\!\!2} \mathcal{M}_b(\hat s)\,,\\
\mathcal{M}_b(\hat s) =\,&\! - \!i \,C_{\!F} g_0^2\!\! \int\! \frac{\mathd^d \ell}{(2 \pi)^d}
\frac{1}{(- \ell^2)^{1 + h} v \!\cdot\! (\ell + k)}\nonumber\\
=\, & \frac{4 C_F g_0^2}{(4\pi)^{2-\varepsilon}} (- \hat s - i 0^+)^{1 - 2 h-2\varepsilon}\,
\frac{\Gamma (2 - h - \varepsilon) \Gamma (2 h - 2 + 2 \varepsilon) }{\Gamma (1 + h)} \,.\nonumber
\end{align}
Combining the two diagrams with the appropriate factor of two for the vertex correction, and taking the discontinuity using Eq.~\eqref{eq:impart} with
$\eta= 2(\varepsilon+ h)$, one obtains the modified one-loop momentum-space bHQET jet function,
which can be converted to position space:
\begin{align}\label{eq:bJET1loop}
m \delta B_n^h (\hat{s}) =\,& -\! \frac{4C_Fg_0^2}{(4\pi)^{2-\varepsilon}} \frac{\Gamma (2 - h -
\varepsilon) \hat{s}^{- 1 - 2 h-2\varepsilon}}{(h + \varepsilon)
\Gamma(1+h) \Gamma (2 - 2 h - 2 \varepsilon)} \,, \\
\delta \tilde B_n^h (x) =\,& \frac{2C_Fg_0^2}{(4\pi)^{2-\varepsilon}} \frac{\Gamma (2 - h - \varepsilon) (i x )^{2(h+\varepsilon)}}
{(1 - 2 h - 2 \varepsilon) (h + \varepsilon)^2\Gamma (h + 1)}\,.\nonumber
\end{align}
To obtain the cusp and non-cusp anomalous dimensions for the bHQET jet function we proceed in position-space. To that end, we identify
${\cal Q} = -ie^{-\gamma_E}/x$ and the 1-loop coefficient $b(\varepsilon,h)$, and obtain the following renormalon master function
\begin{equation}\label{eq:masterB}
G_{\!\tilde B}(\varepsilon,u) = 2 C_{\!F} D (\varepsilon)^{\frac{u}{\varepsilon} - 1} \frac{e^{(\varepsilon-2u)
\gamma_E} \Gamma (2 - u)}{(1 - 2 u) \Gamma (1 + u - \varepsilon)}\,.
\end{equation}
Taking into account the relation $\Gamma_{\!\!\tilde B}=2\Gamma_{\!\rm cusp}$ we find that $2\beta G_{\!\tilde B}(-\beta,0)/\beta_0$ reproduces once more the cusp
anomalous dimension in Eq.~\eqref{eq:cusp-closed}. To obtain the non-cusp anomalous dimension we need the $u$ derivative of $G_{\!\tilde B}(\varepsilon,u)$ at $u=0$:
\begin{align}
&\frac{{\df}G_{\!\tilde B}(-\beta,u)}{{\df} u}
\biggr|_{u = 0} =\frac{C_F (1 +\beta)\Gamma (4 + 2 \beta)}{3 \Gamma (2 + \beta)^3 \Gamma (1 - \beta)} \Biggl\{ 1 -
\psi^{(0)} (1 + \beta)
-\! \frac{1}{\beta} \log\! \Biggl[\! \frac{6\Gamma (2 + \beta)^2 \Gamma (1 -\beta)}{\Gamma (4 + 2\beta)} \!\Biggl] \!\Biggr\}\nonumber\\
&\qquad\qquad\qquad=2 C_F \exp\! \Biggl[\frac{5 \beta}{3} + \sum_{n = 2}
\frac{(-\beta)^n}{n} \Bigl\{\! 1 - 2^{n} -\!\biggl(\frac{2}{3}\biggr)^{\!\!n}\!\! + \zeta_n \bigl[2^n - 3 - (- 1)^n\bigr]\!
\Bigr\}\! \Biggr]\\
&\qquad\qquad\qquad \times\!\! \Biggl[ \frac{8}{3} + \sum_{n = 1} \frac{(-\beta)^n}{n + 1}
\Bigl\{ 2^{n + 1} +\!\biggl(\frac{2}{3}\biggr)^{\!\!n + 1}\!\! - 1 + \zeta_{n + 1} \bigl[3 + n - 2^{n + 1} - (- 1)^n\bigr]\! \Bigr\}\!
\Biggr],\nonumber
\end{align}
where second equality is expressed as a series times the exponential of another series, ideal for a full re-expansion in powers of $\beta$. This expression can be used in
Eq.~\eqref{eq:nocuspPart} together with $G_{\!\tilde B}(-\beta,0) = -G_{Q}(-\beta,0)$ to obtain a closed form for $\gamma_B(\beta)$, as well as the fixed-order coefficients,
with convergence radius $\Delta \beta = 5/2$. We find full agreement with the leading flavor structure of full bHQET results up to
$\mathcal{O}(\alpha_s^3)$~\cite{Fleming:2007xt,Jain:2008gb,Hoang:2015vua}, see last column of Table~\ref{tab:gamma}. From the bHQET consistency condition one can
deduce the following relation between
various non-cusp anomalous dimensions: $\gamma_{H_m}=-2(\gamma_S+\gamma_B)=2(\gamma_H+\gamma_J-\gamma_B)$, where in the second equality we use the
SCET consistency condition to eliminate $\gamma_S$. This relation is exactly verified by our computations at $\mathcal{O}(1/\beta_0)$.
In Fig.~\ref{fig:gammaBHQET} we compare the exact result for $\gamma_B$ with the fixed-order partial sum including up to $\gamma_B^n$. We use
$\alpha_s=0.9$,
but nevertheless find that after adding $6$ or more terms the perturbative result nicely agrees with the exact computation.

To compute the non-logarithmic fixed-order coefficients $a^B_{i, i-1,0}$ of the Fourier-space bHQET jet function, which we define as
\begin{equation}
\tilde B_n = \sum_{i = 1} \beta^i \sum_{j = 0}^{i + 1} \tilde a^B_{i, i-1,j} \log^j (i x e^{\gamma_E} \mu)\,,
\end{equation}
we need to evaluate the renormalon master function at $\beta = 0$:
\begin{figure*}[t!]
\subfigure[]
{\includegraphics[width=0.47\textwidth]{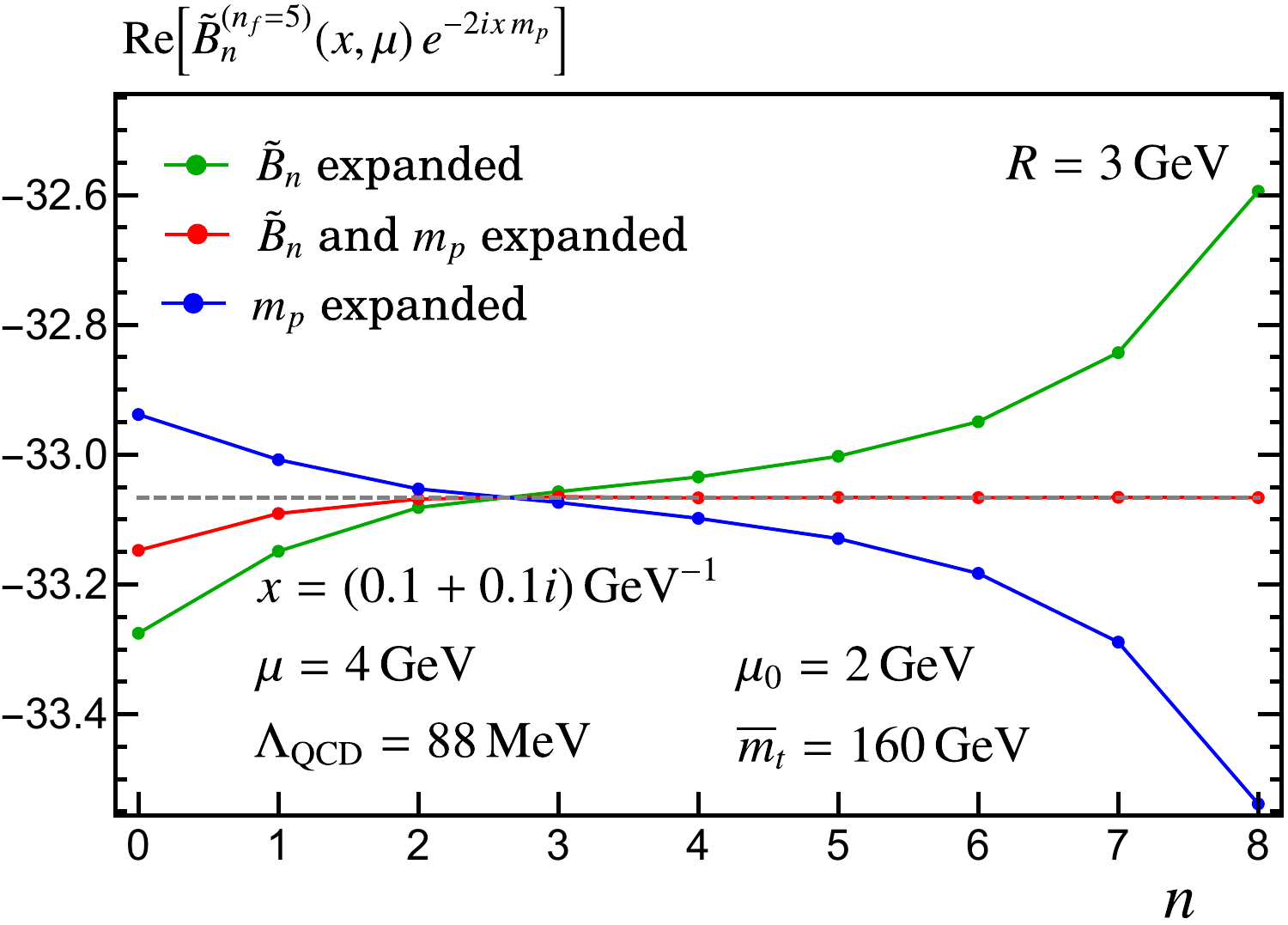}
\label{fig:BRe}}~~~~
\subfigure[]{\includegraphics[width=0.465\textwidth]{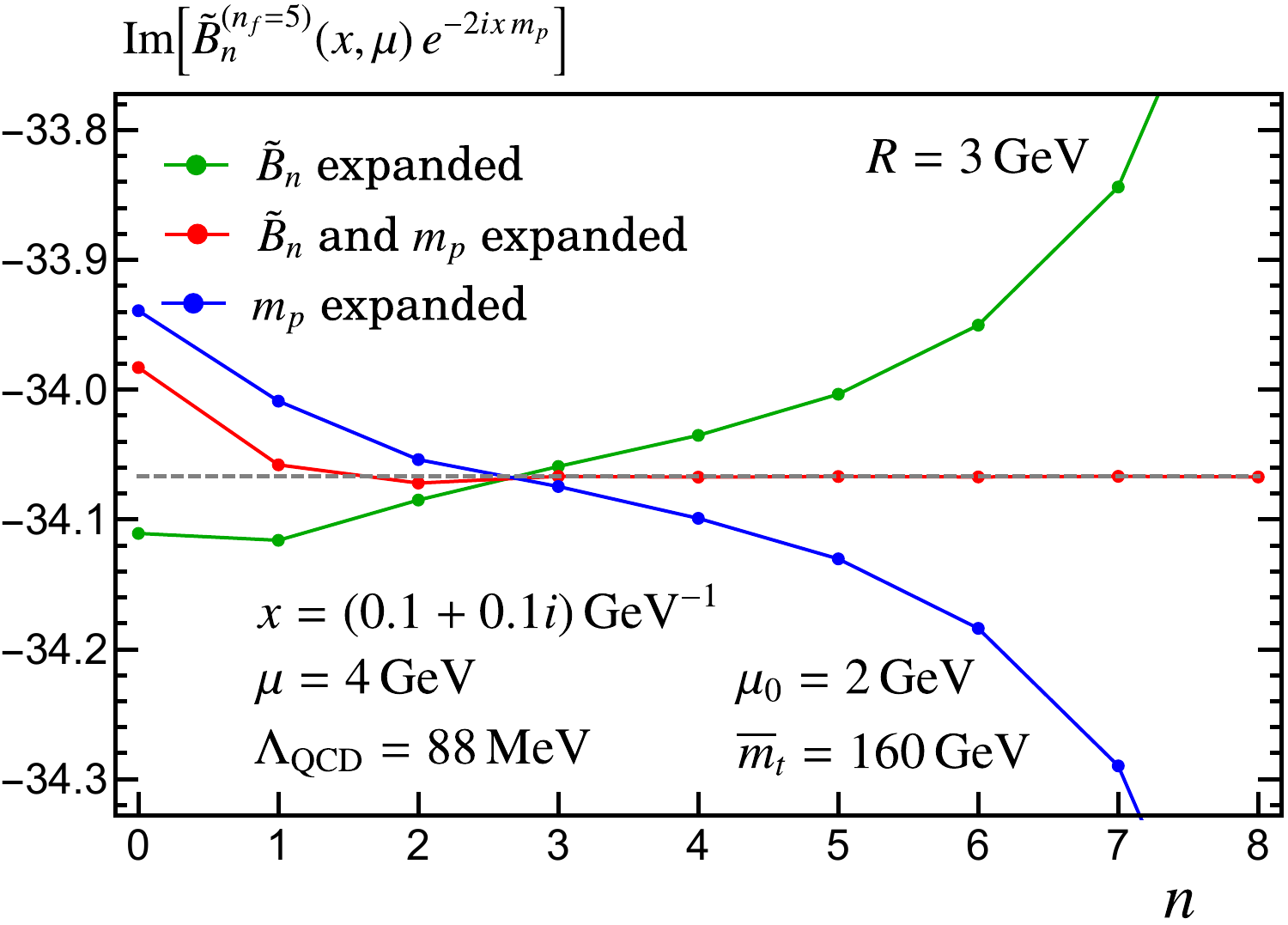}
\label{fig:BIm}}
\caption{Comparison of the fixed-order partial sum (colored dots) and exact results (red dashed line) for the real (left panel) and imaginary (right panel) parts
of the position-space bHQET jet function with renormalon subtraction $\tilde B_n(x,\mu)e^{-2ixm_p}$ in the MSR scheme. The plot uses
\mbox{$x=(0.1 + 0.1 i)\,{\rm GeV}^{-1}$} and $\mu=4\,$GeV. The fixed-order results employ $\mu_0=2\,$GeV and $R=3\,$GeV, and include $n+1$ terms in the
matching condition plus N$^n$LL resummation. We show fixed-order results expanding $\tilde B_n$ only (green), expanding $m_p$ only (blue), and expanding both
(red).} \label{fig:BJet}
\end{figure*}
\begin{equation}
G_{\!\tilde B} (u,0) = \frac{2 C_F e^{\left( \frac{5}{3} - 2 \gamma_E\! \right) u} \Gamma (2 - u)}{(1 - 2 u) \Gamma (1 + u)} =
2 C_F \exp \!\Biggl[\frac{8 u}{3} + \sum_{n = 2} \frac{u^n}{n} \Bigl\{ 2^n - 1 + \zeta_n \bigl[1 - (- 1)^n\bigr] \!\Bigr\} \!\Biggr],
\end{equation}
again provided in a closed form and as the exponential of a series. With this result we correctly reproduce the leading flavor structure of the full theory up to
two loops~\cite{Fleming:2007xt,Jain:2008gb}, see last column of Table~\ref{tab:matEl}. The Borel transform
${\tilde B}_{\!B}(u)\equiv \!\bigl[G_{\!\tilde B}(0,u) - (G_{\!\tilde B})_{0,0} - u (G_{\!\tilde B})_{0,1}\bigr]/u^2$
has simple poles at $u=1/2$ (related to the pole mass ambiguity) and integer values of $u\geq2$. Its asymptotic expansion is therefore expressed
as an isolated term plus an infinite sum:
\begin{equation}
\tilde B_B (u) \asymp - 2C_F\! \Biggl[ \frac{2\,e^{\left( \frac{5}{6} -\gamma_E \right)}}{u - \frac{1}{2}} + \sum_{n =
2} \frac{(- 1)^n (n-1)\,e^{n \left( \frac{5}{3} - 2 \gamma_E \right)}}{n(1 - 2 n) (n!)^2} \,\frac{1}{u - n} \Biggr] .
\end{equation}
The full ambiguity can be written in terms of Bessel functions of the first kind and generalized hypergeometric functions.
The leading ambiguity is, as expected, proportional to $\Lambda_{\rm QCD}$ and takes the following form in position and momentum space:
\begin{equation}\label{eq:delt-Bn}
\delta_{\Lambda}\tilde B(x) = -\frac{4e^\frac{5}{6}C_F}{\beta_0}(i x\Lambda_{\rm QCD})\,,\qquad
\delta_{\Lambda} B(x) = \frac{4 e^\frac{5}{6}C_F}{\beta_0}\frac{\Lambda_{\rm QCD}}{\hat s} \delta(\hat s)\,.
\end{equation}
In position space, except for the factor $ix$, the ambiguity is exactly twice that of $\delta_{\MSBar}(\overline m)$. If one writes the position-space version of
the factorization theorem shown in Eq.~\eqref{eq:factbHQET}, the bHQET jet function {\it effectively} appears in the combination
\mbox{$\hat B_n(x,m_p,\mu) \equiv \tilde B_n(x,\mu) e^{-2 i x m_p}$}, which is free from the leading renormalon ambiguity. Therefore, in order to have a stable perturbative expansion
for $\hat B_n$ one needs to express $m_p$ in a short-distance scheme and expand consistently in powers of $\alpha_s(\mu)$ with $\mu \simeq 1/x\simeq\hat s$. Since the heavy
quark mass is no longer a dynamic scale of the effective theory, one should use a low-scale short-distance scheme whose relation to the pole mass is not proportional to the mass
itself, such as $m^{\rm MSR}$. This avoids having renormalon-subtractions that are much larger than the corresponding fixed-order corrections for $\tilde B_n$, thus
breaking the bHQET power counting. Furthermore, if one chooses $R\sim \mu$ there will be no large logarithms in the renormalon subtraction series. In Fig.~\ref{fig:BJet}
we show how the leading renormalon effectively cancels for the case of the top quark,
both for the real and imaginary parts in panels \ref{fig:BRe} and
\ref{fig:BIm}, respectively. Since the real part of $\tilde B_n$ does not have the $u=1/2$ ambiguity if $x$ is real, we choose a complex number,
but for simplicity keep the renormalization scale $\mu$ real.
When truncating the fixed-order series for $\tilde B_n$ and $\delta_{\rm MSR}$ at any finite order
one gets residual dependence on $R$ and $\mu$.
In the figures we use $n+1$ terms in the fixed-order partial sum, plus N$^n$LL resummation both in $\tilde B_n$ and $m^{\rm MSR}_t$. For this numerical
analysis we choose $\mu_0\neq R\,$. %GeV and $R=3\,$GeV.
If one expands in powers of $\alpha_s(\mu_0)$ only $\tilde B_n$
(green dots) or $m_p$ (blue dots) the asymptotic behavior remains in the series, and it takes more orders to approach the exact value (shown as a gray dashed line). The
way in which the series diverges in each case seems to be almost exactly opposite. When both functions are expanded consistently (red dots), the series appears much
better behaved for large $n$, and converges to the exact result already at three loops.

\section{Jet masses}\label{sec:JetMasses}
As we have seen in Sec.~\ref{sec:bJet}, in the large-$\beta_0$ approximation the leading renormalon of the position-space bHQET jet function is related to that of the pole mass.
This relation holds in full QCD as well and therefore one can use $\tilde B_n$ to define a low-scale short-distance scheme. Using
the exponentiation properties of $\tilde B_n$, in Ref.~\cite{Jain:2008gb} the so-called jet mass, dependent on
$\mu$ and an infrared scale $R$ analogous to that of the MSR mass, was defined as follows
\begin{equation}\label{eq:defJ1}
m_p - m_J(R, \mu) \equiv \delta m_J (R, \mu) = \frac{e^{\gamma_E} R}{2} \frac{\mathd}{\mathd\! \log (i x)} \log
\bigl[\tilde{B}_n (x, \mu)\bigr]_{i x e^{\gamma_E} = 1 / R}\,.
\end{equation}
This idea was adapted in Ref.~\cite{Hoang:2008fs} to define a subtraction scheme for the hemisphere-mass soft-function renormalon. In Ref.~\cite{Bachu:2020nqn}
it was noted that since such subtractions involve a derivative, their asymptotic behavior starts at two loops. Moreover,
for the canonical choice $\mu=R$ the one-loop soft subtraction vanishes. To solve this shortcoming, a new scheme for the soft function not based on derivatives was introduced in
Ref.~\cite{Bachu:2020nqn}. In this section we define a new jet mass following this same spirit. We study both schemes in the large-$\beta_0$ and derive
closed forms for their relation to the pole mass and R-anomalous dimensions.

\subsection{Derivative scheme}
\begin{figure*}[t!]
\subfigure[]
{\includegraphics[width=0.47\textwidth]{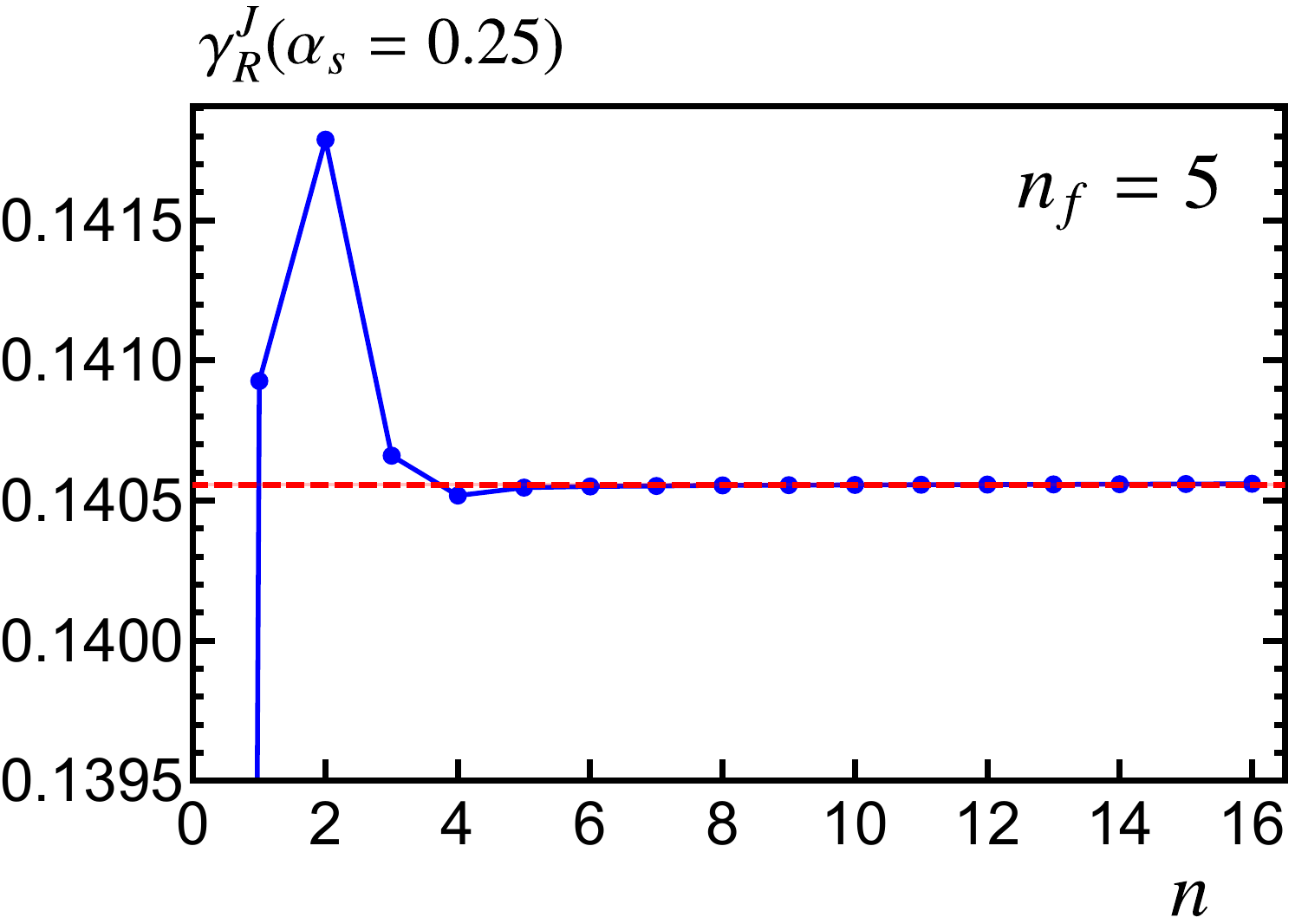}
\label{fig:gammaRJ}}~~~~
\subfigure[]{\includegraphics[width=0.465\textwidth]{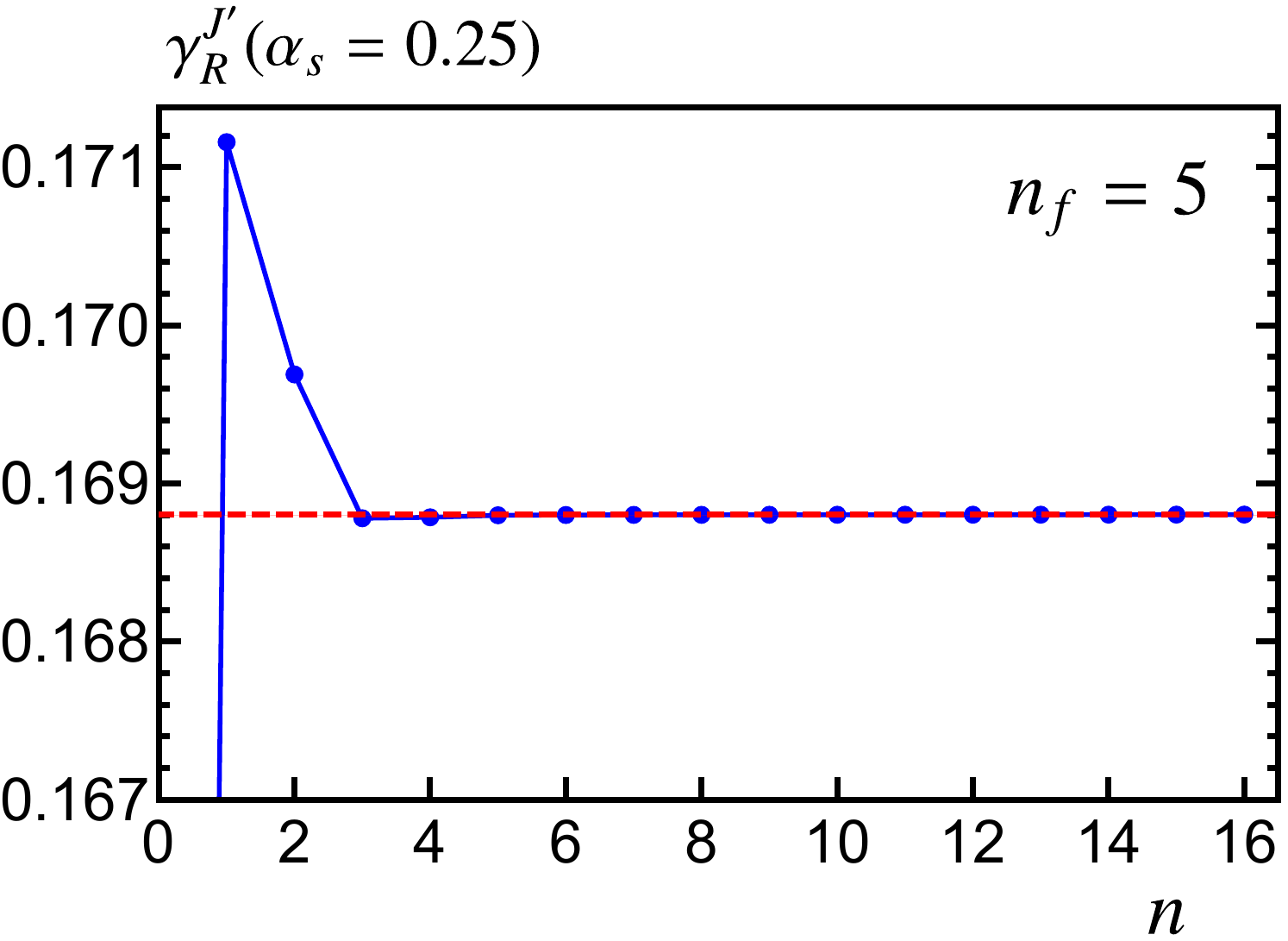}
\label{fig:gammaRJ2}}
\caption{Comparison of the fixed-order partial sum (colored dots) and exact results (red dashed line) for the R-anomalous dimension of the two jet mass schemes:
derivative (left panel) and non-derivative (right panel). In both panels we use and $\alpha_s=0.25$.} \label{fig:gammaRJ12}
\end{figure*}
In the large-$\beta_0$ one has $\log(1+\delta \tilde{B}_n)\simeq \delta \tilde{B}_n$ and the derivative with respect to $\log(ix)$ makes the perturbative series associated
to $m_J (R, \mu)/(Re^{\gamma_E})$ non-cusp type. Indeed, applying Eq.~\eqref{eq:defJ1} to our result in the second line of Eq.~\eqref{eq:bJET1loop}, the modified
one-loop function associated to the jet mass reads
\begin{align}
\frac{\delta m^h_J}{R e^{\gamma_E}} = \frac{2C_Fg_0^2}{(4 \pi)^{2-\varepsilon} } \frac{\Gamma (2 - h - \varepsilon)(R e^{\gamma_E})^{- 2 (h+\varepsilon)}}
{(1 - 2 h - 2 \varepsilon) (h + \varepsilon) \Gamma (1 + h)} \,.
\end{align}
From this result, taking ${\cal Q} = R$, the following renormalon master function is found
\begin{equation}
F_m^J(u,\varepsilon) = 2C_{\!F} D (\varepsilon)^{\frac{u}{\varepsilon} - 1} \frac{e^{(\varepsilon-2u)
\gamma_E} \Gamma (2 - u)}{(1 - 2 u) \Gamma (1 + u - \varepsilon)}\,,
\end{equation}
which is finite for $u\to 0$. From the equation above we can infer that the $\mu$-anomalous dimension of the jet mass is proportional to the universal cusp anomalous dimension.
The evolution in $\mu$ for constant $R$ is therefore given by
\begin{align}
\mu \frac{{\df} m_J(R, \mu)}{{\df} \mu} =\,& -\! Re^{\gamma_E}\, \Gamma_{\rm \! cusp}(\beta)\,,\\[-0.2cm]
m_J(R, \mu_2) - m_J(R, \mu_1) \equiv\,& R\,\Delta^{\mu}_{\rm cusp}(\mu_2,\mu_1) =
\frac{Re^{\gamma_E}}{2}\! \int_{\beta_1}^{\beta_2}\! \frac{{\df}\beta}{\beta^2} \,\Gamma_{\rm \! cusp}(\beta)\,,\nonumber
\end{align}
where in the second line the solution at $\mathcal{O}(1/\beta_0)$ is provided in terms of $\beta_i=\beta(\mu_i)$.
To obtain the fixed-order coefficients of $\delta m_J (R, \mu)$ one needs
\begin{align}
F_m^J(0,u) = \, &2 C_F e^{\left( \frac{5}{3} - 2 \gamma_E \right) u} \frac{\Gamma (2 - u)}{(1 - 2 u) \Gamma (1 + u)}\\
= \, & 2 C_F \exp \!\Biggl[ \frac{8 u}{3} + \!\sum_{n = 2} \frac{u^n}{n} \Bigl\{ 2^n - 1 + \zeta_n \bigl[1 - (- 1)^n \bigr]\! \Bigr\}\! \Biggr],\nonumber
\end{align}
from which one can reproduce the leading flavor structure of the known coefficients for $\delta m_J$ in the full theory up to $\mathcal{O}(\alpha_s^2)$~\cite{Jain:2008gb}.
We can easily find the asymptotic expansion of the corresponding Borel-transform function $m_B^J(u)\equiv\! \bigl[F_m^J(0,u) - F_m^J(0,0)\bigr]/u$:
\begin{equation}
m_B^J(u)\ \asymp 2 C_F\! \Biggl[ -\frac{e^{\frac{5}{6} -
\gamma_E}}{u - \frac{1}{2}} + \sum_{n = 2} \frac{(- 1)^n (n - 1)}{(n!)^2 (2
n - 1)} \frac{e^{\left( \frac{5}{3} - 2\gamma_E \right) n}}{u - n} \Biggr] .
\end{equation}
The leading ambiguity, coming from the $u=1/2$ pole, is independent of $R$ and fully coincides with that of $\delta_{\overline{\rm MS}}(\overline m)$ in Eq.~\eqref{eq:MS-amb},
making $\delta m_J (R, \mu)-\delta_{\overline{\rm MS}}( \overline{m})$ free from the $\mathcal{O}(\Lambda_{\rm QCD})$ renormalon. This renormalon-free difference for
$R=\mu=\overline{m}$ can be used as a matching condition between the jet and $\MSBar$ masses. The R-anomalous dimension of the jet mass is defined along the diagonal
path $\mu=R$:
\begin{equation}
\gamma^J_R(\beta) = -\frac{{\df} m_J(R, R)}{{\df} R}\equiv \frac{1}{\beta_0}\sum_{n=0} \hat \gamma_{R,J}^n\, \beta^{n+1}\,,
\end{equation}
and in the large-$\beta_0$ limit takes the same closed form as Eq.~\eqref{eq:MSRgammaExact} with the obvious substitution $F_{\overline{\rm MS}}\to e^{\gamma_E}F_m^J$,
such that $\gamma^J_R$ does not have a $u=1/2$ ambiguity. Upon expansion in powers of $\beta_R$ one obtains the fixed-order coefficients $\hat \gamma_{R,J}^n$.
In Fig.~\ref{fig:gammaRJ} we compare the exact value of $\gamma_R^J$ (red dashed line), whose ambiguity is too
small to be visible on the plot's scale, with the fixed-order partial sum including up to $\gamma_{R,J}^n$ (blue dots) for a relatively large value of the strong coupling
$\alpha_s=0.25$. The divergent behavior does not manifest for this value within the first $17$ orders shown in the plot.

The solution to this RGE is the same as that explained already in Sec.~\ref{sec:MSR}. In fixed-order perturbation
theory, starting with the value $m_J(\overline{m},\overline{m}) = \overline{m} + \bigl[\delta_{\overline{\rm MS}}( \overline{m}) - \delta m_J (\overline{m}, \overline{m})\bigr]$, one
can obtain the jet mass at the scales $\mu$ and $R$ using the solutions to the corresponding RG equations. Defining $\Delta^{\!R}_J(R_2,R_1)$ as in Eq.~\eqref{eq:Revol-FO}
with the replacement $\gamma_R^n \to \gamma^n_{R,J}$ one proceeds as follows:
\begin{equation}
m_J(R, \mu) = m_J(\overline{m},\overline{m}) + \Delta_J^R(R,\overline{m}) + R\, \Delta^\mu_{\rm cusp}(\mu,R)\,.
\end{equation}
In Fig.~\ref{fig:MSR} we compare the exact value of $m_J(R,\mu)$,
shown as a gray dashed line, with the corresponding computation using the fixed-order partial sum for the matching condition including $n$ terms and R-evolution at N$^n$LL.
We observe that the perturbative result agrees with the exact value obtained with the PV prescription for $3\leq n\leq 10$, and starts diverging for $n>10$.

\begin{figure*}[t!]
\subfigure[]
{\includegraphics[width=0.46\textwidth]{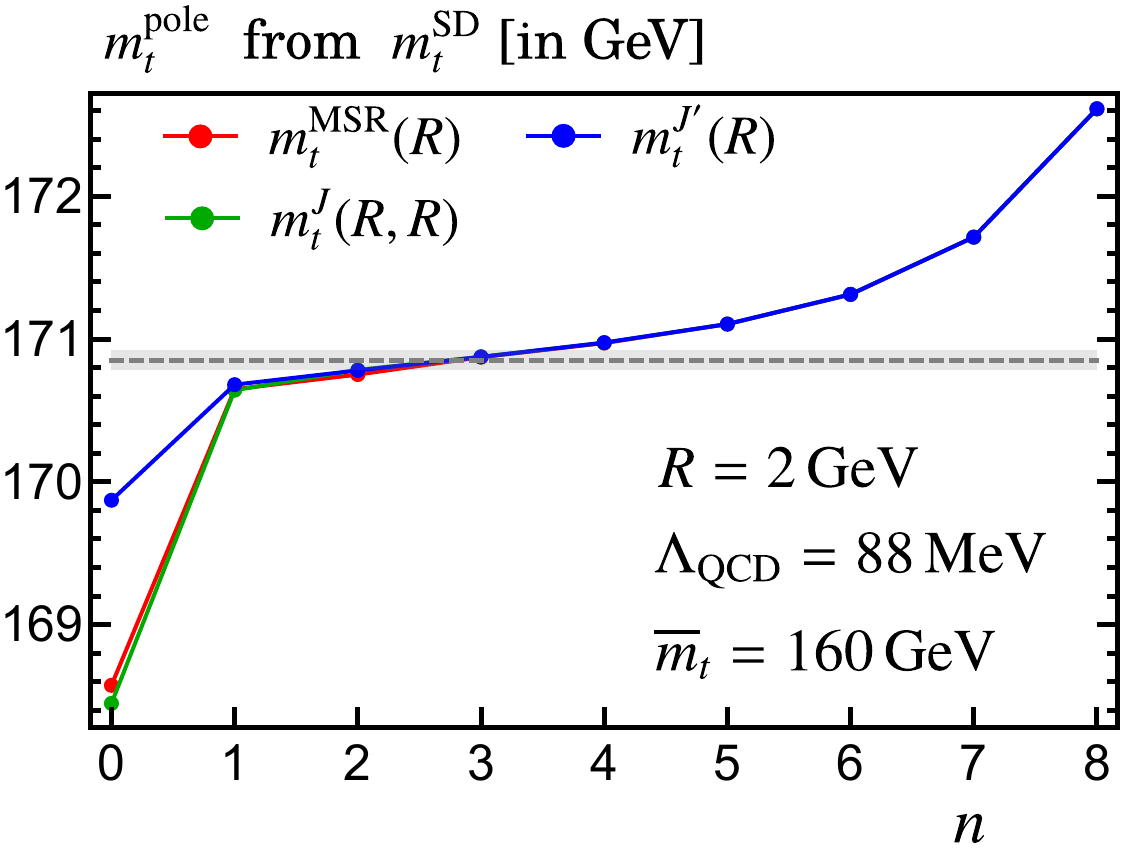}
\label{fig:mtFromMJ}}~~~~
\subfigure[]{\includegraphics[width=0.495\textwidth]{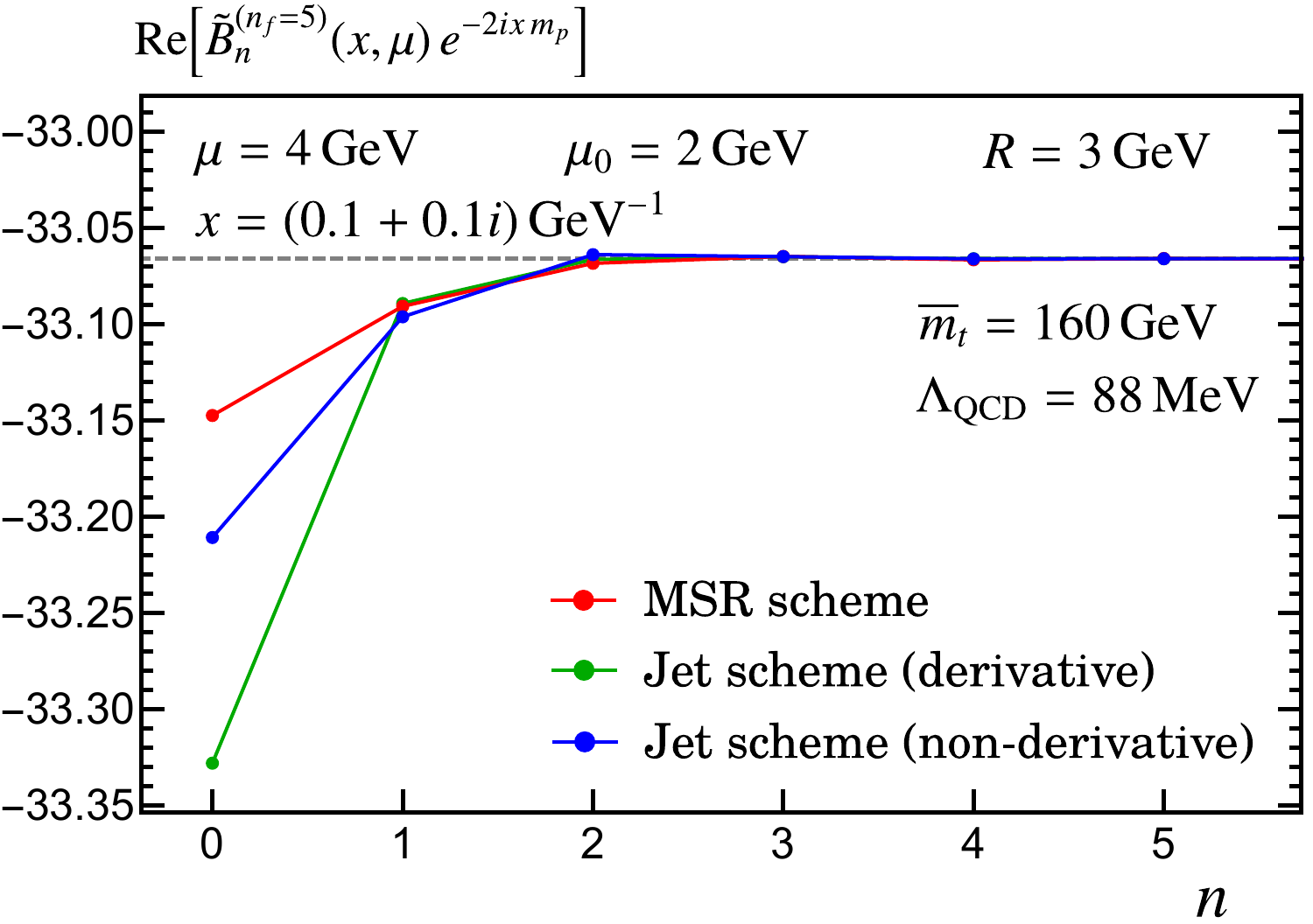}
\label{fig:JB}}
\caption{Left panel: determination of the top quark pole mass from the fixed-order expansion of the MSR mass (red), as well as the derivative (green) and
non-derivative (blue) schemes for the jet mass. We use $R=2\,$GeV in all cases, and for the derivative-scheme jet mass we take $\mu=R$. Right panel:~Comparison of the
fixed-order partial sum (colored dots) and exact results for the real part of the position-space bHQET jet function with renormalon subtraction $\tilde B_n(x,\mu)e^{-2ixm_p}$,
expressing the pole mass in the same schemes as for the left panel with identical color coding, taking $x=(0.1 + 0.1 i)\,{\rm GeV}^{-1}$ and $\mu=4\,$GeV. The fixed-order
expansions have \mbox{$\mu_0=2\,$GeV} and $R=3\,$GeV, and include $n+1$ terms in the matching condition plus N$^n$LL resummation. Both panels: exact results shown as
a gray dashed line with a pale band representing the ambiguity. \label{fig:mJet}}
\end{figure*}

\subsection{Non-derivative scheme}
In this section we consider a low-scale short-distance mass, denoted by $m'_J(R)$, defined form the bHQET jet function which does not depend explicitly on $\mu$.
Based on the prescription introduced in Ref.~\cite{Bachu:2020nqn}, it is defined as
\begin{equation}\label{eq:mJprime}
m_p - m'_J(R) \equiv \delta m'_J (R) = \frac{e^{\gamma_E} R}{2} \log\! \biggl[ \tilde{B}_n\! \biggl(\frac{1}{i R e^{\gamma_E}}, R \biggr) \!\biggr],
\end{equation}
displaying an explicit dependence on the scale $R$. One of the motivations for the scheme definition in Eq.~\eqref{eq:defJ1} was to have a $\mu$-dependent low-scale short-distance
mass. The rationale behind this choice is that, as we have seen, to properly cancel the renormalon of e.g.\ the bHQET jet function one needs to express $\delta m_J(R,\mu)$ in powers
of $\alpha_s(\mu)$, and at first glance it might appear that if the short-distance mass does not depend explicitly on $\mu$ this can only be achieved if we take $R=\mu$. However, as shown
e.g.\ in Ref.~\cite{Bachu:2020nqn} for soft gap subtractions, or in Ref.~\cite{Butenschoen:2016lpz,Mateu:2017hlz} with the MSR mass, one can still use $R\neq \mu$ as long as $\delta m_{J'}(R)$
is expanded in powers of $\alpha_s(\mu)$, what implies that the fixed-order coefficients depend logarithmically on $\mu/R$. As long as $\mu\sim R$ these logarithms are not large.

From the definition in Eq.~\eqref{eq:mJprime}, linearizing again the logarithm at leading order in $1/\beta_0$ and identifying ${\cal Q}=R$ it is trivial to realize that the
master renormalon function for the combination
$2\delta m'_J/(Re^{\gamma_E})$ coincides with $G_{\!\tilde B}$ shown in Eq.~\eqref{eq:masterB}. From the result in Eq.~\eqref{eq:delt-Bn} with the replacement
$ix\to 1/(Re^{\gamma_E})$ one infers that the leading ambiguity for $\delta m'_J (R)$ exactly matches that of $\delta_{\overline{\rm MS}}(\overline m)$ shown in
Eq.~\eqref{eq:MS-amb}. Therefore, the difference $\delta m'_J (R)-\delta_{\overline{\rm MS}}(\overline m)$ is free from the $\mathcal{O}(\Lambda_{\rm QCD})$
renormalon and for $R=\overline m$ can be used as the matching condition to relate $m'_J (\overline m)$ and $\overline m$.

The corresponding R-anomalous dimension is computed as the $R$ derivative acting on $m'_J(R)$, and defining \mbox{$h(\beta)=-2\beta_0\Gamma_{\!\rm cusp}(-\beta)/\beta$}
and $b(\varepsilon) =\! \bigl\{{\df}[G_{\!\tilde B}(\varepsilon,u)-G_{\!\tilde B}(0,u)]/{\df}u\bigr\}_{\!u=0}$ we find the following closed form
\begin{align}\label{eq:gammaJprime}
\gamma^{J'}_R (\beta) = & -\!\frac{{\df} m'_J(R, R)}{{\df} R} \equiv \frac{1}{\beta_0}\sum_{n=0} \hat \gamma_{R,J'}^n\, \beta^{n+1} =
\frac{e^{\gamma_R}}{2\beta_0}\biggl\{ \int_0^{\infty}\! \mathd u\, e^{-\frac{u}{\beta}} (1 - 2 u) \tilde B_B (u) \\
& + \int_{- \beta}^0 \frac{\mathd \varepsilon}{\varepsilon}\, b(\varepsilon)
+ 2 \beta\, b(- \beta) +\! \int_{- \beta}^0 \frac{\mathd \varepsilon}{\varepsilon^2} \log\! \biggl( 1 + \frac{\varepsilon}{\beta}
\biggr) [h (\varepsilon) - \varepsilon h'(0) - h (0)] \nonumber \\
& + 2\!\! \int_{- \beta}^0 \frac{\mathd \varepsilon}{1 + \frac{\varepsilon}{\beta}} \biggl[ \frac{h (\varepsilon) - h(0)}{\varepsilon}
+ \frac{h (- \beta) - h(0)}{\beta} \biggr] \!\Biggr\}.\nonumber
\end{align}
Once more, the factor $(1-2u)$ removes the leading ambiguity from $\gamma^{J'}_R$. To compute the derivative with respect to $\beta$ of the last term in
Eq.~\eqref{eq:Z-res} a regulator has been used such that the logarithm does not diverge at the lower integration limit. After applying the derivative the regulator
can be safely set to zero:
\begin{align}
& \frac{\mathd}{\mathd \beta}\!\! \int_{- \beta}^0\! \frac{\mathd
\varepsilon}{\varepsilon^2} \log\! \biggl(1 + \frac{\varepsilon}{\beta} + \delta\!\biggr) \! \bigl[h (\varepsilon) - \varepsilon h' (0) - h(0)\bigr] \\
& \quad\qquad\qquad\qquad= \log (\delta) \frac{h (- \beta) + \beta h' (0) - h (0)}{\beta^2} -\!
\int_{- \beta}^0 \!\frac{\mathd \varepsilon}{\varepsilon} \frac{h
(\varepsilon) - \varepsilon h' (0) - h (0)}{\beta^2 \bigl( 1 + \delta +
\frac{\varepsilon}{\beta} \bigr)}
\nonumber\\
&\quad\qquad\qquad\qquad = - \frac{1}{\beta^2} \!\int_{- \beta}^0\! \frac{\mathd \varepsilon}{1 +
\frac{\varepsilon}{\beta}}\! \biggl[ \frac{h (\varepsilon) - \varepsilon h'
(0) - h (0)}{\varepsilon} + \frac{h (- \beta) + \beta h' (0) - h (0)}{\beta}
\biggr] \!+\mathcal{O}(\delta)\,. \nonumber
\end{align}
To obtain the last line we write $\log(\delta)=\log(1+\delta) - 1/\beta \int_{-\beta}^0 {\df} \varepsilon/(1+\varepsilon/\beta + \delta)$ and set $\delta = 0$ right away. The terms
proportional to $h'(0)$ are not necessary to make the integrand convergent at the upper integration limit. Furthermore, they cancel one other and therefore do not appear in
last line of Eq.~\eqref{eq:gammaJprime}. Equation~\eqref{eq:gammaJprime} can be integrated analytically in $R$ provided one switches the order of the various double integrals
that appear, recovering the difference of jet masses at two values of $R$. This is an important cross check of our computation.
In Fig.~\ref{fig:gammaRJ2} we perform a comparison of the fixed-order partial sum and exact results for $\gamma^{J'}_R$. The
description and observations are identical to those already presented for Fig.~\ref{fig:gammaRJ}.

Defining $\Delta^{\!R}_{J'}(R_2,R_1)$ in a way analogous to $\Delta^{\!R}_J$ one can obtain the jet mass in fixed-order perturbation theory at any scale $R$ simply
using R-evolution and the matching relation:
$m'_J(R) = \overline{m} + \!\bigl[\delta_{\overline{\rm MS}}( \overline{m}) - \delta m'_J (\overline{m})\bigr] \!+ \Delta_{J'}^R(R,\overline{m})$.
In Fig.~\ref{fig:MSR} we compare the exact value of $m_t^{J'}(R)$, shown as a red dashed line, with the fixed-order partial sum, in green.
Explanations and remarks are identical to those made for the derivative-scheme jet mass.

Comparing Figs.~\ref{fig:gammaRJ12} and \ref{fig:MSRgamma} one concludes that the sub-leading asymptotic behavior for the MSR mass R-anomalous dimension
is more severe than that of the jet masses in either scheme. This might have no effect in practical applications since the value of $\alpha_s$ rarely becomes so large
and at most four perturbative orders are known in full QCD. In any case, this worse perturbative behavior of $\gamma_R$ seems to affect the stability of perturbative
R-evolution when used to determine the MSR and jet masses,
as can be seen in Fig.~\ref{fig:MSR}: if one evolves the various short-distance
masses to a very small scale $R$, the asymptotic behavior for the MSR mass sets in after $7$ orders, while jet masses remain stable until the $12$th loop. On the
other hand, from Fig.~\ref{fig:mtFromMJ} it is hard to conclude that either of them is better suited to estimate the pole mass: starting from the second loop all schemes go hand in
hand. In particular, the MSR (red) and derivative-scheme jet (green) masses are almost on top of each other, while the non-derivative jet mass is closer to the exact value (gray
dashed line with a fade gray band) at one loop. Likewise, one can see in Fig.~\ref{fig:JB}
(with the same color convention) how the three short-scale low-scale masses
successfully remove the leading renormalon in the real part of the position-space bHQET jet function (the imaginary part is not shown as it does not add any
substantial additional information), which uses the same values as in Fig.~\ref{fig:BJet}. Even though the MSR mass appears to reach the exact value earlier,
after the second loop all schemes look almost identical.

\section{Estimate of bHQET jet function at three-loops}\label{sec:estimate}
\begin{table}[t!]
\centering
\begin{tabular}{|c|cccc|}
\hline
$n_\ell$ & $\hat b^\text{1-loop}_3$ & $\hat b^\text{2-loop}_3$ & $\hat b^\text{1-loop}_2$ & $\hat b^{\rm exact}_2$\\
\hline
$3$ & $9000\pm 5200$ & $7300\pm 1600$ & $11.3\pm 5.9$ & $194.20$ \\
$4$ & $7500\pm 4400$ & $5900\pm 1300$ & $10.4\pm 5.4$ & $174.84$ \\
$5$ & $6200\pm 3700$ & $4670\pm 950$ & ~\,$9.6\pm 5.0$ & $155.48$\\
\hline
\end{tabular}
\caption{Numerical values and uncertainties for the logarithm of the bHQET jet function in position space as estimated by Eq.~\eqref{eq:anasymptotic}.\label{tab:bTilde}}
\end{table}
Even though the bHQET jet function has been computed to two-loop accuracy, its cusp and non-cusp anomalous dimensions are known up to $\mathcal{O}(\alpha_s^3)$, which determine
the logarithmic coefficients at this order. Since $\tilde B_n$ depends on a single scale only, the missing non-logarithmic term is a real number. In the previous section we have
determined the leading flavor structure at three-loops, $\alpha_s^3\,n_\ell^2$ while in Ref.~\cite{Jain:2008gb} it was shown that the position-space jet function obeys non-abelian
exponentiation~\cite{Gatheral:1983cz,Frenkel:1984pz}. In particular this encompasses that in the abelian limit (obtained taking $\{n_\ell,C_A\}\to0$) the function $\log(m_p\tilde B_n)$ is one-loop
exact, unambiguously fixing the $C_F^3$ color structure. The pieces which are left to determine are thus $C_F^2 n_\ell$, $C_F C_A n_\ell$, $C_F C_A^2$ and $C_F^2 C_A $, although in
practice one can simply set $C_A=3$ and $C_F=4/3$ and estimate the linear and $n_\ell$-independent terms. The exponentiation theorem and the fact that $\log[\tilde B_n(x,\mu)]-2i x m_p$
is free from the leading $u=1/2$ renormalon can be exploited to estimate the non-logarithmic term. To that end we use a modified version of Eq.~(4.21) in Ref.~\cite{Hoang:2017suc} applied to
the jet mass in the non-derivative scheme defined in the previous section. We start by writing the corresponding perturbative series as
\begin{equation}\label{eq:mJprimeLam}
m_p - m'_J(R) = \frac{e^{\gamma_E} R}{2}\sum_{i=1} {\hat b}_i(\lambda) \biggl[\frac{\alpha_s(\lambda R)}{4\pi}\biggr]^i\,,
\end{equation}
which is formally independent of $\lambda$. The coefficients $\hat b_i\equiv {\hat b}_i(\lambda = 1)$ correspond to the non-logarithmic fixed-order terms of
$\log[\tilde B_n(x,\mu)]$. The $\lambda$-dependent coefficients ${\hat b}_i(\lambda)$ are computed as
\begin{equation}\label{eq:expand-alpha}
{\hat b}_i(\lambda) =\sum_{j=0}^{i-1}\hat c_{i,j}\log^j(\lambda)\,,\quad
\hat c_{i,j} = \frac{2}{j} \sum_{i = j}^{n - 1} i\, \beta_{n - i - 1} \hat c_{i, j - 1} \,,
\end{equation}
with $\hat c_{i,0}\equiv \hat b_i$. From the ${\hat b}_i(\lambda)$ coefficients one computes $\lambda$-dependent $S_k$ coefficients using e.g.\ Eq.~(A.18) of
Ref.~\cite{Hoang:2017suc}. In our case only $S^\lambda_1$ and $S^\lambda_2$ can be determined and accordingly the upper limit on the first sum must be set to $k=2$. In order to
study the convergence of our estimates we will consider setting the upper limit also to $k=1$, which is useful to check if the known two-loop result is reproduced, within uncertainties,
if only $S_1$ is used as input. All in all, the $m$-loop estimate of the coefficient $\hat b_n(\lambda)$ takes the following form:
\begin{equation}
\hat b^m_n(\lambda) = (2\beta_0)^n\! \sum_{k=0}^{\min(m,\,n-1)} S^\lambda_k\!\!
\sum_{\ell=0}^{{\rm min}(n -k-1,\,3)} \!\!g_\ell \biggl(1+\frac{\beta_1}{\beta_0} + k\biggr)_{\!\!n-1-\ell -k}\,,
\label{eq:anasymptotic}
\end{equation}
where the $g_\ell$ coefficients can be computed with the recursive relation provided in Eq.~(A.14) of Ref.~\cite{Hoang:2017suc}.
The benefit of using this formula ---\,rather than the more conventional (4.20) therein\,--- is that it incorporates information beyond leading-renormalon dominance. In particular,
assuming that $m \leq 4$, all coefficients with $n\leq m$ are reproduced exactly. From $b^m_n(\lambda)$ one can obtain an estimate for the canonical
$\hat b^m_n(\lambda=1)\equiv \hat b^m_n$ coefficients simply inverting Eq.~\eqref{eq:expand-alpha} [\,that is, deriving $\hat c^m_{i,j}(\lambda)$ coefficients from the recursive expression
in Eq.~\eqref{eq:expand-alpha} and using the first relation with the replacement $\log(\lambda) \to -\log(\lambda)$\,]. Since the sum over $S^\lambda_k$ in
Eq.~\eqref{eq:anasymptotic} is convergent, truncating it at $k=m$ (rather than the exact value $n-1$) inflicts a small uncertainty that can be quantified directly on $\hat b^{\,j}_n$
varying the parameter $\lambda$ in the range $(1/2,2)$.
\begin{figure*}[t!]
\subfigure[]
{\includegraphics[width=0.54\textwidth]{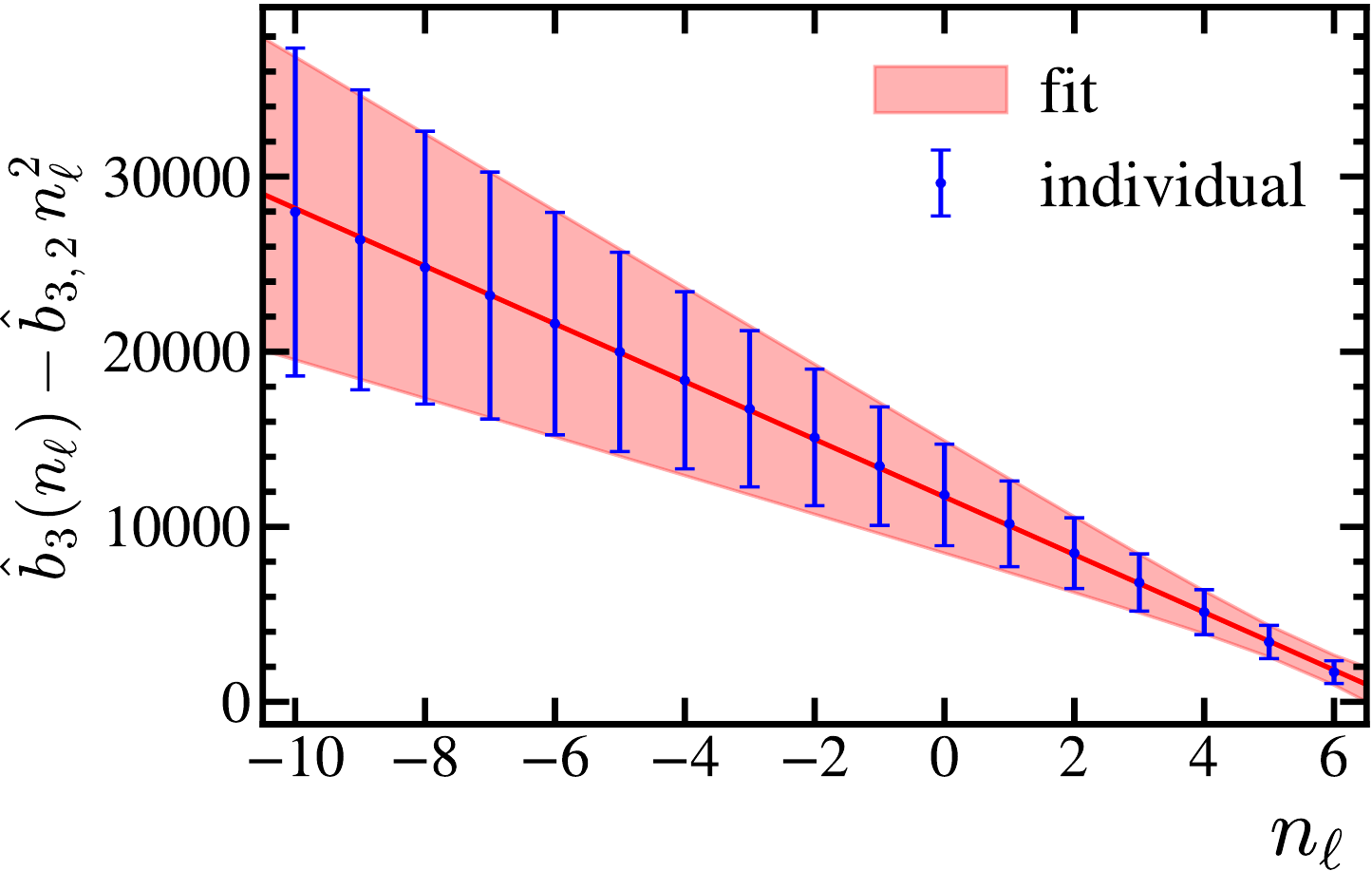}
\label{fig:fit}}~~
\subfigure[]{\includegraphics[width=0.41\textwidth]{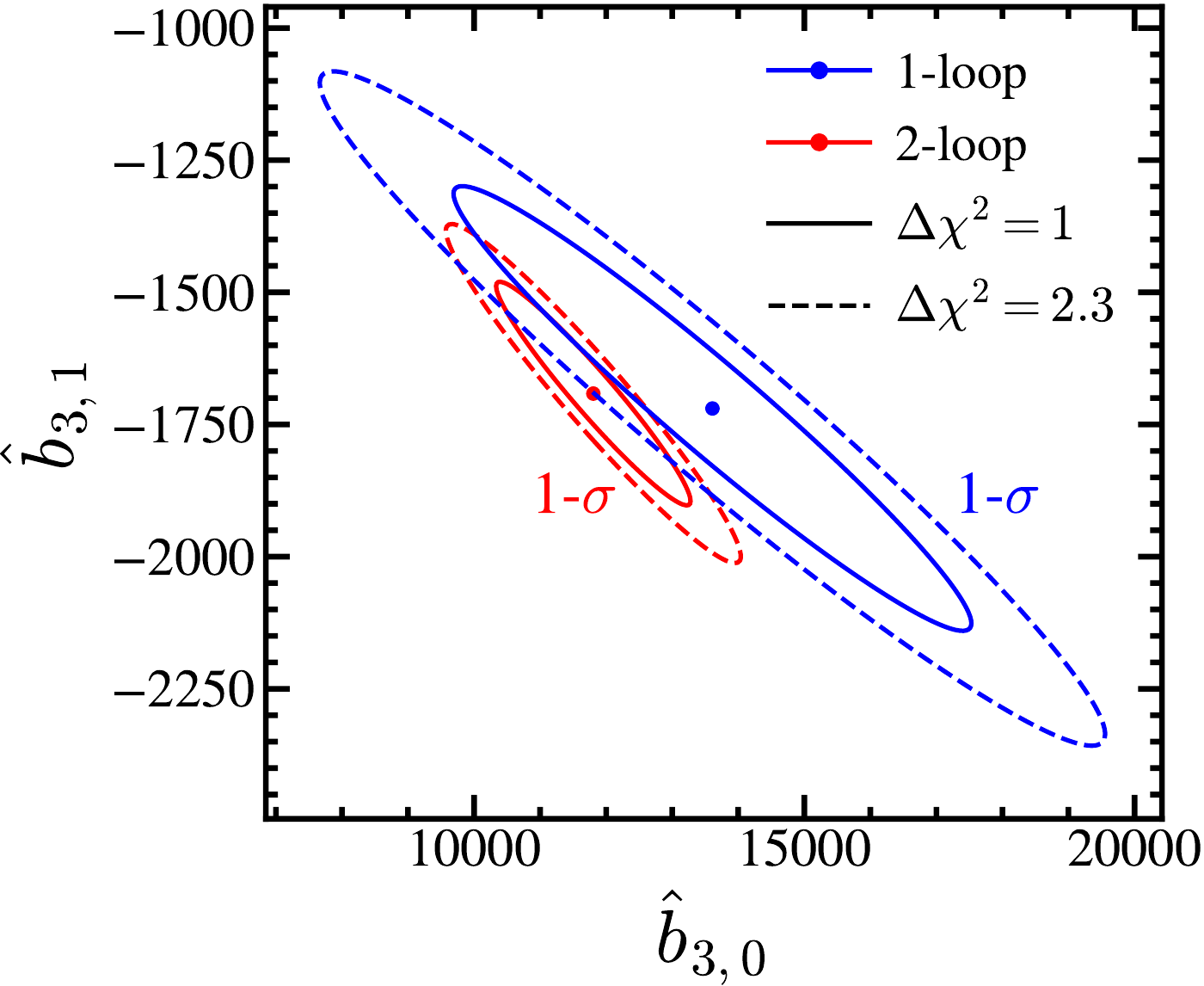}
\label{fig:ellipses}}
\caption{Left panel: Direct estimates for $\hat b_{3}(n_\ell)$
(blue error bars) compared to a fit for $\hat b_{3,0}$ and $\hat b_{3,1}$ with rescaled correlation coefficient (red solid line and pale red band).
Right panel: Error ellipses for the simultaneous determination of the flavor coefficients $\hat b_{3,0}$ and $\hat b_{3,1}$ at 1- (blue) and 2-loops (red).
The correlation coefficients have been individually shrunk such that the uncertainties on $\hat b_{3}(n_\ell=3,4,5)$ match those of a direct determination.
Solid ellipses correspond to 1-$\sigma$ errors in each parameter while dashed ones are 1-$\sigma$ in both dimensions.}
\label{fig:estimate}
\end{figure*}

Let us present our numerical analysis. In first place we fix the active number of flavors to the physically relevant cases for charm, bottom and top $n_\ell=3,4,5$ and directly estimate
$\hat b_3(n_\ell)$ using 1- and 2-loop information. We find that the 2-loop results are contained in the 1-loop uncertainty bands. As an additional check we estimate $\hat b_2(n_\ell)$
using 1-loop information and find that the known exact values are contained within the estimated uncertainties. Our results are summarized in Table~\ref{tab:bTilde}. Next we try to estimate the
coefficients of the various powers of $n_\ell$, which we denote by\footnote{A similar strategy to estimate individual flavor coefficients for the $5$- and $6$-loop corrections to the relation between the pole and $\MSBar$ masses has been used in Ref.~\cite{Kataev:2018gle}.}
\begin{equation}
\hat b_i(n_\ell)=\sum_{k=0}^{i-1} \hat b_{ij}\,n_\ell^j\,.
\end{equation}
To that end we determine the value of $\hat b_3(n_\ell)$ in the range $-10\leq n_\ell \leq 6$, and given that $\hat b_{3,2}$ has been computed in Sec.~\ref{sec:bJet}, least-square minimization is
used to determine the remaining $\hat b_{3,0}$ and $\hat b_{3,1}$. We have checked that varying the $n_\ell$ fit-range boundaries by one (upper limit) or several (lower
limit) units changes the estimated values by an amount much smaller than the uncertainty caused by $\lambda$ variation. Therefore even if included in our final analysis, we will not discuss
this source of error any further.
Considering the individual $\hat b_3(n_\ell)$ errors (obtained by varying $\lambda$) as uncorrelated greatly underestimates the uncertainties of the coefficients. This is expected
since there is of course a strong correlation between the individual uncertainties. One can take the extreme situation in which the errors are considered as $100\%$ correlated by taking the same
value of $\lambda$ for all $n_\ell$ values, determining then $\hat b_{3,0}$ and $\hat b_{3,1}$ as a function of $\lambda$. By varying $\lambda$ in the range specified above one can estimate
the uncertainties of each coefficient and the correlation between them. We find a very strong negative correlation, very close to $100\%$, which is responsible for an $\hat b_3(n_\ell)$
uncertainty in the range $3\leq n_\ell \leq 5$ smaller than that of the direct determination. This is caused by our rough assumption of a $100\%$ correlation in $\lambda$. To correct this deficiency
we keep the individual uncertainties (to which we add in quadrature those coming from the linear regression) of $\hat b_{3,0}$ and $\hat b_{3,1}$ but rescale the correlation between them by a
factor that makes the direct and indirect uncertainties match for the physically relevant values of $n_\ell$, see Fig.~\ref{fig:fit}. This rescaling also makes the 1- and 2-loop estimates of
$\hat b_{3,0}$ and $\hat b_{3,1}$ compatible, as shown in Fig.~\ref{fig:ellipses}. Our numerical results are presented in Table~\ref{tab:bCoef}, where we also shown that using 1-loop input
the two-loop $n_\ell$-independent coefficient is correctly predicted within its uncertainty.

Carrying out the same exercise for the thrust and C-parameter soft function, under the assumption of $u=1/2$ renormalon dominance one obtains less precise results, particularly so for the former.
Since the leading flavor coefficient is not known, we do not attempt to determine the individual $n_f$ coefficients. Following the notation in Refs.~\cite{Abbate:2010xh,Hoang:2014wka} we find
\begin{align}
s_3^{\tau,\,n_\ell=\{4,5,6\}} = &\, \{-1400 \pm 1300, -600 \pm 1300, 100 \pm 1400\}\,, \\
s_3^{\widetilde C,\,n_\ell=\{4,5,6\}} = &\, \{-2000 \pm 500, -1300\pm 240, -670\pm 140\}\, .\nonumber
\end{align}
The thrust result for $n_f=5$ is however not compatible with the determination carried out in Ref.~\cite{Bruser:2018rad}, $s_3^{\tau,\,n_\ell=5} = -10000\pm 2100$, which in turn is based
on the numerical analysis of Ref.~\cite{Monni:2011gb} which uses NNLO binned distributions generated with \texttt{EERAD3}~\cite{Ridder:2014wza}.

\begin{table}[t!]
\centering
\begin{tabular}{|c|ccc|c|cc|}
\hline
order & $\hat b_{3,0}$ & $\hat b_{3,1}$ & correlation & $\hat b_{3,2}$ & $\hat b_{2,0}$ & $\hat b_{2,1}$\\
\hline
$1$ & $13000\pm 8200$ &$-1600\pm 1000$ & $-0.966$ & $50.054$ & $210 \pm 130$ & $-19.364$ \\
$2$ & $11700\pm 3200$ & $-1600 \pm 550$~\, & $-0.961$ & $50.054$ & $252.299$ & $-19.364$ \\
\hline
\end{tabular}
\caption{Numerical values, uncertainties and correlation for the sub-leading flavor structures of the position-space bHQET jet function's exponent coefficients at 3-loops
using at 1- and 2-loop input. For completeness we also shown the known values for $\hat b_{3,2}$ and $\hat b_{2,1}$, and the estimated two-loop coefficient $\hat b_{2,0}$
if 1-loop input is used.\label{tab:bCoef}}
\end{table}

\section{Conclusions}\label{sec:conclusions}
In this article we have adapted the methodology to compute the leading term in the $1/\beta_0$ expansion for series with cusp anomalous dimension. Such series are
ubiquitous in effective field theories for massive and massless jets such as SCET and bHQET. We have found closed expressions for the renormalized series, its
multiplicative renormalization factor as well as their cusp and non-cusp anomalous dimensions. Furthermore, we have solved the corresponding RGE equations in a closed
form in this limit, and shown that the ambiguity does not depend on the renormalization scale. We have provided an optimal algorithm for the principal value prescription used
to compute the inverse Borel integral when there are poles in the integration path.

As a first application, the series that relates the pole and $\MSBar$ masses has been revisited, recovering a known exact from for the $\MSBar$ mass anomalous dimension. We have
derived, for the first time, a closed expressions for the MSR mass R-anomalous dimension $\gamma_R$ in the large-$\beta_0$ limit, and provided an analytic solution for the
corresponding RG equation at any perturbative order. We have investigated how the fixed-order expansion of the series defining this scheme can be used to estimate
a numerical value for the pole mass, finding that for smaller $R$ the series quickly converges to the exact value. On the other hand, the asymptotic behavior of the series only depends
on the renormalization scale $\mu$ at which the strong coupling is evaluated.
The RS mass has also been worked out in this limit, and we have demonstrated that the difference of two such masses for different values of $R$
obeys a perturbative R-evolution equation. Finally, it has been shown that the renormalon normalization constant $N_{1/2}$ as computed with the sum rule derived in
Ref.~\cite{Hoang:2017suc} tends to the large-$\beta_0$ result, which can be obtained from our analytic expressions.

The hard and jet soft functions for massless (in SCET) and massive (in bHQET) quarks have been computed in the large-$\beta_0$ with a modified gluon propagator,
applying the technology developed in this article. Our determination of the universal cusp anomalous dimensions agrees with a previous computation, and our results for the non-cusp
anomalous dimension and matrix elements agree with the leading flavor structure of known results computed in the full theory. Our results for SCET reveal that the leading ambiguities
happen at $u=1$, so they are proportional to $\Lambda_{\rm QCD}^2$. We find that the Borel transform of the SCET hard function has double poles, associated to operators
with non-trivial anomalous dimension in the OPE which depend on $n_f$ at leading order. We also find that its fixed-order expansion converges faster
if $\pi^n$ terms are summed up choosing a complex matching scale for the Wilson coefficient $C_{\!H}$. The SCET jet function, on the other hand, has only two simple poles. Our
computation of the hard and jet functions in bHQET shows that in both cases the leading renormalon is of the most severe type, that is, proportional to $\Lambda_{\rm QCD}$.
We have seen that the fixed-order expansion of the renormalon-free quantity $H_m/m_p^3$ shows nice convergence if the pole mass is expressed in terms of the $\MSBar$ mass.
On the other hand, the fixed-order expansion for the combination $\tilde B_n(x,\mu) e^{-im^2 x}$ appearing in the bHQET factorization theorem, which is also free from the
leading renormalon, behaves well if the pole mass is expressed in terms of the MSR mass (or other R-dependent low-scale short-distance masses). We have estimated the
unknown three-loop non-logarithmic coefficient of $\tilde B_n$ in full QCD with $20\%$ accuracy using renormalon dominance.

The ambiguities found in the EFT computations can have two physical meanings, depending if $\Lambda_{\rm QCD}$ is interpreted as the scale of non-perturbative (hadronic)
physics $\Lambda_{\rm had}$ or as the soft scale $\mu_s$ of SCET or bHQET. For the hard function $H_Q$, since it comes from matching two operators and its definition does
not involve a matrix element (even though, in practice, for its computation matrix elements are taken) it appears natural to interpret $\Lambda_{\rm QCD}$ as the soft scale,
and since $\mu_s/\mu_H\propto \lambda^2$ with $\lambda$ the SCET power-counting parameter, ambiguities are likely related to sub-leading operators in the EFT
power expansion. The jet function is indeed a matrix element and in this case $\Lambda_{\rm QCD}$ could signal either power-suppressed jet functions or effects from
collinear hadronization corrections. In either case, given the relation $\mu_s/\mu_J\propto \lambda$, these are power suppressed contributions. For the $H_m$ hard massive
factor and the bHQET jet function, the leading $u=1/2$ ambiguity is obviously related to the well-understood pole mass renormalon. Higher-order ambiguities can be interpreted
in the same way as just done for $H_Q$ and $J_n$, respectively.

From the position-space bHQET hemisphere jet function one can define the so called jet mass, which involves a derivative of the logarithm of $\tilde B_n$. We have provided
an alternative scheme for a $\mu$-independent jet mass in which no derivative is taken. We have worked out closed expressions
for the relation of these two schemes to the pole mass, showing they have the same leading ambiguity as the corresponding $\MSBar$ series, making possible an
unambiguous matching to $\overline m$. Closed expressions for their R-anomalous dimensions have also been derived, and it has been shown that the two schemes work
similarly to the MSR mass when it comes to removing the renormalon of $\tilde B_n$ or estimating a numerical value for the pole mass in fixed-order perturbation theory.

Our results can be used to derive an exact prediction at leading order in $1/\beta_0$ for the singular structure of massless event-shape distributions in the di-jet limit, as well as
massive event shapes in the peak region, once the large-$\beta_0$ soft function is known. Such computation is more involved than those carried out in this article as it
cannot be cast as the discontinuity of a forward-scattering matrix element. Instead, one has to deal with two-body real-radiation phase-space integrals (when ``cutting'' the quark
bubble), which are similar in complexity to a two-loop integration, with the additional complication of modified gluon propagators. This computation is therefore left to future work.
Our results also suggest that studying the effect of $\pi$ summation in the SCET hard function for a full QCD event-shape distribution is warranted, and we will do so in the near future.
Likewise, it appears important to figure out a practical implementation of the bHQET hard factor renormalon subtraction within the corresponding factorization theorem, and
how this is linked to reparametrization invariance~\cite{Luke:1992cs}, a transformation that connects different orders in the $1/m$ expansion.

\acknowledgments

This work was supported in part by the Spanish MINECO Ram\'on y Cajal program (RYC-2014-16022), the MECD grant PID2019-105439GB-C22, the IFT Centro de
Excelencia Severo Ochoa Program under Grant SEV-2012-0249, the EU STRONG-2020 project under the program H2020-INFRAIA-2018-1, grant agreement
no.\ 824093 and the COST Action CA16201 PARTICLEFACE. NGG is supported by a JCyL scholarship funded by the regional government of Castilla y Le\'on and European Social Fund,
2017 call. We thank D.\ Boito and M.\,V. Rodrigues for useful discussions and pointing us to the right references.

\appendix
\section{Inductive proofs for sums in renormalon series}
\label{sec:AppSum}
In the first part of this appendix it is shown by induction that the sum
\begin{equation}\label{eq:Isum}
I_{i,m}\equiv\sum_{n = 1}^i (- 1)^n \binom{i - 1}{n - 1} n^{m - 1}\,,
\end{equation}
is zero unless $m=0$ or $m\geq i$. We first compute its value for $m=0$
\begin{equation}
I_{i,0} = \sum_{n = 1}^i \frac{(- 1)^n}{n} \binom{i - 1}{n - 1} = \frac{1}{i} \sum_{n
= 1}^i (- 1)^n \frac{i!}{n! (i - n) !} = \frac{1}{i} \biggl[ - 1 + \sum_{n =
0}^i (- 1)^n \binom{i}{n} \biggr] = - \frac{1}{i},
\end{equation}
where in the second equality the definition of binomial is used, in the third a term with $n=0$ has been added and subtracted, and to obtain its final form
we simply use that the sum starting with $n=0$ is $(1 - 1)^n = 0$. To show that $I_{i,m}$ vanishes in the range specified above we define the similar-looking
sum\footnote{To have a more compact form for $J_{i, m}$ we assume that $n^m=1$ even if $n=m=0$. This assumption is also implicitly made in
Eq.~\eqref{eq:ItoJ} when using the binomial expansion for $(1+m)^j$ to cover the case $m=0$.}
\begin{equation}\label{eq:Jdef}
J_{i, m} = 1 + \sum_{n = 1}^{i - 1} (- 1)^n \binom{i - 1}{n} n^m \equiv \sum_{n = 0}^{i - 1} (- 1)^n \binom{i - 1}{n} n^m\,.
\end{equation}
$I_{i,m}$ can be written in terms of $J_{i,m}$ shifting $n \rightarrow n + 1$ in Eq.~\eqref{eq:Isum} and assuming $i,m \geqslant 1$
\begin{align}\label{eq:ItoJ}
I_{i, m} =\,& - \sum_{n = 0}^{i - 1} (- 1)^n \binom{i - 1}{n} (n + 1)^{m - 1} =
- \sum_{n = 0}^{i - 1} (- 1)^n \binom{i - 1}{n} \sum_{j = 0}^{m - 1}
\binom{m - 1}{j} n^j \\
=\,& - \sum_{j = 0}^{m - 1} \binom{m - 1}{j} J_{i, j} \,.\nonumber
\end{align}
Hence we need to show that $J_{i, m} $ vanishes in the range $0 \leqslant m \leqslant i - 1$ for strictly positive values of $i$, or equivalently $J_{i\geq m+1, m\geq 0} = 0$.
Let us start showing that $J_{i\geq 1,0}=0$ and then proceed by induction:
\begin{equation}
J_{i, 0} = \sum_{n = 0}^{i - 1} (- 1)^n \binom{i - 1}{n} = (1 - 1)^{i - 1} = 0.
\end{equation}
Since $J_{0,m}=1$ in general one has $J_{i, 0} = \delta_{i0}$. Let us build a recursion relation for $m > 0$
\begin{align}
J_{i, m} = \,& \sum_{n = 1}^{i - 1} (- 1)^n \frac{(i - 1) ! n^{m - 1}}{(n -
1) ! (i - 1 - n) !} = - (i - 1) \sum_{n = 0}^{i - 2} (- 1)^n \binom{i -2}{n} (n + 1)^{m - 1} \\
= \,& - (i - 1) \sum_{n = 0}^{i - 2} (- 1)^n \binom{i - 2}{n} \sum_{j =
0}^{m - 1} \binom{m - 1}{j} n^j = - (i - 1) \sum_{j = 0}^{m - 1} \binom{m -
1}{j} J_{i - 1, j} \,, \nonumber
\end{align}
where in the first step the definition of binomial is used, in the second we shift $n\to n-1$, pull out the largest factor in $(i-1)!=(i-1)(i-2)!$, and combine it with the
rest of factorials as a new binomial. To obtain the first equality in the second line we use the binomial expansion for $(n + 1)^{m - 1}$.
For $i=1$ the last equality automatically vanishes due to the $(i-1)$ factor multiplying the sum. Since for $i\geq2$ one has $J_{i - 1, 0} = 0$ it is implied that
$J_{i\geq 2, 1} = 0$; likewise, given that for $i\geq32$ both $J_{i - 2} = J_{i - 1} = 0$ it immediately follows that $J_{i\geq3, 2} = 0$. Proceeding in this fashion one concludes
that $J_{i\geq m+1, m} = 0$ for any non-negative $m$, and this proofs $I_{i, m} = 0$ for $1 \leq m \leq i - 1$.

In the second part of this appendix we provide a proof for Eq.~\eqref{eq:sum2}, making use again of recursion relations. Noting that $K_1 = - 1$,
assuming $i \geq 2$ and shifting $n \rightarrow n + 1$ in Eq.~\eqref{eq:sum2} one gets
\begin{align}
K_i =\, & - \sum_{n = 0}^{i - 1} (- 1)^n \binom{i - 1}{n} (n + 1)^{i - 1} =
- \sum_{n = 0}^{i - 1} (- 1)^n \binom{i - 1}{n} \sum_{j = 0}^{i - 1}
\binom{i - 1}{j} n^j \\
=\, &- \sum_{j = 0}^{i - 1} \binom{i - 1}{j} J_{i, j} = \sum_{n = 1}^{i - 1} (- 1)^{n+1} \binom{i - 1}{n} n^{i - 1} = (i -
1) \sum_{n = 1}^{i - 1} \frac{(- 1)^{n-1}(i - 2) ! n^{i - 2}}{(n - 1) ! (i - 1
- n) !} \nonumber\\
=\,&-\! (i -1) \sum_{n = 1}^{i - 1} (- 1)^{n} \binom{i-2}{n-1}n^{i-2}= - (i - 1) K_{i-1} \,, \nonumber
\end{align}
where in the second equality we have used the binomial expansion for $(n + 1)^{i - 1}$ (assuming $n^j=1$ for $n=j=0$);
to get the third equality the definition of $J_{i, j}$ in Eq.~\eqref{eq:Jdef} is used. Our previous result for $J_{i, j}$ (that is, only if $j = i - 1$ it is non-zero) is used to obtain
the second equality in the second line: only the larger value of $j$ in the sum is kept and the definition for $J_{i, i-1}$ is used. To obtain the one-to-last
equality we write the binomial in terms of factorials and pull out the largest factor in both $(i-1)!$ and $n!$; writing the various factorials back as a binomial
one arrives at the last equality. Applying the recursion relation $i-1$ times we obtain the result in Eq.~\eqref{eq:sum2}.

Next we show a result involving the sum of binomials, necessary to obtain the final form in Eq.~\eqref{eq:sum3}. We start by noting that the harmonic
numbers can be written as the following integral
\begin{equation}
\int_0^1 {\df} x\, \frac{1 - x^n}{1 - x} =
\sum_{i = 1}^n \int_0^1 {\df} x\, x^{i - 1} = \sum_{i = 1}^n \frac{1}{i} \equiv H_n \,,
\end{equation}
where in the first step we have used $a^n - b^n= (a-b)\sum_{i=0}^{n-1} a^i b^{n-1-i}$ with $a=1$. To find the relation we seek one switches variables
$x \rightarrow 1 - u$ in the integration above
\begin{align}\label{eq:HnSum}
H_n =\,& \int_0^1 {\df} u \frac{1 - (1 - u)^n}{u} = \int_0^1 \frac{{\df} u}{u} \biggl[ 1 - \sum_{i = 0}^n \binom{n}{i} (- u)^i \biggr] \\
=\,& - \sum_{i = 1}^n \binom{n}{i} (- 1)^i \int_0^1 {\df} u \,u^{i - 1} = - \sum_{i = 1}^n
\binom{n}{i} \frac{(- 1)^i}{i} \,, \nonumber
\end{align}
where we have used the binomial expansion to obtain the second equality, and noticed that the $i=0$ contribution in the sum cancels against the term outside; after integrating the
remaining terms the final expression is obtained.

In this final part of the appendix we show how to convert a sum over harmonic numbers into an integral, that is, we provide a proof for Eq.~\eqref{eq:Harm-sum-id}
\begin{align}\label{eq:Hn2Proof}
&\int_{- \beta}^0 {\df} \varepsilon \log \biggl( \frac{\beta}{\beta + \varepsilon} \biggr) \varepsilon^{i - 1}
= (- \beta)^i \int_0^1 {\df} x \log (x) (1 - x)^{i - 1} \\
&\quad=\ (- \beta)^i \sum_{j = 0}^{i - 1} \binom{i - 1}{j} \int_0^1 {\df}
x \log (x) (- x)^j = (- \beta)^i \sum_{j = 0}^{i - 1} \binom{i -
1}{j} (- 1)^j \biggl[\frac{{\df}}{{\df} \epsilon} \int_0^1 {\df} x \,x^{j +
\varepsilon} \biggr]_{\epsilon = 0} \nonumber\\
&\quad= - (- \beta)^i \frac{{\df}}{{\df} \epsilon} \Biggl[ \sum_{j =
1}^i \binom{i - 1}{j - 1} \frac{(- 1)^j}{j + \epsilon}
\Biggl]_{\epsilon = 0} = (- \beta)^i \sum_{j = 1}^i \binom{i - 1}{j -
1} \frac{(- 1)^j}{j^2} \nonumber\\
&\quad= \frac{(- \beta)^i}{i} \sum_{j = 1}^i \binom{i}{j} \frac{(-
1)^j}{j} = - \frac{(- \beta)^i H_i}{i} \,,\nonumber
\end{align}
where in the first step we switch variables $\varepsilon = -\beta (1 - x)$; to obtain the first expression in the second line we perform the binomial expansion
of $(1 - x)^{i - 1}$; next we write the logarithm in terms of a derivative
and integrate each term in the sum. After carrying out the $\epsilon$ derivative we write the binomial in terms of factorials and use $(i-1)! = i!/i$ and $j^2(j-1)!=j\times j!$
to obtain yet another binomial. Finally, using the result in \eqref{eq:HnSum} we conclude our proof for Eq.~\eqref{eq:Harm-sum-id}.

\section{Alternative derivation of the RGE solution for a series with cusp anomalous dimension}
\label{sec:AppManipulo}
In this appendix we derive Eq.~\eqref{eq:cuspRGE} using the perturbative expansions for the once-subtracted cusp and non-cusp anomalous dimensions,
as well as the twice subtracted cusp:
\begin{align}
\omega ({\cal Q}, \mu) = & -\! \frac{1}{2} \!\int_0^{\beta_{\mu}} \!\!{\df} \beta
\sum_{n = 0} \beta^n\! \biggl( \!\hat\gamma^{n + 1}_A + \frac{\hat\Gamma^{n + 2}_{\!\!A}}{2}
- \frac{\hat\Gamma^{n + 1}_{\!\!A}}{2 \beta_{\cal Q}} \biggr)\! = \!- \frac{1}{2}\! \sum_{n = 0}
\frac{\beta^{n + 1}}{n + 1}\! \biggl(\! \hat\gamma^{n + 1}_A + \frac{\hat\Gamma^{n +
2}_{\!\!A}}{2} - \frac{\hat\Gamma^{n + 1}_{\!\!A}}{2 \beta_{\cal Q}} \biggr) \nonumber\\
= & - \!\frac{1}{\beta_0} \sum_{n = 0} \frac{(- \beta_{\mu})^{n + 1}}{n +
1} \biggl( G_{n + 1, 1} + H_{n + 2} G_{n + 2, 0} - G_{n + 2, 0} -
\frac{G_{n + 1, 0}}{\beta_{\cal Q}} \biggr) \nonumber\\
= & - \frac{1}{\beta_0} \sum_{n = 0} (- \beta_{\mu})^{n + 1} \biggl[
\frac{1}{n + 1} \biggl( G_{n + 1, 1} + H_{n + 1} G_{n + 2, 0} -
\frac{G_{n + 1, 0}}{\beta_{\cal Q}} \biggr)\! - \frac{G_{n + 2, 0}}{n + 2}
\biggr] \nonumber\\
= & - \frac{1}{\beta_0} \sum_{n = 0} \frac{(-
\beta_{\mu})^{n + 1}}{n + 1} \biggl[ G_{n + 1, 1} + H_{n + 1} G_{n + 2, 0} +
\biggl( \frac{1}{\beta_{\mu}} - \frac{1}{\beta_{\cal Q}} \biggr) G_{n + 1, 0}
\biggr] \!- \frac{ G_{1, 0}}{\beta_0} \nonumber\\
= & - \frac{1}{\beta_0} \sum_{n = 0} \biggl\{ \int_0^{\beta_{\mu}}
\frac{{\df} \varepsilon}{\varepsilon} (- \varepsilon)^{n + 1} \biggl[ G_{n +
1, 1} + \biggl( \frac{1}{\beta_{\mu}} - \frac{1}{\beta_{\cal Q}} \biggr) G_{n + 1,
0} \biggr] \nonumber\\
& + \int_{- \beta_{\mu}}^0 \frac{{\df}
\varepsilon}{\varepsilon^2} \log \biggl( 1 + \frac{\varepsilon}{\beta_{\mu}}
\biggr) \varepsilon^{n + 2} G_{n + 2, 0} \biggr\} \!- \frac{G_{1,0}}{\beta_0} \nonumber\\
= &\, \frac{1}{\beta_0} \biggl\{ \int_{- \beta_{\mu}}^0\! \frac{{\df}
\varepsilon}{\varepsilon} \biggl[ \frac{{\df}}{{\df} u}\! \bigl[G (\varepsilon,
u) - G (0, u)\bigr]_{\!u = 0} + \biggl( \frac{1}{\beta_{\mu}} - \frac{1}{\beta_{\cal Q}}
\biggr)\! \bigl[G (\varepsilon, 0) - G (0, 0)\bigr]\! \biggr] \nonumber\\
& + \frac{\hat \Gamma_{\!\!A}^1}{4} -\! \int_{- \beta_{\mu}}^0 \!\frac{{\df}
\varepsilon}{\varepsilon^2} \log \biggl( 1 + \frac{\varepsilon}{\beta_{\mu}}
\biggr) \!\biggl[ G (- \varepsilon, 0) - G (0, 0) - \varepsilon
\frac{{\df} G (\tau)}{{\df} \tau} \biggr|_{\tau = 0} \,\biggr]\! \biggr\}\,,
\end{align}
where each term is integrated in $\beta$ to obtain the second equality; to get the second line we use the results in
Eqs.~\eqref{eq:cuspPart} and \eqref{eq:nocuspPart} for the anomalous dimension coefficients in terms of $G_{i,j}$;
to get to the third line we have used the identity
\begin{equation}
\frac{H_{n + 2} - 1}{n + 1} = \frac{H_{n + 1}}{n + 1} - \frac{1}{n + 2} \,,
\end{equation}
while to obtain the fourth we shift $n\to n-1$ in the sum corresponding to the last term in the third, to which we add (and subtract)
$G_{1,0}/\beta_0$ to make the sum starting at $n=0$ again; to obtain the one-to-last equality $\beta^n/(n+1)$
and $\beta^n H_{n+1}/(n+1)$ are expressed as integrals over $\varepsilon$; finally, summing up the various resulting series the final form is obtained,
in full agreement with Eq.~\eqref{eq:cuspRGE}.

\newpage

\bibliography{EFTs}

\providecommand{\href}[2]{#2}\begingroup\raggedright\begin{thebibliography}{10}

\bibitem{CacciariSalam}
\emph{M.~Cacciari, Jet physics theory}, in proceedings of the 8th International
  Conference on Hard and Electromagnetic Probes of High-energy Nuclear
  Collisions (Hard Probes 2016), Wuhan, China, 22–27, September, 2016.

\bibitem{Bauer:2000ew}
C.~W. Bauer, S.~Fleming and M.~E. Luke, \emph{{Summing Sudakov logarithms in $B
  \to X_s \gamma$ in effective field theory}},
  \href{http://dx.doi.org/10.1103/PhysRevD.63.014006}{\emph{Phys. Rev. D}
  {\bfseries 63} (2000) 014006},
  [\href{https://arxiv.org/abs/hep-ph/0005275}{{\ttfamily hep-ph/0005275}}].

\bibitem{Bauer:2000yr}
C.~W. Bauer, S.~Fleming, D.~Pirjol and I.~W. Stewart, \emph{{An Effective field
  theory for collinear and soft gluons: Heavy to light decays}},
  \href{http://dx.doi.org/10.1103/PhysRevD.63.114020}{\emph{Phys. Rev. D}
  {\bfseries 63} (2001) 114020},
  [\href{https://arxiv.org/abs/hep-ph/0011336}{{\ttfamily hep-ph/0011336}}].

\bibitem{Bauer:2001yt}
C.~W. Bauer, D.~Pirjol and I.~W. Stewart, \emph{{Soft collinear factorization
  in effective field theory}},
  \href{http://dx.doi.org/10.1103/PhysRevD.65.054022}{\emph{Phys. Rev. D}
  {\bfseries 65} (2002) 054022},
  [\href{https://arxiv.org/abs/hep-ph/0109045}{{\ttfamily hep-ph/0109045}}].

\bibitem{Bauer:2001ct}
C.~W. Bauer and I.~W. Stewart, \emph{{Invariant operators in collinear
  effective theory}},
  \href{http://dx.doi.org/10.1016/S0370-2693(01)00902-9}{\emph{Phys. Lett. B}
  {\bfseries 516} (2001) 134--142},
  [\href{https://arxiv.org/abs/hep-ph/0107001}{{\ttfamily hep-ph/0107001}}].

\bibitem{Bauer:2002nz}
C.~W. Bauer, S.~Fleming, D.~Pirjol, I.~Z. Rothstein and I.~W. Stewart,
  \emph{{Hard scattering factorization from effective field theory}},
  \href{http://dx.doi.org/10.1103/PhysRevD.66.014017}{\emph{Phys. Rev. D}
  {\bfseries 66} (2002) 014017},
  [\href{https://arxiv.org/abs/hep-ph/0202088}{{\ttfamily hep-ph/0202088}}].

\bibitem{Collins:1981uk}
J.~C. Collins and D.~E. Soper, \emph{{Back-To-Back Jets in QCD}},
  \href{http://dx.doi.org/10.1016/0550-3213(81)90339-4}{\emph{Nucl. Phys.}
  {\bfseries B193} (1981) 381}.

\bibitem{Korchemsky:1998ev}
G.~P. Korchemsky, \emph{{Shape functions and power corrections to the event
  shapes}},  \href{https://arxiv.org/abs/hep-ph/9806537}{{\ttfamily
  hep-ph/9806537}}.

\bibitem{Korchemsky:1999kt}
G.~P. Korchemsky and G.~Sterman, \emph{{Power corrections to event shapes and
  factorization}},
  \href{http://dx.doi.org/10.1016/S0550-3213(99)00308-9}{\emph{Nucl. Phys.}
  {\bfseries B555} (1999) 335--351},
  [\href{https://arxiv.org/abs/hep-ph/9902341}{{\ttfamily hep-ph/9902341}}].

\bibitem{Korchemsky:2000kp}
G.~P. Korchemsky and S.~Tafat, \emph{{On power corrections to the event shape
  distributions in QCD}},
  \href{http://dx.doi.org/10.1088/1126-6708/2000/10/010}{\emph{JHEP} {\bfseries
  10} (2000) 010}, [\href{https://arxiv.org/abs/hep-ph/0007005}{{\ttfamily
  hep-ph/0007005}}].

\bibitem{Berger:2003iw}
C.~F. Berger, T.~K{\'u}cs and G.~Sterman, \emph{{Event shape / energy flow
  correlations}},
  \href{http://dx.doi.org/10.1103/PhysRevD.68.014012}{\emph{Phys. Rev. D}
  {\bfseries 68} (2003) 014012},
  [\href{https://arxiv.org/abs/hep-ph/0303051}{{\ttfamily hep-ph/0303051}}].

\bibitem{Korchemsky:1994is}
G.~P. Korchemsky and G.~Sterman, \emph{{Nonperturbative corrections in resummed
  cross-sections}},
  \href{http://dx.doi.org/10.1016/0550-3213(94)00006-Z}{\emph{Nucl. Phys.}
  {\bfseries B437} (1995) 415--432},
  [\href{https://arxiv.org/abs/hep-ph/9411211}{{\ttfamily hep-ph/9411211}}].

\bibitem{Lee:2006nr}
C.~Lee and G.~F. Sterman, \emph{{Momentum Flow Correlations from Event Shapes:
  Factorized Soft Gluons and Soft-Collinear Effective Theory}},
  \href{http://dx.doi.org/10.1103/PhysRevD.75.014022}{\emph{Phys. Rev. D}
  {\bfseries 75} (2007) 014022},
  [\href{https://arxiv.org/abs/hep-ph/0611061}{{\ttfamily hep-ph/0611061}}].

\bibitem{Gardi:1999dq}
E.~Gardi and G.~Grunberg, \emph{{Power corrections in the single dressed gluon
  approximation: The Average thrust as a case study}},
  \href{http://dx.doi.org/10.1088/1126-6708/1999/11/016}{\emph{JHEP} {\bfseries
  11} (1999) 016}, [\href{https://arxiv.org/abs/hep-ph/9908458}{{\ttfamily
  hep-ph/9908458}}].

\bibitem{Gardi:2000yh}
E.~Gardi, \emph{{Perturbative and nonperturbative aspects of moments of the
  thrust distribution in $e^+ e^-$ annihilation}},
  \href{http://dx.doi.org/10.1088/1126-6708/2000/04/030}{\emph{JHEP} {\bfseries
  04} (2000) 030}, [\href{https://arxiv.org/abs/hep-ph/0003179}{{\ttfamily
  hep-ph/0003179}}].

\bibitem{Dokshitzer:1995zt}
Y.~L. Dokshitzer and B.~R. Webber, \emph{{Calculation of power corrections to
  hadronic event shapes}},
  \href{http://dx.doi.org/10.1016/0370-2693(95)00548-Y}{\emph{Phys. Lett.}
  {\bfseries B352} (1995) 451--455},
  [\href{https://arxiv.org/abs/hep-ph/9504219}{{\ttfamily hep-ph/9504219}}].

\bibitem{Salam:2001bd}
G.~P. Salam and D.~Wicke, \emph{{Hadron masses and power corrections to event
  shapes}}, \href{http://dx.doi.org/10.1088/1126-6708/2001/05/061}{\emph{JHEP}
  {\bfseries 05} (2001) 061},
  [\href{https://arxiv.org/abs/hep-ph/0102343}{{\ttfamily hep-ph/0102343}}].

\bibitem{Mateu:2012nk}
V.~Mateu, I.~W. Stewart and J.~Thaler, \emph{{Power Corrections to Event Shapes
  with Mass-Dependent Operators}},
  \href{http://dx.doi.org/10.1103/PhysRevD.87.014025}{\emph{Phys. Rev.}
  {\bfseries D87} (2013) 014025},
  [\href{https://arxiv.org/abs/1209.3781}{{\ttfamily 1209.3781}}].

\bibitem{Larkoski:2014wba}
A.~J. Larkoski, S.~Marzani, G.~Soyez and J.~Thaler, \emph{{Soft Drop}},
  \href{http://dx.doi.org/10.1007/JHEP05(2014)146}{\emph{JHEP} {\bfseries 05}
  (2014) 146}, [\href{https://arxiv.org/abs/1402.2657}{{\ttfamily 1402.2657}}].

\bibitem{Dreyer:2018tjj}
F.~A. Dreyer, L.~Necib, G.~Soyez and J.~Thaler, \emph{{Recursive Soft Drop}},
  \href{http://dx.doi.org/10.1007/JHEP06(2018)093}{\emph{JHEP} {\bfseries 06}
  (2018) 093}, [\href{https://arxiv.org/abs/1804.03657}{{\ttfamily
  1804.03657}}].

\bibitem{Hoang:2007vb}
A.~H. Hoang and I.~W. Stewart, \emph{{Designing Gapped Soft Functions for Jet
  Production}},
  \href{http://dx.doi.org/10.1016/j.physletb.2008.01.040}{\emph{Phys. Lett.}
  {\bfseries B660} (2008) 483--493},
  [\href{https://arxiv.org/abs/0709.3519}{{\ttfamily 0709.3519}}].

\bibitem{Fleming:2007qr}
S.~Fleming, A.~H. Hoang, S.~Mantry and I.~W. Stewart, \emph{{Jets from massive
  unstable particles: Top-mass determination}},
  \href{http://dx.doi.org/10.1103/PhysRevD.77.074010}{\emph{Phys. Rev.}
  {\bfseries D77} (2008) 074010},
  [\href{https://arxiv.org/abs/hep-ph/0703207}{{\ttfamily hep-ph/0703207}}].

\bibitem{Webber:1994cp}
B.~R. Webber, \emph{{Estimation of power corrections to hadronic event
  shapes}}, \href{http://dx.doi.org/10.1016/0370-2693(94)91147-9}{\emph{Phys.
  Lett.} {\bfseries B339} (1994) 148--150},
  [\href{https://arxiv.org/abs/hep-ph/9408222}{{\ttfamily hep-ph/9408222}}].

\bibitem{Beneke:1995pq}
M.~Beneke and V.~M. Braun, \emph{{Power corrections and renormalons in
  Drell-Yan production}},
  \href{http://dx.doi.org/10.1016/0550-3213(95)00439-Y}{\emph{Nucl. Phys.}
  {\bfseries B454} (1995) 253--290},
  [\href{https://arxiv.org/abs/hep-ph/9506452}{{\ttfamily hep-ph/9506452}}].

\bibitem{Gardi:2001ny}
E.~Gardi and J.~Rathsman, \emph{{Renormalon resummation and exponentiation of
  soft and collinear gluon radiation in the thrust distribution}},
  \href{http://dx.doi.org/10.1016/S0550-3213(01)00284-X}{\emph{Nucl. Phys.}
  {\bfseries B609} (2001) 123--182},
  [\href{https://arxiv.org/abs/hep-ph/0103217}{{\ttfamily hep-ph/0103217}}].

\bibitem{Eichten:1989zv}
E.~Eichten and B.~R. Hill, \emph{{An Effective Field Theory for the Calculation
  of Matrix Elements Involving Heavy Quarks}},
  \href{http://dx.doi.org/10.1016/0370-2693(90)92049-O}{\emph{Phys. Lett.}
  {\bfseries B234} (1990) 511--516}.

\bibitem{Isgur:1989vq}
N.~Isgur and M.~B. Wise, \emph{{Weak Decays of Heavy Mesons in the Static Quark
  Approximation}},
  \href{http://dx.doi.org/10.1016/0370-2693(89)90566-2}{\emph{Phys. Lett. B}
  {\bfseries 232} (1989) 113--117}.

\bibitem{Isgur:1989ed}
N.~Isgur and M.~B. Wise, \emph{{Weak Transition Form-Factors betwee Heavy
  Mesons}}, \href{http://dx.doi.org/10.1016/0370-2693(90)91219-2}{\emph{Phys.
  Lett.} {\bfseries B237} (1990) 527--530}.

\bibitem{Grinstein:1990mj}
B.~Grinstein, \emph{{The Static Quark Effective Theory}},
  \href{http://dx.doi.org/10.1016/0550-3213(90)90349-I}{\emph{Nucl. Phys.}
  {\bfseries B339} (1990) 253--268}.

\bibitem{Georgi:1990um}
H.~Georgi, \emph{{An Effective Field Theory for Heavy Quarks at Low-energies}},
  \href{http://dx.doi.org/10.1016/0370-2693(90)91128-X}{\emph{Phys. Lett.}
  {\bfseries B240} (1990) 447--450}.

\bibitem{Beneke:1998ui}
M.~Beneke, \emph{{Renormalons}}, {\emph{Phys. Rept.} {\bfseries 317} (1999)
  1--142}, [\href{https://arxiv.org/abs/hep-ph/9807443}{{\ttfamily
  hep-ph/9807443}}].

\bibitem{Grozin:2003gf}
A.~G. Grozin, \emph{{Renormalons: Technical introduction}},
  \href{https://arxiv.org/abs/hep-ph/0311050}{{\ttfamily hep-ph/0311050}}.

\bibitem{Scimemi:2016ffw}
I.~Scimemi and A.~Vladimirov, \emph{{Power corrections and renormalons in
  Transverse Momentum Distributions}},
  \href{http://dx.doi.org/10.1007/JHEP03(2017)002}{\emph{JHEP} {\bfseries 03}
  (2017) 002}, [\href{https://arxiv.org/abs/1609.06047}{{\ttfamily
  1609.06047}}].

\bibitem{Hoang:2008yj}
A.~H. Hoang, A.~Jain, I.~Scimemi and I.~W. Stewart, \emph{{Infrared
  Renormalization Group Flow for Heavy Quark Masses}},
  \href{http://dx.doi.org/10.1103/PhysRevLett.101.151602}{\emph{Phys. Rev.
  Lett.} {\bfseries 101} (2008) 151602},
  [\href{https://arxiv.org/abs/0803.4214}{{\ttfamily 0803.4214}}].

\bibitem{Hoang:2017suc}
A.~H. Hoang, A.~Jain, C.~Lepenik, V.~Mateu, M.~Preisser, I.~Scimemi et~al.,
  \emph{{The MSR mass and the $
  \mathcal{O}\left({\Lambda}_{\mathrm{QCD}}\right) $ renormalon sum rule}},
  \href{http://dx.doi.org/10.1007/JHEP04(2018)003}{\emph{JHEP} {\bfseries 04}
  (2018) 003}, [\href{https://arxiv.org/abs/1704.01580}{{\ttfamily
  1704.01580}}].

\bibitem{Hoang:2009yr}
A.~H. Hoang, A.~Jain, I.~Scimemi and I.~W. Stewart, \emph{{R-evolution:
  Improving perturbative QCD}},
  \href{http://dx.doi.org/10.1103/PhysRevD.82.011501}{\emph{Phys. Rev.}
  {\bfseries D82} (2010) 011501},
  [\href{https://arxiv.org/abs/0908.3189}{{\ttfamily 0908.3189}}].

\bibitem{Jain:2008gb}
A.~Jain, I.~Scimemi and I.~W. Stewart, \emph{{Two-loop Jet-Function and
  Jet-Mass for Top Quarks}},
  \href{http://dx.doi.org/10.1103/PhysRevD.77.094008}{\emph{Phys. Rev.}
  {\bfseries D77} (2008) 094008},
  [\href{https://arxiv.org/abs/0801.0743}{{\ttfamily 0801.0743}}].

\bibitem{Bachu:2020nqn}
B.~Bachu, A.~H. Hoang, V.~Mateu, A.~Pathak and I.~W. Stewart, \emph{{Boosted
  Top Quarks in the Peak Region with N$^3$LL Resummation}},
  \href{https://arxiv.org/abs/2012.12304}{{\ttfamily 2012.12304}}.

\bibitem{Broadhurst:1994se}
D.~J. Broadhurst and A.~G. Grozin, \emph{{Matching QCD and HQET heavy - light
  currents at two loops and beyond}},
  \href{http://dx.doi.org/10.1103/PhysRevD.52.4082}{\emph{Phys. Rev. D}
  {\bfseries 52} (1995) 4082--4098},
  [\href{https://arxiv.org/abs/hep-ph/9410240}{{\ttfamily hep-ph/9410240}}].

\bibitem{Beneke:1994qe}
M.~Beneke and V.~M. Braun, \emph{{Naive nonAbelianization and resummation of
  fermion bubble chains}},
  \href{http://dx.doi.org/10.1016/0370-2693(95)00184-M}{\emph{Phys. Lett. B}
  {\bfseries 348} (1995) 513--520},
  [\href{https://arxiv.org/abs/hep-ph/9411229}{{\ttfamily hep-ph/9411229}}].

\bibitem{Hoang:2021fhn}
A.~H. Hoang, C.~Lepenik and V.~Mateu, \emph{{REvolver: Automated running and
  matching of couplings and masses in QCD}},
  \href{https://arxiv.org/abs/2102.01085}{{\ttfamily 2102.01085}}.

\bibitem{mathematica}
I.~{Wolfram Research}, \emph{{Mathematica Edition: Version 10.0}}.
\newblock Wolfram Research, Inc., Champaign, Illinois, 2014.

\bibitem{Rossum:1995:PRM:869369}
G.~Rossum, \emph{Python Reference Manual}.
\newblock CWI, Amsterdam The Netherlands, (1995) [CS-R9525].

\bibitem{Virtanen:2019joe}
P.~Virtanen et~al., \emph{{SciPy 1.0--Fundamental Algorithms for Scientific
  Computing in Python}},
  \href{http://dx.doi.org/10.1038/s41592-019-0686-2}{\emph{Nature Meth.}
  {\bfseries 17} (2020) 261},
  [\href{https://arxiv.org/abs/1907.10121}{{\ttfamily 1907.10121}}].

\bibitem{oliphant2006guide}
T.~E. Oliphant, \emph{A guide to NumPy}, vol.~1.
\newblock Trelgol Publishing USA, 2006.

\bibitem{PalanquesMestre:1983zy}
A.~Palanques-Mestre and P.~Pascual, \emph{{The $1/N_f$ Expansion of the
  $\gamma$ and $\beta$ functions in QED}},
  \href{http://dx.doi.org/10.1007/BF01212398}{\emph{Commun. Math. Phys.}
  {\bfseries 95} (1984) 277}.

\bibitem{Vermaseren:1997fq}
J.~A.~M. Vermaseren, S.~A. Larin and T.~van Ritbergen, \emph{{The 4-loop quark
  mass anomalous dimension and the invariant quark mass}},
  \href{http://dx.doi.org/10.1016/S0370-2693(97)00660-6}{\emph{Phys. Lett.}
  {\bfseries B405} (1997) 327--333},
  [\href{https://arxiv.org/abs/hep-ph/9703284}{{\ttfamily hep-ph/9703284}}].

\bibitem{Chetyrkin:1997dh}
K.~G. Chetyrkin, \emph{{Quark mass anomalous dimension to $O(\alpha_s^4)$}},
  \href{http://dx.doi.org/10.1016/S0370-2693(97)00535-2}{\emph{Phys. Lett.}
  {\bfseries B404} (1997) 161--165},
  [\href{https://arxiv.org/abs/hep-ph/9703278}{{\ttfamily hep-ph/9703278}}].

\bibitem{Baikov:2014qja}
P.~A. Baikov, K.~G. Chetyrkin and J.~H. K{\"u}hn, \emph{{Quark Mass and Field
  Anomalous Dimensions to ${\cal O}(\alpha_s^5)$}},
  \href{http://dx.doi.org/10.1007/JHEP10(2014)076}{\emph{JHEP} {\bfseries 10}
  (2014) 076}, [\href{https://arxiv.org/abs/1402.6611}{{\ttfamily 1402.6611}}].

\bibitem{Luthe:2016xec}
T.~Luthe, A.~Maier, P.~Marquard and Y.~Schr{\"o}der, \emph{{Five-loop quark
  mass and field anomalous dimensions for a general gauge group}},
  \href{http://dx.doi.org/10.1007/JHEP01(2017)081}{\emph{JHEP} {\bfseries 01}
  (2017) 081}, [\href{https://arxiv.org/abs/1612.05512}{{\ttfamily
  1612.05512}}].

\bibitem{Beneke:1994sw}
M.~Beneke and V.~M. Braun, \emph{{Heavy quark effective theory beyond
  perturbation theory: Renormalons, the pole mass and the residual mass term}},
  \href{http://dx.doi.org/10.1016/0550-3213(94)90314-X}{\emph{Nucl. Phys. B}
  {\bfseries 426} (1994) 301--343},
  [\href{https://arxiv.org/abs/hep-ph/9402364}{{\ttfamily hep-ph/9402364}}].

\bibitem{Ball:1995ni}
P.~Ball, M.~Beneke and V.~M. Braun, \emph{{Resummation of $(\beta_0
  \alpha_s)^n$ corrections in QCD: Techniques and applications to the tau
  hadronic width and the heavy quark pole mass}},
  \href{http://dx.doi.org/10.1016/0550-3213(95)00392-6}{\emph{Nucl. Phys. B}
  {\bfseries 452} (1995) 563--625},
  [\href{https://arxiv.org/abs/hep-ph/9502300}{{\ttfamily hep-ph/9502300}}].

\bibitem{Tarrach:1980up}
R.~Tarrach, \emph{{The Pole Mass in Perturbative QCD}},
  \href{http://dx.doi.org/10.1016/0550-3213(81)90140-1}{\emph{Nucl. Phys.}
  {\bfseries B183} (1981) 384--396}.

\bibitem{Gray:1990yh}
N.~Gray, D.~J. Broadhurst, W.~Grafe and K.~Schilcher, \emph{{Three Loop
  Relation of Quark $\overline{\rm MS}$ and Pole Masses}},
  \href{http://dx.doi.org/10.1007/BF01614703}{\emph{Z. Phys.} {\bfseries C48}
  (1990) 673--680}.

\bibitem{Melnikov:2000qh}
K.~Melnikov and T.~v. Ritbergen, \emph{{The Three loop relation between the
  ${\overline {\rm MS}}$ and the pole quark masses}},
  \href{http://dx.doi.org/10.1016/S0370-2693(00)00507-4}{\emph{Phys. Lett.}
  {\bfseries B482} (2000) 99--108},
  [\href{https://arxiv.org/abs/hep-ph/9912391}{{\ttfamily hep-ph/9912391}}].

\bibitem{Marquard:2016dcn}
P.~Marquard, A.~V. Smirnov, V.~A. Smirnov, M.~Steinhauser and D.~Wellmann,
  \emph{{$\overline{\rm MS}$-on-shell quark mass relation up to four loops in
  QCD and a general SU$(N)$ gauge group}},
  \href{http://dx.doi.org/10.1103/PhysRevD.94.074025}{\emph{Phys. Rev.}
  {\bfseries D94} (2016) 074025},
  [\href{https://arxiv.org/abs/1606.06754}{{\ttfamily 1606.06754}}].

\bibitem{Hoang:2017btd}
A.~H. Hoang, C.~Lepenik and M.~Preisser, \emph{{On the Light Massive Flavor
  Dependence of the Large Order Asymptotic Behavior and the Ambiguity of the
  Pole Mass}}, \href{http://dx.doi.org/10.1007/JHEP09(2017)099}{\emph{JHEP}
  {\bfseries 09} (2017) 099},
  [\href{https://arxiv.org/abs/1706.08526}{{\ttfamily 1706.08526}}].

\bibitem{Pineda:2001zq}
A.~Pineda, \emph{{Determination of the bottom quark mass from the
  $\Upsilon(1S)$ system}},
  \href{http://dx.doi.org/10.1088/1126-6708/2001/06/022}{\emph{JHEP} {\bfseries
  06} (2001) 022}, [\href{https://arxiv.org/abs/hep-ph/0105008}{{\ttfamily
  hep-ph/0105008}}].

\bibitem{Mateu:2018zym}
V.~Mateu, P.~G. Ortega, D.~R. Entem and F.~Fern\'andez, \emph{{Calibrating the
  Na\"\i{}ve Cornell Model with NRQCD}},
  \href{http://dx.doi.org/10.1140/epjc/s10052-019-6808-2}{\emph{Eur. Phys. J.
  C} {\bfseries 79} (2019) 323},
  [\href{https://arxiv.org/abs/1811.01982}{{\ttfamily 1811.01982}}].

\bibitem{Korchemsky:1987wg}
G.~P. Korchemsky and A.~V. Radyushkin, \emph{{Renormalization of the Wilson
  Loops Beyond the Leading Order}},
  \href{http://dx.doi.org/10.1016/0550-3213(87)90277-X}{\emph{Nucl. Phys. B}
  {\bfseries 283} (1987) 342--364}.

\bibitem{Moch:2004pa}
S.~Moch, J.~A.~M. Vermaseren and A.~Vogt, \emph{{The three-loop splitting
  functions in QCD: The non-singlet case}},
  \href{http://dx.doi.org/10.1016/j.nuclphysb.2004.03.030}{\emph{Nucl. Phys.}
  {\bfseries B688} (2004) 101--134},
  [\href{https://arxiv.org/abs/hep-ph/0403192}{{\ttfamily hep-ph/0403192}}].

\bibitem{Henn:2019swt}
J.~M. Henn, G.~P. Korchemsky and B.~Mistlberger, \emph{{The full four-loop cusp
  anomalous dimension in $\mathcal{N}=4$ super Yang-Mills and QCD}},
  \href{http://dx.doi.org/10.1007/JHEP04(2020)018}{\emph{JHEP} {\bfseries 04}
  (2020) 018}, [\href{https://arxiv.org/abs/1911.10174}{{\ttfamily
  1911.10174}}].

\bibitem{Huber:2019fxe}
T.~Huber, A.~von Manteuffel, E.~Panzer, R.~M. Schabinger and G.~Yang,
  \emph{{The four-loop cusp anomalous dimension from the $N=4$ Sudakov form
  factor}}, \href{http://dx.doi.org/10.1016/j.physletb.2020.135543}{\emph{Phys.
  Lett. B} {\bfseries 807} (2020) 135543},
  [\href{https://arxiv.org/abs/1912.13459}{{\ttfamily 1912.13459}}].

\bibitem{vanNeerven:1985xr}
W.~L. van Neerven, \emph{{Dimensional Regularization of Mass and Infrared
  Singularities in two Loop on-shell Vertex Functions}},
  \href{http://dx.doi.org/10.1016/0550-3213(86)90165-3}{\emph{Nucl. Phys.}
  {\bfseries B268} (1986) 453}.

\bibitem{Matsuura:1988sm}
T.~Matsuura, S.~C. van~der Marck and W.~L. van Neerven, \emph{{The Calculation
  of the Second Order Soft and Virtual Contributions to the Drell-Yan
  Cross-Section}},
  \href{http://dx.doi.org/10.1016/0550-3213(89)90620-2}{\emph{Nucl. Phys. B}
  {\bfseries 319} (1989) 570--622}.

\bibitem{Moch:2005id}
S.~Moch, J.~A.~M. Vermaseren and A.~Vogt, \emph{{The Quark form-factor at
  higher orders}},
  \href{http://dx.doi.org/10.1088/1126-6708/2005/08/049}{\emph{JHEP} {\bfseries
  08} (2005) 049}, [\href{https://arxiv.org/abs/hep-ph/0507039}{{\ttfamily
  hep-ph/0507039}}].

\bibitem{Matsuura:1987wt}
T.~Matsuura and W.~L. van Neerven, \emph{{Second Order Logarithmic Corrections
  to the Drell-Yan Cross-Section}},
  \href{http://dx.doi.org/10.1007/BF01624369}{\emph{Z. Phys.} {\bfseries C38}
  (1988) 623}.

\bibitem{Gehrmann:2005pd}
T.~Gehrmann, T.~Huber and D.~Maitre, \emph{{Two-loop quark and gluon form
  factors in dimensional regularisation}},
  \href{http://dx.doi.org/10.1016/j.physletb.2005.07.019}{\emph{Phys. Lett.}
  {\bfseries B622} (2005) 295--302},
  [\href{https://arxiv.org/abs/hep-ph/0507061}{{\ttfamily hep-ph/0507061}}].

\bibitem{Lee:2010cg}
R.~N. Lee, A.~V. Smirnov and V.~A. Smirnov, \emph{{Analytic Results for
  Massless Three-Loop Form Factors}},
  \href{http://dx.doi.org/10.1007/JHEP04(2010)020}{\emph{JHEP} {\bfseries 04}
  (2010) 020}, [\href{https://arxiv.org/abs/1001.2887}{{\ttfamily 1001.2887}}].

\bibitem{Baikov:2009bg}
P.~A. Baikov, K.~G. Chetyrkin, A.~V. Smirnov, V.~A. Smirnov and M.~Steinhauser,
  \emph{{Quark and gluon form factors to three loops}},
  \href{http://dx.doi.org/10.1103/PhysRevLett.102.212002}{\emph{Phys. Rev.
  Lett.} {\bfseries 102} (2009) 212002},
  [\href{https://arxiv.org/abs/0902.3519}{{\ttfamily 0902.3519}}].

\bibitem{Gehrmann:2010ue}
T.~Gehrmann, E.~Glover, T.~Huber, N.~Ikizlerli and C.~Studerus,
  \emph{{Calculation of the quark and gluon form factors to three loops in
  QCD}}, \href{http://dx.doi.org/10.1007/JHEP06(2010)094}{\emph{JHEP}
  {\bfseries 1006} (2010) 094},
  [\href{https://arxiv.org/abs/1004.3653}{{\ttfamily 1004.3653}}].

\bibitem{Abbate:2012jh}
R.~Abbate, M.~Fickinger, A.~H. Hoang, V.~Mateu and I.~W. Stewart,
  \emph{{Precision Thrust Cumulant Moments at $N^3$LL}},
  \href{http://dx.doi.org/10.1103/PhysRevD.86.094002}{\emph{Phys. Rev. D}
  {\bfseries 86} (2012) 094002},
  [\href{https://arxiv.org/abs/1204.5746}{{\ttfamily 1204.5746}}].

\bibitem{Hoang:2014wka}
A.~H. Hoang, D.~W. Kolodrubetz, V.~Mateu and I.~W. Stewart,
  \emph{{$C$-parameter distribution at N$^3$LL' including power corrections}},
  \href{http://dx.doi.org/10.1103/PhysRevD.91.094017}{\emph{Phys. Rev. D}
  {\bfseries 91} (2015) 094017},
  [\href{https://arxiv.org/abs/1411.6633}{{\ttfamily 1411.6633}}].

\bibitem{Lunghi:2002ju}
E.~Lunghi, D.~Pirjol and D.~Wyler, \emph{{Factorization in leptonic radiative
  $B \to \gamma e \nu$ decays.}},
  \href{http://dx.doi.org/10.1016/S0550-3213(02)01032-5}{\emph{Nucl. Phys.}
  {\bfseries B649} (2003) 349--364},
  [\href{https://arxiv.org/abs/hep-ph/0210091}{{\ttfamily hep-ph/0210091}}].

\bibitem{Bauer:2003pi}
C.~W. Bauer and A.~V. Manohar, \emph{{Shape function effects in $B \to X_s
  \gamma$ and $B \to X_ u \ell \bar\nu$ decays}},
  \href{http://dx.doi.org/10.1103/PhysRevD.70.034024}{\emph{Phys. Rev.}
  {\bfseries D70} (2004) 034024},
  [\href{https://arxiv.org/abs/hep-ph/0312109}{{\ttfamily hep-ph/0312109}}].

\bibitem{Becher:2006qw}
T.~Becher and M.~Neubert, \emph{{Toward a NNLO calculation of the $\bar B \to
  X_s \gamma$ decay rate with a cut on photon energy. II. Two-loop result for
  the jet function}},
  \href{http://dx.doi.org/10.1016/j.physletb.2006.04.046}{\emph{Phys. Lett.}
  {\bfseries B637} (2006) 251--259},
  [\href{https://arxiv.org/abs/hep-ph/0603140}{{\ttfamily hep-ph/0603140}}].

\bibitem{Bruser:2018rad}
R.~Br{\"u}ser, Z.~L. Liu and M.~Stahlhofen, \emph{{Three-Loop Quark Jet
  Function}},
  \href{http://dx.doi.org/10.1103/PhysRevLett.121.072003}{\emph{Phys. Rev.
  Lett.} {\bfseries 121} (2018) 072003},
  [\href{https://arxiv.org/abs/1804.09722}{{\ttfamily 1804.09722}}].

\bibitem{Banerjee:2018ozf}
P.~Banerjee, P.~K. Dhani and V.~Ravindran, \emph{{Gluon jet function at three
  loops in QCD}},
  \href{http://dx.doi.org/10.1103/PhysRevD.98.094016}{\emph{Phys. Rev. D}
  {\bfseries 98} (2018) 094016},
  [\href{https://arxiv.org/abs/1805.02637}{{\ttfamily 1805.02637}}].

\bibitem{Fleming:2007xt}
S.~Fleming, A.~H. Hoang, S.~Mantry and I.~W. Stewart, \emph{{Top Jets in the
  Peak Region: Factorization Analysis with NLL Resummation}},
  \href{http://dx.doi.org/10.1103/PhysRevD.77.114003}{\emph{Phys. Rev.}
  {\bfseries D77} (2008) 114003},
  [\href{https://arxiv.org/abs/0711.2079}{{\ttfamily 0711.2079}}].

\bibitem{Leibovich:2003jd}
A.~K. Leibovich, Z.~Ligeti and M.~B. Wise, \emph{{Comment on quark masses in
  SCET}}, \href{http://dx.doi.org/10.1016/S0370-2693(03)00565-3}{\emph{Phys.
  Lett. B} {\bfseries 564} (2003) 231--234},
  [\href{https://arxiv.org/abs/hep-ph/0303099}{{\ttfamily hep-ph/0303099}}].

\bibitem{Hoang:2015vua}
A.~H. Hoang, A.~Pathak, P.~Pietrulewicz and I.~W. Stewart, \emph{{Hard Matching
  for Boosted Tops at Two Loops}},
  \href{http://dx.doi.org/10.1007/JHEP12(2015)059}{\emph{JHEP} {\bfseries 12}
  (2015) 059}, [\href{https://arxiv.org/abs/1508.04137}{{\ttfamily
  1508.04137}}].

\bibitem{Hoang:2008fs}
A.~H. Hoang and S.~Kluth, \emph{{Hemisphere Soft Function at
  $\mathcal{O}(\alpha_s^2)$ for Dijet Production in $e^+e^-$ Annihilation}},
  \href{https://arxiv.org/abs/0806.3852}{{\ttfamily 0806.3852}}.

\bibitem{Butenschoen:2016lpz}
M.~Butenschoen, B.~Dehnadi, A.~H. Hoang, V.~Mateu, M.~Preisser and I.~W.
  Stewart, \emph{{Top Quark Mass Calibration for Monte Carlo Event
  Generators}},
  \href{http://dx.doi.org/10.1103/PhysRevLett.117.232001}{\emph{Phys. Rev.
  Lett.} {\bfseries 117} (2016) 232001},
  [\href{https://arxiv.org/abs/1608.01318}{{\ttfamily 1608.01318}}].

\bibitem{Mateu:2017hlz}
V.~Mateu and P.~G. Ortega, \emph{{Bottom and Charm Mass determinations from
  global fits to $Q\overline{Q}$ bound states at N$^3$\!LO}},
  \href{http://dx.doi.org/10.1007/JHEP01(2018)122}{\emph{JHEP} {\bfseries 01}
  (2018) 122}, [\href{https://arxiv.org/abs/1711.05755}{{\ttfamily
  1711.05755}}].

\bibitem{Gatheral:1983cz}
J.~G.~M. Gatheral, \emph{{Exponentiation of Eikonal Cross-sections in
  Nonabelian Gauge Theories}},
  \href{http://dx.doi.org/10.1016/0370-2693(83)90112-0}{\emph{Phys. Lett. B}
  {\bfseries 133} (1983) 90--94}.

\bibitem{Frenkel:1984pz}
J.~Frenkel and J.~C. Taylor, \emph{{Nonabelian Eikonal Exponentiation}},
  \href{http://dx.doi.org/10.1016/0550-3213(84)90294-3}{\emph{Nucl. Phys. B}
  {\bfseries 246} (1984) 231--245}.

\bibitem{Kataev:2018gle}
A.~L. Kataev and V.~S. Molokoedov, \emph{{Multiloop contributions to the
  on-shell-$ \overline{{\mathrm{MS}}}$ heavy quark mass relation in QCD and the
  asymptotic structure of the corresponding series: the updated
  consideration}},
  \href{http://dx.doi.org/10.1140/epjc/s10052-020-08673-6}{\emph{Eur. Phys. J.
  C} {\bfseries 80} (2020) 1160},
  [\href{https://arxiv.org/abs/1807.05406}{{\ttfamily 1807.05406}}].

\bibitem{Abbate:2010xh}
R.~Abbate, M.~Fickinger, A.~H. Hoang, V.~Mateu and I.~W. Stewart, \emph{{Thrust
  at $N^{3}LL$ with Power Corrections and a Precision Global Fit for
  $\alpha_{s}(mZ)$}},
  \href{http://dx.doi.org/10.1103/PhysRevD.83.074021}{\emph{Phys. Rev. D}
  {\bfseries 83} (2011) 074021},
  [\href{https://arxiv.org/abs/1006.3080}{{\ttfamily 1006.3080}}].

\bibitem{Monni:2011gb}
P.~F. Monni, T.~Gehrmann and G.~Luisoni, \emph{{Two-Loop Soft Corrections and
  Resummation of the Thrust Distribution in the Dijet Region}},
  \href{http://dx.doi.org/10.1007/JHEP08(2011)010}{\emph{JHEP} {\bfseries 08}
  (2011) 010}, [\href{https://arxiv.org/abs/1105.4560}{{\ttfamily 1105.4560}}].

\bibitem{Ridder:2014wza}
A.~Gehrmann-De~Ridder, T.~Gehrmann, E.~W.~N. Glover and G.~Heinrich,
  \emph{{EERAD3: Event shapes and jet rates in electron-positron annihilation
  at order $\alpha_s^3$}},
  \href{http://dx.doi.org/10.1016/j.cpc.2014.07.024}{\emph{Comput. Phys.
  Commun.} {\bfseries 185} (2014) 3331},
  [\href{https://arxiv.org/abs/1402.4140}{{\ttfamily 1402.4140}}].

\bibitem{Luke:1992cs}
M.~E. Luke and A.~V. Manohar, \emph{{Reparametrization invariance constraints
  on heavy particle effective field theories}},
  \href{http://dx.doi.org/10.1016/0370-2693(92)91786-9}{\emph{Phys. Lett. B}
  {\bfseries 286} (1992) 348--354},
  [\href{https://arxiv.org/abs/hep-ph/9205228}{{\ttfamily hep-ph/9205228}}].

\end{thebibliography}\endgroup
\bibliographystyle{JHEP}

\end{document}